\documentclass[longauth]{aa}

\usepackage{gensymb}
\usepackage{graphicx}
\usepackage{txfonts}
\usepackage{natbib}
\bibpunct{(}{)}{;}{a}{}{,} 
\usepackage[breaklinks, colorlinks, citecolor=blue]{hyperref}
\usepackage{bm}
\newcommand{\MUSEp}{{\tt MUSE DRS}}
\newcommand{\DRP}{{\tt DRP}}
\newcommand{\DAP}{{\tt DAP}}
\newcommand{\pPXF}{{\tt pPXF}}
\newcommand{\vorbin}{{\tt vorbin}}
\newcommand{\MAPS}{{\tt MAPS}}
\newcommand{\pymusepipe}{{\tt pymusepipe}}
\newcommand{\mpdaf}{{\tt mpdaf}}
\newcommand{\pypher}{{\tt pypher}}
\newcommand{\astropy}{{\tt astropy}}
\newcommand{\python}{{\tt Python}}
\newcommand{\reproject}{{\tt reproject}}
\newcommand{\numpy}{{\tt numpy}}
\newcommand{\matplotlib}{{\tt matplotlib}}
\newcommand{\scipy}{{\tt scipy}}
\newcommand{\esorex}{{\tt EsoRex}}
\newcommand{\gist}{{\tt gist}}

\newcommand{\oiii}{[O\,\textsc{iii}]}
\newcommand{\nii}{[N\,\textsc{ii}]}
\newcommand{\sii}{[S\,\textsc{ii}]}
\newcommand{\oi}{[O\,\textsc{i}]}
\newcommand{\niion}{[N\,\textsc{i}]}
\newcommand{\hei}{[He\,\textsc{i}]}
\newcommand{\siii}{[S\,\textsc{iii}]}
\newcommand{\oii}{[O\,\textsc{ii}]}
\newcommand{\hii}{H\,\textsc{ii}}
\newcommand{\ha}{H$\alpha$}
\newcommand{\hb}{H$\beta$}

\newcommand{\NaI}{Na\,\textsc{I}}
\newcommand{\kms} {$\mathrm{km s}^{-1}$}

\begin{document}

   \title{The PHANGS-MUSE survey}
   \subtitle{Probing the chemo-dynamical evolution of disc galaxies}
   \titlerunning{PHANGS-MUSE}

   \author{
   Eric~Emsellem\inst{\ref{eso},\ref{lyon}} \and 
   Eva~Schinnerer\inst{\ref{mpia}} \and    
   Francesco~Santoro\inst{\ref{mpia}} \and 
   Francesco~Belfiore\inst{\ref{inaf}} \and 
   Ismael~Pessa\inst{\ref{mpia}} \and 
   Rebecca~McElroy \inst{\ref{usyd}} \and
   Guillermo~A.~Blanc\inst{\ref{uch},\ref{carn}} \and
   Enrico~Congiu\inst{\ref{uch}} \and
   Brent~Groves\inst{\ref{ICRAR},\ref{Canb}} \and    
   I-Ting~Ho\inst{\ref{mpia}} \and 
   Kathryn~Kreckel\inst{\ref{rechen}} \and
   Alessandro~Razza\inst{\ref{uch}} \and
   Patricia~Sanchez-Blazquez\inst{\ref{ucm}} \and 
   Oleg~Egorov\inst{\ref{rechen},\ref{SAI}} \and
   Chris~Faesi\inst{\ref{mpia}} \and
   Ralf~S.~Klessen\inst{\ref{zah},\ref{zw}} \and
   Adam~K.~Leroy\inst{\ref{ohio}} \and 
   Sharon~Meidt\inst{\ref{Gent}} \and
   Miguel~Querejeta\inst{\ref{oan}} \and 
   Erik~Rosolowsky\inst{\ref{alb}} \and
   Fabian~Scheuermann \inst{\ref{rechen}} \and
   %
   %
   Gagandeep~S.~Anand \inst{\ref{UHIfA}} \and
   Ashley~T.~Barnes\inst{\ref{UBonn}} \and
   Ivana~Be\v{s}li\'c\inst{\ref{UBonn}} \and
   Frank~Bigiel\inst{\ref{UBonn}} \and
   Médéric~Boquien\inst{\ref{UA}} \and
   Yixian~Cao\inst{\ref{LAM}} \and 
   M\'elanie~Chevance\inst{\ref{rechen}} \and
   Daniel~A.~Dale\inst{\ref{wyo}} \and
   Cosima~Eibensteiner \inst{\ref{UBonn}} \and
   Simon~C.~O.~Glover\inst{\ref{zah}} \and
   Kathryn~Grasha\inst{\ref{ANU}} \and  
   Jonathan~D.~Henshaw\inst{\ref{mpia}} \and
   Annie~Hughes\inst{\ref{irap}} \and
   Eric~W.~Koch\inst{\ref{CfA}} \and
   J.~M.~Diederik~Kruijssen\inst{\ref{rechen}} \and
   Janice Lee\inst{\ref{gem}}
   Daizhong~Liu \inst{\ref{mpe}} \and
   Hsi-An~Pan\inst{\ref{mpia}} \and
   Jérôme~Pety\inst{\ref{IRAM},\ref{LERMA}} \and
   Toshiki~Saito\inst{\ref{mpia}} \and
   Karin~M.~Sandstrom\inst{\ref{ucsd}} \and
   Andreas~Schruba \inst{\ref{mpe}} \and
   Jiayi~Sun \inst{\ref{ohio}} \and
   David~A.~Thilker\inst{\ref{jhu}} \and
   Antonio~Usero\inst{\ref{oan}} \and
   Elizabeth~J.~Watkins\inst{\ref{rechen}} \and
   Thomas~G.~Williams\inst{\ref{mpia}}
   % ...
   }
   \institute{European Southern Observatory, Karl-Schwarzschild-Stra{\ss}e 2, 85748 Garching, Germany\label{eso} \\
        \email{eric.emsellem@eso.org}
        \and Univ Lyon, Univ Lyon1, ENS de Lyon, CNRS, Centre de Recherche Astrophysique de Lyon UMR5574, F-69230 Saint-Genis-Laval France\label{lyon}
        \and Max-Planck-Institute for Astronomy, K\"onigstuhl 17, D-69117 Heidelberg, Germany\label{mpia}
        \and INAF -- Osservatorio Astrofisico di Arcetri, Largo E. Fermi 5, I-50125, Florence, Italy\label{inaf}
        \and Sydney Institute for Astronomy, School of Physics, Physics Road, The University of Sydney, Darlington 2006, NSW, Australia \label{usyd}
        \and Departamento de Astronom\'ia, Universidad de Chile, Santiago,Chile\label{uch}
        \and Observatories of the Carnegie Institution for Science, Pasadena, CA, USA\label{carn}
        \and International Centre for Radio Astronomy Research University of Western Australia, 7 Fairway, Crawley WA 6009 Australia\label{ICRAR}
        \and Research School of Astronomy and Astrophysics, Australian National University, Canberra, ACT 2611, Australia\label{Canb}
        \and Astronomisches Rechen-Institut, Zentrum f\"ur Astronomie der Universit\"at Heidelberg, M\"onchhofstra{\ss}e 12-14, D-69120 Heidelberg, Germany\label{rechen}
        \and Gemini Observatory/NSF's NOIRLab, 950 N. Cherry Avenue, Tucson, AZ, 85719, USA\label{gem}
        \and Departamento de F\'isica de la Tierra y Astrof\'isica, Universidad Complutense de Madrid, E-28040 Madrid, Spain \label{ucm}
        \and Sternberg Astronomical Institute, Lomonosov Moscow State University, Universitetsky pr. 13, 119234 Moscow, Russia\label{SAI}
        \and Universit\"at Heidelberg, Zentrum f\"ur Astronomie, Institut f\"ur theoretische Astrophysik, Albert-Ueberle-Stra{\ss}e 2, D-69120, Heidelberg, Germany\label{zah}
        \and Universit\"at Heidelberg, Interdisziplin\"ares Zentrum f\"ur Wissenschaftliches Rechnen, Im Neuenheimer Feld 205, D-69120 Heidelberg, Germany\label{zw}
        \and Department of Astronomy, The Ohio State University, 140 West 18th Avenue, Columbus, OH 43210, USA\label{ohio}
        \and Sterrenkundig Observatorium, Universiteit Gent, Krijgslaan 281 S9, B-9000 Gent, Belgium\label{Gent}
        \and Observatorio Astronómico Nacional (IGN), C/Alfonso XII, 3, E-28014 Madrid, Spain\label{oan}        
        \and Department of Physics, University of Alberta, Edmonton, AB T6G 2E1, Canada\label{alb}
        \and Institute for Astronomy, University of Hawaii, 2680 Woodlawn Drive, Honolulu, HI 96822, USA \label{UHIfA}
        \and Argelander-Institut f\"ur Astronomie, Universit\"at Bonn, Auf dem H\"ugel 71, D-53121 Bonn, Germany\label{UBonn}
        \and Centro de Astronomía (CITEVA), Universidad de Antofagasta, Avenida Angamos 601, Antofagasta, Chile\label{UA}
        \and Aix Marseille Univ, CNRS, CNES, LAM (Laboratoire d'Astrophysique de Marseille), F-13388 Marseille, France\label{LAM}
        \and Department of Physics and Astronomy, University of Wyoming, Laramie, WY 82071, USA\label{wyo}        
        \and Research School of Astronomy and Astrophysics, Australian National University, Canberra, ACT 2611, Australia\label{ANU}
        \and Universit\'{e} de Toulouse, UPS-OMP, IRAP, F-31028 Toulouse cedex 4, France\label{irap}
        \and Harvard-Smithsonian Center for Astrophysics, 60 Garden Street, Cambridge, MA 02138, USA\label{CfA}
        \and Max-Planck-Institute for extraterrestrial Physics, Giessenbachstra{\ss}e 1, D-85748 Garching, Germany\label{mpe}
        \and Institut de Radioastronomie Millim\'etrique (IRAM), 300 Rue de la Piscine, F-38406 Saint Martin d'Hères, France\label{IRAM}
        \and LERMA, Observatoire de Paris, PSL Research University, CNRS, Sorbonne Universit\'es, 75014 Paris, France. \label{LERMA}
        \and Center for Astrophysics and Space Sciences, Department of Physics, University of California, San Diego, 9500 Gilman Dr., La Jolla, CA 92093, USA\label{ucsd}
        \and Department of Physics and Astronomy, Johns Hopkins University, Baltimore, MD 21218, USA\label{jhu}
        }

   \authorrunning{MUSE PHANGS team}

   \date{Received July 7, 2021; accepted Dec 8, 2021}

 
  \abstract{We present the PHANGS-MUSE survey, a programme that uses the MUSE integral field spectrograph at the ESO VLT to map 19 massive $(9.4 < \log(M_{\star}/\mathrm{M}_{\odot}) < 11.0)$ nearby ($D \lesssim 20$~Mpc) star-forming disc galaxies. The survey consists of 168 MUSE pointings ($1\arcmin$~by~$1\arcmin$ each) and a total of nearly $15 \times 10^6$ spectra, covering ${\sim}1.5 \times 10^6$ independent spectra. PHANGS-MUSE provides the first integral field spectrograph view of star formation across different local environments (including galaxy centres, bars, and spiral arms) in external galaxies at a median resolution of 50~pc, better than the mean inter-cloud distance in the ionised interstellar medium. This `cloud-scale' resolution allows detailed demographics and characterisations of \hii\ regions and other ionised nebulae. PHANGS-MUSE further delivers a unique view on the associated gas and stellar kinematics and provides constraints on the star-formation history. The PHANGS-MUSE survey is complemented by dedicated ALMA \mbox{CO(2--1)} and multi-band \textit{HST} observations, therefore allowing us to probe the key stages of the star-formation process from molecular clouds to \hii\ regions and star clusters.
 This paper describes the scientific motivation, sample selection, observational strategy, data reduction, and analysis process of the PHANGS-MUSE survey. We present our bespoke automated data-reduction framework, which is built on the reduction recipes provided by ESO but additionally allows for mosaicking and homogenisation of the point spread function. We further present a detailed quality assessment and a brief illustration of the potential scientific applications of the large set of PHANGS-MUSE data products generated by our data analysis framework. The data cubes and analysis data products described in this paper represent {the basis for the first PHANGS-MUSE public data release and are available} in the ESO archive and via the Canadian Astronomy Data Centre.
  }
\maketitle
\section{Introduction}
\label{sec:intro}

The `baryon cycle' -- the collapse of gas to form stars and the subsequent re-injection of matter, energy, and momentum into the interstellar medium (ISM) -- is an intrinsically multi-phase and multi-scale process. Flows of gas in and out of galaxies, as well as internal gas dynamics, connect the small-scale cycle of baryons with the larger galactic and circum-galactic scales. These processes drive both the evolution of galaxies in a large-scale cosmological context, and the still elusive small-scale physics involved in the collapse of gas cores, and the feedback from massive stars \citep{Scannapieco2012, Haas2013, Hopkins2013, Agertz2015, Fujimoto2019}. 

\begin{figure*}
\centering
        \includegraphics[width=0.95\textwidth, trim=0 0 0 0, clip]{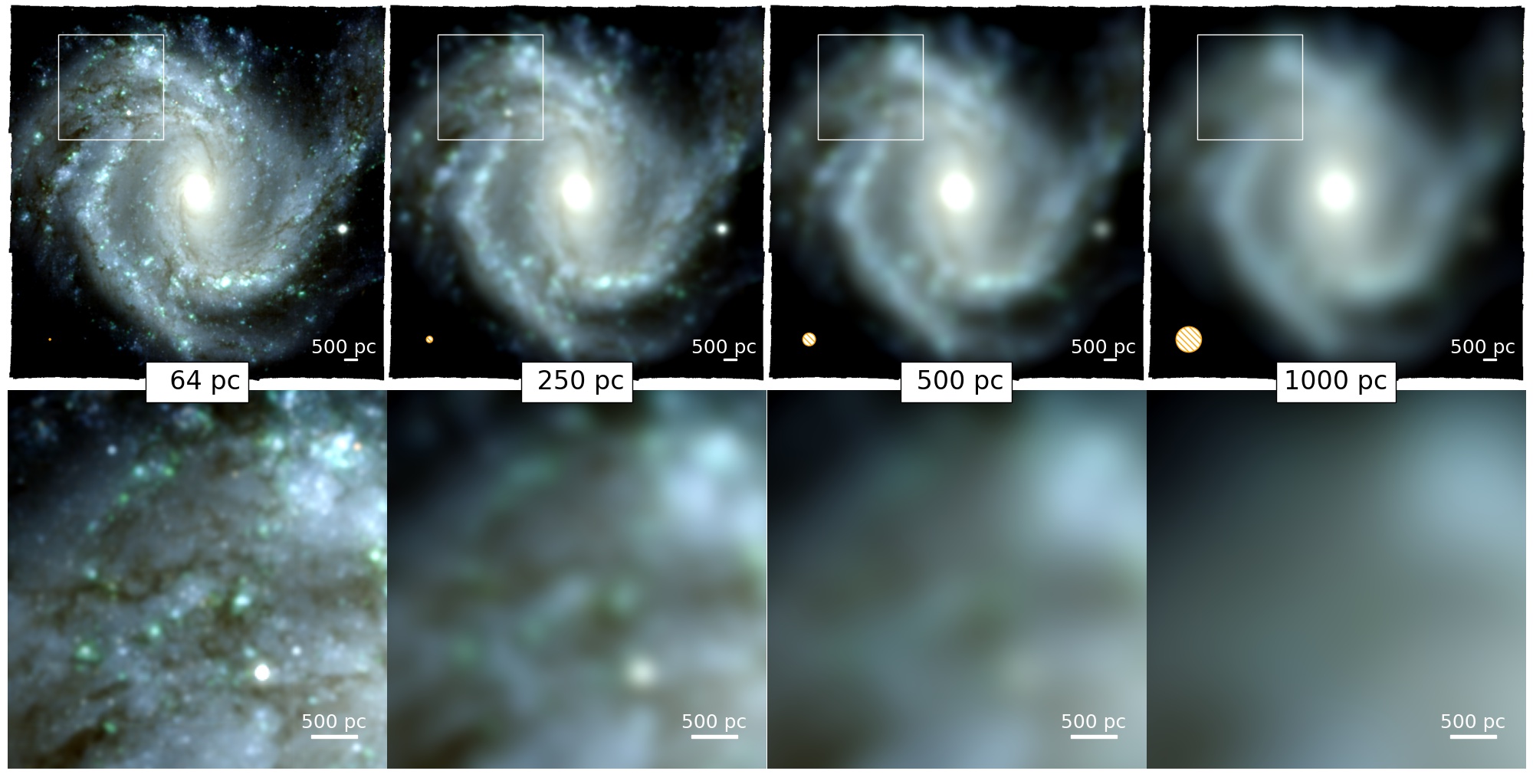}
     \caption{\textbf{From kpc to $\bm{{\sim}100}$~pc scale}, red-green-blue (RGB) colour images (channels derived from reconstructed MUSE mosaic  using the SDSS i, r, and g bands) of NGC\,4303 ($D=17\pm3$~Mpc) at various spatial resolutions. The four columns, from left to right: at 64~pc (the homogenised resolution for our MUSE dataset) and then convolved to 250, 500, and 1000~pc. The size of the beam (FWHM) is provided as a hatched circle at the bottom-left corner of each top row panel. The bottom row shows zoomed-in views of the top row images (area shown as a white rectangle).}
     \label{fig:res}   
\end{figure*}

A key challenge for both observations and theoretical models is to connect the population of galaxies observed across cosmic time with the sub-parsec-scale physics of the star-formation process. Observationally, the cosmological context is addressed by large redshift surveys, while the small-scale physics can be most easily accessed within the Local Group or our own Milky Way. The physical processes associated with star formation are, however, also affected by varying local conditions \citep{Kawamura2009, Colombo2014, Hughes2016, Egusa2018, Hirota2018, Sun2020, Querejeta2021}.

Nearby star-forming galaxies therefore offer a unique viewpoint at the interface of the cosmological and Galactic scales. Several classical studies have been dedicated to detailed multi-wavelength mapping of individual nearby targets \citep[e.g. M51, M33, and M31;][]{Kennicutt2003, de_paz2007, Calzetti2005, Boquien2011, Viaene2014, Corbelli2017, Williams2018}. We have, however, so far lacked a comprehensive multi-tracer campaign covering the entirety of the discs of a representative set of star-forming main-sequence galaxies \citep[where the bulk of today's stars are being formed; e.g.][]{Brinchmann2004} down to their individual star-forming regions and probing the different phases of their ISM.
The target scale is the typical inter-cloud distance scale of a few tens of parsec up to about 100~parsec: it represents the intermediate `structuring size' or `cloud scale' of star-forming galaxies, relating to individual gas clouds, clusters of young stellar objects, and discrete star-forming regions. Already at a few hundred~parsec resolution, most of the structures associated with the gas clouds, stellar clusters, and dusty features are lost (see Fig.~\ref{fig:res}). When exploiting the nearby volume of galaxies up to about 20~Mpc, a scale of 100~pc typically requires arcsecond or sub-arcsecond full width at half maximum (FWHM) beams and, hence, relatively high spatial resolution supported by spectroscopic data over fields of view (FoVs) of a few arcminutes on disc galaxies.

Optical spectroscopy, in particular, is a powerful tool for placing the star-formation process in the context of its galactic host as it can probe the ionised ISM and the stellar backbone, which characterises the local disc environment. It more specifically provides key information pertaining to chemical abundances, stellar mass, star-formation histories, gas, and stellar kinematics. Spectroscopic observations at such scales over the large FoV required to map nearby galaxies have been very challenging due to the lack of suitable instrumentation.

The capabilities offered by integral field spectroscopy (IFS) have, however, greatly evolved over the last 30 years. Different hardware solutions, including fibres, micro-lenses, or advanced slicers, currently allow for a varied set of spatial and spectral samplings, filling factors, FoVs, and overall performance \citep[see e.g.][and references therein]{BaconMonnet2017}. These and other spectroscopic mapping techniques have already begun to be applied to samples of nearby galaxies (e.g. PINGS, \citealt{Rosales-Ortega2010}; VENGA, \citealt{blanc_virus-p_2013}; TYPHOON, M.~Seibert et al. in prep.; SIGNALS, \citealt{Rousseau-Nepton2019} CALIFA \citealt{Sanchez2012}, SAMI, \citealt{croom2012}, MaNGA, \citealt{bundy_overview_2015}). 
The Multi Unit Spectroscopic Explorer (MUSE) at the Very Large Telescope (VLT), with its coverage of most of the optical wavelength range and a FoV of 1~arcmin$^2$, properly sampling the seeing disc, has recently opened a significant new area of parameter space.
In particular, the success and versatility of MUSE lies in a combination of factors, including its high overall performance (${\sim}35$\% peak efficiency around $7000$~\AA; see e.g. MUSE/VLT User's Manual), its multiplexing capabilities ($90{,}000$ spaxels, each with about ${\sim}4000$ spectral pixels), and, most importantly, its robust optical setup (based on advanced slicers and an industrial approach for the building of its 24 spectrographic units) and its dedicated advanced data reduction pipeline \citep[][see also Sect.~\ref{sec:framework}]{Bacon2016, Weilbacher2020}. MUSE is one among just a few integral field units (IFUs) that can deliver a spectro-photometric view of the objects it targets, a key requirement for  being able to robustly derive physical parameters from IFS data.

These combined characteristics make MUSE the ideal instrument for providing, for the first time, extensive mapping of nearby ($D \lesssim 20$~Mpc) star-forming galaxy discs and resolving the mean inter-cloud distance in the ionised ISM (i.e. accessing the cloud scale). This paper presents the realisation of this ambitious goal in the form of the PHANGS-MUSE survey, an ESO Large Programme built on a VLT MUSE pilot project that mapped NGC0628 (Programme IDs: \mbox{1100.B-0651} / PI: E.~Schinnerer; \mbox{095.C-0473} / PI: G.~Blanc; and \mbox{094.C-0623} / PI: K.~Kreckel) to obtain spectrophotometric maps of the ionised gas and stars for 19 nearby galaxies.

The PHANGS-MUSE survey is a key part of the Physics at High Angular Resolution in Nearby Galaxies\footnote{\url{http://www.phangs.org}} (PHANGS) project.
PHANGS aims at obtaining, for the first time, a comprehensive view of the star-formation process across different ISM phases in the range of environments present within a representative sample of nearby, massive, star-forming galaxies \citep[see the detailed discussion about the sample in][]{Leroy2021b}. 
Key goals for PHANGS follow the following scientific threads: (a) infer the timescales of the star-formation process (i.e. molecular cloud lifetimes, feedback timescales, feedback outflow velocities, star-formation efficiencies, and mass loading factors); (b) quantify the importance of the various stellar feedback processes (i.e. ionising radiation, stellar winds, supernova explosions, etc.) in galactic discs, (c) resolve the chemical enrichment and mixing across galaxy discs (in both radial and azimuthal directions), and (d) establish how the clustering of young stars is seeded by and disrupts the structure of the ambient ISM. 

To achieve these goals, observations must map and resolve the individual structures of the star-formation process \citep[with sizes of a few pc to ${\sim}100$~pc; e.g.][]{Sanders1985, Oey2003}, such as (giant) molecular clouds, \hii\ regions, and the resulting star clusters. Sampling the variety of environments (related to e.g. bars, spirals, centres, mass, and dynamical structures) present in nearby galaxies is needed to assess the environmental impact within and among different galaxy discs. In order to link our results to the larger-scale studies of galaxy populations, we have prepared a selection in accordance with the main sequence of star-forming galaxies \citep[e.g.][]{Brinchmann2004}. 

\begin{figure*}
\centering
        \includegraphics[width=0.92\textwidth, trim=0 0 0 0, clip]{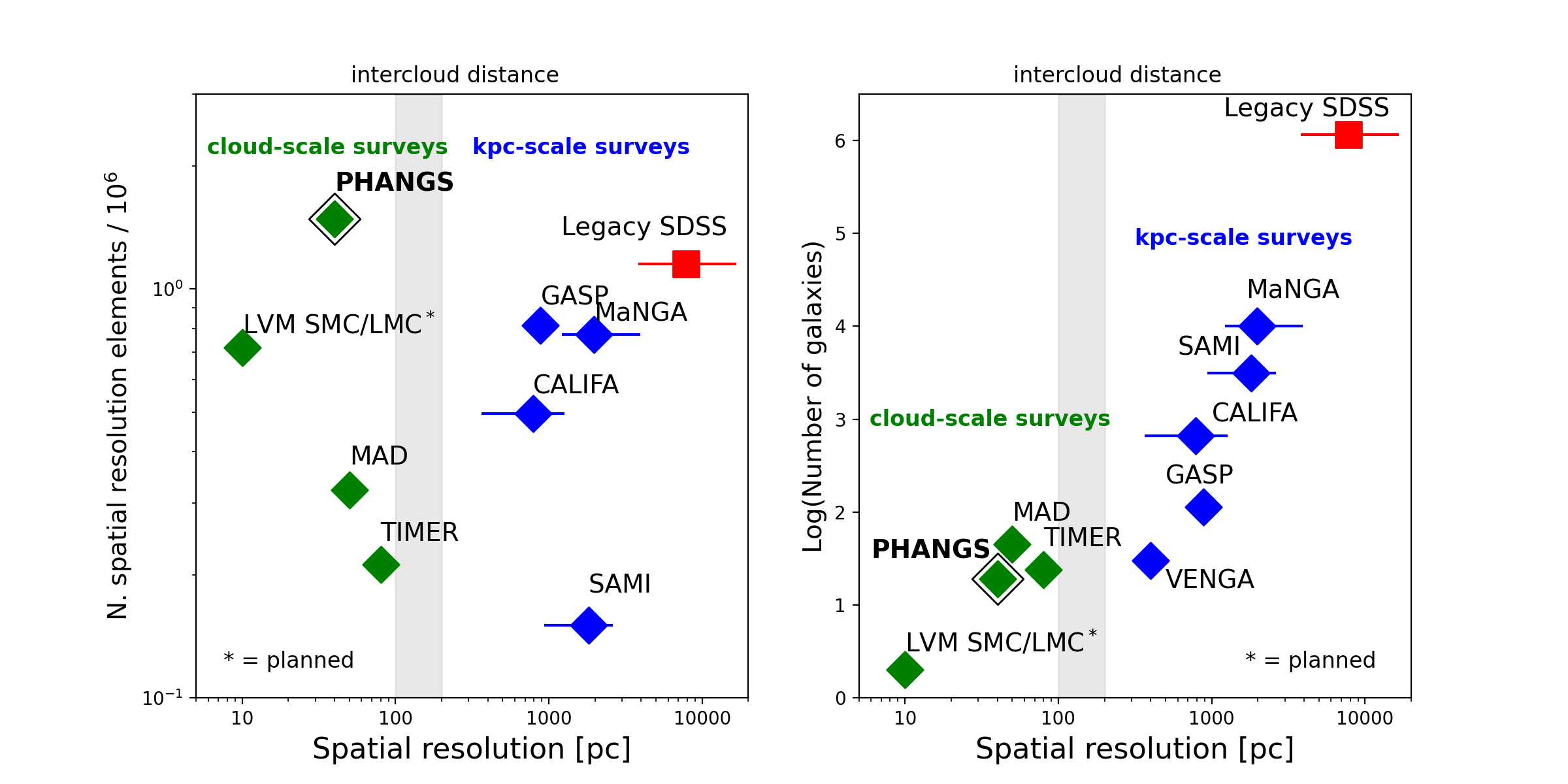}
     \caption{\textbf{Overview of large spectroscopic surveys of nearby galaxies.} \textit{Left}: Large spectroscopic surveys in the plane defined by their spatial resolution (in physical units) and the number of spatial resolution elements surveyed. IFU surveys are shown with diamond symbols (green for those achieving cloud-scale or better resolution and blue if probing at kiloparsec scales). VENGA \citep{blanc_virus-p_2013} is not shown because it features fewer than $10^5$ resolution elements. For reference, the single fibre SDSS survey \citep[][red triangle]{Abazajian2009} is added. PHANGS-MUSE sits in the top-left of this space, ranking highly on both metrics.  
     \textit{Right}: Large spectroscopic surveys in the plane defined by their spatial resolution (in physical units) and the number of galaxies surveyed. PHANGS-MUSE lies on the overall trend line of other IFU surveys. }
     \label{fig:ifu_surveys}   
\end{figure*}

Focusing on the optical spectroscopy aspect, PHANGS-MUSE covers a largely unexplored area of parameter space in terms of number of spectra versus spatial resolution compared to other state-of-the-art spectroscopic studies of nearby galaxies. In Fig.~\ref{fig:ifu_surveys} (\textit{left}) we compare PHANGS-MUSE with several other IFU surveys of local galaxies and with the Legacy (single $3\arcsec$ fibre) Sloan Digital Sky Survey \citep[SDSS][]{strauss_spectroscopic_2002,Abazajian2009}. For each IFU survey, we estimate the number of independent spatial resolution elements as the ratio between the total area surveyed and the area of the point spread function (PSF) FWHM. In this parameter space, PHANGS-MUSE occupies a unique region, combining high spatial resolution with the largest number of spectral elements. PHANGS-MUSE resolves the galactic discs about 1.5 orders of magnitude better than large IFU surveys of the nearby Universe, such as CALIFA, SAMI, and MaNGA \citep{Sanchez2012,croom2012, bundy_overview_2015}, while delivering a factor of ${\sim}2$ more independent spatial resolution elements than MaNGA, the largest of these surveys. 
This comparison highlights the impressive information-gathering power of the MUSE instrument. It also helps contextualise the challenges associated with the processing of the PHANGS-MUSE dataset. PHANGS-MUSE complements the large kiloparsec-scale surveys, such CALIFA, SAMI, and MaNGA, which have observed ${\sim}10^{3} {-} 10^{4}$ galaxies (Fig.~\ref{fig:ifu_surveys}, \textit{right}), and accesses new physics by trading sample size for spatial resolution.

Three other MUSE surveys, the MUSE Atlas of Discs \citep[MAD;][]{Erroz-Ferrer2019}, the Time Inference with MUSE Extragalactic Rings \citep[TIMER;][]{2019MNRAS.482..506G}, and GAs Stripping Phenomena in galaxies with MUSE \citep[GASP;][]{2017ApJ...844...48P}, target nearby star-forming galaxies. TIMER focuses on the impact of bars and active galactic nuclei (AGN) on galaxy evolution, while GASP studies ongoing and past ram pressure stripping events: their samples are therefore focused on addressing specific science questions and are not representative of the population of star-formation main-sequence (SFMS) galaxies. The MAD survey, on the other hand, focused on main-sequence galaxies with $\log(M_\star/M_{\odot}) > 8.5$, selected to be nearby ($z < 0.013$, $D < 55$~Mpc) and moderately inclined ($i<70\degree$), but obtained only one central MUSE pointing per object (and two-pointing mosaics in a few exceptional cases). MAD therefore probes only the inner regions (${\sim}2$~kpc) of nearby galaxies. For more distant targets, it samples a larger fraction of the galactic disc but at coarser spatial resolution (${>}200$~pc), starting to blend structures at the cloud scale (see Fig.~\ref{fig:res}). PHANGS-MUSE is complementary to all these surveys, covering the galactic discs of typical star-forming galaxies at $100$~pc or better resolution.

In this paper {we present the PHANGS-MUSE survey, providing both the global context for this campaign and information pertaining to the associated first public data release (DR1.0). We note that all released PHANGS-MUSE data are available via the ESO Archives (i.e. accessible programmatically using the `PHANGS' data collection flag) as well as from the Canadian Astronomy Data Centre (CADC)}. We start by reviewing the top-level scientific goals of the PHANGS-MUSE programme and present the galaxy sample (Sect.~\ref{sec:samplescgoals}). The MUSE observation strategy is described in Sect.~\ref{sec:obs}. In Sects.~\ref{sec:drs} and~\ref{sec:dap} we detail the data reduction and data analysis pipelines we have developed specifically for this survey and provide further quality assessment in Sect.~\ref{sec:QA}. In Sect.~\ref{sec:datarelease} we briefly describe the relevant internal and public data releases associated with the PHANGS-MUSE dataset. Finally, in Sect.~\ref{sec:keysci} we provide a set of data-demonstration figures to illustrate the potential of the survey, and we present our conclusions in Sect.~\ref{sec:conc}.
\section{PHANGS-MUSE survey: Observational context, sample, and science goals}
\label{sec:samplescgoals}

\subsection{The PHANGS-MUSE galaxy sample}
\label{sec:sample}

\begin{figure*}
\centering
        \includegraphics[width=0.94\textwidth, trim=0 0 0 0, clip]{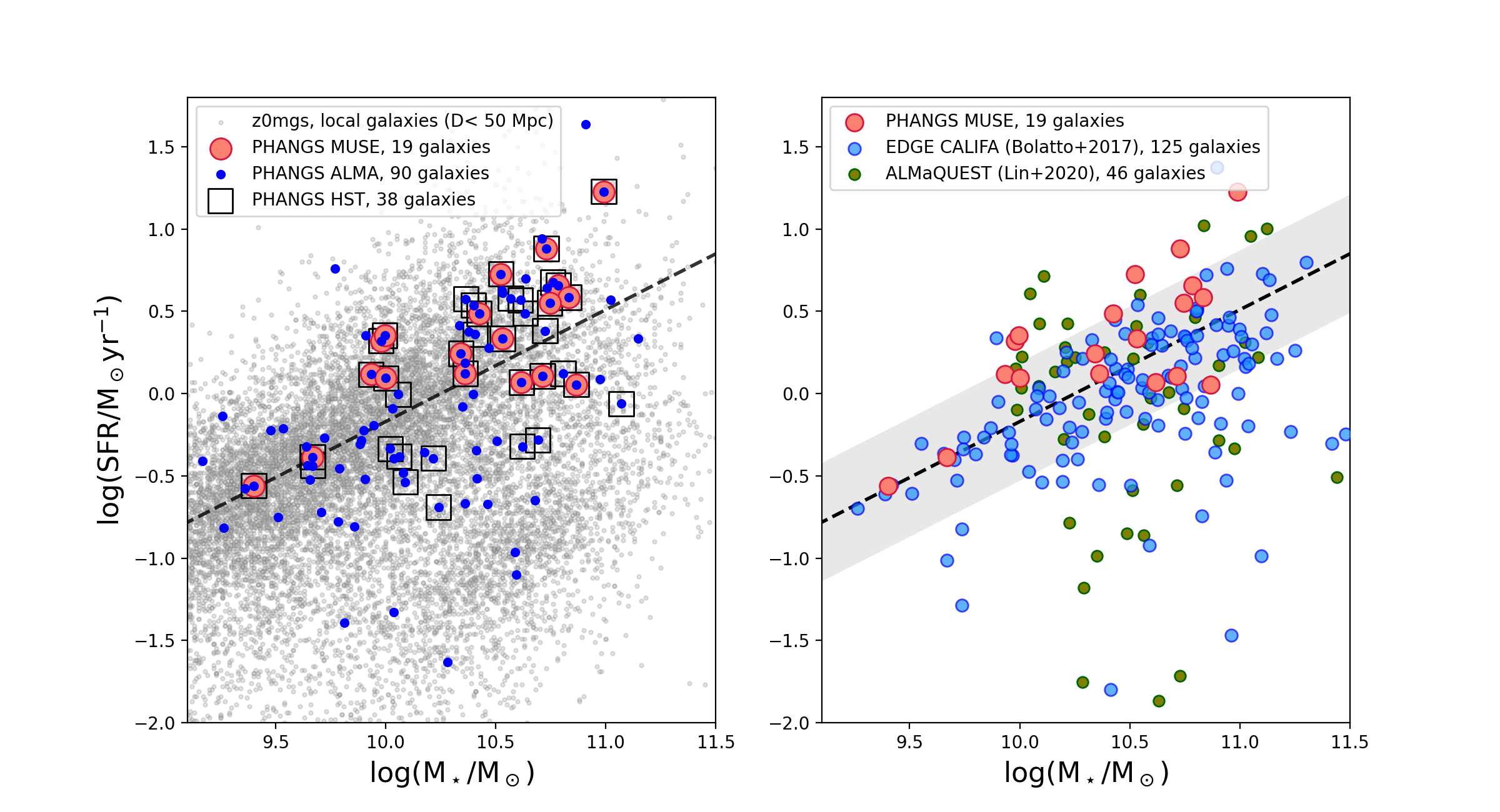}
     \caption{\textbf{ PHANGS-MUSE sample in the $\bm{M_{\star} {-} \mathrm{SFR}}$ plane.} \textit{Left:} PHANGS sample compared with the population of local galaxies from z0MGS \citep[][small grey dots]{Leroy2019}. The large red circles represent the PHANGS-MUSE galaxies. We show the overlap with the ALMA (blue dots) and HST (black empty squares) components of the PHANGS project. The dashed line is the best fit to the SFMS from \citet{Leroy2019}. \textit{Right:} PHANGS-MUSE sample compared to two complementary projects, EDGE-CALIFA \citep{Bolatto2017} and ALMa\-QUEST \citep{Lin_2020}, which also target local galaxies with optical IFS and CO interferometric mapping. The dashed line is the best fit to the SFMS from \cite{Leroy2019} with associated scatter (grey shaded area). }
     \label{fig:sample}   
\end{figure*}

The parent sample of the overall PHANGS programme was originally constructed according to the following four criteria \citep[see details in][]{Leroy2021b}: (i) southern sky accessible, in order to be observable by ALMA and MUSE, with $-75^{\circ}\leq\delta\leq+25^{\circ}$; (ii) nearby ($5~\textrm{Mpc} \leq D \leq 17~\textrm{Mpc}$), in order to probe star-forming regions out to at least an effective radius in a reasonable timescale and simultaneously  provide ${<}100$~pc resolution; (iii) low to moderate inclination ($i < 75^{\circ}$), to limit the effects of extinction and line-of-sight confusion and facilitate the identification of individual star-forming sites;
and (iv) massive star-forming galaxies with $\log(M_{\star}/M_{\odot}) \gtrsim 9.75$ and $\log(\textrm{sSFR}/\textrm{yr}^{-1}) \gtrsim -11$. The mass cut is driven by the desired overlap with ALMA observations, which become increasingly expensive in the low-mass, low-metallicity regime.

These criteria ensure that individual molecular clouds and star-forming regions can be isolated without confusion while the selected galaxies are representative for galaxies where most of the star formation in the local Universe occurs \citep[e.g.][]{Brinchmann2004}. The cuts do not strictly apply to the PHANGS sample described in \citet{Leroy2021b} because of subsequent revisions of the fiducial distance estimates \citep{anand2021}, improved derivation of the mass-to-light ratios (M/Ls) and star-formation rates (SFRs; \citealt{Leroy2019}) and the incorporation of additional galaxies extending the sample in important directions. A subset of 90 galaxies in the PHANGS parent sample has been observed by the PHANGS-ALMA survey, to create wide-field (covering the actively star-forming  disc, roughly $1{-}2\,R_\mathrm{e}$), high-resolution ($\textrm{FWHM}\sim1\arcsec$) \mbox{CO(2--1)} maps resolving the molecular phase into individual molecular clouds \citep{Leroy2021b}. 

As the MUSE effort started at the same time as the ALMA Large Programme, the target selection focused on the 19 targets that were already observed as part of the ALMA pilot projects, or had ALMA archival data of similar characteristics. {More specifically, a MUSE pilot programme \citep[PIs K.~Kreckel and G.~Blanc; see][]{Kreckel2016,Kreckel2017,Kreckel2018} focused on a close nearby face-on target, namely NGC\,628, which was then followed by 16 galaxies observed as part of the ALMA campaign to probe the SFMS, further complemented with targets present in the ESO archive: this led to a sample of 19 nearby systems.} Key properties of the PHANGS-MUSE targets are summarised in Table~\ref{tab:sample}, and the distribution (PHANGS-MUSE objects in red) in the SFR versus stellar mass, $M_{\star}$, plane is shown in Fig.~\ref{fig:sample} (\textit{left}), both in relation to the other PHANGS surveys -- ALMA and the Hubble Space Telescope (\textit{HST}) -- and relative to the main sequence of local ($D<50$~Mpc) star-forming galaxies, as derived by \citet{Leroy2019} via a joint UV+IR analysis. Our sample covers a wide stellar mass range ($9.4 < \log(M_\star/M_\odot) < 11.0$), but is, biased towards high masses (median stellar mass is $\log(M_\star/M_{\odot}) = 10.52$), and towards the upper envelope of the SFMS -- the median SFMS offset is $+0.21$~dex with respect to the $z=0$ Multi-wavelength Galaxy Synthesis (z0MGS) relation (\citealt{Leroy2019}) -- due to the need to adopt early, but sometimes uncertain, measurements of distances and SFR for the full PHANGS sample of star-forming main-sequence galaxies \citep[see details in][]{Leroy2021b}. The PHANGS-MUSE sample does not include any of the Green Valley targets from PHANGS-ALMA and PHANGS-HST. Passive galaxies are excluded from the PHANGS sample by design, although a few passive galaxies have been targeted by PHANGS-ALMA as part of an ancillary programme.

It is useful to compare PHANGS-MUSE with other programmes aiming at obtaining both molecular gas and optical IFU spectroscopy of local galaxies. EDGE-CALIFA \citep{Bolatto2017} and ALMa\-QUEST \citep{Lin_2020}, consisting of 125 and 46 galaxies, respectively, are the only comparable efforts in this category. These surveys provide a more uniform sampling of the main sequence, and extend to the Green Valley, as shown in Fig.~\ref{fig:sample} (\textit{right}). Unlike PHANGS, however, they both observe galaxies at approximately kiloparsec resolution, insufficient to resolve the physics of star formation on the scale of individual clouds. The MUSE targets within the PHANGS sample further have a large set of ancillary data on resolved scales, as emphasised in Sect.~\ref{sec:multi-lambda view}. 

\begin{table*}
\caption{General properties of the PHANGS-MUSE sample.   }
\label{tab:sample}
\centering
\begin{tabular}{lrrrrrrrrrr}
\hline \hline
Name & Distance\tablefootmark{a} & $\mathrm{Log}(M_{\star})$\tablefootmark{b}  & $\rm Log(SFR) $\tablefootmark{b}  & $\rm \Delta_{SFMS} $\tablefootmark{b}  & $\rm R_{25} \tablefootmark{c}$ & PA \tablefootmark{d}  & $i$ \tablefootmark{d} & scale  & PSF \tablefootmark{e} & copt PSF \tablefootmark{f}  \\
     & [Mpc] & [$\rm M_\odot$]  & [$\rm M_\odot$~yr$^{-1}$]     & [dex] & [arcmin]  & [deg]  & [deg] & [pc/arcsec] & [arcsec] & [arcsec] \\
\hline
\hline
IC5332 & 9.0 & 9.67 & -0.39 & 0.01 & 3.0 & 74.4 & 26.9 & 43.7 & 0.72$\pm^{0.08}_{0.12}$ & 0.87 \\
NGC0628 & 9.8 & 10.34 & 0.24 & 0.18 & 4.9 & 20.7 & 8.9 & 47.7 & 0.73$\pm^{0.11}_{0.13}$ & 0.92 \\
NGC1087 & 15.9 & 9.93 & 0.12 & 0.33 & 1.5 & 359.1 & 42.9 & 76.8 & 0.74$\pm^{0.10}_{0.12}$ & 0.92 \\
NGC1300 & 19.0 & 10.62 & 0.07 & -0.18 & 3.0 & 278.0 & 31.8 & 92.1 & 0.63$\pm^{0.18}_{0.13}$ & 0.89 \\
NGC1365 & 19.6 & 10.99 & 1.23 & 0.72 & 6.0 & 201.1 & 55.4 & 94.9 & 0.82$\pm^{0.26}_{0.24}$ & 1.15 \\
NGC1385 & 17.2 & 9.98 & 0.32 & 0.50 & 1.7 & 181.3 & 44.0 & 83.5 & 0.49$\pm^{0.10}_{0.11}$ & 0.77 \\
NGC1433 & 18.6 & 10.87 & 0.05 & -0.36 & 3.1 & 199.7 & 28.6 & 90.3 & 0.65$\pm^{0.18}_{0.14}$ & 0.91 \\
NGC1512 & 18.8 & 10.71 & 0.11 & -0.21 & 4.2 & 261.9 & 42.5 & 91.3 & 0.80$\pm^{0.38}_{0.16}$ & 1.25 \\
NGC1566 & 17.7 & 10.78 & 0.66 & 0.29 & 3.6 & 214.7 & 29.5 & 85.8 & 0.64$\pm^{0.09}_{0.10}$ & 0.80 \\
NGC1672 & 19.4 & 10.73 & 0.88 & 0.56 & 3.1 & 134.3 & 42.6 & 94.1 & 0.72$\pm^{0.17}_{0.08}$ & 0.96 \\
NGC2835 & 12.2 & 10.00 & 0.09 & 0.26 & 3.2 & 1.0 & 41.3 & 59.2 & 0.85$\pm^{0.23}_{0.18}$ & 1.15 \\
NGC3351 & 10.0 & 10.36 & 0.12 & 0.05 & 3.6 & 193.2 & 45.1 & 48.3 & 0.74$\pm^{0.24}_{0.13}$ & 1.05 \\
NGC3627 & 11.3 & 10.83 & 0.58 & 0.19 & 5.1 & 173.1 & 57.3 & 54.9 & 0.77$\pm^{0.21}_{0.10}$ & 1.05 \\
NGC4254 & 13.1 & 10.42 & 0.49 & 0.37 & 2.5 & 68.1 & 34.4 & 63.5 & 0.58$\pm^{0.23}_{0.14}$ & 0.89 \\
NGC4303 & 17.0 & 10.52 & 0.73 & 0.54 & 3.4 & 312.4 & 23.5 & 82.4 & 0.58$\pm^{0.12}_{0.07}$ & 0.78 \\
NGC4321 & 15.2 & 10.75 & 0.55 & 0.21 & 3.0 & 156.2 & 38.5 & 73.7 & 0.64$\pm^{0.45}_{0.18}$ & 1.16 \\
NGC4535 & 15.8 & 10.53 & 0.33 & 0.14 & 4.1 & 179.7 & 44.7 & 76.5 & 0.44$\pm^{0.03}_{0.01}$ & 0.56 \\
NGC5068 & 5.2 & 9.40 & -0.56 & 0.02 & 3.7 & 342.4 & 35.7 & 25.2 & 0.73$\pm^{0.23}_{0.21}$ & 1.04 \\
NGC7496 & 18.7 & 10.00 & 0.35 & 0.53 & 1.7 & 193.7 & 35.9 & 90.8 & 0.79$\pm^{0.03}_{0.17}$ & 0.89 \\
\hline
\end{tabular}
        \tablefoot{
                \tablefoottext{a}{From the compilation of \citet{anand2021}.}
                \tablefoottext{b}{Derived by \citet{Leroy2021b}, using \textit{GALEX} UV and \textit{WISE} IR photometry, following a similar methodology to \cite{Leroy2019}.}
                \tablefoottext{c}{From LEDA \citep{Makarov2014}.}
                \tablefoottext{d}{From \cite{Lang2020}, based on \mbox{CO(2--1)} kinematics.}
                \tablefoottext{e}{FWHM of the Moffat PSF across individual pointing (we report the mean and the minimum and maximum values in the R band).}
                \tablefoottext{f}{FWHM of the Gaussian PSF of the homogenised (copt) mosaic.}
                        }
\end{table*}

\subsection{PHANGS-MUSE in the multi-wavelength context}
\label{sec:multi-lambda view}

\begin{figure*}
\centering
        \includegraphics[width=0.98\textwidth, trim=0 10 30 30, clip]{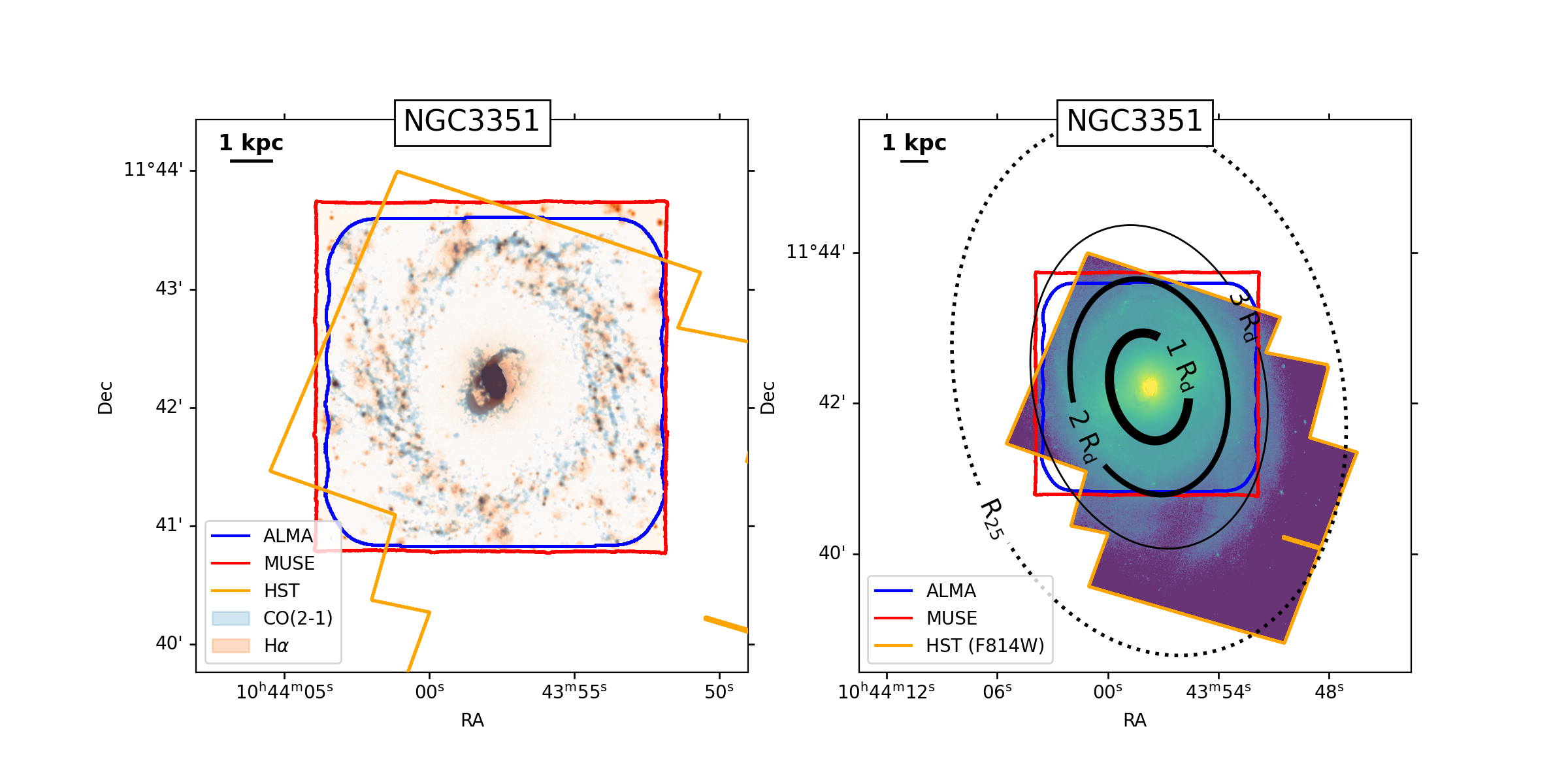}
     \caption{\textbf{Synoptic view of the PHANGS multi-wavelength data using NGC~3351 for illustration.} \textit{Left}: Blue and red contours showing the footprints of the PHANGS-ALMA and PHANGS-MUSE data. Within the respective footprints, we show flux maps for \ha\ from MUSE (light red) and \mbox{CO(2--1)} from ALMA (light blue). The footprint of the HST observations included in the PHANGS-HST data release is also shown as an orange contour.
     \textit{Right}:
     HST imaging (F814W filter) shown in colour. The image covers a larger FoV with respect to the left panel in order to show the entire area imaged by HST. We also indicate several radial metrics: the disc scale length \citep[$R_\mathrm{d}$, as derived in][]{Leroy2021b}  and $R_{25}$. The ALMA and MUSE footprints are also shown, same as in the left panel. We note that all MUSE, ALMA, and HST footprints of the 19 galaxies can be found at \url{https://archive.stsci.edu/hlsp/phangs-hst}.}
     \label{fig:synoptic_3351}   
\end{figure*}

\begin{figure*}
    \centering
    \includegraphics[width=0.99\textwidth]{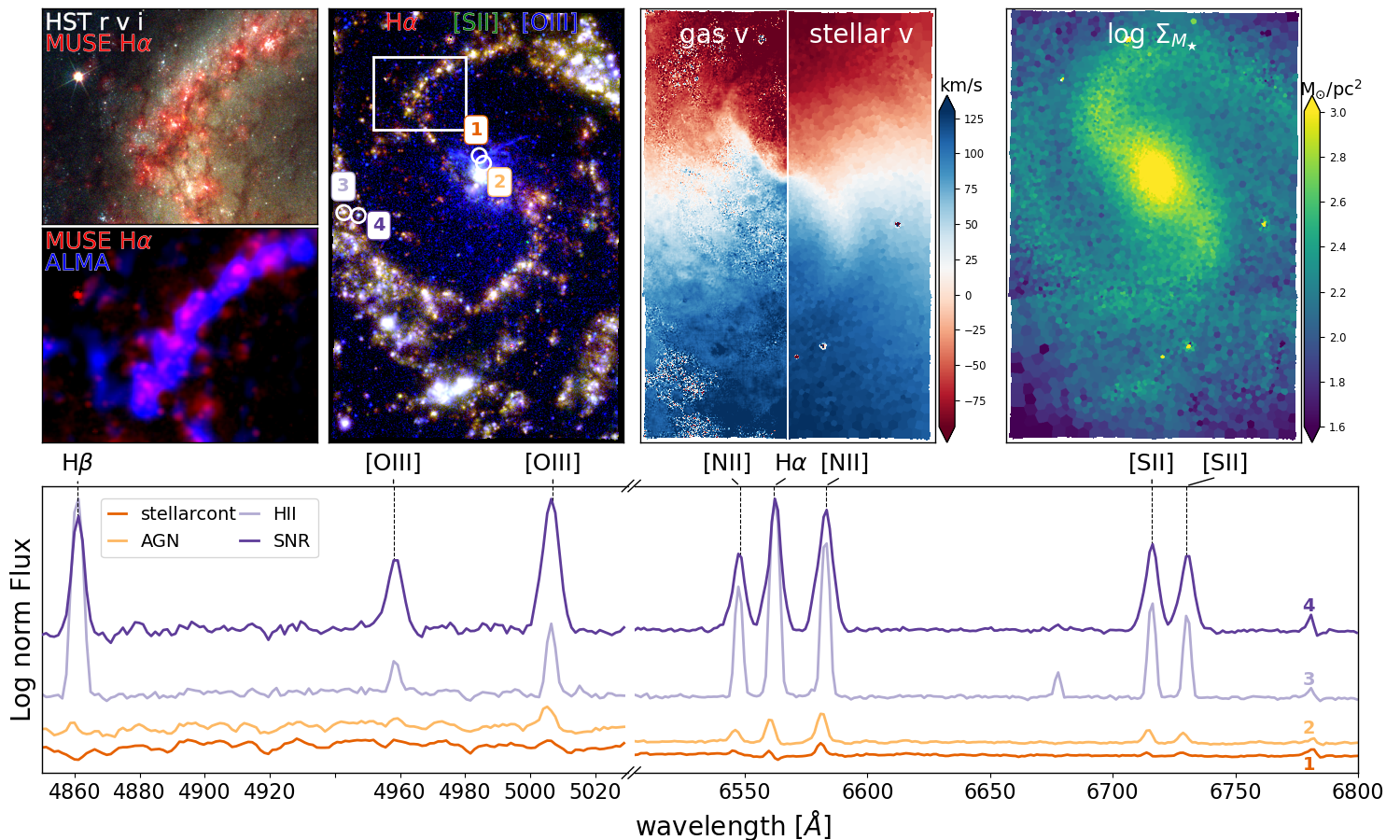}
    \caption{Multi-wavelength, multi-phase view of NGC~4535. The top panels present the stellar and gas distribution and kinematics. Top, second panel from the left: Multi-emission line view (\ha\ in red, \oiii\ in blue, \sii\ in green) tracing the sites of massive star formation along the spiral pattern, with differences in the combined colours highlighting the changes in local physical condition (e.g. abundance, ionisation parameter) and ionising source.Second panel from the left: Extracted spectra (marked as \protect{white} circles) demonstrating the typical characteristics of \protect{four (numbered) regions, namely: 1- dominated by stellar continuum (red); 2- AGN (orange); 3- \hii\ regions (light purple), and 4- supernova remnants (SNR; purple)}. AGN line emission is superimposed on the strong stellar continuum absorption, with distinctive strong \oiii\ and \nii\ emission. In the outer disc, \hii\ regions and SNRs have less contribution from the stellar continuum. Broadened line shapes are apparent in the expanding SNR, in contrast to the narrower \hii\ region line emission, and show  strong \sii\ and \nii\ relative to \ha. Zooming into one section of the spiral arm (white box), spatial offsets between the HST star clusters and ionised gas (left top) and between the ionised gas and ALMA molecular gas (left bottom) demonstrate a  time sequence evolution across the spiral pattern. Centre right: The gas and stellar velocity fields, mapped through the MUSE spectroscopy, highlight deviations from regular rotation and indicate dynamically driven radial flows along the spiral and bar structures. Right: The stellar mass surface density ($\Sigma_{M_\star}$\protect{; see Sect.~\ref{sec:stellarpops}}) highlights the location of these dynamical spiral and bar structures, and provides crucial constraints on the underlying gravitational potential affecting all stellar and gaseous processes in the disc.
    }
    \label{fig:multiwave}
\end{figure*}

The PHANGS programme is built on three main pillars. In addition to the PHANGS-MUSE programme, PHANGS leverages a Large Programme imaging the cold molecular phase in \mbox{CO(2--1)} with ALMA \citep[][PI: E.~Schinnerer]{Leroy2021, Leroy2021b}, and a high-resolution Legacy survey of star clusters and stellar populations using five-band NUV-U-B-V-I imaging with the \textit{HST} \citep[][PI: J.~Lee]{Lee2021}. A fourth pillar is expected in the coming years, as we have also been awarded a James Webb Space Telescope (JWST) Treasury programme (PI: J.~Lee) to image the PHANGS-MUSE sample. This practically means that the PHANGS-MUSE sample of 19 galaxies will ultimately have the full suite of MUSE, ALMA, HST, and JWST data. In Fig.~\ref{fig:synoptic_3351} we give an example of the combined MUSE, ALMA and \textit{HST} coverage for one of the galaxies in our sample, NGC 3351. In Fig.~\ref{fig:multiwave} we provide an illustrative view on the synergy between PHANGS-MUSE, PHANGS-ALMA and PHANGS-HST datasets (top left panels), with the specific power of optical spectroscopy allowed by MUSE, leading superb constraints on stellar and gas kinematics, the distribution and properties of the ionised gas and stellar populations.

This core observational effort is supplemented by a suite of complementary data, including ground-based narrow-band imaging (PHANGS-\ha; A.~Razza et al.\ in preparation, PIs G.~Blanc \& I-T.~Ho), Keck Cosmic Web Imager (KCWI) spectroscopy (PI: K.~Sandstrom), Canada-France-Hawaii Telescope SITELLE \oii\ imaging (PI: A.~Hughes), Russian 6m Fabry-Perot Interferometre spectroscopy (PI: E.~Egorov), atomic hydrogen 21~cm mapping (D.~Utomo et al.\ in preparation, PI: D.~Utomo), far-UV imaging with AstroSAT (E.~Rosolowsky et al.\ in preparation, PI: E.~Rosolowsky), stellar mass maps with corresponding environmental masks \citep{Sheth2010,Querejeta2015,Querejeta2021}, dust maps obtained from archival \textit{Spitzer} and \textit{Herschel} imaging \citep{Kennicutt2003, Kennicutt2011, Clark2018} as reprocessed by J.~Chastenet et al.\ (in preparation), and maps of molecular dense-gas tracers \citep{JimenezDonaire2019}. Tailored numerical work aims to provide the required reference simulations \citep[e.g.][]{Jeffreson2020, Utreras2020}.

\subsection{PHANGS-MUSE survey science goals}
\label{sec:scgoals}

The MUSE observations of our selected nearby galaxies provide observational constraints on the structures (i.e. spirals, bars, centres) that make up the galaxy discs through various measurements by both covering various hosts and spatially resolving those structures. That includes the identification and spectroscopy of individual \hii\ regions, their spatial distributions, brightnesses, metallicities, ionisation properties, and {for a subset of regions} even measurement of weaker temperature-sensitive lines. Those observations also provide detailed optical coverage of the stellar populations, constraining a two-dimensional view of the star-formation history and stellar mass distribution. They represent a unique probe of the gaseous and stellar dynamics via resolved kinematics; and additional components such as dust (extinction via the stellar continuum and Balmer decrement) or AGN (broad lines or high-ionisation emission lines). The PHANGS-MUSE dataset will more specifically inform several key science goals, which we now briefly review in turn.

\medskip
\textbf{A local perspective on scaling relations.}
Scaling relations have often guided observational and theoretical work towards our understanding of star-formation-related processes in galaxies \citep{Kennicutt1998, Brinchmann2004, Salim2007, Bigiel2008, Blanc2009, Leroy2013, CanoDiaz2016, Hsieh2017, Medling2018, Lin_2020, Sanchez2021, Ellison2021, Pessa2021, Querejeta2021}. Basic properties such as stellar, molecular gas, or total gas surface density and SFR surface density have been probed to anchor representative timescales, such as how long it takes to deplete available gas reservoirs on galactic scales \citep{Schmidt1959, Kennicutt1989, Kennicutt1998, Saintonge2011, Cicone2017, Saintonge2017}, as well as more locally on sub-kiloparsec scales \citep{Wong2002, Bigiel2008, Leroy2008, Genzel2010, Schruba2011, Momose2013, Leroy2013,Bolatto2017, Lin_2020,Sorai2019}.

The PHANGS-MUSE dataset provides robust constraints on the SFR, through direct H$\alpha$ flux measurements corrected for extinction (via the Balmer decrement) and contribution from the diffuse gas component. The MUSE spectral coverage also probes gas-phase metallicity tracers, and key age and metallicity-sensitive stellar continuum spectral features, therefore allowing accurate derivations of M/Ls and stellar mass surface densities. The PHANGS dataset allows us to probe scaling relations at the global, kiloparsec-scale or cloud-scale levels, and to investigate the effect of different galactic environments (pressure budget, morphological, dynamical; see Sect.~\ref{sec:SciCase_scaling}). Understanding how scaling relations vary as a function of scale and galactic environment \citep[e.g.][]{Pessa2021} will shine new light on the driving mechanisms for the observed trends, as well as the source of the associated scatter.

\medskip

\textbf{The impact of stellar feedback, in relation to local and global environments.}
All theories and simulations now agree that stellar feedback plays a central role in the self-regulation of the star-formation process across a wide range of galactic environments \citep[e.g.][]{MacLow2004, McKee2007, Ostriker2010, Hopkins2014, Agertz2015, Agertz2016, grisdale_impact_2017, Semenov2018, Fujimoto2019, Semenov2021}. Yet stellar feedback comes in many forms: radiative ionisation and heating, radiation pressure, stellar winds, and supernova explosions. 
These processes affect not just the local (${<}100$~pc) surroundings but can impact on kiloparsec scales and contribute to the pervasive diffuse ionised gas (DIG) observed throughout spiral galaxies \citep{Zurita2000, Haffner2009, Zhang2017}. 
While simulations have made progress considering the combined effects of these feedback processes \citep[e.g.][]{Hopkins2014,Rathjen2021}, observations have lagged behind. Only a few nearby targets, including our Milky Way \citep[e.g.][]{Barnes2020, Olivier2021}, the Magellanic Clouds  \citep{Pellegrini2011,Lopez2011,Lopez2014} or NGC\,300 \citep{McLeod2020}, have seen a careful inventory of the relative strength and location of sources of stellar feedback \citep[see also][]{Chevance2022, Barnes2021}.

The PHANGS-MUSE survey of nearby galaxies will enable the quantitative study of the different forms of stellar feedback (radiative and mechanical) across galactic discs. In combination with the ALMA CO maps, the MUSE data can {probe} the interactions (such as localisation, dynamical and pressure differences) between the warm  ($10^4$~K) and the cold (${<}100$~K) gas reservoir on local and global scales, and potential variations with key galactic parameters. The data can provide measurements of the local balance of input momentum and energy from radiation, stellar winds, and supernovae {constrained via the} HST-derived massive stars and cluster catalogues\citep[][A.~Whitmore et al.\ in press, K.~Larson et al.\ in preparation]{Turner2021}, MUSE-derived \hii\ region properties, and the MUSE-modelled star-formation histories against the restoring forces of gas self-gravity and stellar gravity (derived from the ALMA molecular gas maps and the MUSE stellar mass maps) at a succession of spatial scales \citep[see e.g.][]{Sun2020b, Barnes2021}. We can furthermore model the escape of radiation from individual \hii\ regions and quantify its contribution to the ionisation of the kiloparsec-scale DIG, in combination with hot evolved low-mass stars and other ionising sources \citep{Belfiore2021}. This will give us a local assessment of the impact of individual feedback processes from the scale of individual regions to large parts of galaxies.

\medskip

\textbf{Quantifying the chemical enrichment and mixing in galactic discs.} 
Radial metallicity trends have been observed in galaxy discs for decades, and more recently quantified in the overall population of local galaxies by large IFU surveys (i.e. CALIFA, \citealt{Sanchez2014}; SAMI, \citealt{Croom2021}; MaNGA, \citealt{bundy_overview_2015}). Going beyond the radial trends, measurements of azimuthal variations and small-scale patterns remain poorly constrained, while being crucial to understand the key processes driving the chemical evolution of the ISM \citep[e.g.][]{Zaritsky1994,Sanchez2014,Belfiore2017}, flows of gas (pristine or enriched), and the redistribution of metals from their birth sites to kilo-parsec scales. Tantalising evidence of azimuthal variations in gas-phase oxygen abundance have been obtained by high spatial resolution IFU studies \citep{Sanchez-Menguiano2016,Vogt2017, Kreckel2019} and multi-slit spectroscopy \citep{Berg2015,Croxall2016}. \citet{Ho2017} and \citet{Ho2018}, for example, observed clear azimuthal metallicity variations associated with the spiral arms of NGC~1365 and NGC~2997 in ${\sim}100$~pc resolution pseudo-IFU long-slit data. Yet, it the origin of these azimuthal metallicity variations remains unclear; it is unclear if they are driven by localised self-enrichment and spiral-arm-induced mixing \citep{Ho2017} or by radial flows in the disc \citep{Sanchez-Menguiano2016}. 

Establishing the physical meaning of such variations requires high spatial resolution IFS observations of nearby galaxies, reaching out beyond the brightest \hii\ regions, minimising the contamination by the DIG. It also requires probing various emission lines, isolating individual \hii\ regions and inventorying them. The PHANGS-MUSE data are able to  derive measurements (both from strong-line calibrations, as in \citealt{Kreckel2019}, and from direct electron temperature $T_\mathrm{e}$ determinations, as in \citealt{Ho2019}) of such patterns and their relationships to bars and spiral arms, the key drivers of radial flows in galaxies. By linking such variations with a detailed analysis of bar and spiral arm pattern speeds \citep{Williams2021} and gas flows through our galaxies based on combined MUSE and ALMA data, we will be able to determine the key driver of mixing within galactic discs.

\medskip

\textbf{The role of dynamical regimes on the triggering, boosting or inhibiting of star formation.} Dynamical environments play a key role in the redistribution of the gas reservoir and in setting the efficiency of star formation within discs. Bars, spirals, rings, resonances and central regions \citep[e.g.][]{Verley2007, Sanchez-Blazquez2011, Meidt2013, renaud_environmental_2015, Sun2020, Kretschmer2020, Gensior2020, Henshaw2020} are characterised by different regimes associated with, for example, shear, torques, instabilities, gas flows, compression, or shocks. The PHANGS-MUSE dataset will help us constrain the local star-formation history via spectral fitting techniques, to thus characterise the stellar mass contribution. It will also provide unique leverage on the gravitational potential via the mapping of stellar and gas kinematics, and its various tracers (ionised gas, stellar populations; \citealt{Kalinova2017, Leung2018, Bryant2019, Shetty2020}). The determination of the star-formation history of the stellar disc and its link with the underlying dynamical orbital structure will provide a key constraint to understand the assembly and evolution of stellar discs.
At the same time, we will compare gas and stellar surface densities and kinematics to predictions from equilibrium disc models to assess the scale at which vertical equilibrium \citep[e.g.][]{Ostriker2010, Ostriker2011} and radial disc stability \citep[e.g.][]{Hunter1998, Martin2001, Krumholz2018, Romeo2020} emerge, balancing stellar feedback and gravity. In such a context, the synergy with numerical (hydro-dynamical) simulations will be paramount to further probe the relevant processes and their respective timescales \citep{Utreras2020, Fujimoto2019, Jeffreson2020}.

\textbf{A multi-purpose legacy dataset.} The sensitivity and physical resolution of the PHANGS-MUSE data will additionally allow for further investigation in a number of areas, including, for example, precise distance determination via planetary nebula luminosity functions \citep[][F.~Scheuermann et al.\ in preparation]{Kreckel2017}, and identification of supernova remnants (SNRs) via line ratio diagnostics and line broadening \citep[see e.g.][and references therein]{Kopsacheili2020}. Our science goals are highly complementary to existing kiloparsec resolution IFU studies of hundreds \citep[CALIFA][]{Sanchez2012} or thousands of galaxies (SAMI, \citet{croom2012}; MaNGA, \citet{bundy_overview_2015}) and to future imaging spectroscopy of high-redshift targets, for instance using \textit{JWST} and the Extremely Large Telescope (ELT). The PHANGS-MUSE dataset will serve calibration purposes and act as a training sample to enable accurate physical parameter estimation from lower-resolution observations of massive main-sequence star-forming galaxies. Our overall goal is that PHANGS-MUSE becomes a long-standing legacy dataset, providing the community with a reference for future observational, theoretical and simulation works.

\section{Observations} 
\label{sec:obs}

\subsection{Observing strategy}
\label{sec:obsstrategy}

\begin{figure*}
\centering
        \includegraphics[width=\textwidth]{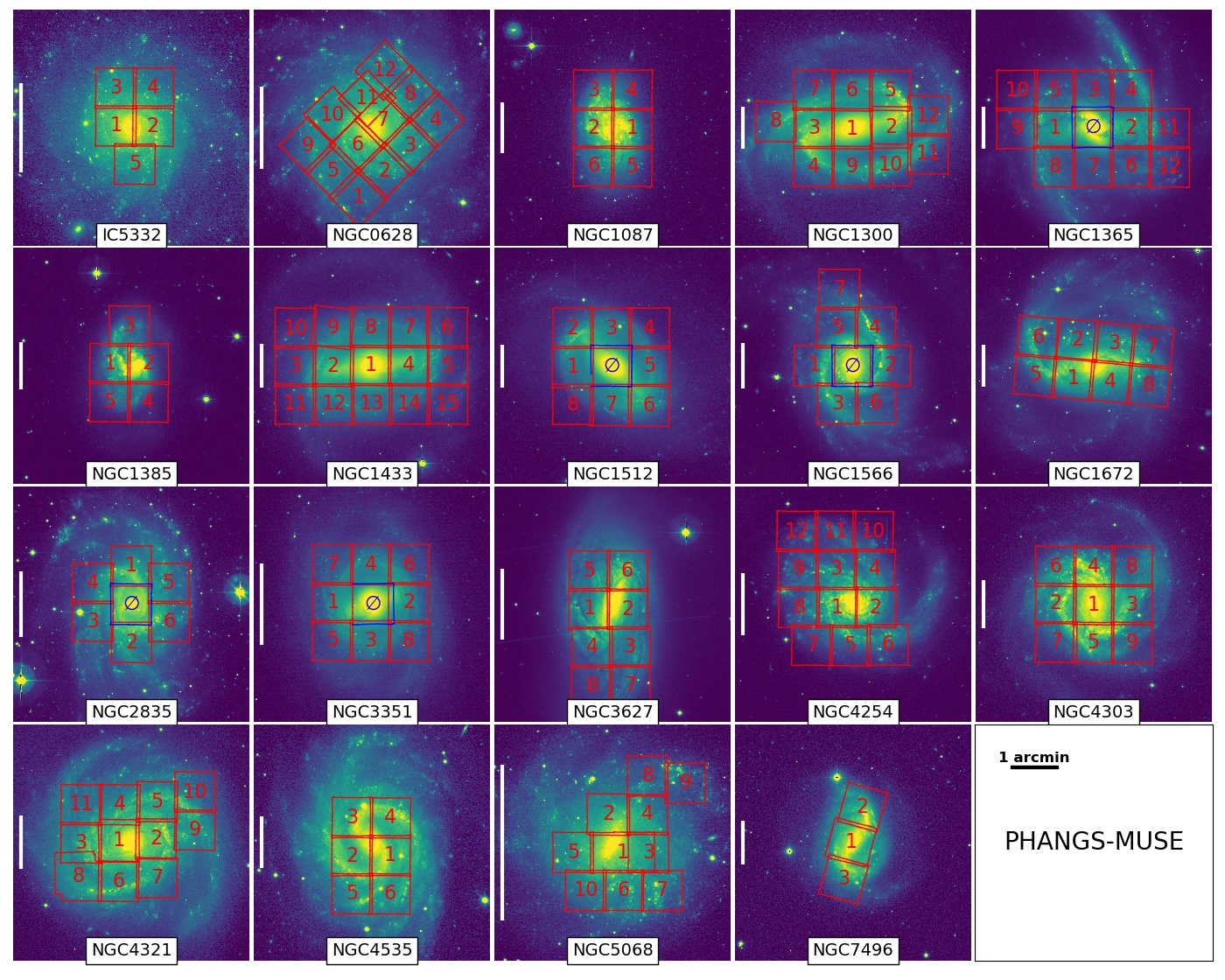}
     \caption{ \textbf{Footprints for the MUSE observations of PHANGS galaxies.} Each panel represents one target of the PHANGS-MUSE sample, with a $5\times5$ arcmin$^2$ FoV from the WFI $R_\mathrm{c}$-band images ($r$-band du~Pont for NGC~7496), and the footprints of the MUSE exposures are overlaid in red. Pointings marked with the $\varnothing$ symbol (in NGC~1365, NGC~1512, NGC~1566, NGC~2835, and NGC~3351) and outlined in blue correspond to observations acquired outside of the main PHANGS campaign but reduced following the same data flow and released as part of PHANGS-MUSE. The vertical white bar on the left side of each panel indicates a scale of $5$~kpc.}
     \label{footprint}
\end{figure*}

The PHANGS-MUSE survey covers a substantial fraction of the star formation within the galactic discs for a sample of 19 nearby star-forming spiral galaxies (see Sect.~\ref{sec:sample}). The footprint of the PHANGS-MUSE survey is shown in Fig.~\ref{footprint}{: it was designed to overlap with the area of sky imaged in \mbox{CO(2--1)} by PHANGS-ALMA, which was itself aimed at encompassing all regions of active star formation inside the disc, including on average 70\% of the WISE3 luminosity across the PHANGS-ALMA sample \citep[see][]{Leroy2021b}}. We designed the mosaics with a preference for a north-south orientation for the individual MUSE pointings. In a few cases, we rotated the position angles of the pointings to best cover the area of the galaxy we wished to map (e.g. NGC~1672 or NGC~7496). 

Individual galaxies are covered by a variable number of pointings depending on their angular size, ranging from 3 (e.g. NGC~7496) to 15 (e.g. NGC~1433), for a total of 168 individual pointings (5 of which were obtained from the ESO archive and had been observed by other programmes). Pointings were placed to have an overlap of $2\arcsec$ ($\sim$two resolution elements, and ${\sim}10$ MUSE spaxels) between adjacent fields to assist in the alignment process for the final mosaics. We further optimised the pointings' gridding, to be as regular as possible to achieve our planned coverage, also for those galaxies that included archival pointings. 

With the PHANGS-MUSE campaign, we aimed at detecting the \hb\ line (at a S/N~$> 5$) on top of the galaxy's stellar continuuum (at S/N~$> 10$). The existing ALMA coverage typically extends to $I$-band surface brightnesses of $\sim 22$~mag~arcsec$^{-2}$. This initially translated to a total of 43 minutes exposure on target (for a dark night, 7 days from the moon, and a typical airmass of 1.25), assuming a 5$\sigma$ detection for \hb\ on individual pointings, a 10$\sigma$ detection on the stellar continuum, and about 20$\sigma$ detection on \ha\ in the interarm regions (assuming a minimum binning of $5\times5$ spaxels, or 1\arcsec${^2}$). Considering the significant spatial variation in stellar populations, extinction, ionised gas content and regimes, this was intended to be a simplified and broad observational strategy. As detailed in Sect.~\ref{sec:em_errors}, we typically detect \hb\ in 50 to 80\% of the individual spaxels throughout our sample (see Fig.~\ref{fig:line_fractions}).

For our Large Programme, each pointing was observed at four different orientations separated by 90 degrees to mitigate the impact of individual MUSE slicers and the effects of the instrumental line spread function (LSF). These exposures were intertwined with two dedicated offset sky pointings, following an Object(O)-Sky(S)-O-O-S-O pattern. For each pointing the total on-source and sky integration times are 43 and 4~minutes, respectively. The entire PHANGS-MUSE Large Programme alone consists of a total telescope time of ${\sim}172$ hours. Observations were carried out using MUSE wide field mode (WFM), using the nominal (non-extended) wavelength range, either in seeing-limited (WFM-noAO) for most of the early acquired data or ground-layer adaptive optics (WFM-AO) mode. The observing campaign was designed for and started with the noAO (without the use of ground-layer adaptive optics) mode, the only mode available at the time. The WFM-AO mode for MUSE was offered in ESO Period 101, one semester after the start of the PHANGS observing runs. After some testing on early PHANGS datasets in terms of the extraction of stellar populations and emission lines information, and considering the small overheads incurred by the AO mode, we decided to implement the systematic use of the WFM-AO mode in 2018 for all targets that had not then yet been started (thus securing a single setup per mosaic). The 168 pointings result in about 15~Million spectra, covering a wavelength range of [$4750{-}9350$~\AA], with the spectral resolution ($\sigma$) going from about $80$~\kms\ (at the blue end) to $35$~\kms\ (at the red end).

A pre-existing set of archival pointings, targeting the centres of some of our target galaxies, were not re-observed, but were reduced together with all other pointings, and included in our mosaics. The central pointing of NGC~3351 was acquired in the course of the MAD survey \citep{Erroz-Ferrer2019}, the central pointing of NGC~1365 was taken from the Measuring Active Galactic Nuclei Under the MUSE Microscope Survey \citep[MAGNUM;][]{Venturi2018} of local AGN, while the central fields for NGC~1512, NGC~1566 and NGC~2835 were observed as part of the TIMER project \citep[][]{2019MNRAS.482..506G}. The observations of NGC~0628 were obtained by the PHANGS collaboration as part of a dedicated pilot programme \citep[PIs K.~Kreckel and G.~Blanc; see][]{Kreckel2016,Kreckel2017,Kreckel2018}.
Overall our mosaics include 168 MUSE pointings: twelve pointings of the pilot target NGC\,0628, 151 pointings executed as part of the PHANGS-MUSE Large Programme, and five archival pointings.

Table~\ref{table:ObsPointings} in Appendix~\ref{sec:contribs} summarises, for each target, the main properties of the pointings such as the sky coordinates, day and time of execution, number of exposures, PSF, and observing mode. Occasionally, due to weather-related or technical issues during the observations, one pointing had to be split into two different observing blocks (OBs).
The observing strategy for the Large Programme consists of a nominal four exposures per pointing. For the prototype NGC~0628, data for each pointing was observed with three exposures instead. For some pointings, we had to discard one or two exposures due to high sky brightness or extreme variability. For some pointings, on the other hand, we obtained one extra exposure, mainly as a result of OB splitting, as mentioned above. Due to a technical problem during the observations of NGC~1385, one of the pointings (i.e.~P03) was observed twice and, therefore, consists of eight science exposures.  

\subsection{Ancillary wide-field imaging data}
\label{sec:ancillary}

In several steps of the reduction and characterisation of the MUSE data (e.g. exposure alignment, PSF measurement), it was necessary to compare our products to reference wide-field imaging data. For this purpose, we used $R_\mathrm{c}$-band images obtained by the PHANGS collaboration with the Wide Field Imager \citep[WFI;][]{Baade1999} on the La Silla's 2.2m MPG/ESO telescope, and $r$-band imaging from the Direct CCD camera of the {100 inch} Las Campanas du~Pont telescope. This imaging data were taken as part of the PHANGS-\ha\ survey, aiming to obtain narrow-band continuum-subtracted \ha\ maps for all PHANGS-ALMA galaxies (A.~Razza et al.\ in preparation).
Eighteen of the PHANGS-MUSE galaxies have been observed with WFI, while NGC~7496 has been observed with Direct CCD on the du~Pont telescope (see Fig.~\ref{footprint} for examples of these data). 

Here we briefly summarise the relevant processing steps for the R-band imaging data. A more detailed description of the observations and reduction steps can be found in A.~Razza et al.\ (in preparation). 
For the R-band imaging we observe each field with a total integration time ranging from 900 to 1200 seconds, and perform a standard reduction. The imaging FoV (WFI: $34^\prime\times33^\prime$, Direct CCD: $8.85^\prime\times8.85^\prime$) is much larger than the MUSE FoV, and also large enough to allow for a robust astrometric and photometric calibration via multiple bright stars. 

Each exposure was astrometrised independently and re-projected into a common grid. Photometric calibration was achieved by performing aperture photometry on a sample of stars and comparing to Gaia DR2 magnitudes \citep{GAIA2018,Riello2018} converted to Johnson-Cousin (WFI) and Gunn (Direct CCD) broadband magnitudes via the appropriate conversions \citep{Evans2018}.
A two-dimensional background was fitted after masking all the sources in the field (galaxy included), and subtracted from all the exposures, which are then combined to obtain the final frames. Typical seeing for the $R$-band observations of the PHANGS-MUSE sample was $0.8\arcsec$, providing a close match to the typical MUSE resolution (listed in Table~\ref{tab:sample}).

\section{Data reduction}
\label{sec:drs}

\begin{figure*}
        \includegraphics[width=\textwidth, trim=0 0 0 0, clip]{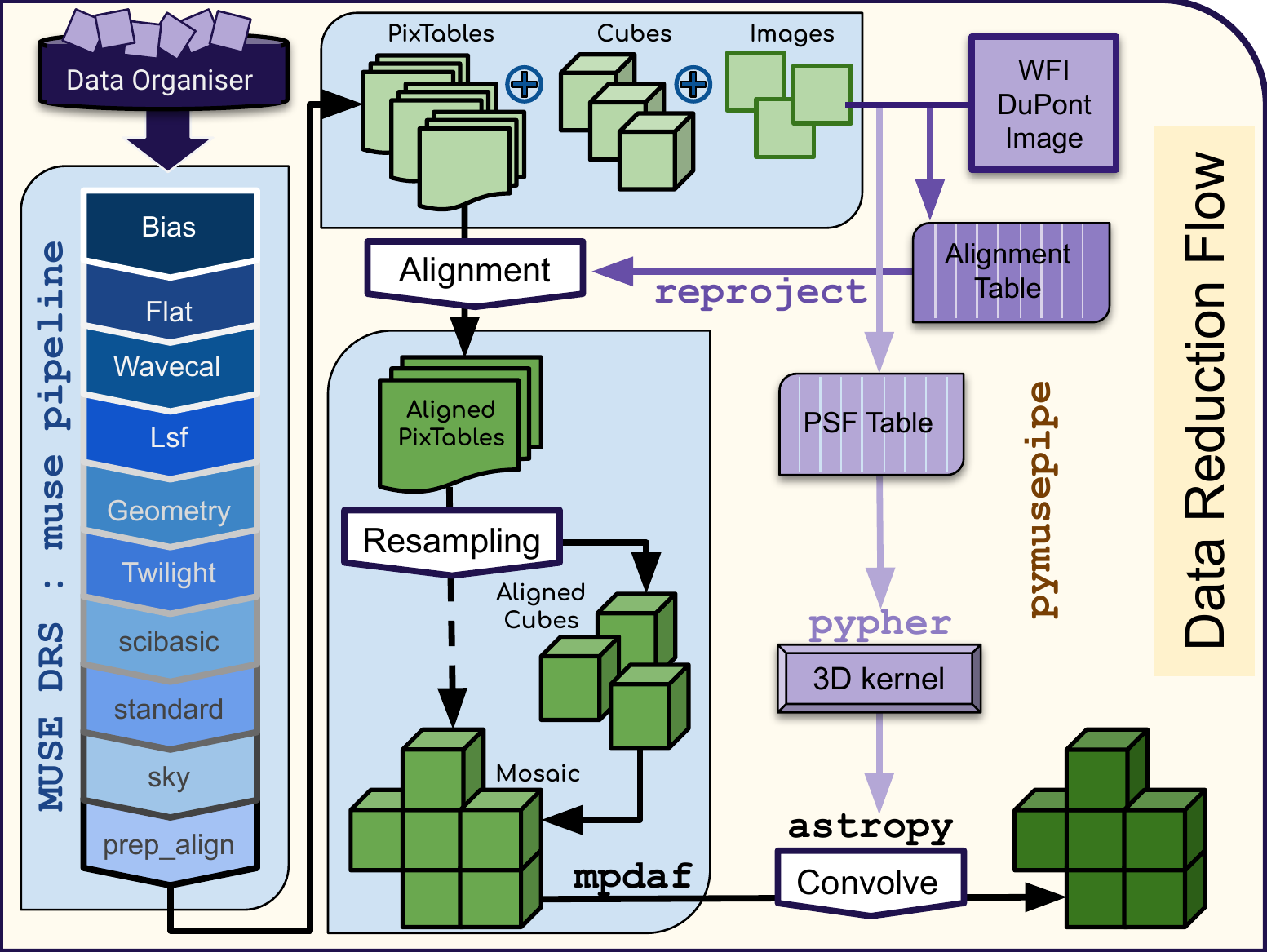}
    \caption{{\bf Schematic data flow representing the PHANGS-MUSE data reduction part (\DRP) of the pipeline.} \pymusepipe\ is used as the global process wrapper. The data flow includes a data organiser that addresses the raw data files, followed by standard MUSE data reduction steps (via \MUSEp), which leads to the first reduced `Pixel Tables' (PixTables) and cubes. These are aligned using WFI or Direct CCD reference images (via \reproject). Aligned individual PixTables are then either resampled on a common grid or directly mosaicked. Optional convolution is performed using the framework from \mpdaf\ and \astropy, while \pypher\ provides the required kernel cube.}
     \label{fig:drs} 
\end{figure*}

MUSE itself is a monolithic instrument consisting of 24 IFUs, each IFU containing an image slicer and spectrograph. The removal of the instrument signature is efficiently addressed via the excellent and advanced MUSE data processing pipeline software (\MUSEp, hereafter) developed within the MUSE consortium \citep{Weilbacher2020}. In order to address, organise, reduce and analyse such a large dataset, we developed a dedicated framework, taking advantage of individual pieces of software or packages, which we describe in the following sections. Figure~\ref{fig:drs} represents a schematic of the data flow for the data reduction part (\DRP) of the PHANGS-MUSE campaign, also flagging the usage of specific packages, with \pymusepipe\footnote{\url{https://pypi.org/project/pymusepipe/}} providing the overall wrapper around these recipes.

\subsection{Framework and software environment}
\label{sec:framework}

We aimed at an almost fully automated pipeline that could be easily tuned to specific needs and potential changes associated with the survey and science goals. A number of constraints were expressed early in the project to satisfy that need, as well as the ability to easily add new acquired datasets, or rerun the entire data flow several times.

This required a modular approach, based on a set of robust packages including the MUSE processing pipeline itself. We adopted \python\ as the main coding and scripting language, exploiting its object-oriented capabilities, its wide community support and the existence of both robust libraries and packages. We now briefly describe the main packages and framework developed or used for the PHANGS-MUSE data.

\paragraph{\pymusepipe\ --}
The \DRP\  is controlled by a newly developed and dedicated \python\ package \pymusepipe, which serves as a wrapper around the main processing steps of the data reduction. \pymusepipe\ includes a simple data organiser and prescriptions for the structure of the data files (but no database per~se), a wrapper around the main functionalities of \MUSEp, accessed via \esorex\ command-line recipes, to remove the instrumental signatures. \pymusepipe\ additionally provides a set of modules supporting the alignment, mosaicking, (two-dimensional and three-dimensional) convolution, and data quality processing. Some of the details pertaining to each module or process are described in the next sections. \pymusepipe\ is thus at the core of the \DRP\  we apply to the MUSE data.

\paragraph{\MUSEp\ --}
The MUSE spectrograph delivers ${\sim}90{,}000$ spectra, each of about $4000$ pixels covering most of the optical wavelength range, over a contiguous FoV of about $1\arcmin \times 1\arcmin$. The raw fits MUSE data thus reflect, via its 24 extensions, the size and shaping of information ending up on the 24 individual detectors. The goal for \MUSEp\ is to address such a complex data and science format, remove the instrument signature, combine and resample exposures for further scientific usage. \MUSEp, developed by the MUSE team \citep{Weilbacher2020}, thus represents a pillar of any data reduction dealing with MUSE data, and follows an approach that minimises the need for resampling steps, using a table-based (PixTable) representation of the data. PixTables encompass the exact origin of the signal on the charge-coupled device and its associated IFU and slice identities. These PixTables can be projected as data cubes onto a given three-dimensional (sky positions and wavelength) grid using given geometric, astrometric and calibration information as derived with \MUSEp.  We are using the latest version of the \MUSEp\ available at the time of writing \mbox{(v2.8.3-1)}.

\paragraph{\mpdaf\ --} The MUSE processed data are natively derived in a PixTable format, which can be projected onto regular three-dimensional data cubes. The PixTables and cubes can themselves be used to reconstruct images in specific filters or extract spectra. \mpdaf\ \citep{Bacon2016} is a \python\ package providing an efficient and user-friendly framework to address such data in a transparent way. \mpdaf\ is thus an important component of \pymusepipe\ where existing \mpdaf\ \python\ classes have been complemented with utility functions (e.g. alignment, convolution, or image reconstruction).

\paragraph{\pypher\ --} The final MUSE mosaics, built either directly from PixTables (via \texttt{muse-scipost} or \texttt{muse-exp\textunderscore combine} in \MUSEp) or from already aligned and resampled data cubes (see Fig.~\ref{fig:drs}), have PSFs that are varying over the spatial FoV and spectral range (Sect.~\ref{sec:psf}). \pypher\footnote{\url{https://pypi.org/project/pypher/}} \citep{Boucaud2016} provides a robust tool to derive kernel cubes feeding a fast-Fourier-transform-based convolution algorithm to homogenise the end-product MUSE data cubes. Given two arbitrary PSF images, the \pypher\ software uses a Wiener filter with a regularisation parameter to compute the convolution kernel needed to move from the input PSF to the output one. The power of such an algorithm is its applicability to general PSFs, expressed analytically or not. We used \pypher\ to move from the wavelength-dependent circular Moffat PSF typical of the MUSE spectrograph, to a wavelength-independent circular Gaussian. The details about the characterisation of the MUSE PSF and the convolution of the data cubes are presented in Sect.~\ref{sec:psf} and~\ref{sec:post-proc}.

\paragraph{General \python\ framework --} We make use of generic but powerful \python\ packages, including \numpy\ \citep{numpy}, \scipy\ \citep{scipy}, \matplotlib\ \citep{matplotlib}, and most importantly \astropy\ \citep{astropy:2018} for its excellent interface with FITS files (\texttt{astropy.io.fits}), and some specific modules including data management tools (e.g. \texttt{astropy.tables}) and units (\texttt{astropy.units}).

\subsection{The data reduction work flow}
In the following, we provide additional details regarding key steps of the MUSE data reduction work flow (\DRP, Fig.~\ref{fig:drs}).

\subsubsection{Data organiser}
\label{sec:orga}

The \pymusepipe\ data organiser relies on reading existing FITS files in a user-defined directory via configuration files. All (compressed or uncompressed) FITS files are scanned and searched for specific keywords (e.g. OBJECT, TYPE, DATE, MODE), which are then used to sort them in dedicated \astropy\ Tables, and stored as attributes of a \pymusepipe\ internal class. Each file is then classified and sorted within predefined categories (including standard stars, flats, biases, arc lamp or illumination exposures, astrometry and geometry calibrations, science and sky exposures), the file structure prescriptions and properties being defined via the \pymusepipe\ configuration module. These are further used for all the subsequent data reduction steps, always prioritising calibration files that are close in time except if otherwise specifically requested by the user (again via a configuration file). The local folder structure is initialised according to the outcome of the file scanning, sorted by file types and categories, following a configuration-dependent \pymusepipe\ hierarchy. The data organiser is both reflected, as mentioned, in the \python\ structure, but also in a set of \astropy\ FITS tables written to disc.

\subsubsection{Instrument signature}
\label{sec:inst}

The MUSE instrument signature is built up and removed using a set of calibration files (biases, flat fields, arc lamps, illumination frames, twilights) acquired soon before or after the main science exposures. These calibrations are processed to derive Master frames including a Master Bias and Flat (via ESO Recipes Execution Tools - \esorex\ - recipes, \texttt{muse\_bias, muse\_flat}), as well as a Trace Table containing the tracing solution for each individual IFU. We note that dark current levels are less than 1~electron per exposure and can be neglected: no dark current correction was applied. 

The wavelength calibration solution (\esorex\ recipe muse\_wavecal) is then derived given a fixed line catalogue. The wavelength solutions are very stable over the 6-year period (October 2014 to December 2020) within which the PHANGS data were accumulated, the distribution of RMS residuals having a median of $0.027$~\AA\ (and an average of $0.027\pm 0.004$~\AA). There is also no detectable difference between the AO and non-AO WFM modes. The next step involved the derivation of the LSF (via \texttt{muse\_lsf}) using the same arc lamps used for the wavelength calibration. Static calibrations provided with the raw data were used both for the geometry and astrometry corrections: hence we did not run \esorex\ recipes \texttt{muse\_geometry} and \texttt{muse\_astrometry}. We should emphasise that other joint astrometry/geometry files, calculated via the MUSE-WISE system \citep[implemented as part of the MUSE Guaranteed Time Observations;][]{MuseWise2015}, may be more closely following the time varying geometric distortions in the instrument: we will carry out such tests and address potential implications in future data {releases} (see Sect.~\ref{sec:datarelease}). The full three-dimensional illumination correction and the relative through-put for each IFU was derived via an illumination correction (\texttt{muse\_twilight}) using twilight sky flat frames. For PHANGS data taken before March~11, 2017, vignetting correction was included as recommended in the MUSE User's Manual \footnote{\url{https://www.eso.org/sci/facilities/paranal/instruments/muse/doc/ESO\-261650\_7\_MUSE\_User\_Manual.pdf}}. Each offset sky exposure was used via the \esorex\ recipe \texttt{muse\_create\_sky} to produce a sky spectrum, {which is then associated with proximate individual exposures minimising the time difference between the exposures on the offset sky and on target}. The main steps covered by this data flow are represented on the left-hand side of Fig.~\ref{fig:drs}, and powered via \MUSEp\ routines using \esorex\ shell piped commands. 

Two key stages needed specific attention and proved challenging in the processing of PHANGS data: the sky subtraction for extended sources, and the mosaicking of final data cubes, including astrometric and absolute flux calibrations.

\subsubsection{Satellite trails}

A few individual exposures included satellite trails producing bright streaks partially or fully crossing the MUSE FoV. These trails were removed from individual exposures by using manually-defined masks. These masks follow a slit-like geometry, with widths between 9 and 12 MUSE spaxels and lengths covering the individual trails detected on reconstructed images before being fed into the mosaicking module. This trail correction step was applied to the first science exposure of pointing~4 for NGC\,1365, the third exposure of pointing~3 for NGC\,1672, and (a partial trail in) the first exposure of pointing 4 for NGC\,3351. Those three trail-masks were fed into the mosaicking module, selecting out the corresponding pixels of the PixTables before proceeding.

\subsubsection{Alignment and mosaicking}
\label{sec:align}

\begin{figure*}
    \centering
        \includegraphics[width=0.81\textwidth, trim=0 0 0 0, clip]{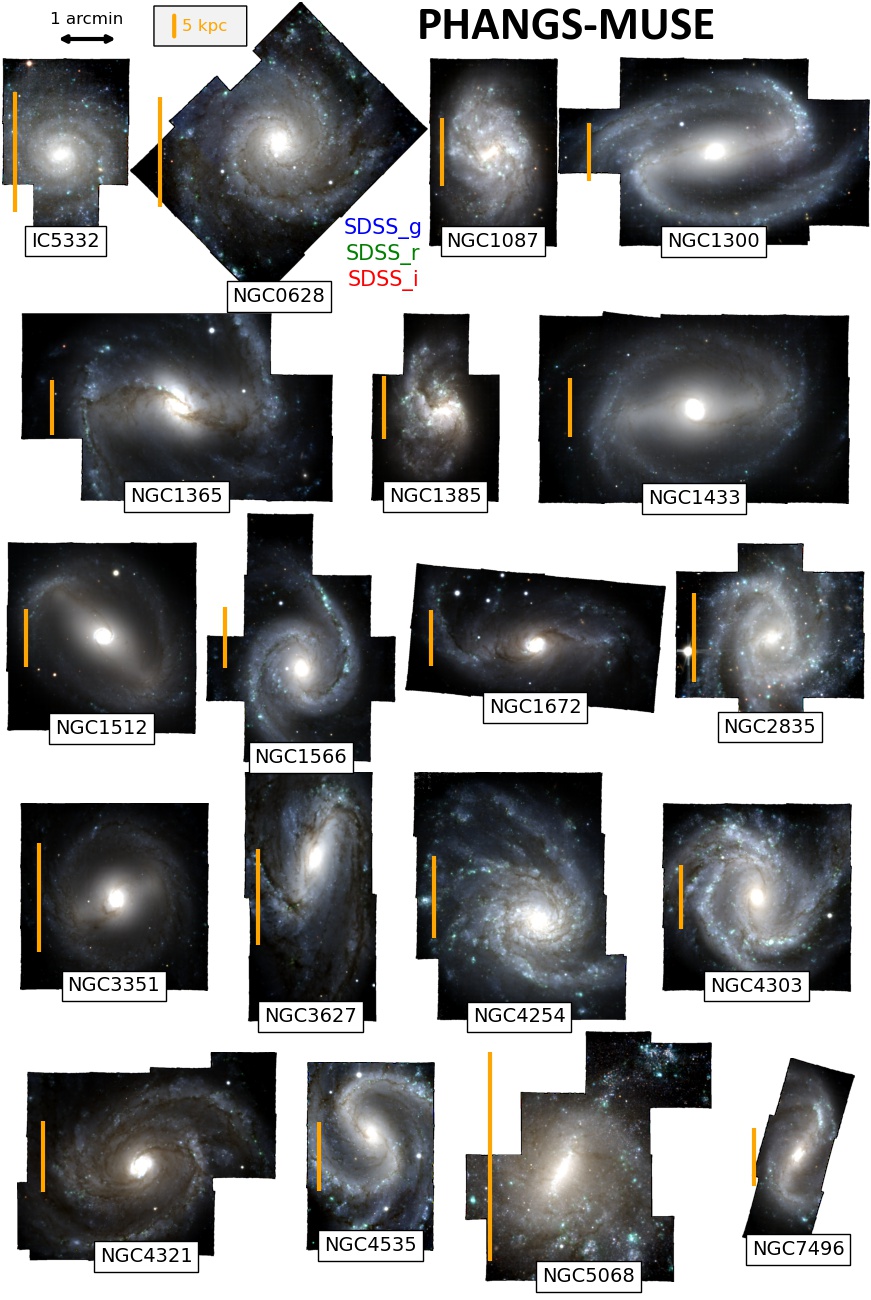}
     \caption{{\bf PHANGS-MUSE reconstructed images}. RGB images combining the three $i$, $r$, and $g$ (red, green, and blue channels, respectively) SDSS-band reconstructed maps of the 19 targets that are full mosaics from the PHANGS-MUSE Large Programme (including the pilot galaxy NGC\,0628) -- produced as part of the \DRP+\DAP\ flow. The scale in arcminutes, which roughly corresponds to the extent of one single MUSE pointing, is shown at the top left of the figure as a labelled black arrow, while the orange vertical bar on the left side of each panel represents the distance-dependent 5~kpc scale for each target.}
     \label{RGBcol}   
\end{figure*}

\MUSEp\ provides basic alignment capabilities, for example via cross-correlation techniques and detection of point sources. For extended objects, like targets in the PHANGS sample, the preliminary relative offsets provided by \MUSEp\ \texttt{muse\_exp\_align} are unfortunately not robust enough to deliver sub-spaxel accuracy, due to the lack of bright point sources. We are also seeking an absolute astrometric solution for each individual exposure that optimises the mosaicking and the comparison with miscellaneous PHANGS datasets (e.g. \textit{HST}, ALMA). We therefore decided to match the astrometry of each individual exposure with the $R$-band imaging acquired in the course of the PHANGS project (Sect.~\ref{sec:ancillary} and A.~Razza et al.\ in preparation). We used the dedicated alignment module in \pymusepipe\ to connect each individual exposure with the WFI and du~Pont images. MUSE images reconstructed from individual PixTables via the WFI (or du~Pont) filter curve are compared, given a first guess relative offset, via a set of associated contours and image plots. This step is manual in the sense that the user checks and either confirms or fine-tunes the offset via the \python\ interface. This process leads to a somewhat subjective assessment of the relative alignment, including a pixel offset (in both $X$ and $Y$) and a potential rotation angle. To increase robustness, this step is performed by at least two (sometimes three) members of the PHANGS team. Visual inspections confirmed that the relative astrometry is well within one fourth of a spaxel (namely, $0\farcs05$): any change larger than $0.1$ spatial pixel in an individual exposure is easily identified visually and leads to strongly asymmetric residuals (especially around point-like sources) when dividing the reference and MUSE images. 

We identified a few issues associated with the geometric and astrometric solutions provided via predefined MUSE calibrations. About 20\% of all exposures exhibit a global small but still significant rotation between $0.1$ and $0.3$ degrees with respect to the $R$-band images, with no apparent correlation with RA, DEC or time when the target was observed. This residual rotation is also corrected simultaneously via the \pymusepipe\ alignment module, as a free parameter set by the user: such a rotation is also confirmed by at least two different users (sometimes three).

\subsubsection{Sky background subtraction}
\label{sec:skyback}

{As mentioned in Sect.~\ref{sec:inst}, we associate each target exposure with the sky spectrum built from the offset sky exposure closest in time. The sky subtraction itself is included in the \texttt{muse\_scipost} processing step, thus using an associated continuum spectrum, and always re-fitting the sky emission lines 
(via the default \texttt{skymethod='model'} option).}
We used the reference $R$-band images to {further} constrain any residual (sky) background resulting from an imperfect sky subtraction, or global flux normalisation discrepancy (per exposure). Assuming that the $R$-band reference image has zero background and the correct absolute flux normalisation, and that the flux in the MUSE reconstructed image represents a linear function of the true flux (involving a normalisation constant plus a background), we can write
\begin{equation}
    \begin{split}
        \mathrm{Flux}^{R}(x,y)
        & = a \times \mathrm{MUSE}^{R}_{1}(x,y) \\
        & = a \times \left(\mathrm{MUSE}^{R}_\mathrm{raw}(x,y) + \mathrm{Sky} - \mathrm{Sky}_{1} \right)\, +\, b
    \end{split}
,\end{equation}
where $\mathrm{Sky}$ is a constant representing the true sky background for that specific exposure, $\mathrm{Sky}_{1}$ is another constant representing the actual value removed during the initial sky subtraction process, and $a$ and $b$ are constants representing a linear regression representation of the $\mathrm{Flux}_{R\textrm{-band}}$ versus the MUSE reconstructed image. A perfect sky subtraction and normalisation would lead to $a=1$ and $b=0$. We then use the fitted $a$ value as a normalisation correction, and $b$ to fix the sky contribution by applying $\mathrm{Sky} = \alpha \times \mathrm{Sky}_1$ where $\alpha = 1 - b / \left( a \cdot \mathrm{Sky_1} \right)$. Hence, knowing $a$ and $b$ as well as $\mathrm{Sky}_1$, the value of the sky continuum integrated within the reference image filter, we derive a correction for the sky normalisation that yields a linear regression where $b=0$. The \pymusepipe\ package implements this approach as an option, using the recorded linear regression $a$ and $b$ values. The regression itself is performed via an orthogonal distance regression (ODR) comparing the reference and MUSE reconstructed images after noise filtering and binning: we use bins of $15\times15$ spaxels ($3\arcsec \times 3\arcsec$) to minimise the impact of unresolved structure in the comparison. 

We find that the distribution of the scaling factors $a$ over the full set of PHANGS-MUSE exposures is well fit by a skew-normal distribution with location 0.99, scale of 0.06 and shape parameter $\alpha$ (related to the skewness) of 1.5 (the best Gaussian fit has a mean of 1.03 and sigma of 0.046). Only 28 (respectively, 27) exposures out of 676 have scaling factors lower than 0.9 (respectively, higher than 1.2). The distribution of background values ($b$) resembles a Gaussian function centred on 0 with a FWHM of about 1.4 
(in units of $10^{-20}$~erg~cm$^{-2}$~s$^{-1}$~$\AA^{-1}$), with a small tail towards positive values.
It is important to note that the sky re-normalisation only acts within the $R$-band filter, assuming that the reference image is background free. Since the reference MUSE sky exposure may result in a reference sky spectrum that is not necessarily an exact representation of the actual sky on the MUSE science exposure, this could lead to a colour variation and, hence, to an over- or under-subtraction of the sky that depends on wavelength (see Sect.~\ref{sec:overlap}). 

\subsubsection{Point spread function}
\label{sec:psf}

\begin{figure}
\centering
        \includegraphics[width=0.45\textwidth, trim=0 0 0 0, clip]{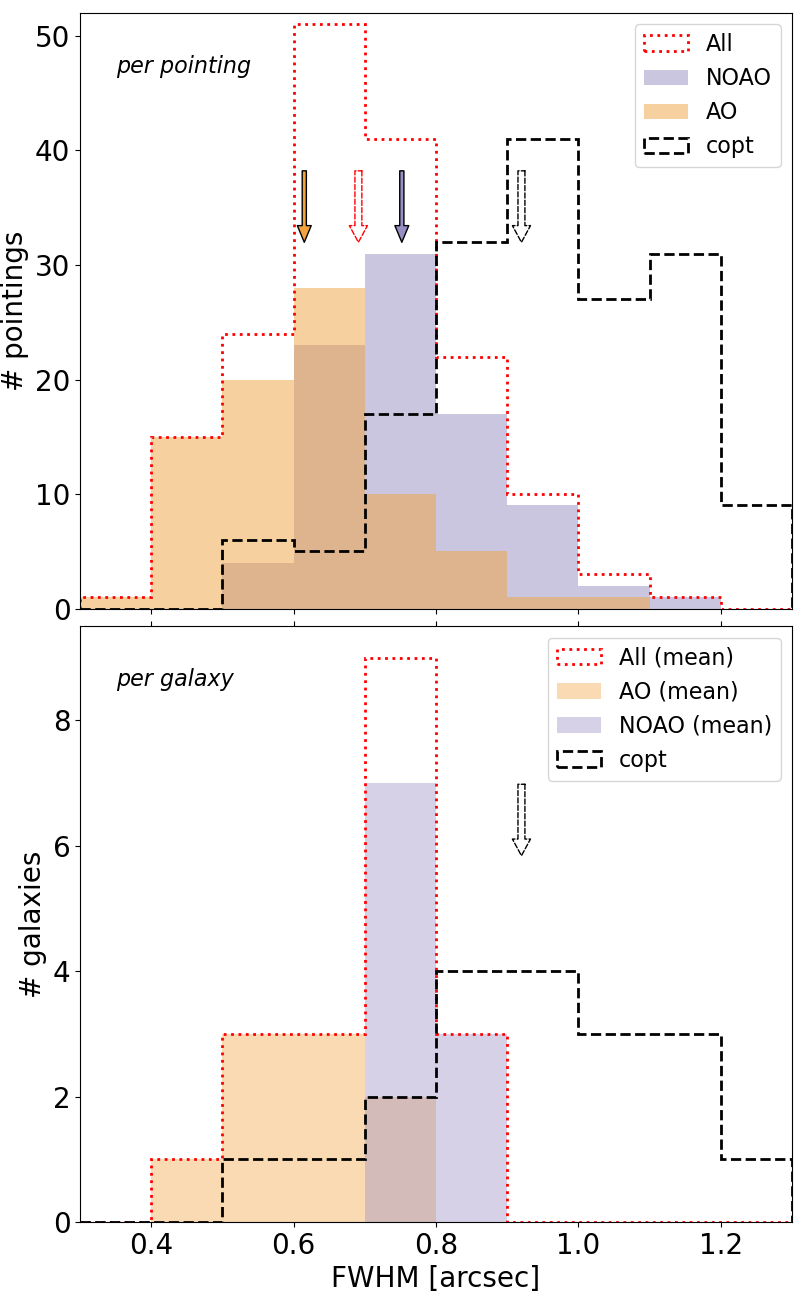}
        \caption{Distribution of the FWHM of the MUSE PSF. The top panel shows the FWHM per pointing as measured by the algorithm described in Sect.~\ref{sec:psf}. The histograms respectively identify the pointings observed in AO mode (orange) and noAO mode (purple). The red dotted histogram represents the distribution for all existing pointings, while the dashed black histogram shows the distribution of PSFs of the copt data cubes (i.e. after the convolution with the worst observed PSF for each galaxy). The bottom panel shows the average FWHM per galaxy and the convolved FWHM (again, using the worst PSF in each target as the baseline), with the same colour scheme as in panel~(a). The arrows indicate the median value of each sub-sample (AO, noAO, all exposures, convolved) with the associated colour scheme.
        }
     \label{fig:psf} 
\end{figure}

As emphasised in Fig.~\ref{footprint}, our sample galaxies extend well beyond a single MUSE FoV. Covering a significant fraction of the galactic discs (typically well beyond 3 effective radii), each target has been observed using from 3 to 15 MUSE individual pointings, each of them observed at different times and with different sky conditions.
The characteristics of the PSF thus naturally vary across a given mosaic. Some of the science goals of the PHANGS project require a homogeneous spatial resolution throughout each galaxy disc. Moreover, a good characterisation of the local PSF at a given location and wavelength within the mosaicked data cube is a key step for the exploitation of such a dataset.

We have assumed that the MUSE PSF can be described at all wavelengths as a circular Moffat function \citep{Fusco2020}, parameterised by its core width and power index. Both quantities are related to the seeing and general atmospheric conditions, combined with the instrument performance during the observations.
The core width serves as a proxy for the width of the PSF, while the power index relates to the relative amplitude and shape of the PSF wings.  The core size $R$ and the Moffat FWHM are connected via the following relation:
\begin{equation}
\label{eq:moffat_fwhm}
\mathrm{FWHM} = 2 \cdot R \sqrt{2^{1 / n} - 1}~,
\end{equation}
where $R$ is the core width and $n$ is the power index of the Moffat function.
For a Moffat function, a small power index corresponds to more prominent wings (i.e. as compared with a Gaussian of the same core size). As the power index increases, the Moffat function converges towards a Gaussian profile.

Previous observations with the MUSE WFM have suggested that the MUSE PSF can be considered as a constant over the FoV of the instrument for a single exposure \citep[e.g.][]{Serre2010,Bacon2017,Fusco2020}. However, the MUSE spectrographs deliver a PSF whose width varies significantly with wavelength. In order to characterise the PSF of our data cubes, we therefore made use of a four parameter function: a reference core width (or FWHM), a power index at reference wavelength, and a first order polynomial (two parameters) to describe the rate of change with wavelength.
These can be directly measured by fitting a Moffat function to a point-like source (i.e. a~star) at different wavelengths for each pointing.

In practice, only a fraction of the observed pointings (${\sim}40\%$) include a star bright enough to robustly recover the four parameters describing the PSF. We thus implemented an alternative method, namely an adaptation of the algorithm presented in \citet{Bacon2017}. The principle relies on a minimisation of the difference between a reference image with a known PSF, and a MUSE reconstructed image, after spatial cross-convolution. Such an algorithm requires
(i) a reference image with a known (or measured) PSF, which covers a significant fraction of the region of interest (i.e. the FoV of the MUSE pointing) and
 (ii) an image of the MUSE pointing extracted using the same pass-band as the reference image.

As mentioned in Sect.~\ref{sec:ancillary}, we used the $R$-band images obtained either with WFI at the 2.2m MPG/ESO telescope in La~Silla or with the du~Pont telescope at the Las Campanas Observatory (see A.~Razza et al.\ in preparation) as reference images for the PSF measurement process. The PSFs of the reference $R$-band images were consistently measured using a set of bright stars throughout their large FoV, and assuming a Moffat two-dimensional profile (which has been shown to be a good representation of their PSF; see A.~Razza et al.\ in preparation).

Within our PHANGS-MUSE pointings we never cover galaxy-free regions that can serve to directly measure the sky background. Since the galaxy emission is highly structured, providing a robust and independent estimate of the power index of the Moffat using such an automatic algorithm has proven quite challenging. Furthermore, we only have a single reference image per galaxy. This means that we can only measure the PSF of the MUSE reconstructed image as a luminosity-weighted value corresponding to the $R$~bandpass.\footnote{The situation may change with the addition of multi-band HST images, but this is beyond the scope of the present release.}

We therefore restricted the minimisation process to a single-parameter optimisation, namely the FWHM of the PSF at $6483.58$~\AA, the reference wavelength of the \textit{R}-band WFI filter (but see Sect.~\ref{sec:future}). As a first approximation, we followed \citet{Bacon2017} and assumed that the FWHM of the PSF linearly decreases as a function of wavelength, with a slope of $-3.0 \times 10^{-5}\ \mathrm{arcsec}/$\AA\ (MUSE team, private communication), while the power index is fixed at the values of $n=2.8$ and $2.3$ for the no-AO and AO modes, respectively \citep[see e.g.][]{Fusco2020}. While we cannot directly confirm the robustness of such a rule as applied to our dataset, due to the lack of isolated bright stars in the MUSE pointings, preliminary checks with a few sources appear to be consistent with that assumption.

Our specific implementation for the PHANGS-MUSE data cubes consists of the following steps: (i) the reference image is converted to the same flux units as the MUSE image; (ii) the reference image is re-projected to match the MUSE gridding, and the two images are aligned using their respective World Coordinate System (WCS) coordinates (and their cross-normalisation checked); (iii) the MUSE image is convolved with a model of the measured PSF of the reference image;
 (iv) the reference image is convolved with a given model of the MUSE PSF, with its FWHM as a free parameter; and (v) foreground stars identified using Gaia-DR2 catalogue are masked. The last two steps (convolution of re-gridded narrowband image and masking) are used as input for a least-square optimisation process.

We note that the WFI PSF was estimated by building an effective PSF using a few bright sources in the image already re-gridded to resemble the MUSE pointing (FoV and sampling).
Finally, while in \citet{Bacon2017} the stars are masked before the convolution, and the minimisation is performed in Fourier space, we decided to mask the stars after the convolution and to keep the computation in real space. Due to the complexity of the background, masking the stars before the convolution step would create potentially strong edge effects that could drive the minimisation towards incorrect values. Masking after the convolution and deriving the difference in the image space allow us to avoid this issue, at the expense of having a more time-consuming algorithm. Table~\ref{table:ObsPointings} summarises the properties of each pointing, including the FWHM of the PSF measured with the algorithm described in this section, and Fig.~\ref{fig:psf} provides histograms of the measured distribution for the FWHM measured on individual pointings and on completed mosaics for all targets. The minimum, maximum and median values of the measured FWHM are $0\farcs38$, $1\farcs18$ and $0\farcs69$, respectively, with about 80\% and 97\%  of all pointings having FWHM smaller than $0\farcs8$ and $1\arcsec$, respectively (only 3 out of 168 having FWHM between $1\arcsec$ and $1\farcs2$).\footnote{This is to be compared with the prior requirement for the observational setup (ESO Phase 2) of a seeing better than 0\farcs8.}

\subsubsection{Line spread function}
\label{sec:lsf}

The LSF is in principle not affected by the observing conditions and thus solely determined by the instrument characteristics. MUSE is a relatively robust instrument and the LSF is stable enough that it does not need to be characterised for every exposure or pointing, as we did for the PSF \citep{Bacon2017}. The LSF is observed to change slightly over the FoV, but usually the variation is small enough \citep[$< 0.05$~\AA;][]{Husser2016} that, for our purposes, it can be considered constant.
However, it does change significantly as a function of wavelength, and a good knowledge of its behaviour is critical, for example, to measure the stellar and gas velocity dispersion (via absorption and emission lines, respectively).
The MUSE LSF can be roughly approximated by a Gaussian profile \citep{Bacon2017} whose FWHM changes as a function of wavelength. The FWHM varies between about 3~\AA\ towards the blue end of the spectrum ($4800$~\AA) and $2.4$~\AA\ at $\sim 7500$~\AA\ \citep[Fig.~2 from][]{Bacon2017}. Over the whole range, this variation can be described by a second order polynomial. In this work, we will make use of Eq.~8 of \citet{Bacon2017}:
\begin{equation}
    \label{eq:lsf}
    \mathrm{FWHM} \, (\lambda \, [\text{\AA}]) = 5.866\times10^{-8} \lambda^2 - 9.187\times10^{-4}\lambda + 6.040~.
\end{equation}
This was shown to represent a fair approximation of the variation in the LSF with wavelength: those variations were measured to have a scatter of about 1 to 3\%, representing an average of $0.05$~\AA\ over the full MUSE spectral range \citep[see e.g.][]{Bacon2017, Emsellem2019}.

\subsubsection{Post processing}
\label{sec:post-proc}

At this stage of the data reduction, mosaicked data cubes whose astrometry and background have been calibrated to match those of the reference $R$-band images were computed. We refer to these mosaics as `native' (for native spatial resolution): the variation in the PSF over the field and as a function of wavelength is not corrected (see Sect.~\ref{sec:psf}). The native data cubes have the advantage of having the highest spatial resolution possible with the given observations, while the PSF variation may impair robust measurements throughout the FoV or depending on wavelength. In addition to the set of native resolution mosaics, we produce data cubes with homogenised PSFs, labelled as the `copt' (for convolved, optimised) dataset.

The homogenisation procedure first requires a measure of the input MUSE PSF, as described in Sect.~\ref{sec:psf}. Our target PSF is a circular two-dimensional Gaussian whose FWHM is constant as a function of wavelength and position within each individual mosaic. A Gaussian target PSF was selected to simplify further post-processing, including convolution to coarser spatial resolutions.
We make use of a direct convolution scheme with a three-dimensional kernel (`kernel cubes') representing the transfer function from the original to the target PSF. Each individual MUSE exposure is addressed independently, while all exposures from a given pointing adopt the same common pointing-dependent PSF. Measuring the PSF at the pointing level leads to a more robust outcome, while the PSF variations between exposures of a given pointing become largely irrelevant due to the linearity of the convolution scheme. We note that we perform the convolution at the individual exposure level to maximise the number of individual cubes that end up being combined during the final mosaicking step, hence optimising the robustness of the rejection of spurious pixels.

The PSF homogenisation proceeds as follows: First, for each individual pointing, a three-dimensional model of the best-fit Moffat PSF is created, with constant power index and with a FWHM varying as a function of wavelength as described in Sect.~\ref{sec:psf}.

Second, a three-dimensional model of the target PSF with constant FWHM at all wavelengths is created, assuming a FWHM strictly larger than the worst value measured within the mosaic (position, wavelength). More specifically, we use the worst FWHM value and add $0\farcs2$ in quadrature, which is a reasonable compromise to reach a robust Gaussian profile at all positions. 

Third, the convolution kernel cube is created via the \python\ package \pypher\ (see Sect.~\ref{sec:framework}).
Fourth, the resulting \pypher\ three-dimensional kernel is fed into a fast-Fourier-transform two-dimensional convolution scheme, for each individual MUSE exposure belonging to the given pointing, addressing each wavelength slice independently, thus reducing the computational time needed for the operation. The spectral variance is also propagated slice by slice via the relation $\sigma^2_{\mathrm{copt}} = \sigma^2 \otimes \mathit{ker}^2$, as derived from the usual principles of propagation of uncertainties. $\sigma^2_{\mathrm{copt}}$ is the variance of the convolved cube, $\sigma^2$ is the variance of the native cube, \textit{ker} is the kernel and $\otimes$ denotes a convolution.

Fifth, the convolved data cubes go through a step of binary erosion\footnote{This applies mathematical morphology algorithms for image processing; see e.g. \texttt{binary\_erosion} in \texttt{scipy.ndimage}.} to lower the impact of edge effects during the convolution. We do not keep track of the correlation that this procedure generates between individual spaxels in the convolved cube. Finally, PSF-homogenised mosaics are built from the individually convolved data cubes, via the same combination algorithm used to create the native data cubes.

We should emphasise that the convolution process leads to high levels of spaxel-to-spaxel correlations, which in principle should be reflected in covariance terms: we are ignoring those in subsequent analyses. The variances reported from the convolved cubes should therefore be used with caution when binning the data to larger spaxel sizes.

The resulting copt mosaics are further processed to create additional lower-resolution mosaics, where the spatial resolution is set to either a fixed physical scale (in parsec) or to a fixed on-sky resolution (in arcseconds). All native and convolved data cubes are analysed in the same way, {using a dedicated PHANGS-MUSE Data Analysis Pipeline (\DAP)}, as described in the next section (Sect.~\ref{sec:dap}).

\subsubsection{Data reduction products}
\label{sec:drsprod}

The data reduction flow described in the previous sections delivers a set of data products in various forms, including data cubes and images. They include measurements of the PSFs, listed in Table~\ref{table:ObsPointings}, alignment tables in FITS or ASCII formats including the applied RA and DEC offsets for each individual exposures, as well as the flux normalisation (Sect.~\ref{sec:skyback}). The final products that feed the \DAP\  (see next section) include the full mosaic data cubes for each galaxy target (native, and copt; see Sect.~\ref{sec:post-proc}). We also generate a set of reconstructed broadband images for an extensive set of given filters ($i_\mathrm{SDSS}$, $r_\mathrm{SDSS}$, $g_\mathrm{SDSS}$ etc; see Fig.~\ref{RGBcol}).

\section{The data analysis pipeline}
\label{sec:dap}

\begin{figure*}
\centering
        \includegraphics[width=0.9\textwidth, trim=0 0 0 0, clip]{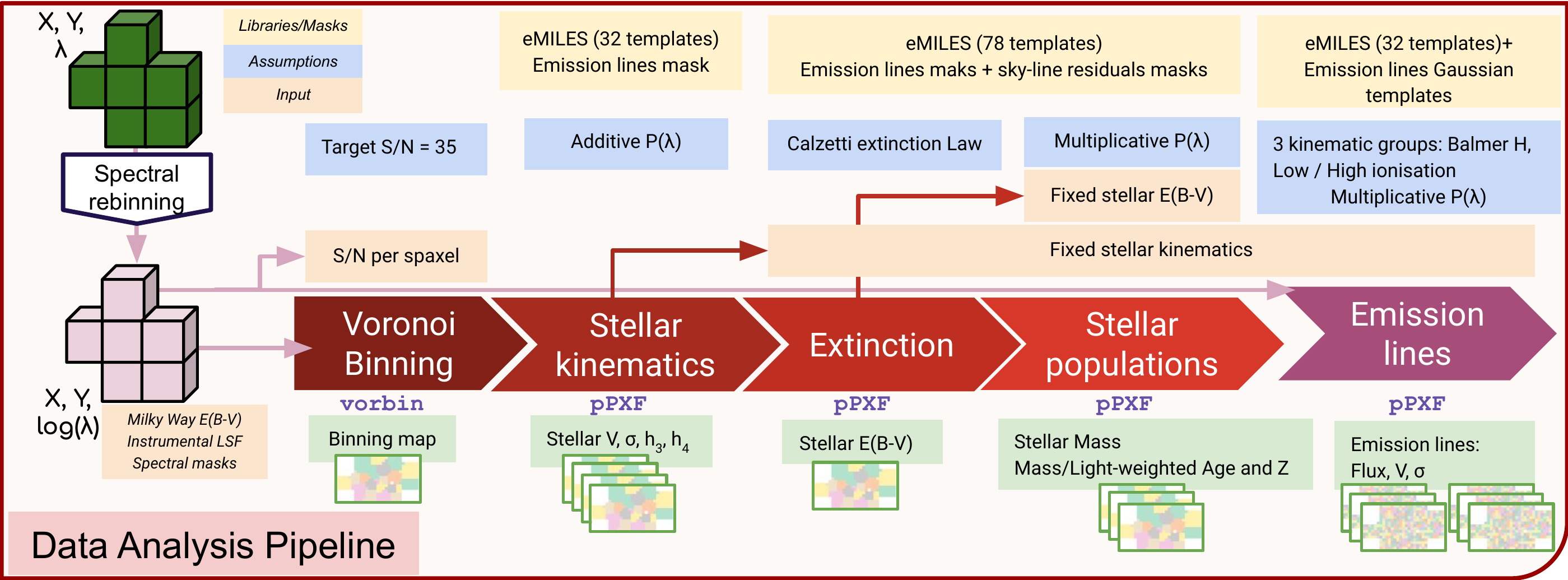}
     \caption{Analysis flow for the \protect{PHANGS-MUSE \DAP}. After spectral re-binning using a $\mathrm{ln}(\lambda)$ prescription, the resulting mosaicked cubes are used for the extraction of the stellar kinematics, stellar extinction, and stellar population maps, using an appropriate adaptive binning scheme, while the emission line maps (including the corresponding gas kinematics) are derived from the originally sampled cubes. The main information pertaining to the stellar libraries, spectral masks, assumptions and fixed input for each step are mentioned in the upper part of the figure, while the data products are illustrated in its lower part. See Sect.~\ref{sec:dap_flow} for further details.}
     \label{fig:DAPflow}
\end{figure*}

\subsection{Requirements and design}
\label{sec:dap_requirements}

The aim of the {\DAP} for PHANGS-MUSE is to generate high-level data products (i.e. fluxes, kinematics, etc), from the stellar continuum and absorption lines, as well as from the ionised gas emission lines. Several software tools have been developed to this aim in recent years, especially to process the data associated with large IFU surveys: CALIFA, MaNGA, SAMI, and TIMER. An incomplete list of notable tools includes Pipe3D \citep{Sanchez2016a, Sanchez2016b}, the MaNGA \DAP\  \citep{Belfiore2019, Westfall2019}, LZIFU \citep{Ho2016}, and \gist\ \citep{bittner2019}. 

We impose several requirements to the software framework for the analysis of the PHANGS-MUSE data, namely: (i) integrate an adaptive spatial binning scheme, to ensure a minimum signal-to-noise ratio required for a robust measurement of observable quantities;
(ii) perform a well-tested and robust extraction of physical quantities, including gas and stellar kinematics, and stellar population properties;
(iii) parallelise the spectral fitting over multiple cores, to allow for efficient processing of ${\sim}10^6$ spectra (possibly several times, when iterating over subsequent data releases);
(iv) be sufficiently modular to allow us to implement changes and/or replace individual modules, without affecting the structure of the code; and (v) support the analysis of IFU datasets from other instruments and surveys, to compare with publicly available high-level products, and therefore benchmark the output of our code against best practices in the field.

We judged these requirements to correspond closely to the philosophy behind the \gist\ code\footnote{\url{https://abittner.gitlab.io/thegistpipeline}} \citep{bittner2019}, which has a module-based structure and supports parallelisation for the fitting stages. We therefore adopted \gist\ as the starting point for our pipeline environment.

The modular structure of \gist\ allowed us to easily replace several of its constituent modules with algorithms more closely aligned to our goals with PHANGS-MUSE. In particular, with respect to the public implementation of \gist, we made changes to virtually every module, and replaced the emission line spectral fitting and the stellar population analysis routines with ones written by members of our team (F.~Belfiore and I.~Pessa). The version of the code used for the analysis of PHANGS-MUSE is publicly available\footnote{\url{https://gitlab.com/francbelf/ifu-pipeline}}.

Our pipeline implementation, which we refer to as \DAP\ is described in detail in the next subsections. Several of these pipeline-level software tools share core pieces of software to perform spatial binning and spectral fitting. Two such modules stand out for their wide applicability: \vorbin\ \citep{cappellari_adaptive_2003} and \pPXF\ \citep{cappellari_parametric_2004}. These were originally developed for the pioneering IFU work performed as part of the SAURON/ATLAS$^\mathrm{3D}$ surveys \citep{de_zeeuw_sauron_2002,cappellari_atlas3d_2011}, and subsequently updated and upgraded \citep{Cappellari2017}. They address the key tasks of binning two-dimensional data to reach a specific S/N level, and to fit the stellar continuum (and, optionally, the gas emission lines) with a non-negative linear combination of templates. Below we briefly described these modules, which we utilised in the analysis of the PHANGS-MUSE data.

\paragraph{\vorbin\ --} This is a robust and broadly used package to adaptively bin data cubes along the two spatial dimensions. The method uses Voronoi tessellations \citep{cappellari_adaptive_2003}, an optimal solution being found via an iterative process constrained by a given parameter representing the targeted signal to noise. For the present datasets, we used the estimate of the signal-to-noise ratio per spaxel as direct input. 

\paragraph{\pPXF\ --} The extraction of the stellar and gas kinematics, and information pertaining to the stellar population content of spectra, has been popularised via a set of excellent pieces of software that exploit spectral features in various ways. \pPXF\ \citep{cappellari_parametric_2004,Cappellari2017} is an intensively tested and robust algorithm to perform direct pixel fitting of spectra making use of spectral template libraries. The generic fitting module of \pPXF\ is extremely flexible and supports a wide range of applications, including the simultaneous fitting of absorption and emission lines and the extraction of non-parametric star-formation histories, with the possibility to add multiple kinematic components and generic constraints (e.g. line flux ratios). The \DAP\ consists of several modules wrapping \pPXF\ for specific applications.

\subsection{The data analysis flow}
\label{sec:dap_flow}

In this section we describe the analysis flow developed within the \DAP\ going from the preparatory to the main computational steps delivering specific data products. The \DAP\ workflow consists of a set of modules running in series, some depending on the outcome of previous steps (e.g. derivation of the stellar kinematics is needed as input to the emission lines fitting module). Figure~\ref{fig:DAPflow} contains a schematic representation of \DAP, including the main input and output parameters, libraries and constraints. Each of the individual \DAP\ modules writes to disc a set of intermediate output files, which can be useful to re-run specific modules in case of a failure, and also contains a more extensive set of outputs that may be of interest for specialised analysis. The key set of physical parameters produced by the \DAP\ are then consolidated into a main output file, described in Sect.~\ref{sec:dap_output}.

\subsubsection{Preparatory steps}
\label{sec:dap_prep}

The \DAP\ requires as input a configuration file, which specifies certain pipeline parameters, the location of the input IFU data, the galaxy's redshift (or systemic velocity), and the Galactic extinction at its position. 

The data are read via a bespoke input module, which is instrument-specific (tailored to MUSE in our case), but may be replaced in order to process data from different instruments. By appropriately changing the parameters in the configuration files and writing a new input module, users can easily process data from other surveys. 
For PHANGS-MUSE, we utilise a common data input model for both the WFM-noAO and WFM-AO observations. The wavelength range corresponding to the AO gap is automatically masked by \MUSEp\ in the case of AO observations. This mask is propagated by the \DAP. We assume the LSF given by Eq.~\ref{eq:lsf}. The systemic velocity of each galaxy is taken from \cite{Lang2020}, who derived these from an analysis of the PHANGS-ALMA CO(2--1) kinematic maps. The MUSE data are corrected for foreground Galactic extinction, using the \citet{Cardelli1989} extinction law and the $E(B-V)$ values from \citet{Schlafly2011}.

Since we chose to produce fully reduced MUSE data cubes on a linear wavelength axis,\footnote{The \MUSEp\ allows one to reduce the data on a logarithmically sampled wavelength axis when needed, which would save one re-binning step in the spectral fitting process. Our team, however, decided to maintain the default linear sampling that has been used for most existing MUSE data published in the literature to date.} 
we perform a resampling of the data on a logarithmic (natural log) wavelength axis, as required for input to \pPXF\ using a channel size of 50~\kms. This channel size is sufficient to Nyquist sample the LSF of the MUSE data with more than two pixels for $\lambda < 7000$~\AA, but inevitably over-samples it at the blue edge of the MUSE wavelength range. As discussed further below, we fit the wavelength range $4850{-}7000$~\AA\ to avoid strong sky residuals in the redder part of the MUSE wavelength range.

\subsubsection{Spatial binning}
\label{sec:binning}

\begin{figure*}
\centering
        \includegraphics[width=0.99\textwidth, trim=0 0 0 0, clip]{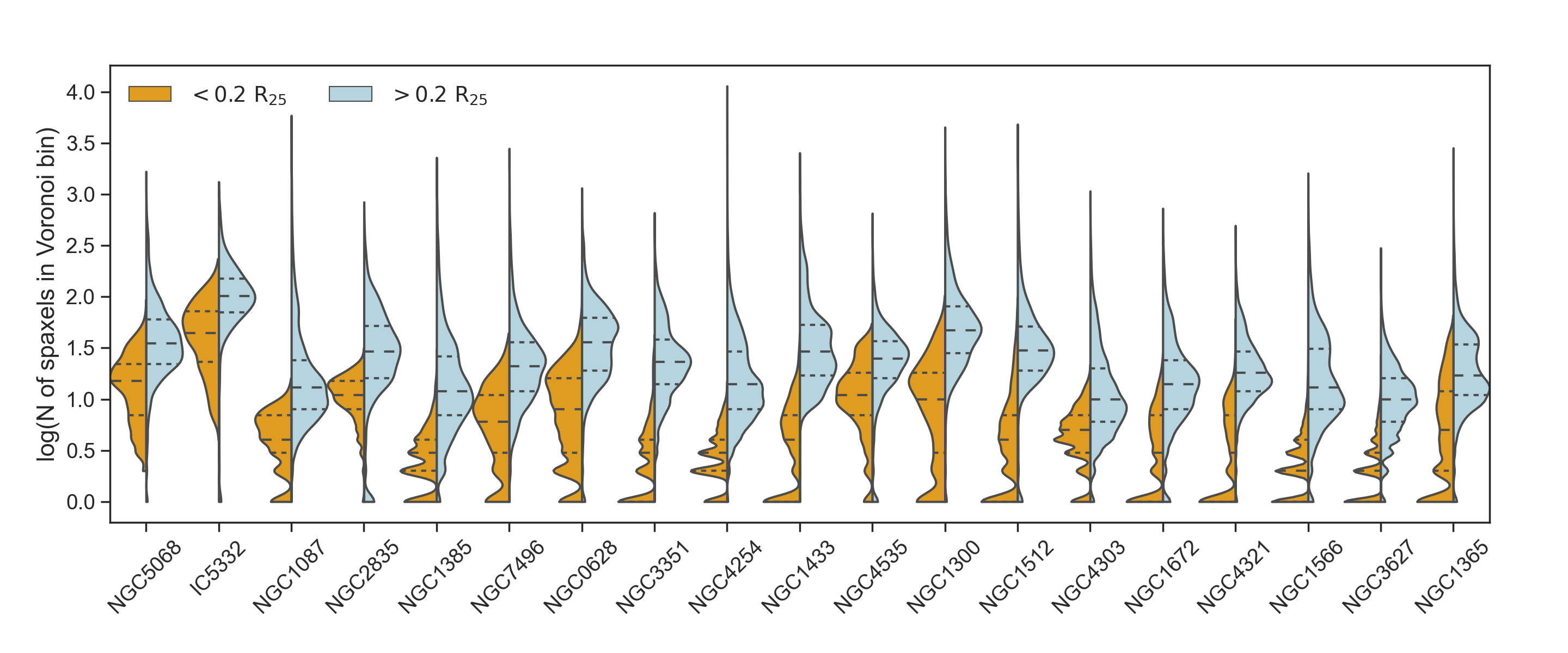}
     \caption{Histograms (violin plots) of the distribution of Voronoi bin sizes for the galaxies in our sample. In orange we show bins within $0.2\,R_{25}$ and in blue those at larger galactocentric radii. Galaxies are ordered from left to right by stellar mass. The median bin size outside the central regions ($R> 0.2\,R_{25}$) is between 10 and 100 spaxels, roughly corresponding to circularised radii of $0.4\arcsec$ and $1.1\arcsec$. The 25$^{\rm th}$, 50$^{\rm th}$, and 75$^{\rm th}$ percentiles of each distribution are marked with dashed lines. }
     \label{fig:violin_bins}   
\end{figure*}

As extensively discussed in the literature, an accurate and unbiased determination of the stellar kinematics and stellar population properties from full spectral fitting requires a minimum signal-to-noise ratio \citep[S/N;][]{Johansson2012, Westfall2019}. The \DAP\ computes the continuum S/N in the $5300{-}5500$~\AA\ wavelength range, using the noise vector from the reduction pipeline. 

The data are then Voronoi binned to a target S/N of 35 making use of \vorbin. This S/N level is used to determine both the stellar kinematics and the stellar population properties. The value of 35 was chosen to ensure that the relative uncertainty in the stellar mass measurement, which we estimated via Monte Carlo realisations of the errors (see Sect.~\ref{sec:stellarpops} below), stays below 15\%. For comparison, the MaNGA data are re-binned to a $\mathrm{S/N}=10$ to determine the stellar kinematics by the publicly available run of the dedicated analysis pipeline \citep{Westfall2019}. We opted for a higher S/N threshold for two reasons: a) we aimed to keep the same Voronoi bins for both the stellar kinematics analysis and the determination of stellar population properties via full spectral fitting, which generally require higher S/N, b) a S/N target of 35 still generated bins that are generally of small size, comparable with the scale of the PSF, except for the edges of the maps where bins can be significantly larger. 

In Fig.~\ref{fig:violin_bins}, we show histograms of the number of spaxels included in a Voronoi bin across our sample of galaxies. The histograms are smoothed via a kernel density estimator for ease of visualisation. Bins within $0.2\,R_{25}$ are shown in orange, while those at larger radii are presented in blue. Galaxies are ordered by stellar mass from left to right. The 25$^{\rm th}$, 50$^{\rm th}$, and 75$^{\rm th}$ percentiles of each distribution are marked with dashed lines. The figure confirms that the median bin size outside the central regions ($R > 0.2\,R_{25}$) is between 10 and 100 spaxels, corresponding to circularised radii $0.4\arcsec$ and $1.1\arcsec$ (assuming bins are roughly circular). The two lowest-mass galaxies in our sample (NGC~5068 and IC~5332), together with NGC~2835, exhibit lower surface brightness levels even in the central regions, and therefore require larger bin sizes. Other galaxies show a significant number of bins consisting of one, or a few spaxels (which appear as isolated peaks in the figure due to the log-scaling of the y-axis and the smoothing of the distribution). We also note that there exists a tail of bins with a very large number of spaxels (${>}1000$), but these are very uncommon and correspond to the outermost regions.

\begin{figure*}
\centering
        \includegraphics[width=0.99\textwidth, trim=0 0 0 0, clip]{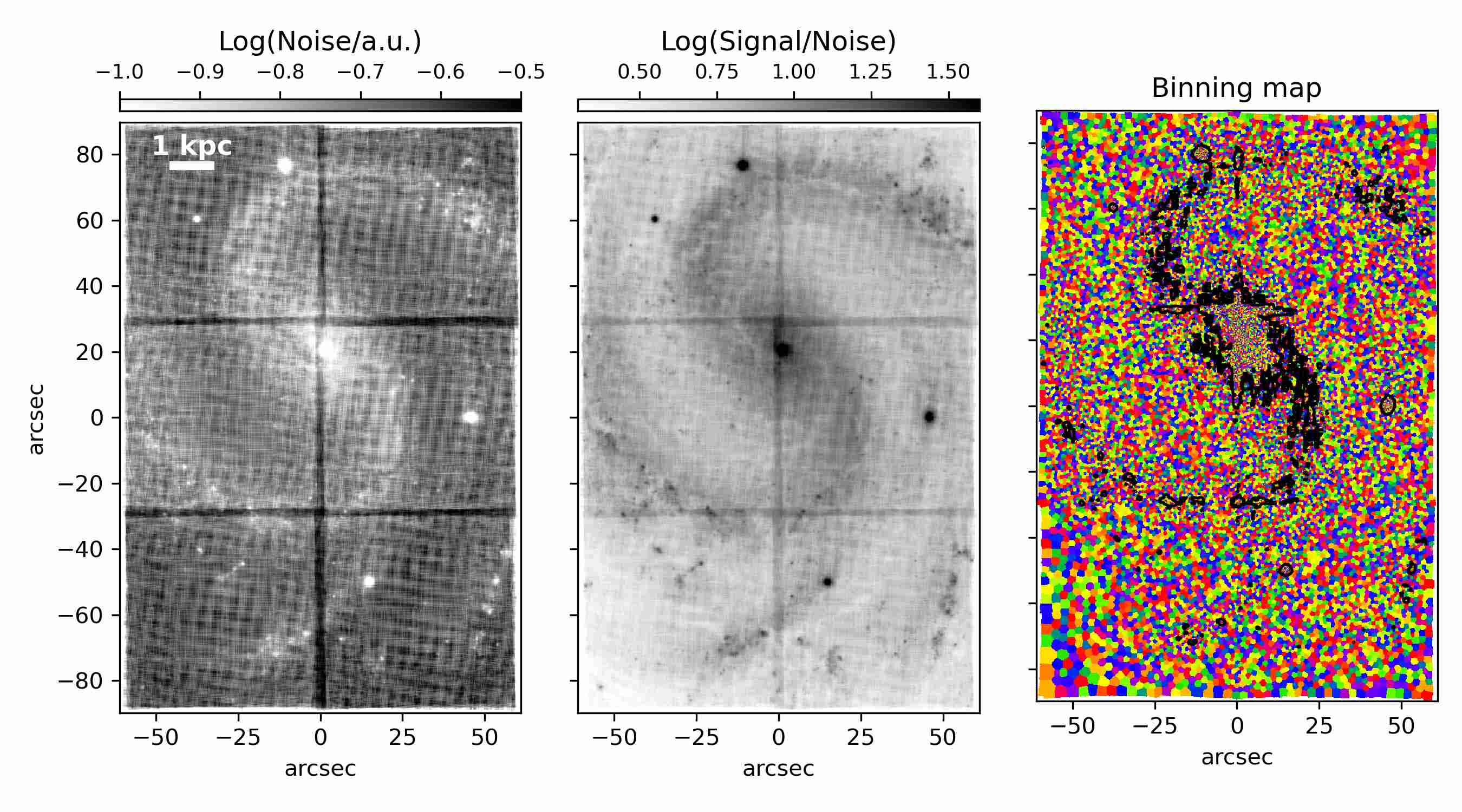}
     \caption{Maps of the average noise (left panel), S/N (middle panel), and binning map (right panel) for NGC~4535. The noise and S/N maps are computed by averaging the pipeline noise and flux over the $5300{-}5500$~\AA\ wavelength range. The stripes dividing the surveyed area into six squared subregions correspond to the overlap regions of the six MUSE pointings obtained for this galaxy. The noise map also shows an evident cross-hatch pattern within individual pointings due to the cube-generating resampling step in the MUSE \DRP\  when combining exposures with different rotation angles. This behaviour is also visible in the S/N map but does not significantly affect the results of the binning process. The binning map shows the result of the Voronoi binning procedure with target $\mathrm{S/N} = 35$. The black contour shows the $\mathrm{S/N} = 12$ level on individual spaxels. In this target only the galaxy centre (and a few foreground stars) have $\mathrm{S/N} > 35$ in individual spaxels, which are therefore left unbinned.}
     \label{fig:noise}   
\end{figure*}

The large size of our MUSE mosaics, well beyond the size of images intended for use with the \vorbin\ package by their authors, meant that we had to limit the size of an optimisation loop in the \vorbin\ code in order to ensure convergence.
When calculating the errors for the binned spectra, we assume the MUSE spaxels to be independent. This is not strictly correct, because of the inevitable resampling of the raw data in the data-cube-generation process. In short, spaxels nearly congruent with a single detector pixel will have almost no covariance, while spaxels whose flux originates from several pixels at the detector level will have errors that are strongly correlated with their nearest neighbours  \citep[see Sect.~4.6 of][]{Weilbacher+20}. In our native data cubes, this effect is visible as weak horizontal and vertical bands (the two orientations are due to the set of $90\deg$ rotations we perform) in the noise maps. An example map of the average noise vector and the resulting S/N distribution in the $5300{-}5500$~\AA\ wavelength range are shown in  Fig.~\ref{fig:noise}. The striping pattern we are referring to should not be confused with the much more evident change in the noise properties of the data in the overlap region of the different MUSE pointings, which corresponds to a real reduction in the noise due to longer effective exposure times. Within mosaics where one pointing consists of fewer than the nominal four exposures (e.g. because one exposure was discarded for quality control reasons) the noise is also consequently higher. 

In Fig.~\ref{fig:noise}, we also show the resulting Voronoi binning map, demonstrating that the small-scale noise striping pattern does not have a visible effect on the resulting Voronoi bins. In this work, we therefore neglect the issue of small-scale spatial covariance in the MUSE data.

The \DAP\ supports the determination of emission line properties for two different binning schemes: either the same Voronoi bins as the stellar kinematics or for single spaxels. For the first public PHANGS-MUSE data {release (DR1.0, hereafter; see Sect.~\ref{sec:datarelease})}, the emission line properties are derived for single spaxels because the H$\alpha$ emission is detected at the single-spaxel level across most maps.

We also tested the pipeline with different binning schemes. Two such implementations, optimised for the study of \hii\ regions and the DIG, respectively, are discussed in \cite{Santoro2021} and \cite{Belfiore2021}.

\subsubsection{Stellar kinematics}
\label{sec:stkin}

\begin{figure}
\centering
        \includegraphics[width=0.48\textwidth, trim=0 0 0 0, clip]{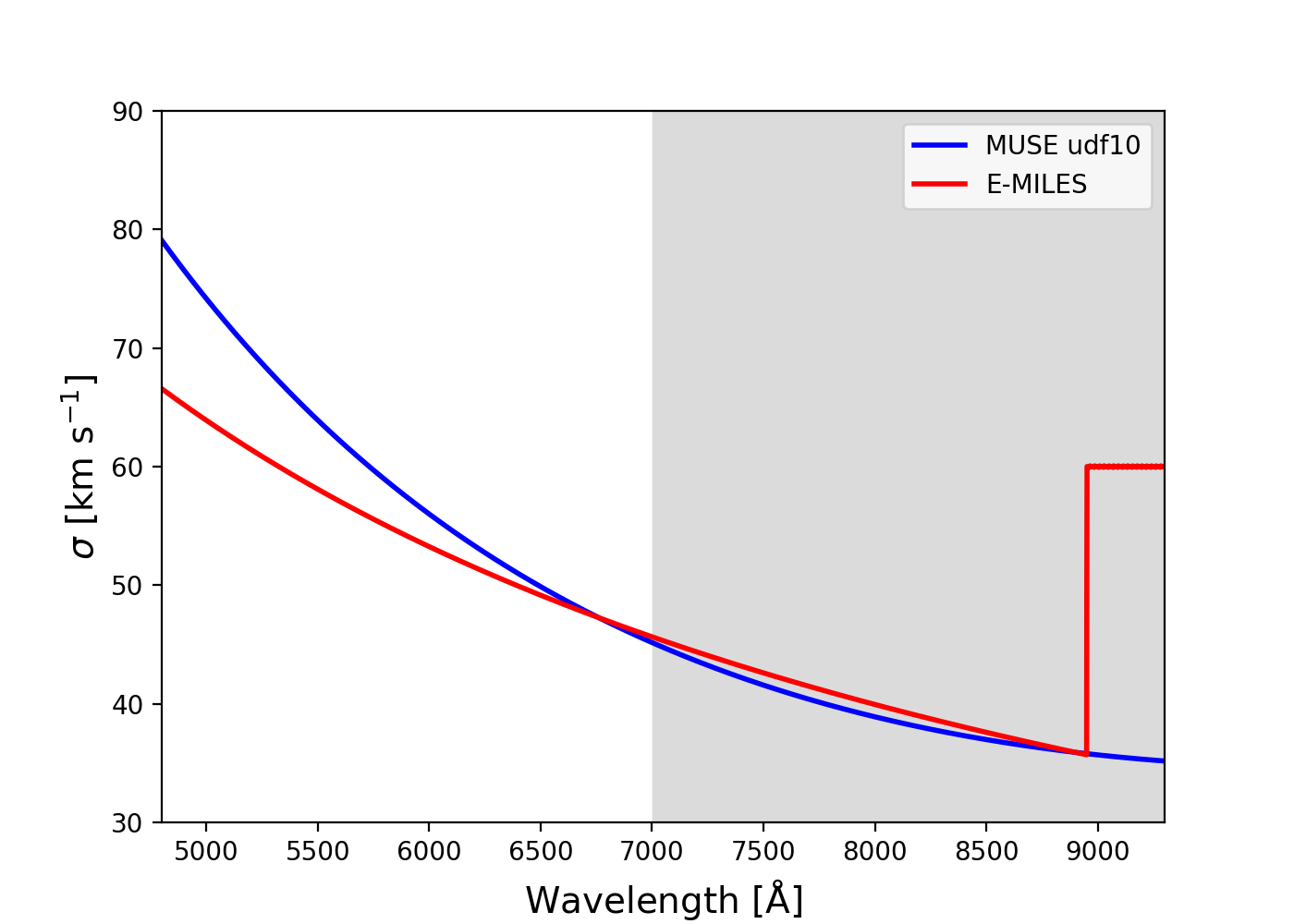}
 
     \caption{Comparison of the spectral resolution of the MUSE data (Eq.~\ref{eq:lsf}, taken from \citealt{Bacon2017}) and the E-MILES stellar templates. The greyed-out part of the wavelength range ($\lambda > 7000$~\AA) was excluded from the fit in this \protect{first DR1.0 PHANGS-MUSE data release}.}
     \label{fig:lsf_emiles}   
\end{figure}

The stellar kinematics are derived using \pPXF, following the same procedure as in \gist. 
Briefly, to fit the stellar continuum we use E-MILES simple stellar population models \citep{Vazdekis2016}, generated with a \citet{Chabrier2003} initial mass function, BaSTI isochrones \citep{Pietrinferni2004}, eight ages ($0.15{-}14$~Gyr, logarithmically spaced in steps of $0.22$~dex) and four metallicities ($\mathrm{[Z/H]} = [-1.5, -0.35, 0.06, 0.4]$), for a total of 32 templates. 
We again fit the wavelength range $4850{-}7000$~\AA\ in order to avoid strong sky residuals in the redder part of the MUSE wavelength range. The regions around the expected positions of ionised gas emission lines are masked. The mask width is taken to be $\pm400$~\kms\ of the systemic velocity of the galaxy. This mask width is found to be appropriate for the range of velocities and dispersions present in the PHANGS data. In MUSE-noAO observations the region around the \NaI~D absorption doublet is also masked, because of the potential ISM contribution. Finally, we mask the region around the bright sky lines at the observed wavelengths of $5577$, $6300$ and $6363$~\AA.

The spectral resolution of E-MILES\footnote{Taken from \url{http://www.iac.es/proyecto/miles/pages/spectral-energy-distributions-seds/e-miles.php}} is higher than that of the MUSE data within the wavelength range we are considering (see Fig.~\ref{fig:lsf_emiles}), although at ${\sim}7000$~\AA\ E-MILES is expected to have virtually the same spectral resolution as MUSE. The templates are therefore convolved to the spectral resolution of the data, using an appropriate wavelength-dependent kernel. No convolution of the data is performed where the E-MILES resolution is worse than or comparable to that of the MUSE data (beyond about $6750$~\AA).
Velocities are computed with respect to the systemic velocity of the galaxy. We fitted four moments of the line-of-sight velocity distribution (i.e. velocity, velocity dispersion, $h3$ and $h4$). To derive the stellar kinematics we make use of additive Legendre polynomials (12$^{\rm th}$ order, in the spectral direction), and no multiplicative polynomials. Polynomials are found to be advantageous in the derivation of stellar kinematics with \pPXF\ \citep{Westfall2019}; the choice between additive and multiplicative is purely dictated by computing efficiency in this step.

The code is able to perform a Markov chain Monte Carlo (MCMC) based error estimate for kinematic parameters (a feature inherited from \gist), but this step was not performed for the current data release (DR1.0). The errors reported for the kinematic parameters are therefore formal errors, as given by \pPXF.
Examples of kinematic maps obtained from our data are shown in Sect.~\ref{sec:SciCase_stellarpops}.

\subsubsection{Stellar populations}
\label{sec:stellarpops}

The stellar population module also employs \pPXF\ in a sequence of steps optimised for use with our MUSE data. We do not use the original \gist\ code for this module, because we moved away from `regularisation', as explained below. The stellar population analysis is performed on the Voronoi-binned data (i.e. the same spectra used for the stellar kinematics determination).

As in the derivation of stellar kinematics, we used E-MILES stellar population models {using the same setup as for the stellar kinematics (see Sect.~\ref{sec:stkin})}, but we increased the number of templates to 78, with ages = [0.03, 0.05, 0.08, 0.15, 0.25, 0.40, 0.60, 1.0, 1.75, 3.0, 5.0, 8.5, 13.5] Gyr and [Z/H] = [$-1.49$, $-0.96$, $-0.35$, $+0.06$, $+0.26$, $+0.4$]. We fit the same wavelength range as used for the stellar kinematics determination.
We retain the previously derived stellar kinematics parameters by fixing them in the \pPXF\ fit, and mask the regions around emission lines, as in Sect.~\ref{sec:stkin}. Additionally, we mask further regions of the spectrum particularly affected by sky line residuals (wavelengths within ranges [$5861.1{-}5911.7$], [$6528.8{-}6585.1$] and [$6820.0{-}6990.9$]~\AA). 
The fit is performed in two stages. In the first stage we determine the extinction of the stellar continuum, as parametrised by a \citet{calzetti2001} extinction curve. In the second iteration the stellar extinction is kept fixed, and multiplicative polynomials of 12$^{\rm th}$ order, with a mean of one, are used to correct for residual inaccuracies in the relative flux calibration. This two-tiered procedure was demonstrated to be necessary to overcome inaccuracies in the sky continuum subtraction, which can cause subtle changes in the spectral shape. To the best of our knowledge, such residuals do not have a large effect on the derived stellar population properties, except for the $E(B-V)$ of the stellar component, as discussed in Sect.~\ref{sec:extinction}. Additive polynomials cannot be used as they both modify the absorption line equivalent widths (EWs) and affect our measurements of stellar mass surface density.

Given the degeneracy that exists in spectral fitting (i.e. between attenuation, metallicity, and age), the use of regularisation (i.e. along metallicity and age axes) when trying to recover star-formation histories has been promoted for certain situations \citep[e.g.][]{Cappellari2017}. Given the average spatial resolution of ${\sim}50$~pc, our data resolve individual star-forming regions with significant contributions from very young stars, resulting in strong variations in the spatial distribution of stellar ages across the galactic discs. Extensive tests showed that forcing a fixed level of regularisation on our dataset leads to star-formation histories with strong biases, which themselves vary strongly from region to region and are difficult to control. While we could envision a scheme to provide a more controlled bias, this is beyond the scope of the present {PHANGS-MUSE DR1.0 public release}. We therefore decided to rely on Monte Carlo simulations to estimate the uncertainty in the recovered stellar population parameters, and use un-regularised fitting. For each spectrum, we performed 20 Monte Carlo iterations. In each iteration, we added to the input spectrum Gaussian noise with a mean of zero and a standard deviation corresponding to the error vector at each wavelength bin. The uncertainties of stellar population parameters were calculated as the standard deviation of their distributions produced by the Monte Carlo realisations. This is meant as a first-order estimate of the true uncertainties. 
We only ran Monte Carlo realisations during the second step of the fitting procedure, once the stellar extinction has been fixed. Hence, no error was computed for the stellar $E(B-V)$. The output of \pPXF\ is a vector with the weights of the templates that, when linearly combined, best reproduce the observed spectrum. These weights represent the fraction of the total stellar mass born with a given age and metallicity,  and they can be used to produce the final maps, including stellar mass surface densities, and both light- or mass-weighted ages and metallicities. The stellar mass surface density maps include both, the contributions from live stars and remnants.
For each pixel, the average age and metallicity are computed as
\begin{equation}
\label{eq:age}
 \langle \log \mathrm{age} \rangle = \frac{\Sigma_{i}\, \log(\mathrm{age}_{i})\,w_{i}}{\Sigma_{i}\,w_{i}}
\end{equation}
and
\begin{equation}
\label{eq:z}
 \langle \mathrm{[Z/H]} \rangle = \frac{\Sigma_{i}\, \mathrm{[Z/H]}_{i}\, w_{i}}{\Sigma_{i}\, w_{i}}~,
\end{equation}
where age$_{i}$ and [Z/H]$_{i}$ correspond to the age and metallicity of each template, and $w_{i}$ is its corresponding weight in the linear combination. To convert mass-weighted quantities to light-weighted quantities, we use the M/L of each template in the $V$-band. We compute luminosity-fraction weights as
\begin{equation}
w_{i}^{\rm LW} = \frac{w_{i}}{(M/L_{V})_{i}}~,
\end{equation}
where $w_{i}^{\rm LW}$ corresponds to the luminosity-fraction weight of a given template, $w_{i}$ its mass-fraction weight, and $(M/L_{V})_{i}$ correspond to its M/L in the V-band. We can use these luminosity-fraction weights to calculate light-weighted properties, following Eqs.~\ref{eq:age} and~\ref{eq:z}. It is possible to use the stellar population weights we derive to produce maps of, for example, stellar mass in different age ranges (e.g. young, intermediate, and old stars), or to study the age versus metallicity relation within individual regions. Such analyses are beyond the scope of the present paper. 

\subsubsection{Emission lines}
\label{sec:emlines}

\begin{table*}
        
        \centering

    \caption{Wavelengths and ionisation potential of the relevant ion for each emission line. All lines are corrected for the Milky Way foreground contribution. Wavelengths are taken from the
    National Institute of Standards and Technology 
    (NIST; \url{https://physics.nist.gov/PhysRefData/ASD/lines_form.html}), and are Ritz wavelengths in air except for the H~Balmer lines, in which case we use the `observed' wavelength in air as reported in NIST. The
    \DAP\ string name is used to identify the correct extension in the \MAPS\ files. Ionisation
    potentials are taken from \cite{Draine2011}.  Lines redder than
    ${\sim}7000$~\AA\ are currently not fitted.}
        
        \begin{tabular}{l l l c c}
                \hline
                line name & Wavelength & String ID & Ionisation potential & Fixed ratio \\
                 & (air) [\AA| & & [eV] & \\
                \hline
                \multicolumn{5}{c}{Hydrogen Balmer lines} \\
                \hline
                \hb\ & 4861.35 & \texttt{HB4861} & 13.60 & no \\
                \ha\ & 6562.79 & \texttt{HA6562} & 13.60  & no\\
                \hline
                \multicolumn{5}{c}{Low-ionisation lines} \\
                \hline
        \niion$\lambda$5197 & 5197.90  & \texttt{NI5197} & --- & no\\
                \niion$\lambda$5200 & 5200.26  & \texttt{NI5200} & --- & no \\
                \nii$\lambda$5754  &  5754.59  &\texttt{NII5754} & 14.53 & no\\
                \oi$\lambda$6300   & 6300.30  & \texttt{OI6300} & --- & no\\
                \oi$\lambda$6364   & 6363.78   &  \texttt{OI6363} & --- & 0.33 \oi$\lambda$6300 \\
                \nii$\lambda$6548  &  6548.05  & \texttt{NII6548} & 14.53 & 0.34 \nii$\lambda$6584\\
                \nii$\lambda$6584  &  6583.45 & \texttt{NII6583}  & 14.53 & no \\
                \sii$\lambda$6717  & 6716.44   & \texttt{SII6716} & 10.36 & no \\
                \sii$\lambda$6731  & 6730.82  & \texttt{SII6730} & 10.36 & no\\
                \hline
                \multicolumn{5}{c}{High-ionisation lines} \\
                \hline
                \oiii$\lambda$4959 & 4958.91  & \texttt{OIII4958} & 35.12 & 0.35 \oiii$\lambda$5007 \\ 
                \oiii$\lambda$5007 & 5006.84   & \texttt{OIII5006} &35.12 & no\\ 
                \hei$\lambda$5876                   &  5875.61  & \texttt{HeI5875} & 24.58  & no\\
                \siii$\lambda$6312 & 6312.06  & \texttt{SIII6312} & 23.34 & no\\
                \hline
        \end{tabular}
        
        \label{table_ems}
\end{table*}

Emission lines are fitted by performing an independent call to \pPXF, where emission lines are treated as additional Gaussian templates, and the stellar continuum is fitted simultaneously. We do not use the module provided with \gist, based on the \texttt{gandalf} implementation \citep{sarzi_sauron_2006}. This choice mirrors the philosophy of the MaNGA \DAP, and is motivated by the greater flexibility of the \pPXF\ implementation and the extensive testing and experience documented in \cite{Belfiore2019} and \cite{Westfall2019}. Some of the code we use to interface with \pPXF\ in this fitting stage was adapted directly from the MaNGA \DAP, and makes use of the analytical Fourier transform implemented in version~$>6$ of \pPXF\ \citep{Cappellari2017}.

The fits are performed on individual spaxels, fixing the stellar velocity moments to the values obtained during the stellar kinematics fitting step within the associated Voronoi bin (Sect.~\ref{sec:stkin}). We tested the effect of leaving the stellar kinematics free and find largely identical results for spaxels with large S/N in the continuum. We use the same set of 32 stellar templates as in Sect.~\ref{sec:stkin}.

The kinematic parameters of the emission lines (velocity and velocity dispersion) are tied together within three different groups, as follows: (i) hydrogen Balmer lines: \ha, \hb; 
(ii) low-ionisation lines: \oi$\lambda\lambda$6300,64, \niion $\lambda\lambda$5197,5200, \nii$\lambda\lambda$6548,84, \nii$\lambda$5754, 
and \sii$\lambda\lambda$6717,31; and (iii) high-ionisation lines: \hei$\lambda$5875, \oiii$\lambda\lambda$4959,5007, and \siii$\lambda$6312.

We tie the intrinsic (astrophysical) velocity dispersion within each kinematic group, prior to convolution with the instrumental LSF. The measured velocity dispersion of lines belonging to the same kinematic group is therefore different, but the difference is that required to bring the intrinsic velocity dispersion into agreement, given the assumed LSF. Because the MUSE LSF changes with wavelength, this is an important effect to take into account. 

Initial testing showed that the kinematics of the \oiii\ line is sufficiently different from that of the Balmer lines to require an independent kinematic component. On the other hand, we found that leaving the kinematics of the \hb\ line free generates a large number of non-physical Balmer decrements (\ha/\hb) at low S/N. The definition of a third group of low-ionisation metal lines may support specific science cases focusing on the DIG, where such low-ionisation lines are prevalent with respect to the hydrogen Balmer lines. Alternative tying strategies  can be trivially implemented by changing simple keywords in a configuration file of the \DAP. 

During the emission lines fit, \pPXF\ is run with  eight order multiplicative Legendre polynomials, but no additive polynomials (polynomials are only applied to the stellar continuum templates). The use of additive polynomials would be inappropriate in this fitting stage as they modify the EW of the Balmer absorption lines, and therefore potentially introduce non-physical corrections to the hydrogen Balmer line fluxes. The use of a different set of polynomials for this fitting stage with respect to the stellar kinematics fitting stage, in addition to the different S/N of the continuum (going from bins to single spaxels) and the absence of a mask in the wavelength regions around emission lines contribute to creating subtle differences in the best-fit continua generated in the two, complementary fitting stages. The effects of the different polynomials treatment and of masking versus simultaneous fitting of the emission lines are discussed in detail in \citet[][their Sect.~5.2]{Belfiore2019}, who find that the effect of polynomials can be large (${\sim}10$\% of the emission line flux), especially for the high-order Balmer lines (not present, however, in the MUSE spectral range). They find, moreover, that the effects of masking versus simultaneous fitting of the emission lines are only evident in the regions of Balmer absorption, causing small systematic changes in the fluxes of hydrogen Balmer lines ($<2$\% for H$\alpha$). In light of these findings we prefer to refit the continuum in the emission lines fitting stage. 

We note that alternative approaches to this problem are possible. In the Pipe3d CALIFA analysis, for example, the best-fit continuum from the bins used for the stellar kinematics extraction is simply re-scaled according to the median flux in the constituent spaxels, so that the spectrum that is subtracted at the spaxel level is left unchanged \citep{Sanchez2016a}. In the SAMI public data release, a new fit is performed at the spaxel level, as done here, but to limit the impact of degeneracies only those stellar templates with non-zero weights in the parent Voronoi bin are retained in the fit of the individual spaxels \citep{croom2012}. In general, the effect of restricting the template library in this way are small if the stellar populations are reasonably uniform within the Voronoi bin \citep[][specifically their Sect.~7.4.1]{Westfall2019}. At the spatial resolution of PHANGS this may not be always the case, since the distribution of young stellar populations is stochastic on small scales. We prefer, therefore, not to restrict the range of templates used, also in light of the fact that we use a small set of templates to start with. Aside from the correction for foreground Milky Way extinction, applied by the \DAP\ as part of its preliminary steps, no correction for dust is applied within the target galaxy. 

\begin{figure*}
\centering
        \includegraphics[width=0.95\textwidth, trim=0 0 0 0, clip]{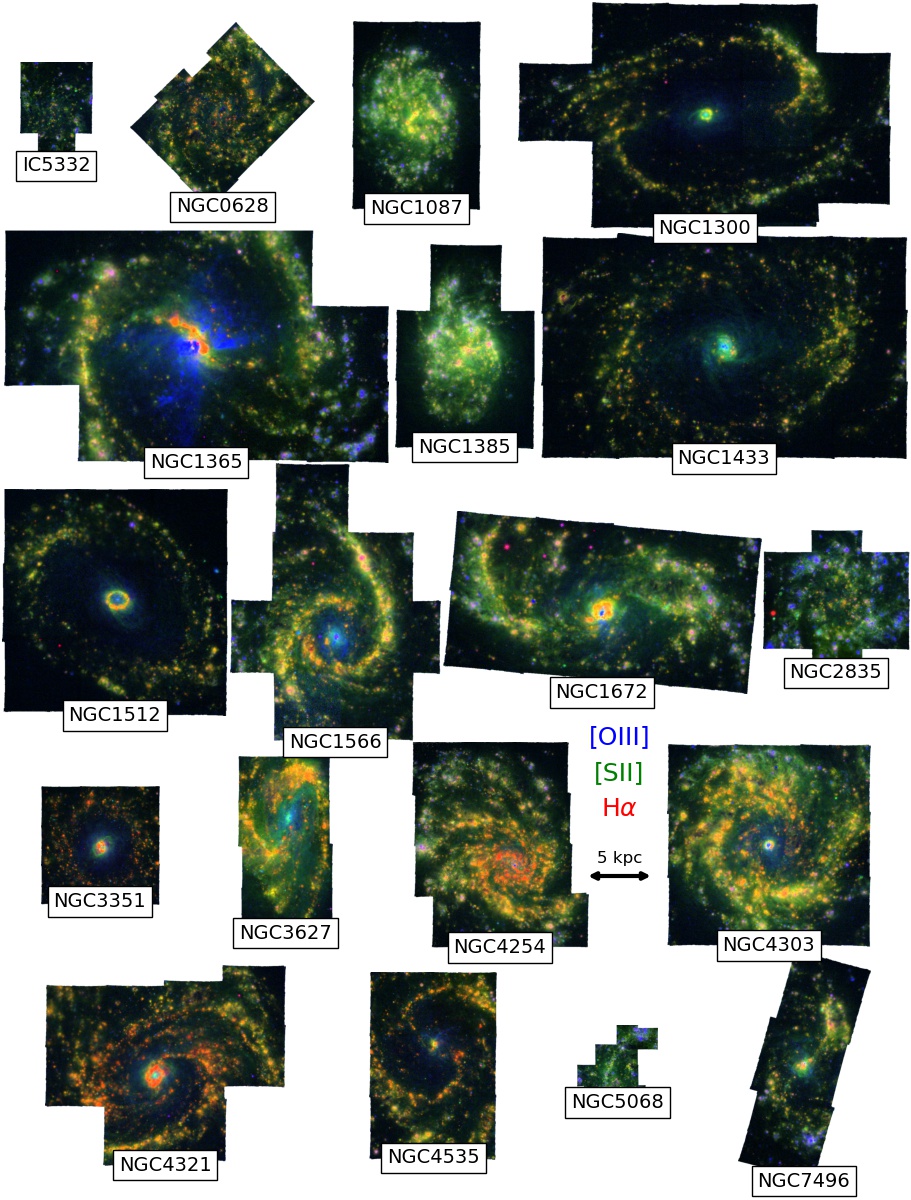}
     \caption{RGB images of the 19 targets in the MUSE sample obtained by combining the emission line maps for \sii, \ha\, and \oiii\ (green, red, and blue channels, respectively). A 5~kpc double black arrow indicates the fixed physical scale for all panels, just below the RGB labels respectively associated with each emission line map.}
     \label{fig:3emicol}
\end{figure*}

An overview of the line maps for \ha, \sii\ and \oiii\ is presented for the full PHANGS-MUSE sample in Fig.~\ref{fig:3emicol}. The emission line science enabled by MUSE is unprecedented, due to the remarkable sensitivity and wide wavelength coverage of the instrument. When applied to nearby galaxies, we have the further advantage of achieving the $50{-}100$~pc physical scales necessary to isolate individual ionising sources from each other and from the surrounding diffuse ionised emission. Our wide coverage across the galaxy discs enables us to connect these small scales to the large kiloparsec-scale relations determined in the literature (see also Sect.~\ref{sec:SciCase_scaling}).

\subsubsection{Output files}
\label{sec:dap_output}

The \DAP\ output that we make publicly available consists of two-dimensional maps of parameters and physical properties of interest. These maps are consolidated in the so-called \MAPS\ file, a multi-extension FITS file. {The full set of maps currently contains 133 extensions}. These \MAPS\ employ the exact same pixel grid and world coordinate system as the mosaic data cubes produced by the reduction pipeline. The full list of extensions of \MAPS\ is described in Table~\ref{tab:maps}. The extensions include two-dimensional maps of stellar kinematics, ionised gas fluxes and kinematics, and stellar-population-related (stellar mass and age) measurements. In the first public PHANGS-MUSE DR1.0 release, all maps pertaining to the stellar and ionised gas fluxes and kinematics are available, so far excluding the stellar populations that need further attention (see Sect.~\ref{sec:youngbias}). \DAP\ also produces many intermediate products that are not part of the {first public data PHANGS-MUSE DR1.0 release}, but could be useful to expert users for specific science cases (e.g. the individual weights of the best-fit stellar templates, which can be used to compute the mass fraction within specified age bins). We welcome requests for these products from interested members of the scientific community. Depending on the feedback our team will receive, future {public} data releases may include a larger set of products.

\begin{table*}
\caption{List of \texttt{FITS} extensions included in the \MAPS\ file, the main output of the PHANGS-MUSE \DAP. Each extension is a two-dimensional map on the same WCS as the mosaic data cube. We list the extension names and a brief description of the map associated with that extension. }              
\label{tab:maps}      
\centering    
\begin{tabular}{l l}
\hline
\hline
Extension name &  Description\\
\hline
\hline
    \multicolumn{2}{c}{Binning} \\
\hline
        \texttt{ID} &  unique ID for each spaxel \\
        \texttt{FLUX} & white-light image\\
        \texttt{SNR} &  continuum S/N for individual spaxels \\
        \texttt{SNRBIN} &  continuum S/N for each Voronoi bin \\
        \texttt{BIN\_ID} &  unique ID for each Voronoi bin, unbinned spectra have bin IDs of $-1$ \\
\hline
        \multicolumn{2}{c}{Stellar kinematics} \\
        \multicolumn{2}{c}{\texttt{HN$^{\#}$\_STARS} = higher order Gauss-Hermite velocity moment, if available (e.g. \texttt{H3\_STARS, H4\_STARS})}\\
\hline
        \texttt{V\_STARS} & stellar velocity [$\rm km  \ s^{-1}$], after subtracting the systemic velocity\\
        \texttt{FORM\_ERR\_V\_STARS} & formal velocity error [$\rm km \ s^{-1}$]\\
        \texttt{ERR\_V\_STARS} & MCMC-calculated error for velocity (currently not available) [$\rm km  \ s^{-1}$]\\
        \texttt{SIGMA\_STARS} & stellar velocity dispersion [$\rm km \ s^{-1}$]\\
        \texttt{FORM\_ERR\_SIGMA\_STARS} & formal sigma error [$\rm km  \ s^{-1}$] \\
        \texttt{ERR\_SIGMA\_STARS} & MCMC-calculated error for sigma (if available) [$\rm km \ s^{-1}$]\\
        \texttt{HN$^{\#}$\_STARS} & higher order moments of the stellar LOSVD (when available) \\
        \texttt{FORM\_ERR\_HN$^{\#}$\_STARS}  & formal errors in the high-order moments \\
        \texttt{ERR\_HN$^{\#}$\_STARS} & MCMC errors for higher-order moments (not yet available)  \\
\hline
        \multicolumn{2}{c}{Stellar populations} \\
\hline
        \texttt{STELLAR\_MASS\_DENSITY}  &  stellar mass surface density [$\rm M_\odot \ pc^{-2}$] \\
        \texttt{STELLAR\_MASS\_DENSITY\_ERR}  &  error in the above [$\rm M_\odot \ pc^{-2}$] \\
        \texttt{AGE\_MW}  & log(Age/yr), where the Age is mass-weighted \\
        \texttt{AGE\_MW\_ERR}  & error in the above quantity\\
        \texttt{Z\_MW}  &  mass-weighted [Z/H] \\
        \texttt{Z\_MW\_ERR}  &  error in the above quantity \\
        \texttt{AGE\_LW}  &  log(Age/yr), where the Age is luminosity-weighted (V-band)  \\
        \texttt{AGE\_LW\_ERR}  &  error in the above\\
        \texttt{Z\_LW}  & luminosity-weighted (V-band) [Z/H]  \\
        \texttt{Z\_LW\_ERR}  & error in the above quantity  \\
        \texttt{EBV\_STARS}  & $E(B-V)$ of the stellar component [mag] \\
\hline
        \multicolumn{2}{c}{Emission lines} \\
        \multicolumn{2}{c}{ \texttt{*emline} = emission line string id listed in Table~\ref{table_ems} }\\
\hline
        \texttt{BIN\_ID\_LINES}  & unique bin for emission lines, these are individual spaxels in the current DR2 
        \\
        \texttt{CHI2\_TOT}  & The $\chi^2$ over the full fitted wavelength range. 
        \\
        \texttt{*emline\_FLUX}  & emission line flux [$\rm 10^{-20} \ erg \ s^{-1} \ cm^{-2}  \ spaxel^{-1}$] \\
        \texttt{*emline\_FLUX\_ERR}  & emission line flux error [$\rm 10^{-20} \ erg s^{-1} cm^{-2} \ spaxel^{-1}$] \\
        \texttt{*emline\_VEL}  & emission line velocity [$\rm km \ s^{-1}$] \\
        \texttt{*emline\_VEL\_ERR} & emission line velocity error [$\rm km \ s^{-1}$]\\
        \texttt{*emline\_SIGMA} & emission line velocity dispersion [$\rm km \ s^{-1}$] \\
        \texttt{*emline\_SIGMA\_ERR} & emission line velocity dispersion error [$\rm km \ s^{-1}$] \\
        \texttt{*emline\_SIGMA\_CORR} & instrumental velocity dispersion at the position of the line [$\rm km \ s^{-1}$]\\
\hline
\end{tabular}
\end{table*}

We also emphasise that there is a difference in the way the velocity dispersions for gas and stars are reported. The stellar velocity dispersion is reported as the astrophysical dispersion, corrected for instrumental broadening, while the velocity dispersion of the lines is reported as the observed velocity dispersion. To obtain the astrophysical velocity dispersion for the emission lines, the relevant instrumental dispersion at the wavelength of the emission line contained in the \texttt{SIGMA\_CORR} extension should be subtracted from the observed value in quadrature. Since the kinematics of the emission lines are tied within the groups defined in Sect.~\ref{sec:emlines} there is a large amount of redundancy in the velocities and the astrophysical velocity dispersions (i.e. these quantities are not independent). We nonetheless decided to maintain the redundancy in the data products files, to ensure a format robust against changes in the line-tying scheme. Users should be warned, however, that in the current implementation there are only three independent velocity and velocity dispersions for the emission lines, corresponding to the kinematic groups defined above. 

\subsection{Star masks}
\label{sec:starmask}

Depending on the area mapped and Galactic latitude, we can have up to tens of foreground stars in a single PHANGS-MUSE mosaic. Using the Gaia all-sky catalogue we can localise and mask these contaminants. However, the Gaia catalogue also includes bright stellar clusters, and even some bright galaxy centres, that we do not wish to mask. We therefore follow a series of steps to create our final foreground star masks. These {masks will be publicly released together with catalogues extracted from miscellaneous PHANGS datasets, including the MUSE data, in the near future}.

We use the Gaia DR2 archive \citep{GAIA2018}, searching for all sources within an $8\arcmin$ aperture centred on each galaxy, allowing therefore for stars up to the edges of our MUSE mosaic FoVs. We then define a circular aperture for each source that attempts to capture all spaxels impacted significantly by the source. The radius of the aperture is defined as a linear function of the Gaia $g$-band magnitude, empirically calibrated to the MUSE white-light image. We set as the minimum aperture size the worst PSF in the mosaic. 

For each source, we determine the maximum and mean of the calcium triplet CaT EWs at heliocentric (i.e. zero) velocity of the spaxels contained within the aperture (i.e. the sum of three bands centred at the features as compared to median flux over $\sim$~$i$-band). We also determine the maximal stellar velocity difference between the spaxels within the aperture and the median stellar velocity in the unaffected surrounding spaxels representing the typical galaxy velocity. These quantities are then used to classify each source as a Galactic (i.e. star) or extragalactic object (e.g. cluster).

We label each Gaia DR2 object as a star if the velocity difference is large (${>}50$~\kms), and the CaT EW is greater than a determined threshold (CaT EW mean $> 0.03$~\AA\ and maximum CaT $> 0.05$~\AA). These empirically defined thresholds were determined by examining several spectra of Gaia sources and classifying these by eye. These thresholds were found to prevent almost all galaxy centres, bright clusters and \hii\ regions from being automatically masked. Some faint stars ($g-\mathrm{mag} > 19.5$), however, are missed by these criteria. 

Other contaminants that may affect the stellar population fits (i.e. diffraction spikes from very bright stars, background galaxies) are not masked. We also note that some faint stars have been found not to be in the Gaia catalogue, meaning that they will be missed by our masking procedure. NGC~1566 has its centre masked due to its type~1 AGN (i.e. exhibiting strong and broad emission lines as well as thermal emission).

We note that the spectral cubes or analysis data product \MAPS\ that we provide as part of the public data release {(PHANGS-MUSE DR1.0)} are not automatically masked at the positions of the foreground stars we have identified. The users are, on the other hand, left to judge the suitability of these masks depending on their science goals. We caution, however, that most two-dimensional maps of physical parameters (e.g. emission line fluxes, stellar masses) will be biased or incorrect at the position of foreground stars, because of the incorrect spectral fitting (see Sect.~\ref{sec:spectral_fits}).

\section{Quality assessment}
\label{sec:QA}

To validate our final data products of the PHANGS-MUSE sample we derive several statistical measures within the million spectra of our Large Programme, and also cross-compare with existing data on our galaxies. In this section we present quality tests performed both on the final mosaicked cubes, and the high-level data products of the data analysis pipelines.

\subsection{Quality of the mosaicked data cubes}
\label{sec:dap_quality}

\subsubsection{Validation of the photometric calibration}

\begin{figure}
    \centering
    \includegraphics[width=0.5\textwidth]{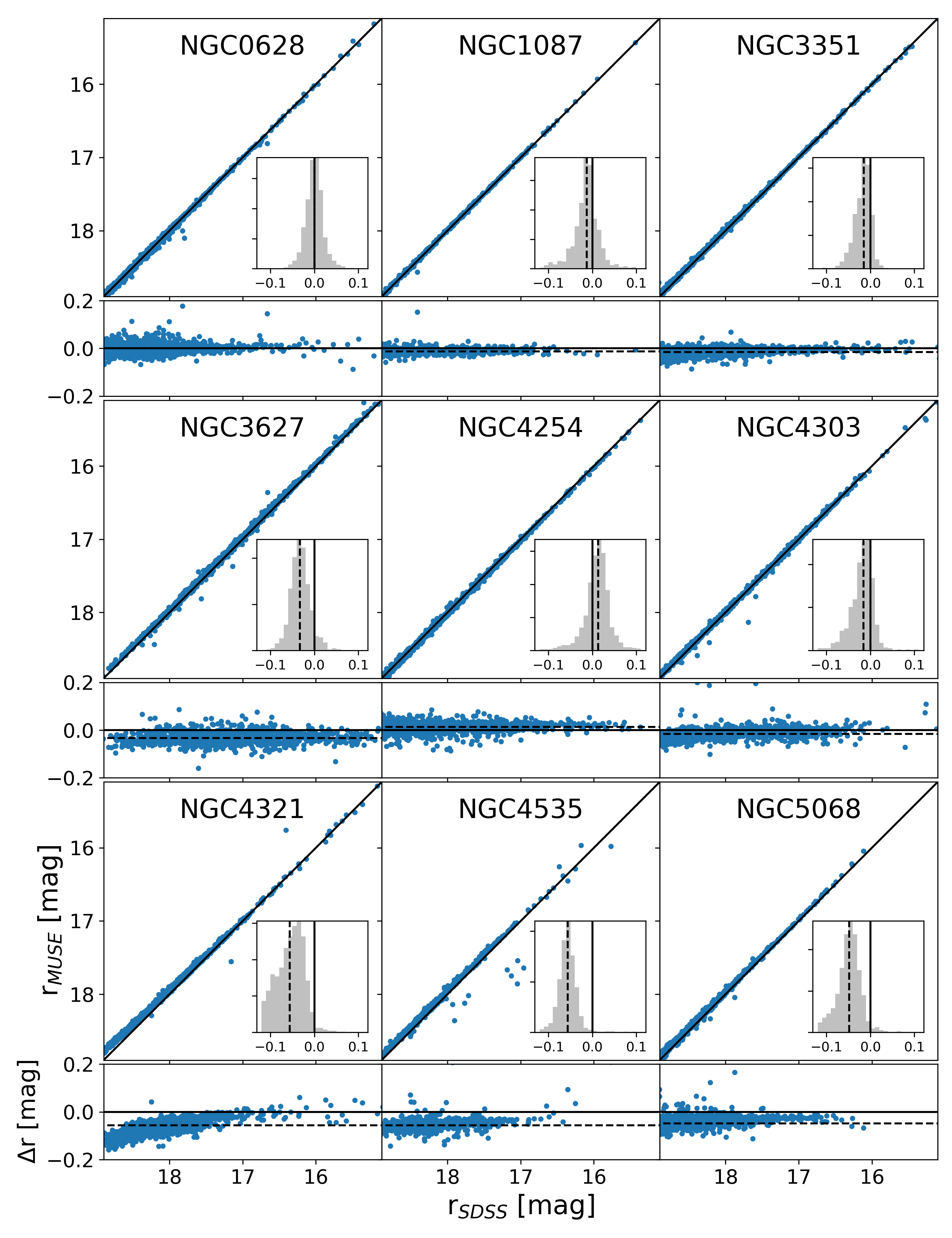}
    \caption{Comparison of $r$-band magnitudes measured over $5\arcsec \times 5\arcsec$ apertures within MUSE synthetic images ($r_\mathrm{MUSE}$) and SDSS images ($r_\mathrm{SDSS}$) for the nine galaxies with SDSS imaging available. \protect{Each panel is separated into a main one-to-one comparison plot and an additional plot below showing the magnitude difference ($\Delta r_\mathrm{MUSE - SDSS}$) versus the SDSS value. Insets in each panel show the histograms of the magnitude difference, with the median offset indicated with a dashed line. Solid lines indicate the identity line (or the 0 value for the magnitude difference).} Across this sample of galaxies, the median offset ranges from $-0.06$ to $0.01$~mag. Typical scatter within any galaxy is ${\sim}0.04$~mag. 
    \label{fig:sdss_compare}}
\end{figure}

To validate the overall photometric calibration of the cubes, we compare synthetic broadband images against existing SDSS images for the nine galaxies that lie within the SDSS legacy survey footprint (Fig.~\ref{fig:sdss_compare}). To construct the synthetic $r$-band images we apply the Sloan $r$-band filter curve to the MUSE spectral cubes. We use the SDSS Science Archive Server Mosaic tool\footnote{\url{https://dr12.sdss.org/mosaics}} in order to construct large-scale $r$-band images that cover the MUSE mosaic footprint. We then sample both the SDSS image and our MUSE image with uniformly spaced $5\arcsec\times5\arcsec$ apertures. This aperture size was chosen to be large enough to mitigate any differences in seeing or astrometric offsets between the two datasets. Across this sample of galaxies, the median photometric offset ranges from $-0.06$ to $0.01$~mag. NGC~4321 in particular shows a skewed distribution, corresponding to a larger offset for the apertures with lower fluxes, which could potentially be attributed to sky subtraction residuals in either dataset. Overall, the median offset of the galaxies in this sample is $-0.01$~mag, corresponding to slightly brighter magnitudes in the MUSE image. This is roughly consistent with the SDSS photometric calibration uncertainty \citep{Padmanabhan2008}. Typical scatter within any galaxy is ${\sim}0.04$~mag. 

\subsubsection{Validation of the absolute astrometric calibration}
\label{sec:dap_validastro}

As discussed in Sect.~\ref{sec:align}, not every MUSE pointing includes enough bright stars to perform an accurate determination of the astrometry, or to robustly check it a posteriori. This was the motivation to rely on the full extended (stellar and gaseous) emission from the galaxy, and on its comparison with independent ground-based imaging covering much larger FoVs. Still, across our mosaics there are a total of 96 stars that can be overall used to partly validate the absolute and relative astrometric calibrations of the PHANGS-MUSE survey. The only two galaxies that do not contain any foreground stars within the FoV covered by MUSE are NGC~1385 and NGC~7496.

To validate the astrometric solution of the MUSE data, we compared the positions of stars in the MUSE mosaics (as defined in Sect.~\ref{sec:starmask}) both with their WFI and du~Pont broadband locations and with their Gaia DR1 locations. We first used the \textsc{photutils} routine \textsc{IRAFStarFinder} \citep{photutils2020} on our MUSE mosaics to accurately determine the centroid of each object classified as a star in our stellar masks.
When comparing the MUSE positions with the broadband positions (measured with the same procedure), we obtain $\Delta\mathrm{RA} = 0.026\arcsec\pm 0.047\arcsec$ and $\Delta\mathrm{Dec} = -0.013\arcsec\pm 0.044\arcsec$: such values are observed consistently across the full PHANGS-MUSE sample and are well within the accuracy expected from our alignment routine, representing only from about 1/5$^{\rm th}$ to 1/20$^{\rm th}$ of a MUSE spaxel size. This confirms that our alignment process is robust, yielding a good astrometric calibration using an intermediate imaging dataset. We note that the relative astrometry between the WFI ground-based imaging and Gaia DR1 is consistent with a null offset, and a scatter significantly better than 100~mas.

We more directly compared the MUSE and Gaia DR1 positions for stars in common and actually find a non-zero offset of $\Delta\mathrm{RA} = -0.011\arcsec\pm 0.067\arcsec$ and $\Delta\mathrm{Dec} = -0.10\arcsec\pm 0.09\arcsec$. Such an offset is basically zero in Right Ascension, but represents about half a MUSE spaxel in declination, which appears significant. When using the same small subset of stars to compare WFI and Gaia DR1 star positions, we consistently find a similar offset. We should emphasise that the stars in the MUSE FoVs represent much fewer than 1\% of the total number of stars used for the astrometric solution of WFI versus Gaia DR1.  Hence, if we now consider stars in the broadband FoV that are not covered by MUSE, but still represent more than 99\% of all available stars, we measure a shift of less than 1~mas, with scatters of 70~mas and 88~mas in RA and Dec, respectively.

All this suggests that we have a robust absolute and relative astrometric solution at the level of a MUSE sub-spaxel. There is, however, a clear residual offset with respect to Gaia DR1, observed only for stars that overlap with the PHANGS-MUSE galaxy targets. Such a residual shift may be due to systematics of the Gaia solutions associated with the presence of an extended emission background, or could be more generally driven by the difficulty to derive centroids of stars superimposed on bright emission. In any case, addressing this issue pertains to the WFI dataset and a more detailed analysis will be included in A.~Razza et al.\ (in preparation) when presenting the imaging data.

\subsubsection{Validation of the wavelength solution}
\label{sec:dap_validlambda}
As noted in \citet{Weilbacher+20}, temperature variations between the daytime arc lamp-based wavelength solution and nighttime observations can result in a zero-point shift of the wavelength calibration of up to 1/10$^{\rm th}$ of the MUSE spectral resolution. As a validation of our zero-point calibration, we compare the \ha\ velocity centroids, tracing the ionised gas kinematics, with the molecular gas kinematics, as traced by PHANGS-ALMA observations of the CO emission \citep{Leroy2021b}. While these two different gas phases are not necessarily co-spatial, they are predominantly constrained to the mid-plane of the disc and can reasonably be expected to globally trace the same kinematics. Here we used the first moment maps from the CO observations, and convert them to a heliocentric reference frame using the optical velocity definition. To minimise uncertainties and ensure we exclude most of the DIG, we require a $\mathrm{S/N} > 20$ on the \ha\ line.  We find the \ha\ velocities are systematically offset by $0.6$~km~s$^{-1}$ towards smaller values compared to the CO, corresponding to 1/100$^{\rm th}$ of a MUSE channel, with ${\sim}10$~km~s$^{-1}$  scatter. This systematic agreement provides increased confidence in our overall calibration accuracy, while the variations could reflect differences in the physical impact of feedback processes on different phases of the ISM.

\subsubsection{Sky subtraction}
\label{sec:overlap}
\begin{figure*}
\centering
        \includegraphics[width=0.8\textwidth]{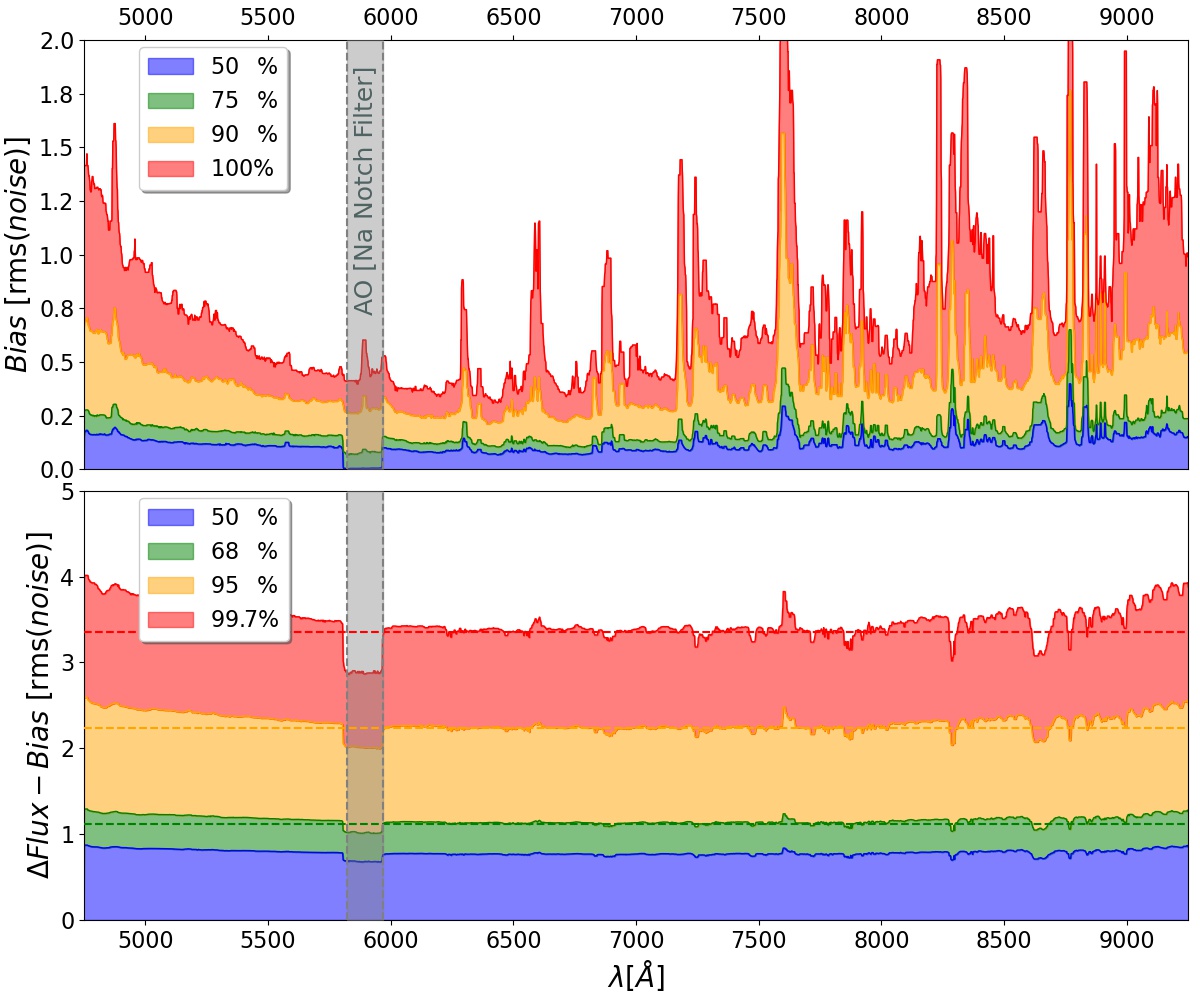}
     \caption{Median offset and percentiles of the difference between spectra of the same regions from overlapping pointings.
     The statistics have been computed from a set of 249 regions covering the full PHANGS-MUSE sample, and the resulting percentile vectors have been filtered to make it legible (keeping the sky line residuals visible). {\rm Top panel}: Percentiles of the distribution of the bias level, normalised by the typical (overlap-region-averaged) noise level. Ninety percent of the spectra have a \protect{normalised} bias that is typically between 30 and 50\% of the noise level, while a small fraction show up at levels of $60{-}120$\% of the noise, specifically in the blue or red part of the spectrum. We note that although beyond $7000$~\AA\ residuals are heavily contaminated by sky line residuals, the pipeline still constructs a roughly correct noise vector. {\rm Bottom panel}: Percentiles (50, 68.3, 95.5\%, and 99.7\%) of the distribution of differences normalised by the individual spectra noise level, after subtraction of a wavelength-independent median bias offset (see text). The dashed (respectively, green, yellow, and red) lines show values of $1.12$ (12\% above~1), $2.24,$ and $3.36$, showing that the noise level is slightly underestimated (by about 12\%). A trend is visible towards the redder and bluer end of the wavelength coverage. }
     \label{fig:compover}   
\end{figure*}

As described in Sect.~\ref{sec:skyback}, the sky continuum spectrum subtracted from individual exposures is computed from dedicated sky exposures. Unfortunately, the overall sky brightness as well as the detailed properties of the sky spectrum generally vary between and during exposures on a timescale of a few minutes. We therefore expect systematic differences (shape, absorption features) between the actual sky continuum in a given science exposure and the sky continuum extracted from the closest sky exposure. To control for any overall shift in the normalisation of the sky brightness, we apply to our MUSE data a normalising factor, deduced from a reference broadband image as described in Sect.~\ref{sec:skyback}. This normalisation, however, cannot address any wavelength-dependent changes in the sky spectrum, for example slope or global shape. Using our MUSE data, we can estimate the impact of these changes by comparing either spectra from individual exposures within the same pointing (assuming sky variations are the main source of variations once any astrometric shifts have been corrected) or spectra from overlapping pointings. Here we favoured the latter approach, as we can use pointings made of several merged exposures, thus optimising the S/N of the evaluated spectra. In this section, we briefly report on such a systematic comparison.

We identified all overlapping regions between different MUSE pointings throughout our sample (as a reminder, such regions are generally about $2\arcsec$ wide; see Sect.~\ref{sec:skyback}). For each such overlap region we selected a set of $n_\mathrm{s}$ spaxels and form all $n_\mathrm{s}$ pairs of corresponding spectra (Spec$_\mathrm{P1}^i$, Spec$_\mathrm{P2}^i$,  $i \in [1 - n_\mathrm{s}]$), where P1 and P2 refer to a generic Pointing~1 and Pointing~2. We discarded pairs of spaxels located at the edges of the pointings using a binary erosion process, thus focusing on spectra within the central part of the overlapping region. We further discard overlap regions containing fewer than $500$ spaxels. The process finally generated a set of $249$ overlap regions with a minimum number of common spaxels.

For each overlap region, we then derived the median of all the (paired) differences ${\rm Spec}^\mathrm{med}_\mathrm{P1/P2} = {\rm median}({\rm Spec}_\mathrm{P1}^i - {\rm Spec}_\mathrm{P2}^i)$: this represents the median `bias' offset between P1 and P2 as a function of wavelength. We compare the median bias to the typical noise level in the spectra over a given overlap region, taken as the square root of the average variance of each pair ($\langle{\rm Var}_\mathrm{P1/P2}\rangle = \sum_i ({\rm Var}_\mathrm{P1}^i + {\rm Var}_\mathrm{P2}^i)/2$). The result is illustrated in the top panel of Fig.~\ref{fig:compover} where we see the median bias typically represents about 10\% (respectively, 15, 30, 50\%) of the noise level for 50\% (respectively, 75, 90, 100\%) of the spectra. There is a very significant trend towards the blue and red part of the wavelength coverage (with biases up to 150\% of the typical noise level). As expected, we also note that sky line residuals have a significant impact on this budget.

We then evaluated the residual difference between pairs of spectra, after removal of that median bias, as compared to the noise level (i.e. the variance of those spectra). This exercise is aimed at testing the reliability of the noise variance delivered by our data flow for individual pointings. This is illustrated in the bottom panel of Fig.~\ref{fig:compover} where we present some percentiles of its distribution, over the 249 overlap regions. Assuming a standard normal distribution of the noise in the spectra, we would expect the 68.3, 95.5 and 99.7\% percentiles of the absolute value of the differences to correspond to 1, 2, and 3 times the standard deviation. The figure shows that we generally underestimate the noise level by about 12\% (the three coloured dashed lines in the bottom panel corresponding to values of 1.12, 2.24, and 3.36). We also observe a clear trend with higher values towards the blue and red ends of the spectra, where the discrepancy goes from, for example, 12\% to ${\sim}30$\%. This is consistent with previous estimates; for example, \citet{Bacon2017} quote a re-scaling of the variance by a factor of about 1.3, which they attribute to the impact of the interpolation process and its impact on the noise covariance.

Overall, we conclude that we may slightly under-estimate the noise level by 10 to 30\% when using the derived variances (and ignoring the covariance terms), and that the bias due to improper sky continuum subtraction is present, but negligible for most of the spectra, but can be significant for about 10 to 20\% of them, especially towards the blue end of the MUSE wavelength range.

By construction, there are no offsets in the broadband colour reconstructed images of individual pointings. We confirm that we observe no systematic differences between adjacent pointings using such broadband filters, a good a posteriori check of our implementation. This is, however, not necessarily true for colours. The spectral dependence of the median bias suggests that the shape of the sky continuum spectrum used for the sky background subtraction process may sometimes depart from the true one. We interpret jumps in the stellar extinction maps (see Fig.~\ref{fig:example_ebv}) as a direct consequence of that discrepancy. Fixing such an issue would require a spectrally dependent correction of the reference sky spectrum itself. This may be addressed by using photometric reference points (e.g. \textit{HST} imaging) in several bands (as opposed to the single \textit{Rc}-band used here), but it is beyond the scope of the present release {(DR1.0)}.

\begin{figure*}
\centering
        \includegraphics[width=0.8\textwidth]{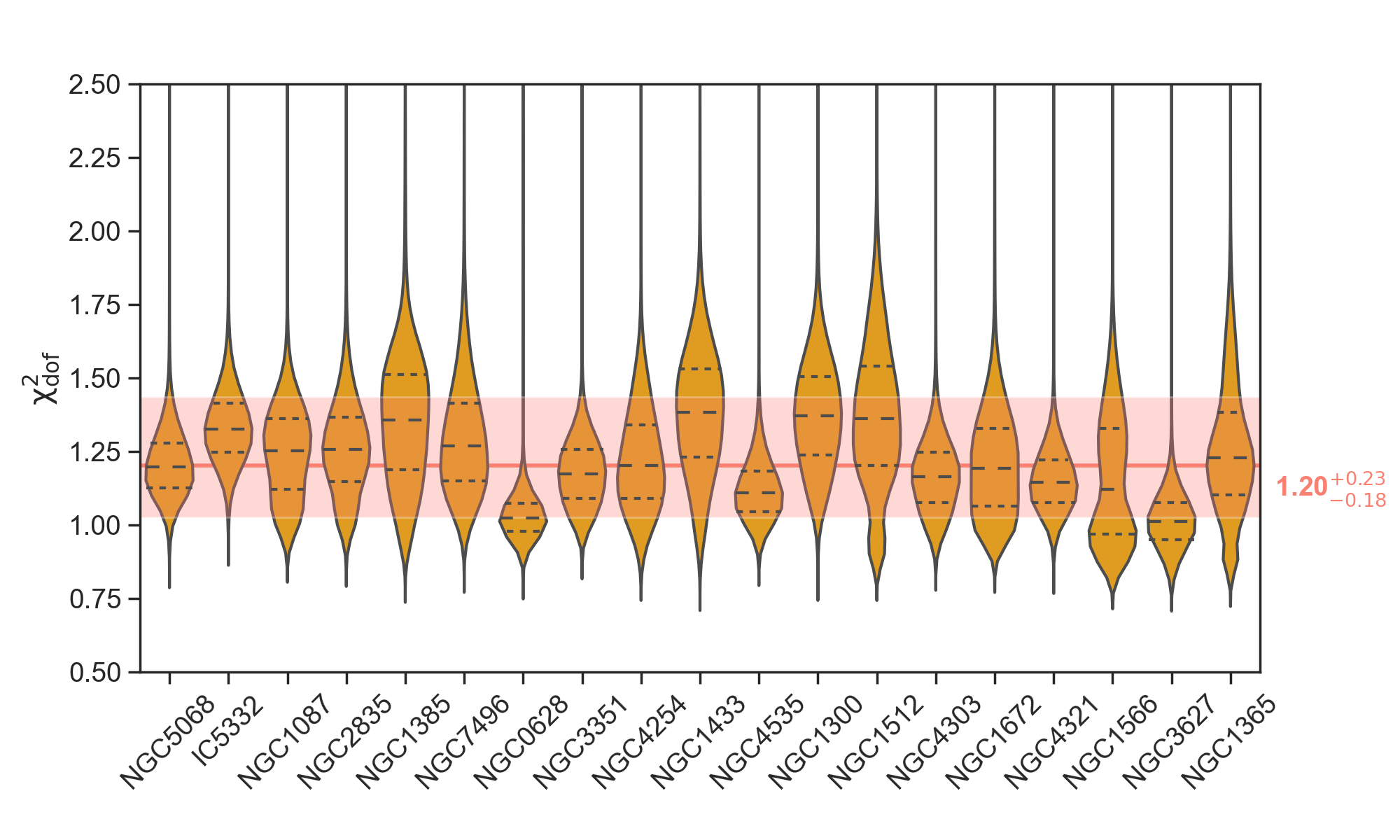}
     \caption{Histograms of the distributions of reduced chi-square, $\chi^2_{\rm dof}$, of the resulting spectral fits for all the spaxels in  all the galaxies in our sample. The reduced $\chi^2$ is calculated over the wavelength range fitted by the \DAP\ ($4850{-}7000$~\AA). The red horizontal line and shaded area corresponds to the median (16$^{\rm th}$ and 84$^{\rm th}$) percentiles of the distribution across all galaxies, corresponding to $\chi^2_{\rm dof} = 1.20 ^{+0.23}_{-0.18}$.}
     \label{fig:chi2_violin}   
\end{figure*}

\subsection{Quality assessment of the spectral fitting and analysis products}

In the following subsections we provide a brief assessment of the reliability of our spectral fitting procedure carried out by the \DAP. 
We focus on a few specific questions: the reliability of the error vectors and consequent $\chi^2$ of the spectral fits, the errors and detectability of emission lines, and the comparisons of our derived stellar masses with those obtained from $3.6~\mu$m imaging. Detailed discussion of the SFR (and extinction corrections) derived from our data and comparison with UV+IR estimates of SFR are presented in F.~Belfiore et al.\ (in preparation), and in a summarised form in \citet{Leroy2021b}, and will therefore not be repeated here.

\subsubsection{Overall quality of the spectral fits}
\label{sec:spectral_fits}

In order to validate the overall performance of the \DAP\ we investigate the quality of spectral fits in terms of the {reduced $\chi^2$} ($\chi^2_{\rm dof}$). If the errors provided by the pipeline are correct, we expect $\chi^2_{\rm dof} \sim 1$ for good spectral fits. Vice versa, assuming that the spectral fits are correct, we may consider to re-scale the average value of $\sqrt{\chi^2_{\rm dof}}$ as a correction factor to apply to the error vectors to bring them in good agreement with the residuals. 
In Fig.~\ref{fig:chi2_violin} we show histograms of the {reduced} $\chi^2_{\rm dof}$ distributions over the full fitted wavelength range fitted by the \DAP\ ($4850{-}7000$~\AA) for the galaxies in our sample. Galaxies are ordered by stellar mass from left to right. The 25$^{\rm th}$, 50$^{\rm th}$, and 75$^{\rm th}$ of each distribution are marked with dashed lines. The red horizontal line corresponds to the median value (16$^{\rm th}$ and 84$^{\rm th}$ percentiles shaded) across all galaxies $\chi^2_{\rm dof} = 1.20 ^{+0.23}_{-0.18}$. The median {reduced} $\chi^2_{\rm dof}$ value for individual galaxies ranges from $1.0$ to~$1.4$. This demonstrates that our spectral fits lead to residuals in good agreement with the error vectors provided by the pipeline. In all galaxies, however, a very small number of spaxels show much larger $\chi^2_{\rm dof}$ (only 0.1\% of spaxels have $\chi^2_{\rm dof}> 3$). We can gain more insight into these extreme regions by locating them spatially within our galaxies.
\begin{figure*}
\centering
        \includegraphics[width=0.99\textwidth]{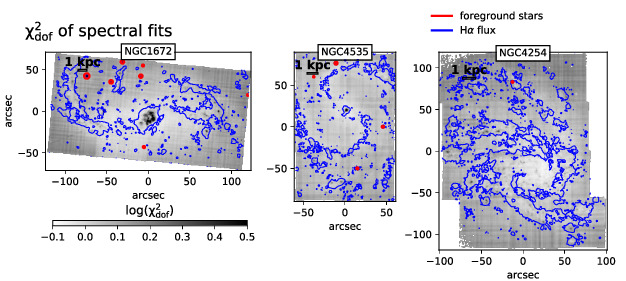}
     \caption{Maps of the $\chi^2_{\rm dof}$ for three example galaxies \protect{at the individual spaxel level}. We show in blue contours the \ha\ surface brightness and in red the regions masked as foreground stars. The $\chi^2_{\rm dof}$ maps demonstrate the quality of our spectral fits $\chi^2_{\rm dof} \sim 1$, except in a few regions of bright H$\alpha$ emission (e.g. the central regions in NGC~1672 and NGC~4535). A few regions of higher $\chi^2_{\rm dof}$ can also be seen in the middle of spiral arms, corresponding to the centres of bright \hii\ regions. Foreground stars appear as small spots of very high $\chi^2_{\rm dof}$ and are emphasised with red contours.}
     \label{fig:chi2_maps}   
     \includegraphics[width=0.99\textwidth]{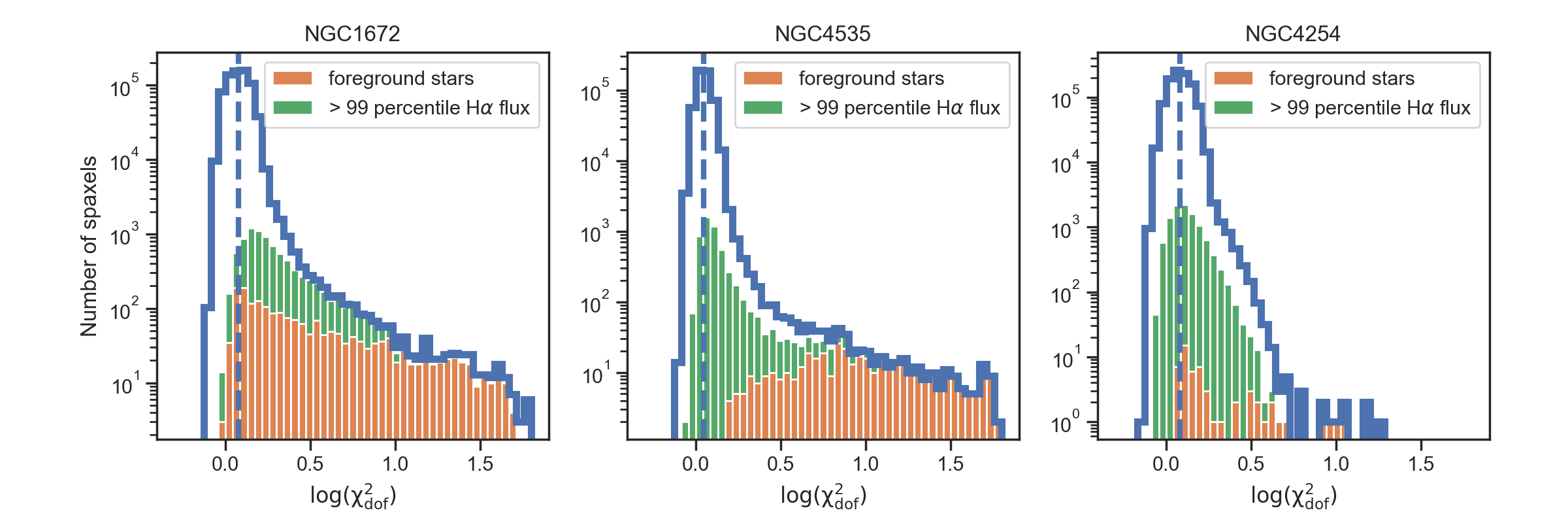}
     \caption{Histogram of the distributions of $\chi^2_{\rm dof}$ (in blue) for the same example galaxies as in Fig.~\ref{fig:chi2_maps}. The median $\chi^2_{\rm dof}$ is shown as a blue vertical dashed line. We show as a bar plot in orange and green the stacked histograms of pixels belonging to our foreground star masks (orange) and to the brightest line-emitting regions (${>}99^{\rm th}$ percentile of the \ha\ emission, in green). Because the bar plots are stacked, their sum must be smaller than the histogram of all spaxels in blue. This allows us to visually assess that the vast majority of high-$\chi^2_{\rm dof}$ pixels are due to foreground stars and bright line-emitting regions.}
     \label{fig:example_chi2} 
\end{figure*}

In Fig.~\ref{fig:chi2_maps} we show example {reduced} $\chi^2_{\rm dof}$ maps for three galaxies in our sample. As expected, the median {reduced} $\chi^2_{\rm dof}$ is very close to~1 (1.2, 1.1, and 1.2 for NGC~1672, NGC~4535, and NGC~4254, respectively). We also see the expected pointing to pointing variations (due to the slight differences in the noise levels across pointings) and the cross-hatch pattern, which has already been discussed in Sect.~\ref{sec:binning} (see Fig.~\ref{fig:noise}). It is also evident that a few localised regions have much higher {reduced} $\chi^2_{\rm dof}$ than the median. One such class of deviant regions is represented by foreground stars, which are enclosed in red contours in Fig.~\ref{fig:chi2_maps} (as defined by our foreground star masks; see Sect.~\ref{sec:starmask}). In addition, regions of very bright line emission, both in the middle of spiral arms (traced by the \ha\ emission, shown with blue contours) and the galaxy centres (in NGC~1672 and NGC~4535 in particular) show enhanced {reduced} $\chi^2_{\rm dof}$. We interpret this as a consequence of non-Gaussianity in the bright emission lines (see Sect.~\ref{sec:em_errors}), causing strong residuals at their specific wavelengths.

Finally, we observe (especially in NGC~4254) a trend of increasing $\chi^2_{\rm dof}$ with galactocentric distance. This effect is likely the consequence of the presence of sky residuals in the spectra. These residuals contribute to the $\chi^2_{\rm dof}$ in regions of low surface brightness, while in regions of higher surface brightness the Poissonian errors from the continuum become dominant. Therefore, at high surface brightness levels pipeline errors are a better representation of the model residuals, resulting in {reduced} $\chi^2_{\rm dof}$ {closer to unity}.

We investigate further whether foreground stars and bright emission lines can explain all the high $\chi^2_{\rm dof}$ spaxels by presenting in Fig.~\ref{fig:example_chi2} a histogram of the $\chi^2_{\rm dof}$ distributions for our three example galaxies (blue histograms), {computing such a quantity at the individual spaxel level}. We plot the $\rm log(\chi^2_{\rm dof})$ on the x-axis and employ a logarithmic y-axis to emphasise the small tail of spaxels at high $\chi^2_{\rm dof}$ values. We also present as stacked barplots the distributions of pixels associated with foreground stars (orange) and bright line emission (parametrised here as the 1\% of spaxels with brightest \ha\ emission, green). Since the orange and green barplots are stacked on each other, their total height would reach the blue histogram if all the high-$\chi^2_{\rm dof}$ spaxels fall into these two categories. Overall, we see that for NGC~1672 and NGC~4535 we explain virtually all the spaxels $\chi^2_{\rm dof}>3$ with either foreground stars or bright line emission. The absence of foreground stars in NGC~4254 creates a $\chi^2_{\rm dof}$ with a much more limited tail, which is overall largely explained by bright line emission.

We conclude noting that the residuals from our spectral fits are in excellent agreement with the pipeline noise. If we take the median values of $\chi^2_{\rm dof} = 1.20$ and assume that our fits are good, then we would conclude that the noise vectors provided by the pipeline ought to be corrected upwards by 10\% on average, in excellent agreement with the 12\% estimates obtained from the overlap region statistics in Sect.~\ref{sec:overlap}. Such a level of agreement can be considered extremely satisfactory, and highlights the robustness of the error vectors provided the \MUSEp. Spaxels with highly deviant $\chi^2_{\rm dof}$ values are explained as either foreground stars, which can be masked efficiently by the masks we provide, or regions of bright line emission, where the emission line residuals dominated the overall spectral $\chi^2_{\rm dof}$.

\subsubsection{Reliability of errors for emission line fluxes}
\label{sec:em_errors}

We obtain emission line errors directly from the output of \pPXF,\, which obtains them as the output of its non-linear fitting stage, performed via Levenberg-Marquardt minimisation, using a \python\ version of the well-established \texttt{mpfit} routine, originally written in IDL by Markwardt (2009). Several authors have found that when fitting emission lines with single Gaussian the formal errors obtained from this procedure are accurate to within a few percent \citep{Ho2016, Belfiore2019}. We performed our own simple recovery simulations, where noise is added to a model spectrum consisting of Gaussian emission lines, to assess the validity of the errors under such idealised conditions, and find, in agreement with previous work, that the flux is recovered with no bias and flux errors are correct to within better than 5\% percent at all flux levels.

Users interested in errors for weak lines may wish to consider the underestimation of the noise vector of ${\sim}12$\%, discussed in Sects.~\ref{sec:overlap} and~\ref{sec:spectral_fits}. Weak lines that are found in regions of enhanced sky residuals should also be considered carefully. A subtler effect affects our estimates of the errors for very strong emission lines. It is found empirically that for $\mathrm{S/N} > 30$, residuals from the Gaussian fit are generally larger than the errors, leading to an increase in $\chi^2$ of the fit within the core of the line as a function of increasing S/N \citep[see Fig.~3 in][]{Belfiore2019}. 


\begin{figure*}
    \centering
    \includegraphics[width=7in]{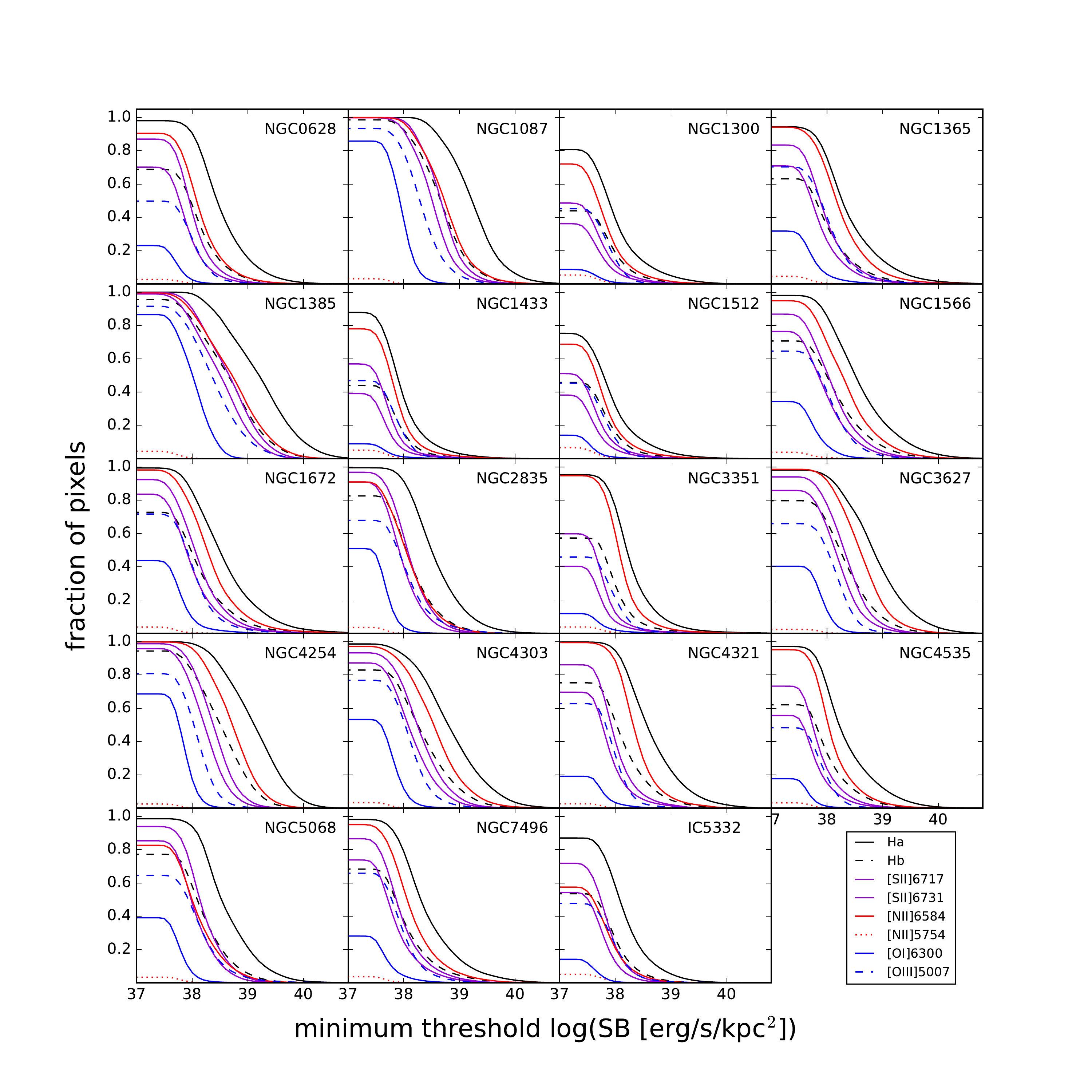}
    \caption{Fraction of 0.2\arcsec\ spaxels inside of $0.5\,R_{25}$ that have $3\sigma$ detections above a given surface brightness threshold (SB) for a representative sample of emission lines. H$\alpha$ is detected in upwards of 95\% of all pixels in most galaxies, with a lower ${\sim}80$\% fraction in some strongly barred systems (e.g. NGC~1300, NGC~1433, and NGC~1512).  Typical 3$\sigma$ flux sensitivity in H$\alpha$ is 
    $4{-}7 \times 10^{37}$ erg~s$^{-1}$~kpc$^{-2}$ ($3{-}7 \times 10^{-19}$ erg~s$^{-1}$~cm$^{-2}$ per 0.2\arcsec\ spaxel).
    H$\beta$ is typically detected in $50{-}80$\% of spaxels.  Low-ionisation lines (\nii6584, \sii6717, and \sii6731) are detected in $60{-}95$\% of spaxels, while the high-ionisation \oiii5007 line emission is less common ($50{-}70$\% of pixels). For contrast, the faint auroral \nii5754 emission line is detected at a 3$\sigma$ level in ${\sim}5$\% of spaxels (though fewer than 1\% are detected at 5$\sigma$).}
    \label{fig:line_fractions}
\end{figure*}

\subsubsection{Emission line detection thresholds}
\label{sec:elthreshold}
We demonstrate the sensitivity of our data by plotting the fraction of pixels above a given surface brightness threshold, for a representative sample of emission lines mapped within $0.5\,R_{25}$ and at $\mathrm{S/N} > 3$ (Fig.~\ref{fig:line_fractions}). This provides a more quantitative illustration than multi-line maps (Fig.~\ref{fig:3emicol}), which we hope may be of use to the reader in proposal and project planning. Our brightest line, \ha, is detected in ${\sim}95$\% of all pixels for most galaxies. Typical $3\sigma$ flux sensitivity in \ha\ is $4{-}7 \times 10^{37}$ erg~s$^{-1}$~kpc$^{-2}$ ($3{-}7 \times 10^{-19}$ erg~s$^{-1}$~cm$^{-2}$ per pixel). \hb\ is detected in $50{-}80$\% of pixels, which has implications for our ability to perform matched-resolution extinction corrections based on the Balmer decrement (see also Sect.~\ref{sec:SciCase_emline}). Low-ionisation lines (\nii6584, \sii6717, and \sii6731) are detected in $60{-}95$\% of pixels, while the high-ionisation \oiii5007 line emission is less common ($50{-}70$\% of pixels). The faint auroral \nii5754 emission line is detected at a 3$\sigma$ level in ${\sim}5$\% of pixels (though fewer than 1\% are detected at ${>}5\sigma$). Even moderate spatial binning can help significantly with the detection of low-surface-brightness emission and fainter auroral lines \citep[e.g.][]{Santoro2021,Belfiore2021}.

\subsubsection{Stellar masses}
\label{sec:s4g}

\begin{figure*}
    \centering
    \includegraphics[width=\textwidth]{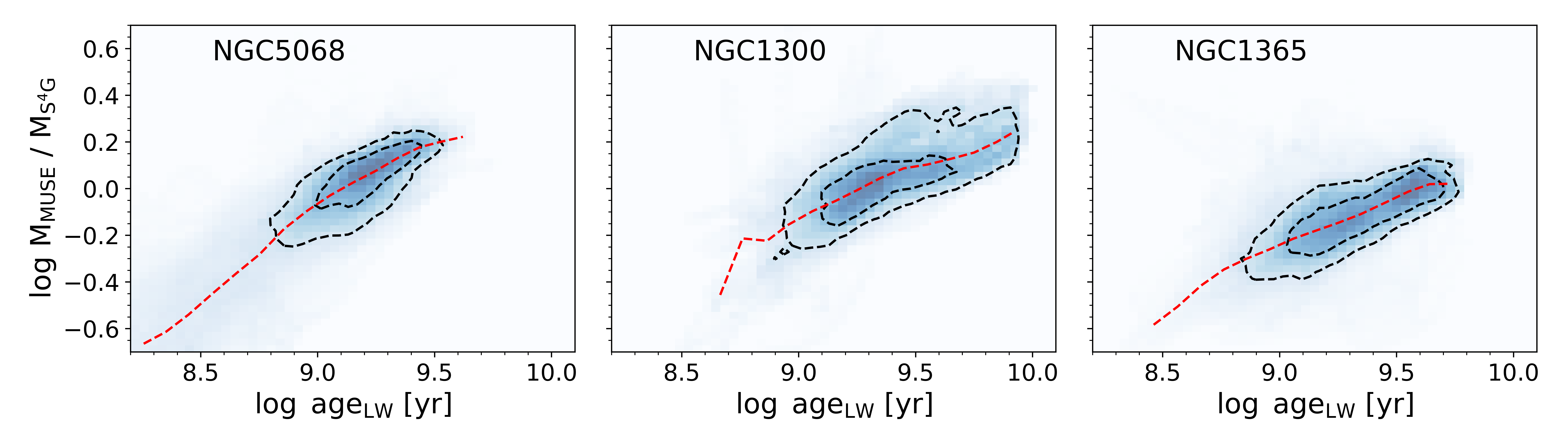}
    \caption{Two-dimensional histogram of the ratio between the stellar mass surface density derived from full spectral fitting from MUSE (Sect.~\ref{sec:stellarpops}) and archival stellar mass maps from S$^4$G, based on \textit{Spitzer} IRAC maps \citep{Querejeta2015}, as a function of the light-weighted age of the underlying stellar population in each pixel, derived from the full spectral fitting. We show three galaxies in our sample, spanning the stellar mass range of the PHANGS-MUSE sample (in ascending order from left to right). The figure shows good agreement between the MUSE and the S$^4$G mass maps in pixels dominated by stellar populations older than ${\sim}4$~Gyr. However, in regions hosting younger populations, the masses derived from the near-IR data are systematically larger, possibly due to the contribution of AGB stars to the near-IR flux. The contours mark the iso-probability curves where the probability density drops below $40\%$ and $10\%$ of the maximum. The dashed red line shows the median mass ratio at a given luminosity-weighted age.}
    \label{fig:MUSE-S4G-MassRatio}
\end{figure*}

We now compare the values of stellar mass surface density derived from our stellar population analysis (Sect.~\ref{sec:stellarpops}) with those derived by  \citet{Querejeta2015} using the $3.6~\mu$m and $4.5~\mu$m IRAC bands from the \textit{Spitzer} Survey of Stellar structure in Galaxies \citep[S$^4$G;][]{Sheth2010}.
The S$^4$G work uses the independent component analysis (ICA) presented in \citet{Meidt2014} to separate the contributions of the stellar and the dust emission to the IRAC fluxes and  to derive a relation that allows the M/L to be obtained using the $[3.6] {-} [4.5]$ colour, assuming that the contribution to the stellar light at those wavelengths is dominated by an old population with an almost constant M/L ratio. 

We convolved the MUSE maps to an angular resolution of $1.5\arcsec$, consistent with that from the S$^4$G maps, and computed, for each pixel, the ratio $M_{\rm MUSE} / M_{\rm S^4G}$, where $M_{\rm MUSE}$ and $M_{\rm S^4G}$ correspond to the stellar mass derived from our MUSE data and S$^4$G data, respectively. On average, we found that $M_{\rm MUSE}$ values are about 30$\%$ smaller than $M_{\rm S^4G}$, but the differences are strongly dependent on the age of the underlying stellar population estimated from the full spectral fitting.

This can be seen in Fig.~\ref{fig:MUSE-S4G-MassRatio}, where we show galaxies spanning the stellar mass range probed by PHANGS-MUSE, NGC~5068 being the least massive object in our sample, NGC~1300 close to the median mass, and NGC~1365 being the most massive galaxy. We show $\log(M_{\rm MUSE}/M_{\rm S^4G})$ as a function of the light-weighted age of each pixel. 
 As can be seen, both methods agree when the light is dominated by old stars, but start to diverge when the mean light-weighted age is lower than ${\sim}4$~Gyr. A similar trend was reported by \citet{deAmorim2017} using CALIFA data.  This is not surprising, as the S$^4$G calibration assumes that the stellar flux in the near-IR is  dominated by old stars. However, in stellar populations with ages of $1{-}2$~Gyr, the contribution of Asymptotic Giant Branch (AGB) stars to the near-IR flux can be dominant (leading to a different M/L than that appropriate for and old population) and all the pixels with a mean light-weighted age below ${\sim}4$~Gyr have a contribution of stars in this age range. It is worth mentioning that the age dependence of the mass difference is much stronger when a constant M/L ratio is applied to the original 3.6~$\mu$m image (i.e. the ICA has some effect in compensating for the excess of light in regions that host young stellar populations) but does not completely remove the age trend.
 
\begin{figure}
    \centering
    \includegraphics[width=\columnwidth]{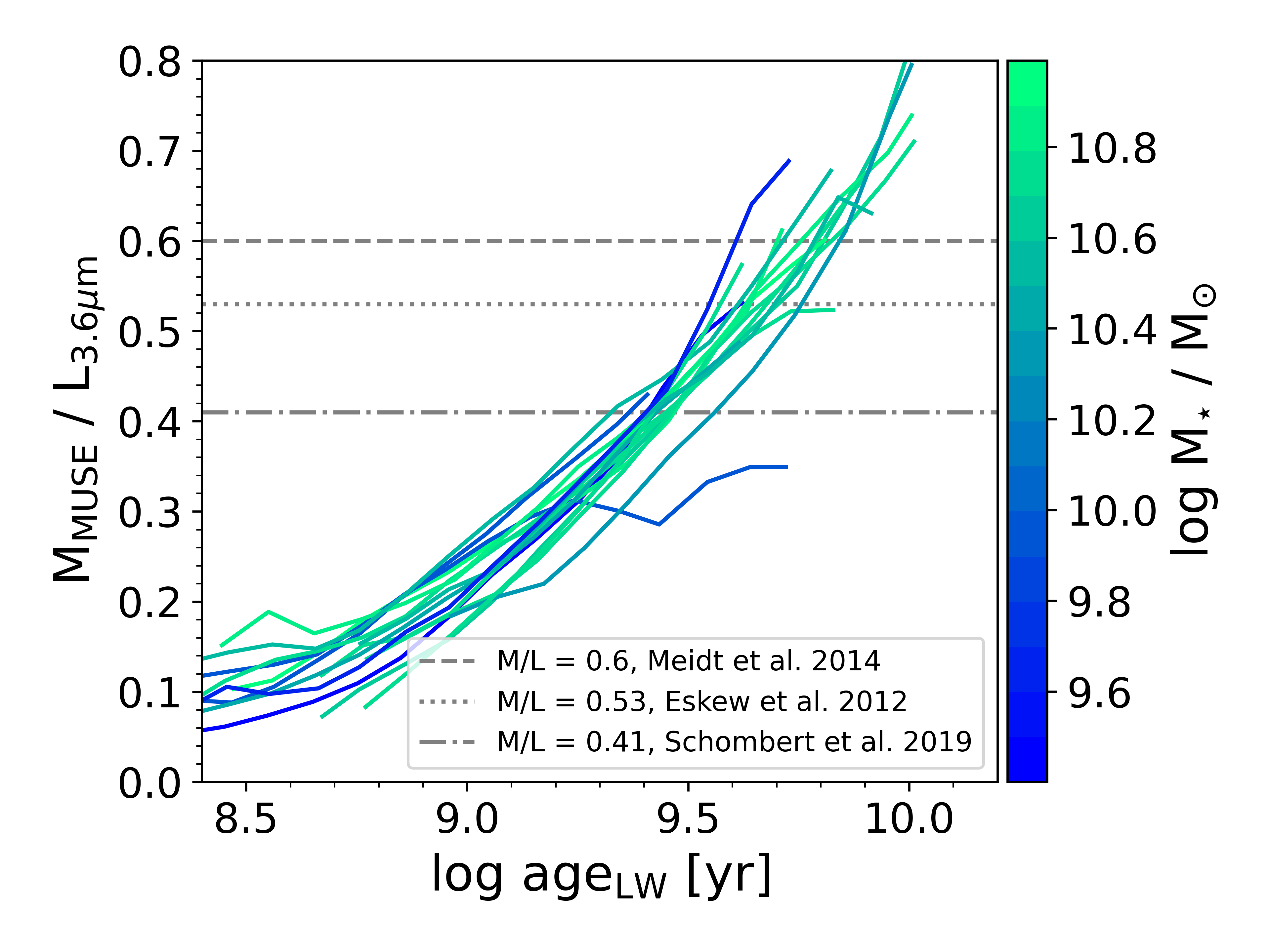}
    \caption{Age dependence of the $\mathrm{M}_\mathrm{MUSE}/\mathrm{L}_{3.6\,\mu\mathrm{m}}$ ratio \protect{(in M$_{\odot}$/L$_{\odot}$)} for the galaxies in our sample, where $\mathrm{L}_{3.6\,\mu\mathrm{m}}$ corresponds to the luminosity of a pixel in the original IRAC $3.6~\mu$m image (not the ICA version from S$^4$G). Each line represents a galaxy, coloured by its total stellar mass. The figure shows a positive trend, with pixels hosting older stellar populations having larger M/L ratios. The horizontal lines mark different values adopted in the literature (dot-dashed line $ \mathrm{M/L}=0.41$, \citealt{Schombert2019}; dotted line, $\mathrm{M/L}=0.53$ \citealt{Eskew2012}; dashed line $\mathrm{M/L}= 0.6$, \citealt{Meidt2014}).}
    \label{fig:MUSE-S4G-M2L-calib}
\end{figure}

 Figure~\ref{fig:MUSE-S4G-M2L-calib} shows the median $\mathrm{M/L}_{3.6\,\mu\mathrm{m}}$ ratio of a pixel, at a given light-weighted age, calculated as the ratio $\mathrm{M}_\mathrm{MUSE} / \mathrm{L}_{3.6\,\mu\mathrm{m}}$, where $\mathrm{L}_{3.6\mu\mathrm{m}}$ corresponds to the luminosity of a pixel in the original IRAC $3.6~\mu$m image (not the ICA version from S$^4$G), in solar units. Each line represents a galaxy, coloured by its total stellar mass. The typical dispersion in the M/L ratio measurement at a given age across galaxies is ${\sim}0.03$. The figure shows little variations of this trend among galaxies, with no obvious correlation with total stellar mass. 
 The horizontal lines in Fig.~\ref{fig:MUSE-S4G-M2L-calib} correspond to $\mathrm{M/L}$ values commonly adopted in the literature. The value of \citet{Meidt2014} was intended to be applied to the  dust-corrected images (i.e. after applying the ICA to remove dust emission), and is therefore higher than the other values from the literature and closer to the value expected for a very old stellar population. 
 
 \cite{Leroy2021b} compare the MUSE mass maps presented here with their \textit{GALEX}+ \textit{WISE} version from z0MGS \citep[following][]{Leroy2019}.
 As an alternative to the ICA procedure, they use a M/L that scales with \textit{GALEX} + \textit{WISE} colours to compensate for the impact of specific SFR on the $3.6~\mu$m or $3.4~\mu$m flux, in the range of $0.2 < \mathrm{M/L} < 0.5$. Our mass maps agree well with this range of values. However, they also find an offset of $0.08$~dex between their masses inferred from \textit{GALEX} + \textit{WISE} imaging and our MUSE-derived mass maps. The offset is found to be roughly independent of specific SFR, which can be understood as a proxy for stellar age, and is therefore likely associated with other systematic differences in the stellar population modelling (e.g. differences in the Single Stellar Populations -- SSP -- models).  

\subsection{Known systematic errors in the stellar population maps}

\subsubsection{Imperfect sky subtraction: Effect on stellar extinction}
\label{sec:extinction}

\begin{figure*}
        \includegraphics[width=0.99\textwidth, trim=0 0 0 0, clip]{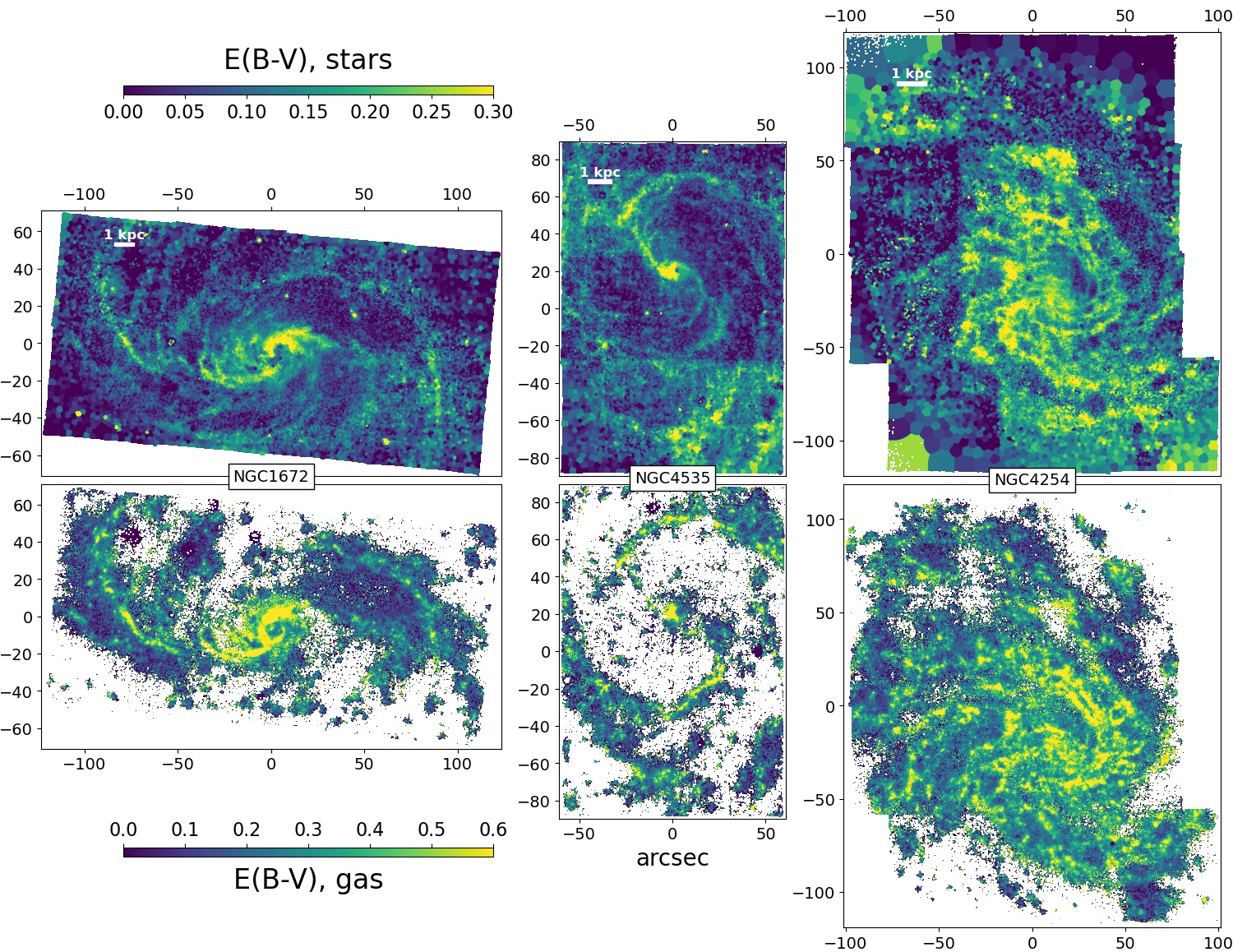}
     \caption{Stellar and gas $E(B-V)$ for three example galaxies. The stellar ${E(B-V)}_\mathrm{stars}$ is determined from the Voronoi binned data cubes used for stellar population analysis, while the gas ${E(B-V)}_\mathrm{gas}$ is computed from the Balmer decrement, derived from the spaxel-level analysis of the ionised gas emission lines (only regions with $\mathrm{S/N} > 4$ in the line emission are considered). The colour bars for the gaseous $E(B-V)$ is stretched by a factor of two to account for the average ratio ${E(B-V)}_\mathrm{stars} \sim 0.5\, {E(B-V)}_\mathrm{gas}$. ${E(B-V)}_\mathrm{stars}$ traces similar structures as ${E(B-V)}_\mathrm{gas}$, although jumps are evident when comparing the average level of some MUSE pointings to that of their neighbours (e.g. the bottom-right pointing in NGC~4535 and the three left-most pointings in NGC~4254).}
     \label{fig:example_ebv}   
\end{figure*}

As discussed earlier, the stellar $E(B-V)$ map shows clear systematic jumps between different MUSE pointings. These jumps are caused by spectral differences in the sky continuum subtraction across adjacent pointings in the mosaic (see Sect.~\ref{sec:skyback}). We use the overlap regions among adjacent MUSE pointings to quantify the impact of the different continuum levels on the recovered stellar extinction. We computed the stellar $E(B-V)$ for this set of $2 \times 249$ spectra (two per overlap region), extracted from the individual overlap regions. For each pair of spectra, we calculated $\Delta{E(B-V)}$, as the absolute difference in the extinction between the two. We therefore measured $249$ values of $\Delta{E(B-V)}$ across our full sample obtaining a median value (as well as the 68, 95 and 99.7$^{\rm th}$ percentiles, respectively) of $\Delta{E(B-V)} = 0.037$ ($0.058$, $0.133$ and $0.261$, respectively). This test implies than, although most of the pointings are relatively homogeneous with respect to their neighbours, with $\Delta{E(B-V)} < 0.04$, a fraction of them show larger spectral differences that lead to systematic offsets in the derived stellar extinction, up to $\Delta{E(B-V)} \approx 0.3$, in agreement with the pointing-to-pointing jumps visible in Fig.~\ref{fig:example_ebv}. Fixing such an issue would require the usage of different sky continuum reference spectra for individual exposures, something that is envisioned for future {public} data releases (see Sect.~\ref{sec:future}).

\subsubsection{Systematic errors in the stellar population fits at young ages}
\label{sec:youngbias}

While examining the maps associated with the stellar population fitting (see Sect.~\ref{sec:stellarpops}), we noticed the presence of low metallicity values ($\mathrm{LW~[Z/H]} < -1.3$) in a few regions encompassing very young stellar clusters ($\mathrm{LW~age} < 400$~Myr). Such low metallicity values would be inconsistent with an internal and progressive chemical enrichment of the ISM \citep[e.g.][]{Ho2017}. This suggests that the fitting process converges towards a misleading local minimum, the bluest available stellar template, constrained by the youngest age bin (30~Myr) of the implemented template library. These low-metallicity regions usually coincide with strong \ha\ emission, indicating that these young clusters coincide with active star formation. In addition to the lack of younger templates (due to the low number of young and metal-poor stars in the E-MILES library), contributions from nebular emission as well as unmasked emission lines are expected to further impact the $\chi^2$ minimisation in such `young' regions. A visual inspection of the spectral fits for some associated spaxels revealed a systematic overestimation of the stellar continuum at wavelengths bluer than ${\sim}5100$~\AA. We therefore deem these age and metallicity measurements (as well as $E(B-V)$) unreliable. This issue has already been reported in several studies \citep[e.g.][]{Carrillo2020, Bittner2020}, who similarly reported unexpected young and metal-poor regions.

\begin{figure*}
        \includegraphics[width=0.99\textwidth, trim=0 0 0 0, clip]{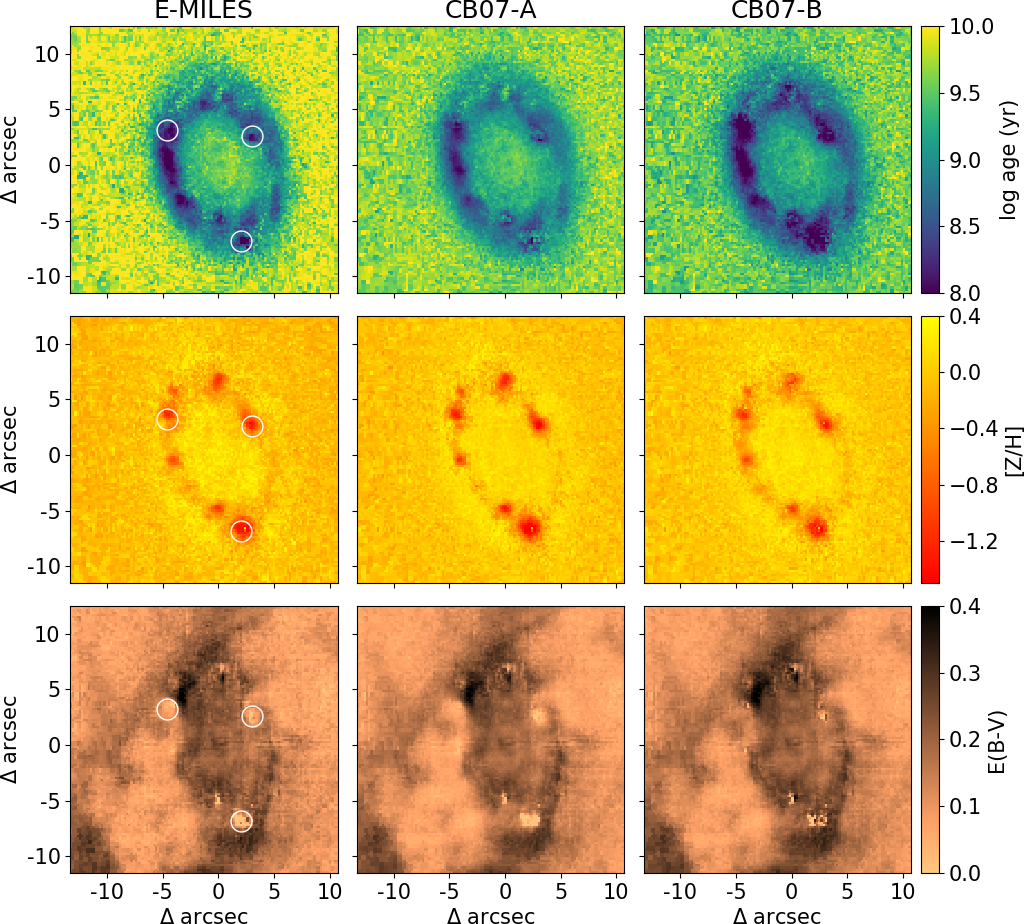}
     \caption{Age (top row), metallicity (middle row), and extinction (bottom row) of the inner star-forming ring of NGC~3351 obtained using three different sets of templates: E-MILES (left column), CB07, excluding templates younger than 30~Myr (CB07-A; central column), and CB07 including young templates (CB09-B; right column). Adding younger templates to our template grid slightly alleviates the problem of extremely metal-poor and young regions in the sense that the young clusters are younger and mildly less metal-poor. However, the improvement is marginal, and these regions are still significantly more metal-poor than the surrounding pixels. Panels in the last row show, additionally, an abnormally low extinction in the young and metal-poor regions \protect{(marked as white circles in the left column)}. This is likely a consequence of the same issue, as it is partially improved when younger templates are included. Overall, adding templates younger than 30~Myr to our age-metallicity grid does not provide a solution to the issue of young and extremely metal-poor regions.}
     \label{fig:test_young_templates}   
\end{figure*}

We explored whether adding younger templates to our age-metallicity grid is sufficient or not to overcome this issue. To this end, we used SSP models from the Bruzual \& Charlot evolutionary population synthesis database (Charlot and Bruzual 2007, private communication; CB07), computed assuming a Padova 1994 isochrone and a \citet{Chabrier2003} IMF. In order to account for differences driven by using a different stellar library, we defined two sets of CB07 templates: CB07-A, with five metallicity bins ([Z/H] = [$-1.7$, $-0.7$, $-0.4$, $0$, $0.4$]) and 16~age bins, log-spaced from 30~Myr to 20~Gyr (i.e. with the same low-age limit as E-MILES), and CB07-B, with the same metallicity bins, but with a larger age coverage, including 25~age bins ranging from 1~Myr to 20~Gyr .
Figure~\ref{fig:test_young_templates} presents the output LW age, LW [Z/H] and stellar $E(B-V)$, obtained with each different set of templates (E-MILES, CB07-A, and CB07-B) for the nuclear star-forming ring of NGC~3351.As expected, adding younger templates has an impact, in the sense that the young clusters are younger and mildly less metal-poor. However, the improvement is marginal, the derived metallicities for the above-mentioned regions being still significantly lower than for their surrounding spaxels. The associated low measured $E(B-V)$ is also likely driven by the same mechanism, biasing the blue end of the best-fit spectrum. We conclude here that including templates younger than 30~Myr unfortunately does not solve the degeneracy. A robust solution is beyond the reach of the present paper, and would require a broader consideration, examining a combined set of factors such as nebular emission (continuum and lines), template mismatch due to the currently poor observational constraints in the young and metal-poor regime, and potential sky-subtraction residuals.

\section{Data versions}
\label{sec:datarelease}

\subsection[]{The first PHANGS-MUSE public data release \protect{(DR1.0)}}
\label{sec:dr1}
PHANGS-MUSE is a legacy dataset, with wide applications beyond those pursued by the PHANGS team. To this end, we provide public data releases of science-ready pipeline products, including both the fully reduced mosaics {(at native, copt)}, and the \textsc{DAP} high-level data products (at the same two resolution levels). The science-ready data cubes and data products {are provided through two main links, via the ESO Archive tool\footnote{\url{http//archive.eso.org/scienceportal/home}} and also via the CADC portal, where the PHANGS-ALMA dataset is also available.

The first public PHANGS-MUSE data release (DR1.0) includes the datasets} and data products as described in Sects.~\ref{sec:drsprod} and~\ref{sec:dap_output}. {Maps associated with the stellar population analysis will be made available in subsequent releases, as well as other convolved datasets (e.g. 15\arcsec).} The {products associated with DR1.0} were available as an internal PHANGS data release (with minor modifications), and have been already exploited in the course of several studies (published or submitted at the time of this writing), including \cite{Leroy2021b, Pessa2021, Turner2021, Williams2021, Barnes2021, Beslic2021, Williams2022,Belfiore2021}.
An \hii\ region catalogue based on the PHANGS-MUSE data is presented \cite{Santoro2021}.

\subsection{Early versions of the PHANGS-MUSE data}
\label{sec:dr0}

Results from the PHANGS-MUSE survey have already been published using preliminary versions of the \DRP, coupled with different analysis methods. We briefly summarise here the differences between these early versions of the reduced PHANGS data and that presented in the {first} public PHANGS-MUSE data release (Sect.~\ref{sec:dr1}).

An initial sample of eight galaxies (IC\,5332, NGC\,0628, NGC\,1087, NGC\,1672, NGC\,2835, NGC\,3627, NGC\,4254, NGC\,4535) was reduced manually using {\tt esorex} pipeline recipes (i.e. without the {\tt pymusepipe} software framework). Stellar continuum fitting and emission line subtraction was performed using LZIFU \citep{Ho2016}. Further details on the data reduction and data analysis are provided in \cite{Kreckel2019}. This data version was used in associated early PHANGS papers \citep{Kreckel2017, Kreckel2018, Ho2019, Herrera2020, Kreckel2020}. Both the astrometric accuracy and the absolute photometric calibration were not yet fully developed for this data version. Absolute flux calibration was determined only for NGC~0628, and we find an overall agreement in the H$\alpha$ line flux between the older maps and the current \DAP\ products to within 2\% for this galaxy. Data from the remaining galaxies was used only to determine line ratios, and pixel-to-pixel comparisons between those maps and our latest version of the \DRP\ and \DAP\ find they underestimate line fluxes by up to 10\% at $\mathrm{S/N} > 5$ in the H$\alpha$ line flux, due to non-photometric observing conditions. Line ratios show significantly better systematic agreement, to within 3\% at $\mathrm{S/N} > 5$ in \nii/H$\alpha$. \hii\ region catalogues derived from both versions have astrometric agreement of $0.3\arcsec$ when cross-matching the unresolved objects.

Finally, the \DRP\ and \DAP\ described in this paper were used, in a preliminary form, for the development of tools internally to the PHANGS collaboration. The associated preliminary dataset for NGC\,0628 was specifically used by \citet{Andrews2021}. The main limitation of these early data versions is in the astrometric calibration, and it did not significantly impact the results in that paper.

\subsection{Future developments}
\label{sec:future}
\begin{figure*}
    \centering
    \includegraphics[width=7in]{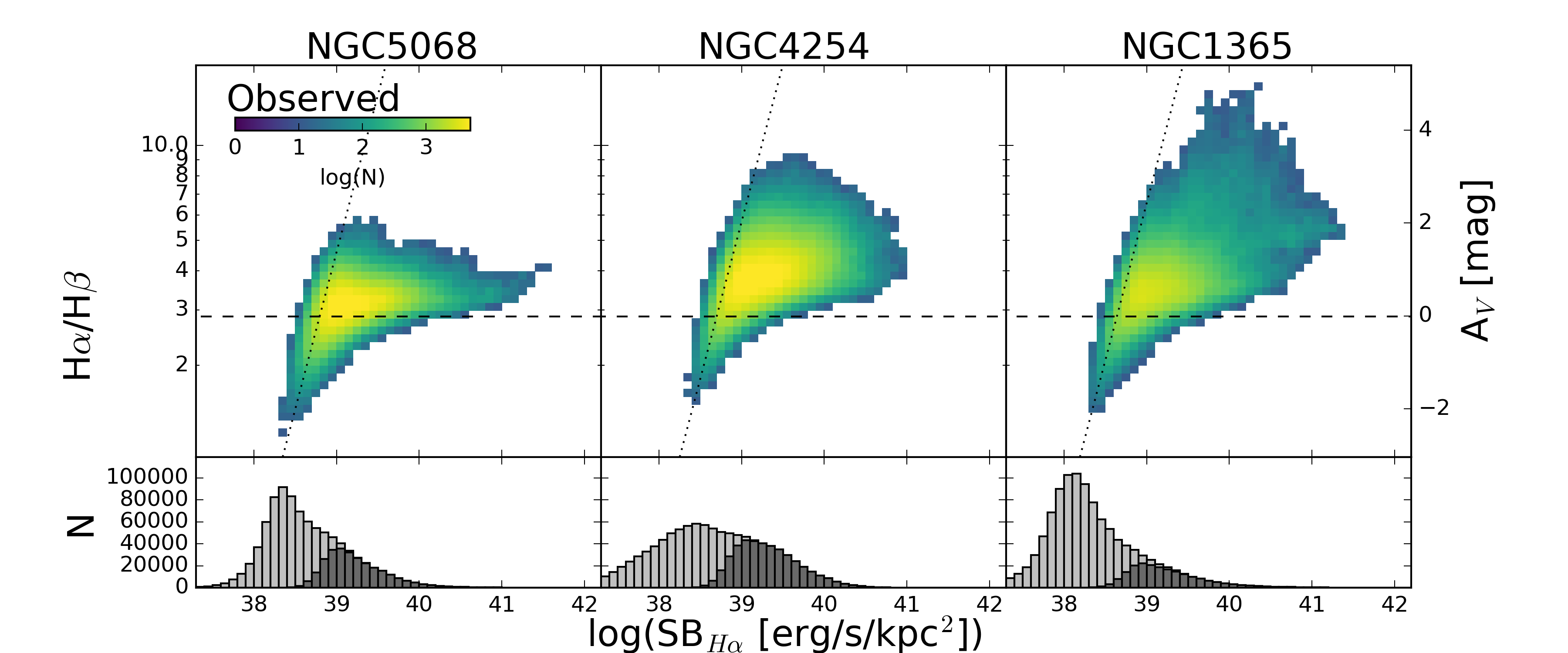}
    \includegraphics[width=7in]{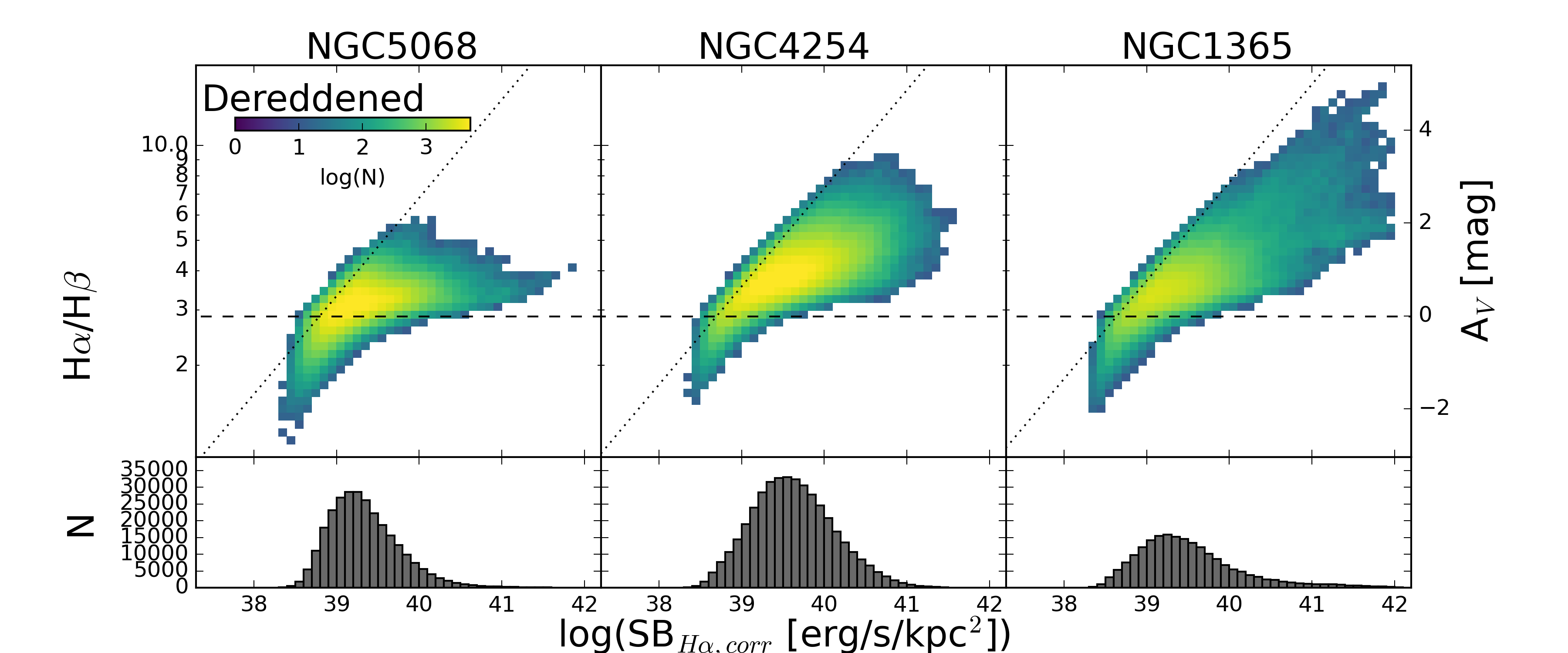}
    \caption{Distribution of pixel-based \ha/\hb\ line ratios as a function of observed (top) and extinction corrected (bottom) \ha\ surface brightness (SB$_{\mathrm{H}\alpha}$) for three galaxies of our sample (see text). We employ a S/N cut of 10 for both lines. Most pixels have \ha/\hb\ $> 2.86$ (dashed line, as predicted by case~B recombination for $T_\mathrm{e} = 10^4$~K, $n_\mathrm{e} = 100$~cm$^{-3}$), and those that have ratios below this value are consistent within the errors. Because of this, when de-reddening \ha\ we apply no correction for those pixels with H$\alpha$/H$\beta$ $<$ 2.86. 
    At low \ha\ surface brightness, our \hb\ sensitivity sets an upper threshold on the recoverable value of \ha/\hb. The value of this threshold for the median \hb\ error is indicated in all panels by the dotted lines.    Histograms of SB$_{\mathrm{H}\alpha}$ are included for each set of plots for pixels that meet our S/N cut (in black) and those that do not (in grey). A significant number of pixels are excluded by requiring a $\mathrm{S/N} > 10$ detection of \hb. A wide range of extinctions is observed, with no obvious correlation with SB$_{\mathrm{H}\alpha}$. This demonstrates the need for matched resolution constraints on the dust extinction. Within PHANGS-MUSE, the use of the Balmer decrement provides a unique opportunity to perform robust, extinction-corrected SFR measurements across the bulk of the galaxy disc.
    }
    \label{fig:av}
\end{figure*}

This paper describes the data reduction and data analysis steps employed as a baseline in producing science-ready data products. However, as discussed throughout the paper, we are aware of several limitations of the current analysis, which we plan to address accordingly.
Our planned next steps are as follows.
{

\em Improved sky subtraction.} This both concerns the sky continuum and the sky emission lines. The former will be better constrained via a refined estimate of the sky continuum contribution using overlapping exposures and external constraints from multi-band HST images when available. For the latter, we would consider a principle component analysis (PCA) approach, similar to what has been developed by other groups \citep[e.g.][]{Soto2016}. We hope this would result in significantly reduced stellar ${E(B-V)}$ residuals between pointings (see Fig.~\ref{fig:av}) within the mosaic, and further scientific exploitation of the red end of the MUSE spectral coverage.

{\em Improved geometric transformation.} The data release described in the present paper relies on fixed geometric and astrometric MUSE calibration files, as delivered with the raw datasets. We would examine whether a time-varying set of calibration files \citep[provided via e.g. the MuseWise system implemented by the MUSE GTO team;][]{MuseWise2015} could improve the registration of individual MUSE exposures of the same FoV.
   
   {\em Improved absolute flux zero-point calibration and astrometric registration.} We would make use of an improved photometry and astrometry for the ground-based du~Pont and Direct CCD imaging, and further exploit the existing PHANGS-HST imaging when possible.

\textit{Use of the auto-calibration functionalities in the \MUSEp}\ to improve on the residual imprints left by the slicer-to-slicer variations (flat-fielding). While this is clearly a secondary and low-amplitude effect, the exploitation of all blank sky exposures during the night associated with individual exposures seems to give reasonable results and a significant improvement on test exposures. It still requires pre-calculated `AUTOCAL\_FACTORS' with carefully defined sky masks and, hence, goes beyond the present data release.
  
  {\em Application of alternative stellar population templates.} This would include a systematic study of the choice of alternative stellar population templates and its impact on inferred properties. In particular, we would assess whether the identification of anomalously low-metallicity young stellar population associated with high \ha\ EWs is physical or reflects limitations in the templates currently employed. 
    
    {\em Multi-component Gaussian emission line fits.} Secondary components or non-Gaussian profiles are sometimes visible for the emission lines in the MUSE data. Those may be associated with, for example, small-scale outflows, complex dynamical regimes (e.g. shocks) or (spectrally, spatially) unresolved structures. A systematic identification of spectra significantly impacted by our default assumption of single Gaussian profiles for each individual emission line is one of the listed improvements to be added to a future data release \citep[see e.g.][]{Henshaw2020b}. 

\section{Key science enabled by PHANGS-MUSE} 
\label{sec:keysci}

The PHANGS-MUSE survey enables a number of key science goals (see Sect.~\ref{sec:scgoals}). In the following subsections we demonstrate the richness of the PHANGS-MUSE dataset.

\subsection{Mapping of star-formation rate}
\label{sec:SciCase_scaling}

Dust proves an obstacle to inferring the direct physical conditions, as it acts to both extinguish and redden the stellar and nebular light. With PHANGS-MUSE, we can directly parameterise the effect of dust on nebular emission lines through the Balmer decrement (traced by the ratio of \ha/\hb). Figure~\ref{fig:av} demonstrates the range of \ha/\hb\ line ratios that we observe at $\mathrm{S/N} > 10$ for three galaxies in our sample, as a function of \ha\ surface brightness. These three galaxies represent the lowest stellar mass (NGC~5068), an average stellar mass flocculent morphology (NGC~4254) and the highest stellar mass, AGN-dominated, strongly barred morphology (NGC~1365). The majority of pixels are above the canonical \ha/\hb\ of 2.86 (applicable for Case~B recombination, assuming $T_\mathrm{e} = 10^4$~K and $n_\mathrm{e} = 100$~cm$^{-3}$), revealing the impact of dust reddening across these spiral galaxies. Those pixels with lower values are predominantly consistent with 2.86 within uncertainties, even accounting for our fairly high $\mathrm{S/N} > 10$ cut. 

As also reflected in Fig.~\ref{fig:line_fractions}, a large fraction of pixels have significant detections in \ha, but are not detected in \hb\ (Fig.~\ref{fig:av}, histogram). For pixels where both emission lines are detected, we can apply an extinction correction at matched resolution. Applying an \cite{ODonnell1994} extinction law, and applying no correction for \ha/\hb $<$2.86, the observed Balmer decrements correspond to extinctions up to $A_V \approx 4$~mag in these galaxies, which is the typical maximum detectable extinction value for our sample. Most pixels are consistent with $A_V \approx 1{-}2$~mag, typical for many face-on galaxies \citep{Kennicutt1998}. In NGC~1365, where the central AGN contributes to the high \ha\ intensity pixels, the assumed intrinsic \ha/\hb$ = 2.86$ is probably not appropriate, resulting in an overestimation of $A_V$ for these pixels. It is worth emphasising that applying a constant intrinsic Balmer decrement might be restrictive as the Balmer decrement is sensitive to temperature: other approaches and methodologies exist to estimate the extinction correction via the emission lines, that is, allowing for a varying \ha/\hb\ ratio via the fitting of photoionisation models \citep[see e.g.][]{Mingozzi2020, Pellegrini2020}.

Applying an extinction correction imposes a correlation between \ha/\hb\ and extinction-corrected \ha\ surface brightness. Due to our H$\beta$ detection limit, we cannot determine an extinction correction for highly extincted pixels at low \ha\ surface brightness (represented by the dotted line). The small number of points in excess of this upper limit reflect slight variations in the \hb error map.  
Accounting for that, we see no clear correlation between observed \ha\ brightness and dust extinction. The uniformity in \ha\ pixel-based surface brightness distribution between these three quite different galaxies reflects the uniformity in the \hii\ region luminosity functions, tracing the underlying photoionisation physics powering the majority of the brightest emission \citep{Santoro2021}. The PHANGS-MUSE dataset thus builds on previous work \citep[e.g.][]{Kreckel2018} and represents the basis for robust SFR maps using matched resolution extinction correction \citep{Pessa2021}.

\subsection{Emission line diagnostics}
\label{sec:SciCase_emline}

\begin{figure*}
    \centering
    \includegraphics[width=0.9\textwidth]{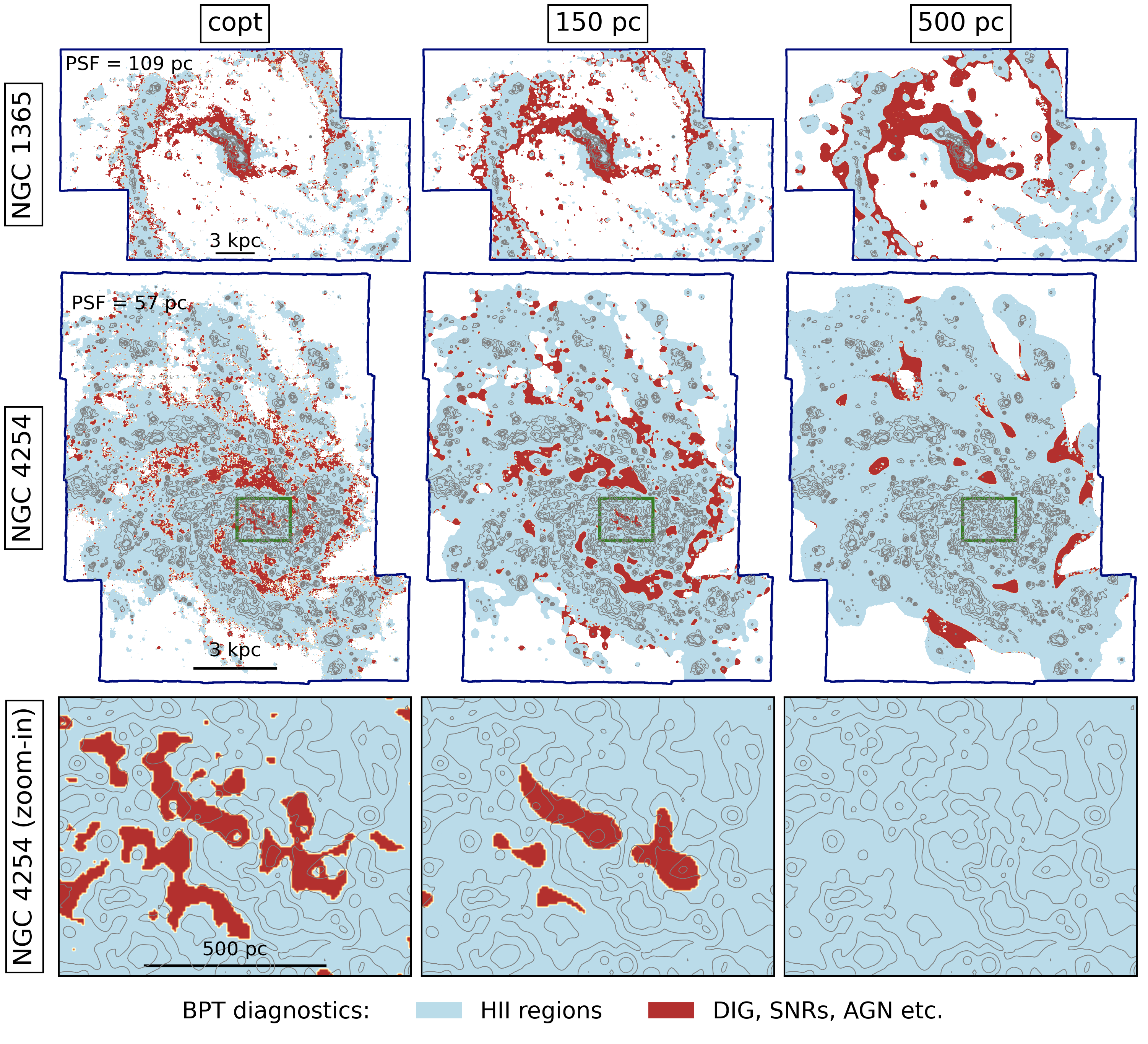}
    \caption{Mapping how the pixel-based BPT classifications change as a function of physical resolution for  NGC\,1365 and NGC\,4254. We convolve line maps of each target from copt resolution to 150~pc and 500~pc scales and apply the \cite{kauffmann_host_2003} \nii/\ha\ demarcation to distinguish photoionisation (blue) from harder ionising sources (shocks and AGN, in red). At high spatial resolution, the DIG also displays line ratios consistent with such shock models \citep{Belfiore2021}. H$\alpha$ contours at the copt resolution (left panels) are shown at each scale for reference. In NGC\,4254, as we convolve to larger spatial scales we see that the luminosity weighting results in a larger area of the map appear consistent with photoionisation. This is more apparent in a zoomed-in view (bottom panel). Individual isolated sources with shock excitation (likely SNRs) and the imprint of the DIG are no longer distinguishable in such disc galaxies as we move to the largest 500~pc scales. In contrast, the strong, extended AGN contamination in NGC~1365 can still be recovered. These figures demonstrate the power of diagnostic line ratios in distinguishing ionisation sources, as well as the importance of high (${<}150$~pc resolution) imaging to obtain an accurate census of ionising sources in the disc. }
    \label{fig:bpt}
\end{figure*}

As seen in Fig.~\ref{fig:line_fractions}, the detection of \ha\ emission across nearly all pixels in our mosaics is complemented by a high (${>}50$\%) detection fraction for a wide array of high and low-ionisation species of oxygen, nitrogen and sulphur atoms even at the level of individual pixels. With binning, we can achieve near complete detection of these strong emission lines across our mosaics, providing new insights into not just the brightest regions but also into the physical conditions impacting the diffuse gas \citep{Belfiore2021}. In addition, we achieve direct detection of weak, temperature-sensitive auroral emission lines across the ${\sim}5$\% of pixels that correspond to the centres of the brightest \hii\ regions \citep{Ho2019}. {While those detections may be biased towards higher-metallicity regions (as for the metallicity-dependent \nii$\lambda$5754 line), line ratios such as \nii$\lambda$5754/\nii$\lambda$6548 and \siii$\lambda$6312/\siii$\lambda$9058 represent relevant insights on the local conditions.} This wealth of emission lines enables a variety of diagnostics that constrain the physical properties of the ionised ISM, including the temperature, density, abundances, pressure, and dust extinction \citep{Kreckel2019, Kreckel2020, Barnes2021}.

Well-established line diagnostics \citep[BPT diagrams, named after Baldwin, Philipps and Terlevich]{Baldwin1981} can also distinguish the contribution of different sources of ionisation, for example SNRs, shocks, and planetary nebulae. Many of those sources act on ${<}100$~pc scales and are less luminous than neighbouring photoionised \hii\ regions. As such, they are distinguishable only when sufficiently high spatial resolution observations are obtained.  When binning to larger scales, the intrinsic luminosity weighting means that the line ratios often become dominated by the brighter photoionised \hii\ regions, and all spectral information on other ionising sources is lost. 

Figure~\ref{fig:bpt} demonstrates how BPT classifications (here using the \citealt{kauffmann_host_2003} demarcation as an example) appear to change with changing spatial resolution, from our native $50{-}100$~pc scales, to 150~pc and 500~pc scales. In the two galaxies shown, the diagnostic maps indicate widespread photoionisation due to star-formation activity (in blue) in addition to ionisation from harder ionising sources (in red) such as shocks and AGN. Diffuse ionised gas also exhibits line ratios consistent with shock excitation, though the exact explanation for these line ratios is still unclear \citep{Zhang2017}. In NGC~1365 the strong central AGN clearly imprints a signal both in the galaxy centre as well as out to large kiloparsec scales \citep{Venturi2018}. In NGC~4254, a flocculent star-forming disc, the red regions are likely due to a combination of DIG emission and SNRs.

Convolving our maps from the copt resolution ($50{-}100$~pc) to 150~pc and 500~pc scales, we see that the number and distribution of ionising sources changes with decreasing spatial resolution. Nearly all individual sources in NGC~4254 are lost at 500~pc. However, the impact of the central AGN on NGC\,1365 is recoverable even at this lowest physical resolution. These figures demonstrate the importance of our high ($<100$~pc) spatial resolution approach in constructing a complete census of ionising sources in the discs of these galaxies. 

These images also demonstrate that while weaker emission lines often cannot be detected in pixel-scale maps  certain regions; for example, in the outer part of the disc, a detection can be recovered with rather modest binning.  In this sense, spatial binning of the original data cubes holds great promise for providing diagnostics for the full disc, and offers leverage for the study of the pervasive DIG \citep{Belfiore2021}.

The multi-wavelength, multi-phase and multi-tracer approach, as adopted by the PHANGS project and illustrated in Fig.~\ref{fig:multiwave}, has the potential to provide a key link for further dissecting the baryon cycle, probing those complex gaseous plus stellar galactic ecosystems, and placing quantitative constraints on the feedback physics that regulates the chemical and dynamical evolution of the ISM \citep[][]{Chevance2022, Kim2021, Barnes2021}. PHANGS-MUSE more specifically enables us to catalogue individual \hii\ regions \citep[][]{Kreckel2019,Santoro2021} along with other ionising sources (Planetary Nebulae -- PNe --, SNRs; F.~Scheuermann et al., in preparation), crucially informing us about the impact of the individual young clusters catalogued by \textit{HST} \citep{Turner2021}, and the evolving parent Giant Molecular Clouds \citep[GMCs][A.~Hughes et al. in preparation]{Sun2020b,Sun2020,Rosolowsky2021}.   
This cross-observatory approach will provide new insights into stellar feedback \citep{Herrera2020} and provide fertile testing grounds for future \textit{JWST} studies of heavily embedded star formation. 

\subsection{Star-formation histories and dynamical environments}
\label{sec:SciCase_stellarpops}

\begin{figure*}
   \centering
    \includegraphics[width=0.85\textwidth]{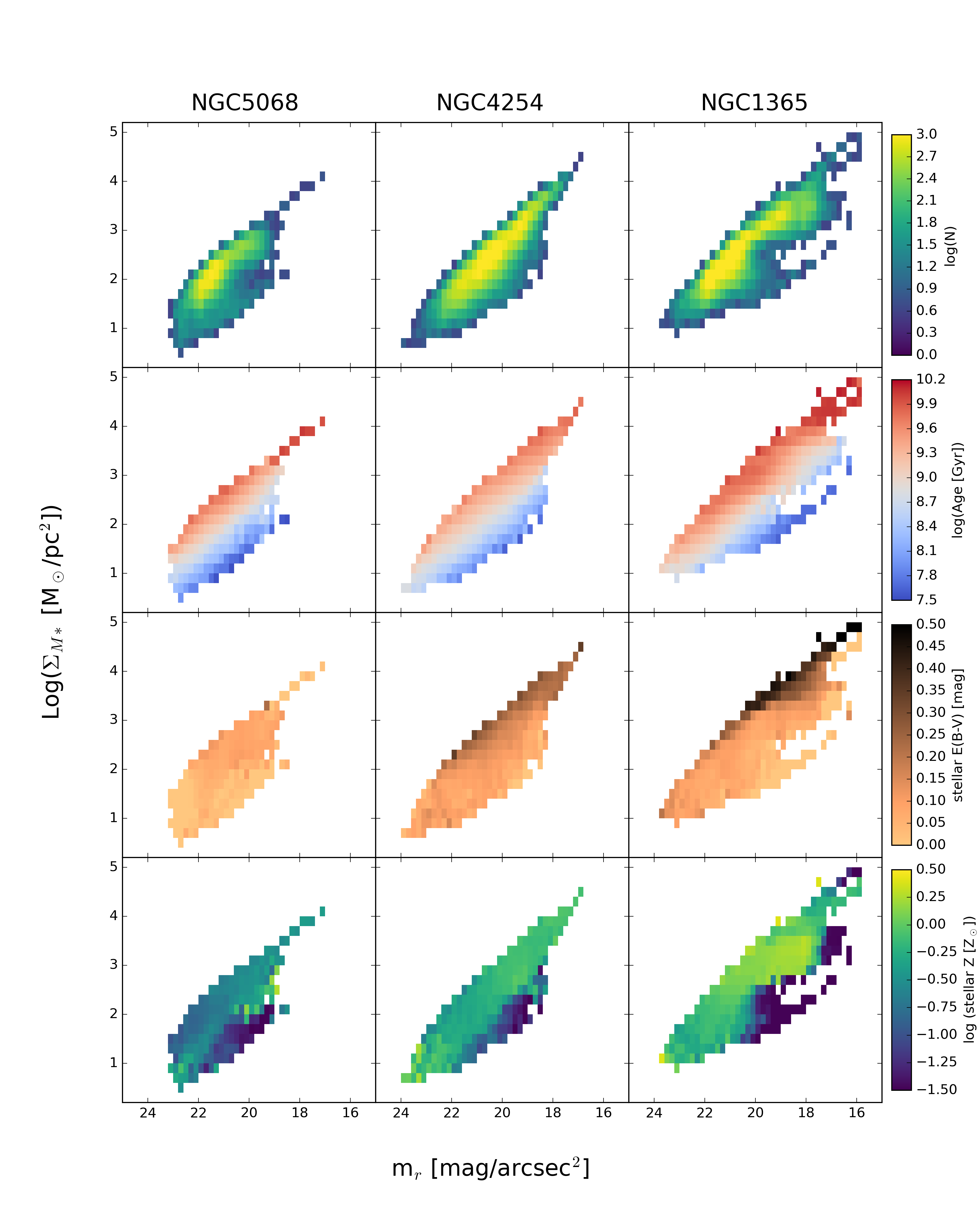}
 \caption{Stellar mass surface density ($\Sigma_{M_\star}$) computed from the MUSE spectroscopy as a function of $r$-band surface brightness ($m_r$) for three representative galaxies (see caption of Fig.~\ref{fig:av} and the associated text). All measurements are performed on Voronoi binned regions that reach at least a stellar continuum $\mathrm{S/N} = 35$. For all histograms, bins are required to contain at least three measurements. Top row: Binned distribution revealing a clear correlation between mass and light, but with up to an order of magnitude scatter (the colour scale, in $\log(N),$ being set up by the point number density). Second row: Colouring bins by the median luminosity-weighted stellar age.\ We demonstrate that the spread is dominated by variations in stellar age. Third row: Colouring bins by the median stellar reddening $E(B-V)$ revealing a secondary dependence, reflecting the long known degeneracy between dust extinction and stellar age. Bottom row: Colouring bins by the median stellar metallicity illustrating the global scaling with mass as well as additional local variations with both luminosity and stellar mass surface density throughout the disc. We note that very low metallicity values may correspond to an unsolved degeneracy due to bright and very young stellar clusters (see Sect.~\ref{sec:youngbias}). As demonstrated in this set of figures, PHANGS-MUSE enables us to infer stellar masses and ages, breaking the long-standing degeneracy between reddening due to dust and reddening due to an ageing stellar population. }
 \label{fig:m2l}
\end{figure*}

Turning to the stellar continuum, MUSE data enable us to infer stellar masses and ages, breaking the long-standing degeneracy between reddening due to dust and reddening due to an ageing stellar population, going well beyond what is possible from broadband colours. Figure~\ref{fig:m2l} reflects the power of our stellar spectroscopic fits presenting the multi-dimensionality of the PHANGS-MUSE data products. While it is clear that our inferred stellar masses trace the broadband light (top row), the order-of-magnitude spread reflects predominantly the range in stellar ages (second row). This correlation qualitatively dominates over variations driven by stellar dust reddening (third row) or changes in stellar abundance (bottom row). Ongoing work explores how these high-resolution (${\sim}100$~pc) constraints on the stellar population relate to traditional scaling relations, such as the resolved stellar main sequence \citep{Pessa2021}. It also provides essential input when carrying out comprehensive modelling of ionising sources in the disc, which in the literature has been suggested to contain a non-negligible contribution from hot evolved main-sequence stars \citep{Zhang2017,Belfiore2021}.

One other crucial part of the PHANGS-MUSE science case relies on  mapping wide areas in order to sample a variety of dynamical environments \citep[spiral arm, bar, centre][]{Querejeta2021}. These trace the individual impact of the galactic potential on the stars and the ionised gas, as well as perturbations from non-axisymmetry (e.g. bar and spiral structures) in the disc (see Fig.~\ref{fig:multiwave}). 

\begin{figure*}
    \centering
    \includegraphics[width=0.95\textwidth]{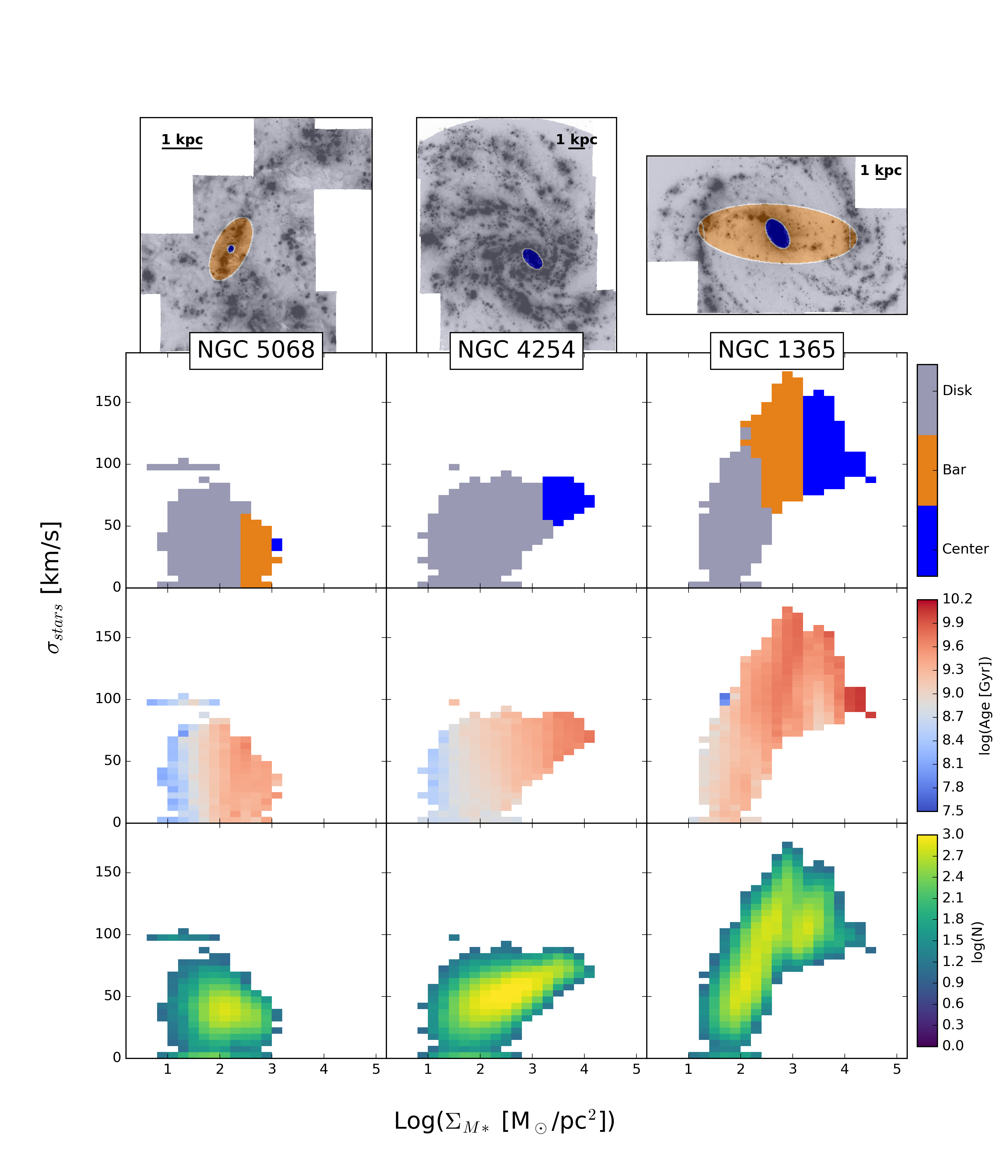}
    \caption{Changes in stellar velocity dispersion reflected by their dynamical environment. These three galaxies represent galaxies at the lowest stellar mass (NGC~5068), an average stellar mass flocculent morphology (NGC~4254), and the highest stellar mass, AGN-dominated, strongly barred morphology (NGC~1365). Top: Each galaxy broken down into its centre (blue), bar (orange),  and disc (grey) environments from \citet{Querejeta2021}. Bottom: Distribution of  values from the individual Voronoi bins plotted (see Sect.~\ref{sec:dap_requirements}), comparing the stellar velocity dispersion ($\sigma_{\rm stars}$) to the stellar mass surface density ($\Sigma_{M_\star}$). We further colour-code each part of the histogram by the environment (second row), the median luminosity-weighted stellar age (third row), or the number density (fourth row). The disc-dominated galaxies (NGC~5068 and NGC~4254) show moderate $\sigma_{\rm stars}$ with an increase at high $\Sigma_{M_\star}$, corresponding to the central environment, and younger stellar ages at lower $\Sigma_{M_\star}$ (in the spiral arm and outer disc). NGC~1365 presents a more complicated star-formation history, with older stellar ages dominating across the bar associated with increased stellar dispersions.  In this set of figures, we see how our PHANGS-MUSE data allow us to trace the individual impact of the galactic potential on the stars, as well as perturbations from non-axisymmetry (e.g. bar and spiral structures) in the disc. 
   }
    \label{fig:kin2}
\end{figure*}

As an illustration, Fig.~\ref{fig:kin2} shows the pixel-based distribution of stellar velocity dispersion as a function of stellar mass surface density ($\Sigma_{M_\star}$) for three of our galaxies (top row). We apply  environmental masks \citep{Querejeta2021} that classify each target into distinct morphological environments: centre, bar, and disc. Environments and galaxy morphologies are shown directly at the top of Fig.~\ref{fig:kin2}, while each bin is colour-coded by the median environment (second row). The highest dispersions are associated with high $\Sigma_{M_\star}$, and map to the locations morphologically classified as centre- or bar-dominated, while disc-dominated regions generally exhibit lower velocity dispersions (as expected). We also find good correspondence between stellar ages, mass surface density and dynamics. Colouring each bin by the median luminosity-weighted age (centre row) reveals an expected correlation between younger ages and lower $\Sigma_{M_\star}$ (e.g. at larger radii) in the most disc-dominated galaxies (NGC~5068 and NGC~4254). NGC~1365 presents a more complicated star-formation history, with older stellar ages dominating across the bar. While we find good consistency between expectations based on morphology and the dynamical structure revealed by MUSE kinematics, it also provides new insight into the nature of various dynamical structures. For example, we detect a turn-over in the stellar velocity dispersion towards the centre of NGC~1365, at the highest densities, suggesting a cold, and potentially older, central dynamical structure.

Figures~\ref{fig:m2l} and~\ref{fig:kin2} are meant as a first glimpse at the richness of the PHANGS-MUSE dataset, emphasising the power and beauty (Fig.~\ref{fig:multiwave}) of two-dimensional spectroscopic mapping of star-forming galactic discs, and the scientific potential that will be developed in future studies by our team and the community via the public data releases.

\section{Summary}
\label{sec:conc}

In this paper we present a detailed account of the PHANGS-MUSE survey, which delivers MUSE/VLT IFS in the optical for a set of 19 nearby star-forming galactic discs. PHANGS-MUSE both covers and resolves a significant fraction of the optical disc, and in particular maps the region enclosing most of its observed molecular gas content (as detected via the \mbox{CO(2--1)} line). The 19 galaxies of the PHANGS-MUSE sample are part of 38 systems surveyed by PHANGS-HST, all drawn from the parent sample of 90 PHANGS-ALMA galaxies, and will be targeted for eight band 2--21 $\mu$m imaging as part of the PHANGS-JWST Cycle~1 Treasure programme. This makes such a campaign a unique tool for addressing the physics of various tracers pertaining to the onset of star formation, stellar evolution, and feedback.

We provide a description of the main science objectives for PHANGS-MUSE. They include a refreshed and resolved view on scaling relations, probing the impact of feedback and its relation to the local environment, better understanding the chemical history of galactic discs, and determining the intricate link between dynamical regimes and star formation. The PHANGS-MUSE data can be further exploited on many scientific fronts, representing a huge legacy programme for the scientific community.

Compared to other  spectroscopic surveys of galaxies,  PHANGS-MUSE stands out for the sheer number of spectra observed. With millions of independent spatial elements, PHANGS-MUSE covers a rich variety of galactic local environments, all at a typical intrinsic resolution of 50~pc. Compared to other two-dimensional IFU galaxy surveys (e.g. MaNGA, SAMI, and CALIFA), we target many fewer galaxies but in far greater detail, achieving a qualitatively different physical resolution. Compared to planned surveys of the Milky Way and Magellanic Clouds (e.g. \mbox{SDSS-V}/LVM), we more uniformly sample the $z=0$ galaxy population. And thanks to the multi-wavelength coverage by HST and ALMA at matched or better resolution, we have an unparalleled multi-wavelength view of the baryon cycle within nearby massive star-forming main-sequence galaxies.

We detail the pipeline flows associated with the data reduction and analysis, demonstrating a robust albeit challenging setup to address our ambitious PHANGS-MUSE dataset. We emphasise the modularity of our setup, strongly motivated by the need to re-process and re-analyse the entire set of raw and reduced data several times, driven by, for example, algorithmic improvements and tests. The reduction pipeline is heavily based on the \MUSEp\ routines, wrapped up via the \pymusepipe\ package, while for our \DAP\ we further tuned a PHANGS-specific package inherited from the \gist\ package, focusing on the use of, for example, \pPXF\ for all major fitting stages (stellar kinematics, emission lines, and stellar populations). Challenges on the data reduction side include the sky subtraction, accurate astrometric solution for all individual exposures, and post-processed homogenisation of the PSF for each mosaic: these are all addressed via specific and simple algorithms, sometimes supported by existing software (e.g. \pypher\ or \mpdaf). The demanding steps for the data analysis are mostly associated with the individual binning schemes and the extraction of the star-formation-history (and associated extinction) maps.

We then provide quantitative measures of the quality of the MUSE-PHANGS data when it comes to wavelength, astrometry, and photometric calibrations. We also give a detailed account of the quality of the spectral fits that drive the delivered science maps, demonstrating that the fits are overall of excellent quality, while regions heavily contaminated by foreground stars or reflecting the presence of very young sources may need further attention.

The outcome of the data reduction and analysis processes are formatted into a PHANGS-MUSE public data release, namely DR1.0, which includes: full native data cubes of the 19 galaxy mosaics, encompassing both the data spectra and their estimated variances, their reconstructed images for a list of standard broad- and medium-band filters, and a large set of derived maps representing extracted information pertaining to the stellar and gas kinematics, emission line gas, stellar and gas extinction, and stellar populations. Earlier versions of the PHANGS-MUSE dataset were distributed internally and exploited in a series of published works. Going beyond PHANGS-MUSE DR1.0, subsequent data releases will potentially include a number of key improvements concerning, for example, the photometry, astrometry, and parameter fitting.

We close the presentation of the PHANGS-MUSE survey by illustrating the science potential of this new observational window on nearby star-forming discs. We provide brief accounts of the \ha\ pixel-based surface brightness distribution and how dust correction impacts the associated distribution of line fluxes and ratios (Sect.~\ref{sec:SciCase_scaling}), the importance of resolving regions with different emission line regimes (Sect.~\ref{sec:SciCase_emline}), and the variation in gas physical states, dynamical states, and star-formation histories among various dynamical environments within discs (Sect.~\ref{sec:SciCase_stellarpops}). These represent only a glimpse of the richness of the PHANGS-MUSE dataset: we believe that its (first and subsequent) public releases will help in developing a better understanding of how the baryon cycle proceeds and how it connects with nearby galaxy disc hosts and their dynamical structures.

\begin{acknowledgements}
This work was carried out as part of the PHANGS collaboration. 
ES, FS, RME, TGW acknowledge funding from the European Research Council (ERC) under the European Union's Horizon 2020 research and innovation programme (grant agreement No. 694343).

SCOG and RSK acknowledge financial support from the German Research Foundation (DFG) via the collaborative research centre (SFB 881, Project-ID 138713538) ``The Milky Way System'' (subprojects A1, B1, B2, and B8). They also acknowledge funding from the Heidelberg Cluster of Excellence ``STRUCTURES'' in the framework of Germany's Excellence Strategy (grant EXC-2181/1, Project-ID 390900948) and from the European Research Council via the ERC Synergy Grant ``ECOGAL'' (grant 855130).

FBi, ATB, IB acknowledge funding from the European Research Council (ERC) under the European Union's Horizon 2020 research and innovation programme (grant agreement No.726384/Empire).

CE acknowledges funding from the Deutsche Forschungsgemeinschaft (DFG) Sachbeihilfe, grant number BI1546/3-1.

EJW acknowledges support from the Deutsche Forschungsgemeinschaft (DFG, German Research Foundation) - Project-ID 138713538 -- SFB 881 (``The Milky Way System'', subproject P2).

JMDK and MC gratefully acknowledge funding from the Deutsche Forschungsgemeinschaft (DFG) through an Emmy Noether Research Group (grant number KR4801/1-1) and the DFG Sachbeihilfe (grant number KR4801/2-1) and from the European Research Council (ERC) under the European Union's Horizon 2020 research and innovation programme via the ERC Starting Grant MUSTANG (grant agreement number 714907).

E.C. acknowledges support from ANID project Basal AFB-170002.

KK, OE, and FS gratefully acknowledge funding from the German Research Foundation (DFG) in the form of an Emmy Noether Research Group (grant number KR4598/2-1, PI: Kreckel). EW acknowledges support from the DFG via SFB 881 ``The Milky Way System'' (Project-ID 138713538; sub-project P2).

MB gratefully acknowledges support by the ANID BASAL project FB210003.

JPe acknowledges support by the Programme National ``Physique et Chimie du
Milieu Interstellaire'' (PCMI) of CNRS/INSU with INC/INP, co-funded by CEA
and CNES.
Based on observations collected at the European Southern Observatory under ESO programmes 1100.B-0651, 095.C-0473, and 094.C-0623 (PHANGS-MUSE; PI: Schinnerer), as well as 094.B-0321 (MAGNUM; PI: Marconi), 099.B-0242, 0100.B-0116, 098.B-0551 (MAD; PI: Carollo) and 097.B-0640 (TIMER; PI: Gadotti).
This paper makes use of data gathered with the 2.5 meter du~Pont located at Las Campanas Observatory, Chile, and data based on observations carried out at the MPG 2.2m telescope on La Silla, Chile.
JCL acknowledges support for PHANGS-HST (program number 15654), provided through a grant from the Space Telescope Science Institute under NASA contract NAS5-26555. PHANGS-HST is based on observations made with the NASA/ESA Hubble Space Telescope, and obtained from the data archive at STScI. STScI is operated by the Association of Universities for Research in Astronomy, Inc. under NASA contract NAS 5-26555.
\end{acknowledgements}

\bibliographystyle{aa}
\bibliography{library}

\begin{thebibliography}{190}
\expandafter\ifx\csname natexlab\endcsname\relax\def\natexlab#1{#1}\fi

\bibitem[{{Abazajian} {et~al.}(2009){Abazajian}, {Adelman-McCarthy},
  {Ag{\"u}eros}, {Allam}, {Allende Prieto}, {An}, {Anderson}, {Anderson},
  {Annis}, {Bahcall}, {Bailer-Jones}, {Barentine}, {Bassett}, {Becker},
  {Beers}, {Bell}, {Belokurov}, {Berlind}, {Berman}, {Bernardi}, {Bickerton},
  {Bizyaev}, {Blakeslee}, {Blanton}, {Bochanski}, {Boroski}, {Brewington},
  {Brinchmann}, {Brinkmann}, {Brunner}, {Budav{\'a}ri}, {Carey}, {Carliles},
  {Carr}, {Castander}, {Cinabro}, {Connolly}, {Csabai}, {Cunha}, {Czarapata},
  {Davenport}, {de Haas}, {Dilday}, {Doi}, {Eisenstein}, {Evans}, {Evans},
  {Fan}, {Friedman}, {Frieman}, {Fukugita}, {G{\"a}nsicke}, {Gates},
  {Gillespie}, {Gilmore}, {Gonzalez}, {Gonzalez}, {Grebel}, {Gunn},
  {Gy{\"o}ry}, {Hall}, {Harding}, {Harris}, {Harvanek}, {Hawley}, {Hayes},
  {Heckman}, {Hendry}, {Hennessy}, {Hindsley}, {Hoblitt}, {Hogan}, {Hogg},
  {Holtzman}, {Hyde}, {Ichikawa}, {Ichikawa}, {Im}, {Ivezi{\'c}}, {Jester},
  {Jiang}, {Johnson}, {Jorgensen}, {Juri{\'c}}, {Kent}, {Kessler}, {Kleinman},
  {Knapp}, {Konishi}, {Kron}, {Krzesinski}, {Kuropatkin}, {Lampeitl},
  {Lebedeva}, {Lee}, {Lee}, {French Leger}, {L{\'e}pine}, {Li}, {Lima}, {Lin},
  {Long}, {Loomis}, {Loveday}, {Lupton}, {Magnier}, {Malanushenko},
  {Malanushenko}, {Mandelbaum}, {Margon}, {Marriner}, {Mart{\'\i}nez-Delgado},
  {Matsubara}, {McGehee}, {McKay}, {Meiksin}, {Morrison}, {Mullally}, {Munn},
  {Murphy}, {Nash}, {Nebot}, {Neilsen}, {Newberg}, {Newman}, {Nichol},
  {Nicinski}, {Nieto-Santisteban}, {Nitta}, {Okamura}, {Oravetz}, {Ostriker},
  {Owen}, {Padmanabhan}, {Pan}, {Park}, {Pauls}, {Peoples}, {Percival}, {Pier},
  {Pope}, {Pourbaix}, {Price}, {Purger}, {Quinn}, {Raddick}, {Re Fiorentin},
  {Richards}, {Richmond}, {Riess}, {Rix}, {Rockosi}, {Sako}, {Schlegel},
  {Schneider}, {Scholz}, {Schreiber}, {Schwope}, {Seljak}, {Sesar}, {Sheldon},
  {Shimasaku}, {Sibley}, {Simmons}, {Sivarani}, {Allyn Smith}, {Smith},
  {Smol{\v{c}}i{\'c}}, {Snedden}, {Stebbins}, {Steinmetz}, {Stoughton},
  {Strauss}, {SubbaRao}, {Suto}, {Szalay}, {Szapudi}, {Szkody}, {Tanaka},
  {Tegmark}, {Teodoro}, {Thakar}, {Tremonti}, {Tucker}, {Uomoto}, {Vanden
  Berk}, {Vandenberg}, {Vidrih}, {Vogeley}, {Voges}, {Vogt}, {Wadadekar},
  {Watters}, {Weinberg}, {West}, {White}, {Wilhite}, {Wonders}, {Yanny},
  {Yocum}, {York}, {Zehavi}, {Zibetti}, \& {Zucker}}]{Abazajian2009}
{Abazajian}, K.~N., {Adelman-McCarthy}, J.~K., {Ag{\"u}eros}, M.~A., {et~al.}
  2009, \apjs, 182, 543

\bibitem[{{Agertz} \& {Kravtsov}(2015)}]{Agertz2015}
{Agertz}, O. \& {Kravtsov}, A.~V. 2015, \apj, 804, 18

\bibitem[{{Agertz} \& {Kravtsov}(2016)}]{Agertz2016}
{Agertz}, O. \& {Kravtsov}, A.~V. 2016, \apj, 824, 79

\bibitem[{{Anand} {et~al.}(2021){Anand}, {Lee}, {Van Dyk}, {Leroy},
  {Rosolowsky}, {Schinnerer}, {Larson}, {Kourkchi}, {Kreckel}, {Scheuermann},
  {Rizzi}, {Thilker}, {Tully}, {Bigiel}, {Blanc}, {Boquien}, {Chandar}, {Dale},
  {Emsellem}, {Deger}, {Glover}, {Grasha}, {Groves}, {Klessen}, {Kruijssen},
  {Querejeta}, {S{\'a}nchez-Bl{\'a}zquez}, {Schruba}, {Turner}, {Ubeda},
  {Williams}, \& {Whitmore}}]{anand2021}
{Anand}, G.~S., {Lee}, J.~C., {Van Dyk}, S.~D., {et~al.} 2021, \mnras, 501,
  3621

\bibitem[{{Andrews} {et~al.}(2021){Andrews}, {Jencson}, {Van Dyk}, {Smith},
  {Neustadt}, {Sand}, {Kreckel}, {Kochanek}, {Valenti}, {Strader}, {Bersten},
  {Blanc}, {Bostroem}, {Brink}, {Emsellem}, {Filippenko}, {Folatelli},
  {Kasliwal}, {Masci}, {McElroy}, {Milisavljevic}, {Santoro}, \&
  {Szalai}}]{Andrews2021}
{Andrews}, J.~E., {Jencson}, J.~E., {Van Dyk}, S.~D., {et~al.} 2021, \apj, 917,
  63

\bibitem[{{Astropy Collaboration} {et~al.}(2018){Astropy Collaboration},
  {Price-Whelan}, {Sip{\H{o}}cz}, {G{\"u}nther}, {Lim}, {Crawford}, {Conseil},
  {Shupe}, {Craig}, {Dencheva}, {Ginsburg}, {Vand erPlas}, {Bradley},
  {P{\'e}rez-Su{\'a}rez}, {de Val-Borro}, {Aldcroft}, {Cruz}, {Robitaille},
  {Tollerud}, {Ardelean}, {Babej}, {Bach}, {Bachetti}, {Bakanov}, {Bamford},
  {Barentsen}, {Barmby}, {Baumbach}, {Berry}, {Biscani}, {Boquien}, {Bostroem},
  {Bouma}, {Brammer}, {Bray}, {Breytenbach}, {Buddelmeijer}, {Burke},
  {Calderone}, {Cano Rodr{\'\i}guez}, {Cara}, {Cardoso}, {Cheedella}, {Copin},
  {Corrales}, {Crichton}, {D'Avella}, {Deil}, {Depagne}, {Dietrich}, {Donath},
  {Droettboom}, {Earl}, {Erben}, {Fabbro}, {Ferreira}, {Finethy}, {Fox},
  {Garrison}, {Gibbons}, {Goldstein}, {Gommers}, {Greco}, {Greenfield},
  {Groener}, {Grollier}, {Hagen}, {Hirst}, {Homeier}, {Horton}, {Hosseinzadeh},
  {Hu}, {Hunkeler}, {Ivezi{\'c}}, {Jain}, {Jenness}, {Kanarek}, {Kendrew},
  {Kern}, {Kerzendorf}, {Khvalko}, {King}, {Kirkby}, {Kulkarni}, {Kumar},
  {Lee}, {Lenz}, {Littlefair}, {Ma}, {Macleod}, {Mastropietro}, {McCully},
  {Montagnac}, {Morris}, {Mueller}, {Mumford}, {Muna}, {Murphy}, {Nelson},
  {Nguyen}, {Ninan}, {N{\"o}the}, {Ogaz}, {Oh}, {Parejko}, {Parley}, {Pascual},
  {Patil}, {Patil}, {Plunkett}, {Prochaska}, {Rastogi}, {Reddy Janga},
  {Sabater}, {Sakurikar}, {Seifert}, {Sherbert}, {Sherwood-Taylor}, {Shih},
  {Sick}, {Silbiger}, {Singanamalla}, {Singer}, {Sladen}, {Sooley},
  {Sornarajah}, {Streicher}, {Teuben}, {Thomas}, {Tremblay}, {Turner},
  {Terr{\'o}n}, {van Kerkwijk}, {de la Vega}, {Watkins}, {Weaver}, {Whitmore},
  {Woillez}, {Zabalza}, \& {Astropy Contributors}}]{astropy:2018}
{Astropy Collaboration}, {Price-Whelan}, A.~M., {Sip{\H{o}}cz}, B.~M., {et~al.}
  2018, \aj, 156, 123

\bibitem[{{Baade} {et~al.}(1999){Baade}, {Meisenheimer}, {Iwert}, {Alonso},
  {Augusteijn}, {Beletic}, {Bellemann}, {Benesch}, {B{\"o}hm}, {B{\"o}hnhardt},
  {Brewer}, {Deiries}, {Delabre}, {Donaldson}, {Dupuy}, {Franke}, {Gerdes},
  {Gilliotte}, {Grimm}, {Haddad}, {Hess}, {Ihle}, {Klein}, {Lenzen}, {Lizon},
  {Mancini}, {M{\"u}nch}, {Pizarro}, {Prado}, {Rahmer}, {Reyes}, {Richardson},
  {Robledo}, {Sanchez}, {Silber}, {Sinclaire}, {Wackermann}, \&
  {Zaggia}}]{Baade1999}
{Baade}, D., {Meisenheimer}, K., {Iwert}, O., {et~al.} 1999, The Messenger, 95,
  15

\bibitem[{{Bacon} {et~al.}(2017){Bacon}, {Conseil}, {Mary}, {Brinchmann},
  {Shepherd}, {Akhlaghi}, {Weilbacher}, {Piqueras}, {Wisotzki}, {Lagattuta},
  {Epinat}, {Guerou}, {Inami}, {Cantalupo}, {Courbot}, {Contini}, {Richard},
  {Maseda}, {Bouwens}, {Bouch{\'e}}, {Kollatschny}, {Schaye}, {Marino},
  {Pello}, {Herenz}, {Guiderdoni}, \& {Carollo}}]{Bacon2017}
{Bacon}, R., {Conseil}, S., {Mary}, D., {et~al.} 2017, \aap, 608, A1

\bibitem[{{Bacon} \& {Monnet}(2017)}]{BaconMonnet2017}
{Bacon}, R. \& {Monnet}, G. 2017, {Optical 3D-Spectroscopy for Astronomy}
  (Weinheim, Germany: Wiley-VCH Verlag GmbH \& Co. KGaA)

\bibitem[{{Bacon} {et~al.}(2016){Bacon}, {Piqueras}, {Conseil}, {Richard}, \&
  {Shepherd}}]{Bacon2016}
{Bacon}, R., {Piqueras}, L., {Conseil}, S., {Richard}, J., \& {Shepherd}, M.
  2016, {MPDAF: MUSE Python Data Analysis Framework}

\bibitem[{{Baldwin} {et~al.}(1981){Baldwin}, {Phillips}, \&
  {Terlevich}}]{Baldwin1981}
{Baldwin}, J.~A., {Phillips}, M.~M., \& {Terlevich}, R. 1981, \pasp, 93, 5

\bibitem[{{Barnes} {et~al.}(2021){Barnes}, {Glover}, {Kreckel}, {Ostriker},
  {Bigiel}, {Belfiore}, {Be{\v{s}}li{\'c}}, {Blanc}, {Chevance}, {Dale},
  {Egorov}, {Eibensteiner}, {Emsellem}, {Grasha}, {Groves}, {Klessen},
  {Kruijssen}, {Leroy}, {Longmore}, {Lopez}, {McElroy}, {Meidt}, {Murphy},
  {Rosolowsky}, {Saito}, {Santoro}, {Schinnerer}, {Schruba}, {Sun}, {Watkins},
  \& {Williams}}]{Barnes2021}
{Barnes}, A.~T., {Glover}, S.~C.~O., {Kreckel}, K., {et~al.} 2021, \mnras, 508,
  5362

\bibitem[{{Barnes} {et~al.}(2020){Barnes}, {Longmore}, {Dale}, {Krumholz},
  {Kruijssen}, \& {Bigiel}}]{Barnes2020}
{Barnes}, A.~T., {Longmore}, S.~N., {Dale}, J.~E., {et~al.} 2020, \mnras, 498,
  4906

\bibitem[{{Belfiore} {et~al.}(2017){Belfiore}, {Maiolino}, {Tremonti},
  {S{\'a}nchez}, {Bundy}, {Bershady}, {Westfall}, {Lin}, {Drory}, {Boquien},
  {Thomas}, \& {Brinkmann}}]{Belfiore2017}
{Belfiore}, F., {Maiolino}, R., {Tremonti}, C., {et~al.} 2017, \mnras, 469, 151

\bibitem[{{Belfiore} {et~al.}(2021){Belfiore}, {Santoro}, {Groves},
  {Schinnerer}, {Kreckel}, {Glover}, {Klessen}, {Emsellem}, {Blanc}, {Congiu},
  {Barnes}, {Boquien}, {Chevance}, {Dale}, {Kruijssen}, {Leroy}, {Pan},
  {Pessa}, {Schruba}, \& {Williams}}]{Belfiore2021}
{Belfiore}, F., {Santoro}, F., {Groves}, B., {et~al.} 2021, \aap, in press;
  arXiv e-prints, arXiv:2111.14876

\bibitem[{{Belfiore} {et~al.}(2019){Belfiore}, {Westfall}, {Schaefer},
  {Cappellari}, {Ji}, {Bershady}, {Tremonti}, {Law}, {Yan}, {Bundy}, {Shetty},
  {Drory}, {Thomas}, {Emsellem}, \& {S{\'a}nchez}}]{Belfiore2019}
{Belfiore}, F., {Westfall}, K.~B., {Schaefer}, A., {et~al.} 2019, \aj, 158, 160

\bibitem[{{Berg} {et~al.}(2015){Berg}, {Skillman}, {Croxall}, {Pogge},
  {Moustakas}, \& {Johnson-Groh}}]{Berg2015}
{Berg}, D.~A., {Skillman}, E.~D., {Croxall}, K.~V., {et~al.} 2015, \apj, 806,
  16

\bibitem[{{Be{\v{s}}li{\'c}} {et~al.}(2021){Be{\v{s}}li{\'c}}, {Barnes},
  {Bigiel}, {Puschnig}, {Pety}, {Herrera Contreras}, {Leroy}, {Usero},
  {Schinnerer}, {Meidt}, {Emsellem}, {Hughes}, {Faesi}, {Kreckel}, {Belfiore},
  {Chevance}, {den Brok}, {Eibensteiner}, {Glover}, {Grasha},
  {Jimenez-Donaire}, {Klessen}, {Kruijssen}, {Liu}, {Pessa}, {Querejeta},
  {Rosolowsky}, {Saito}, {Santoro}, {Schruba}, {Sormani}, \&
  {Williams}}]{Beslic2021}
{Be{\v{s}}li{\'c}}, I., {Barnes}, A.~T., {Bigiel}, F., {et~al.} 2021, \mnras,
  506, 963

\bibitem[{{Bigiel} {et~al.}(2008){Bigiel}, {Leroy}, {Walter}, {Brinks}, {de
  Blok}, {Madore}, \& {Thornley}}]{Bigiel2008}
{Bigiel}, F., {Leroy}, A., {Walter}, F., {et~al.} 2008, \aj, 136, 2846

\bibitem[{{Bittner} {et~al.}(2019){Bittner}, {Falc{\'o}n-Barroso}, {Nedelchev},
  {Dorta}, {Gadotti}, {Sarzi}, {Molaeinezhad}, {Iodice}, {Rosado-Belza}, {de
  Lorenzo-C{\'a}ceres}, {Fragkoudi}, {Gal{\'a}n-de Anta}, {Husemann},
  {M{\'e}ndez-Abreu}, {Neumann}, {Pinna}, {Querejeta},
  {S{\'a}nchez-Bl{\'a}zquez}, \& {Seidel}}]{bittner2019}
{Bittner}, A., {Falc{\'o}n-Barroso}, J., {Nedelchev}, B., {et~al.} 2019, \aap,
  628, A117

\bibitem[{{Bittner} {et~al.}(2020){Bittner}, {S{\'a}nchez-Bl{\'a}zquez},
  {Gadotti}, {Neumann}, {Fragkoudi}, {Coelho}, {de Lorenzo-C{\'a}ceres},
  {Falc{\'o}n-Barroso}, {Kim}, {Leaman}, {Mart{\'\i}n-Navarro},
  {M{\'e}ndez-Abreu}, {P{\'e}rez}, {Querejeta}, {Seidel}, \& {van de
  Ven}}]{Bittner2020}
{Bittner}, A., {S{\'a}nchez-Bl{\'a}zquez}, P., {Gadotti}, D.~A., {et~al.} 2020,
  \aap, 643, A65

\bibitem[{{Blanc} {et~al.}(2009){Blanc}, {Heiderman}, {Gebhardt}, {Evans}, \&
  {Adams}}]{Blanc2009}
{Blanc}, G.~A., {Heiderman}, A., {Gebhardt}, K., {Evans}, Neal~J., I., \&
  {Adams}, J. 2009, \apj, 704, 842

\bibitem[{Blanc {et~al.}(2013)Blanc, Weinzirl, Song, Heiderman, Gebhardt,
  Jogee, Evans~II, Bosch, Luo, Drory, Fabricius, Fisher, Hao, Kaplan, Marinova,
  Vutisalchavakul, \& Yoachim}]{blanc_virus-p_2013}
Blanc, G.~A., Weinzirl, T., Song, M., {et~al.} 2013, The Astronomical Journal,
  145, 138

\bibitem[{{Bolatto} {et~al.}(2017){Bolatto}, {Wong}, {Utomo}, {Blitz}, {Vogel},
  {S{\'a}nchez}, {Barrera-Ballesteros}, {Cao}, {Colombo}, {Dannerbauer},
  {Garc{\'\i}a-Benito}, {Herrera-Camus}, {Husemann}, {Kalinova}, {Leroy},
  {Leung}, {Levy}, {Mast}, {Ostriker}, {Rosolowsky}, {Sandstrom}, {Teuben},
  {van de Ven}, \& {Walter}}]{Bolatto2017}
{Bolatto}, A.~D., {Wong}, T., {Utomo}, D., {et~al.} 2017, \apj, 846, 159

\bibitem[{{Boquien} {et~al.}(2011){Boquien}, {Calzetti}, {Combes}, {Henkel},
  {Israel}, {Kramer}, {Rela{\~n}o}, {Verley}, {van der Werf}, {Xilouris}, \&
  {HERM33ES Team}}]{Boquien2011}
{Boquien}, M., {Calzetti}, D., {Combes}, F., {et~al.} 2011, \aj, 142, 111

\bibitem[{{Boucaud} {et~al.}(2016){Boucaud}, {Bocchio}, {Abergel}, {Orieux},
  {Dole}, \& {Hadj-Youcef}}]{Boucaud2016}
{Boucaud}, A., {Bocchio}, M., {Abergel}, A., {et~al.} 2016, \aap, 596, A63

\bibitem[{Bradley {et~al.}(2020)Bradley, Sip{\H o}cz, Robitaille, Tollerud,
  Vin{\'{\i}}cius, Deil, Barbary, Wilson, Busko, G{\"u}nther, Cara, Conseil,
  Bostroem, Droettboom, Bray, Bratholm, Lim, Barentsen, Craig, Pascual, Perren,
  Greco, Donath, de~Val-Borro, Kerzendorf, Bach, Weaver, D'Eugenio, Souchereau,
  \& Ferreira}]{photutils2020}
Bradley, L., Sip{\H o}cz, B., Robitaille, T., {et~al.} 2020, astropy/photutils:
  1.0.0

\bibitem[{{Brinchmann} {et~al.}(2004){Brinchmann}, {Charlot}, {White},
  {Tremonti}, {Kauffmann}, {Heckman}, \& {Brinkmann}}]{Brinchmann2004}
{Brinchmann}, J., {Charlot}, S., {White}, S.~D.~M., {et~al.} 2004, \mnras, 351,
  1151

\bibitem[{{Bryant} {et~al.}(2019){Bryant}, {Croom}, {van de Sande}, {Scott},
  {Fogarty}, {Bland-Hawthorn}, {Bloom}, {Taylor}, {Brough}, {Robotham},
  {Cortese}, {Couch}, {Owers}, {Medling}, {Federrath}, {Bekki}, {Richards},
  {Lawrence}, \& {Konstantopoulos}}]{Bryant2019}
{Bryant}, J.~J., {Croom}, S.~M., {van de Sande}, J., {et~al.} 2019, \mnras,
  483, 458

\bibitem[{Bundy {et~al.}(2015)Bundy, Bershady, Law, Yan, Drory, MacDonald,
  Wake, Cherinka, Sánchez-Gallego, Weijmans, Thomas, Tremonti, Masters,
  Coccato, Diamond-Stanic, Aragón-Salamanca, Avila-Reese, Badenes,
  Falcón-Barroso, Belfiore, Bizyaev, Blanc, Bland-Hawthorn, Blanton,
  Brownstein, Byler, Cappellari, Conroy, Dutton, Emsellem, Etherington,
  Frinchaboy, Fu, Gunn, Harding, Johnston, Kauffmann, Kinemuchi, Klaene,
  Knapen, Leauthaud, Li, Lin, Maiolino, Malanushenko, Malanushenko, Mao,
  Maraston, McDermid, Merrifield, Nichol, Oravetz, Pan, Parejko, Sanchez,
  Schlegel, Simmons, Steele, Steinmetz, Thanjavur, Thompson, Tinker, van~den
  Bosch, Westfall, Wilkinson, Wright, Xiao, \& Zhang}]{bundy_overview_2015}
Bundy, K., Bershady, M.~A., Law, D.~R., {et~al.} 2015, The Astrophysical
  Journal, 798, 7

\bibitem[{{Calzetti}(2001)}]{calzetti2001}
{Calzetti}, D. 2001, \pasp, 113, 1449

\bibitem[{{Calzetti} {et~al.}(2005){Calzetti}, {Kennicutt}, {Bianchi},
  {Thilker}, {Dale}, {Engelbracht}, {Leitherer}, {Meyer}, {Sosey}, {Mutchler},
  {Regan}, {Thornley}, {Armus}, {Bendo}, {Boissier}, {Boselli}, {Draine},
  {Gordon}, {Helou}, {Hollenbach}, {Kewley}, {Madore}, {Martin}, {Murphy},
  {Rieke}, {Rieke}, {Roussel}, {Sheth}, {Smith}, {Walter}, {White}, {Yi},
  {Scoville}, {Polletta}, \& {Lindler}}]{Calzetti2005}
{Calzetti}, D., {Kennicutt}, R.~C., J., {Bianchi}, L., {et~al.} 2005, \apj,
  633, 871

\bibitem[{{Cano-D{\'\i}az} {et~al.}(2016){Cano-D{\'\i}az}, {S{\'a}nchez},
  {Zibetti}, {Ascasibar}, {Bland-Hawthorn}, {Ziegler}, {Gonz{\'a}lez Delgado},
  {Walcher}, {Garc{\'\i}a-Benito}, {Mast}, {Mendoza-P{\'e}rez},
  {Falc{\'o}n-Barroso}, {Galbany}, {Husemann}, {Kehrig}, {Marino},
  {S{\'a}nchez-Bl{\'a}zquez}, {L{\'o}pez-Cob{\'a}}, {L{\'o}pez-S{\'a}nchez}, \&
  {Vilchez}}]{CanoDiaz2016}
{Cano-D{\'\i}az}, M., {S{\'a}nchez}, S.~F., {Zibetti}, S., {et~al.} 2016,
  \apjl, 821, L26

\bibitem[{{Cappellari}(2017)}]{Cappellari2017}
{Cappellari}, M. 2017, \mnras, 466, 798

\bibitem[{{Cappellari} \& {Copin}(2003)}]{cappellari_adaptive_2003}
{Cappellari}, M. \& {Copin}, Y. 2003, \mnras, 342, 345

\bibitem[{{Cappellari} \& {Emsellem}(2004)}]{cappellari_parametric_2004}
{Cappellari}, M. \& {Emsellem}, E. 2004, \pasp, 116, 138

\bibitem[{Cappellari {et~al.}(2011)Cappellari, Emsellem, Krajnović, McDermid,
  Serra, Alatalo, Blitz, Bois, Bournaud, Bureau, Davies, Davis, Zeeuw,
  Khochfar, Kuntschner, Lablanche, Morganti, Naab, Oosterloo, Sarzi, Scott,
  Weijmans, \& Young}]{cappellari_atlas3d_2011}
Cappellari, M., Emsellem, E., Krajnović, D., {et~al.} 2011, Monthly Notices of
  the Royal Astronomical Society, 416, 1680

\bibitem[{{Cardelli} {et~al.}(1989){Cardelli}, {Clayton}, \&
  {Mathis}}]{Cardelli1989}
{Cardelli}, J.~A., {Clayton}, G.~C., \& {Mathis}, J.~S. 1989, \apj, 345, 245

\bibitem[{{Carrillo} {et~al.}(2020){Carrillo}, {Jogee}, {Drory}, {Kaplan},
  {Blanc}, {Weinzirl}, {Song}, \& {Luo}}]{Carrillo2020}
{Carrillo}, A., {Jogee}, S., {Drory}, N., {et~al.} 2020, \mnras, 493, 4094

\bibitem[{{Chabrier}(2003)}]{Chabrier2003}
{Chabrier}, G. 2003, \pasp, 115, 763

\bibitem[{{Chevance} {et~al.}(2022){Chevance}, {Kruijssen}, {Krumholz},
  {Groves}, {Keller}, {Hughes}, {Glover}, {Henshaw}, {Herrera}, {Kim}, {Leroy},
  {Pety}, {Razza}, {Rosolowsky}, {Schinnerer}, {Schruba}, {Barnes}, {Bigiel},
  {Blanc}, {Dale}, {Emsellem}, {Faesi}, {Grasha}, {Klessen}, {Kreckel}, {Liu},
  {Longmore}, {Meidt}, {Querejeta}, {Saito}, {Sun}, \& {Usero}}]{Chevance2022}
{Chevance}, M., {Kruijssen}, J.~M.~D., {Krumholz}, M.~R., {et~al.} 2022,
  \mnras, 509, 272

\bibitem[{{Cicone} {et~al.}(2017){Cicone}, {Bothwell}, {Wagg}, {M{\o}ller}, {De
  Breuck}, {Zhang}, {Mart{\'\i}n}, {Maiolino}, {Severgnini}, {Aravena},
  {Belfiore}, {Espada}, {Fl{\"u}tsch}, {Impellizzeri}, {Peng}, {Raj},
  {Ram{\'\i}rez-Olivencia}, {Riechers}, \& {Schawinski}}]{Cicone2017}
{Cicone}, C., {Bothwell}, M., {Wagg}, J., {et~al.} 2017, \aap, 604, A53

\bibitem[{{Clark} {et~al.}(2018){Clark}, {Verstocken}, {Bianchi}, {Fritz},
  {Viaene}, {Smith}, {Baes}, {Casasola}, {Cassara}, {Davies}, {De Looze}, {De
  Vis}, {Evans}, {Galametz}, {Jones}, {Lianou}, {Madden}, {Mosenkov}, \&
  {Xilouris}}]{Clark2018}
{Clark}, C.~J.~R., {Verstocken}, S., {Bianchi}, S., {et~al.} 2018, \aap, 609,
  A37

\bibitem[{{Colombo} {et~al.}(2014){Colombo}, {Hughes}, {Schinnerer}, {Meidt},
  {Leroy}, {Pety}, {Dobbs}, {Garc{\'\i}a-Burillo}, {Dumas}, {Thompson},
  {Schuster}, \& {Kramer}}]{Colombo2014}
{Colombo}, D., {Hughes}, A., {Schinnerer}, E., {et~al.} 2014, \apj, 784, 3

\bibitem[{{Corbelli} {et~al.}(2017){Corbelli}, {Braine}, {Bandiera},
  {Brouillet}, {Combes}, {Druard}, {Gratier}, {Mata}, {Schuster}, {Xilouris},
  \& {Palla}}]{Corbelli2017}
{Corbelli}, E., {Braine}, J., {Bandiera}, R., {et~al.} 2017, \aap, 601, A146

\bibitem[{{Croom} {et~al.}(2012){Croom}, {Lawrence}, {Bland-Hawthorn},
  {Bryant}, {Fogarty}, {Richards}, {Goodwin}, {Farrell}, {Miziarski}, {Heald},
  {Jones}, {Lee}, {Colless}, {Brough}, {Hopkins}, {Bauer}, {Birchall}, {Ellis},
  {Horton}, {Leon-Saval}, {Lewis}, {L{\'o}pez-S{\'a}nchez}, {Min}, {Trinh}, \&
  {Trowland}}]{croom2012}
{Croom}, S.~M., {Lawrence}, J.~S., {Bland-Hawthorn}, J., {et~al.} 2012, \mnras,
  421, 872

\bibitem[{{Croom} {et~al.}(2021){Croom}, {Owers}, {Scott}, {Poetrodjojo},
  {Groves}, {van de Sande}, {Barone}, {Cortese}, {D'Eugenio}, {Bland-Hawthorn},
  {Bryant}, {Oh}, {Brough}, {Agostino}, {Casura}, {Catinella}, {Colless},
  {Cecil}, {Davies}, {Drinkwater}, {Driver}, {Ferreras}, {Foster},
  {Fraser-McKelvie}, {Lawrence}, {Leslie}, {Liske}, {L{\'o}pez-S{\'a}nchez},
  {Lorente}, {McElroy}, {Medling}, {Obreschkow}, {Richards}, {Sharp}, {Sweet},
  {Taranu}, {Taylor}, {Tescari}, {Thomas}, {Tocknell}, \&
  {Vaughan}}]{Croom2021}
{Croom}, S.~M., {Owers}, M.~S., {Scott}, N., {et~al.} 2021, \mnras, 505, 991

\bibitem[{{Croxall} {et~al.}(2016){Croxall}, {Pogge}, {Berg}, {Skillman}, \&
  {Moustakas}}]{Croxall2016}
{Croxall}, K.~V., {Pogge}, R.~W., {Berg}, D.~A., {Skillman}, E.~D., \&
  {Moustakas}, J. 2016, \apj, 830, 4

\bibitem[{{de Amorim} {et~al.}(2017){de Amorim}, {Garc{\'\i}a-Benito}, {Cid
  Fernandes}, {Cortijo-Ferrero}, {Gonz{\'a}lez Delgado}, {Lacerda}, {L{\'o}pez
  Fern{\'a}ndez}, {P{\'e}rez}, \& {Vale Asari}}]{deAmorim2017}
{de Amorim}, A.~L., {Garc{\'\i}a-Benito}, R., {Cid Fernandes}, R., {et~al.}
  2017, \mnras, 471, 3727

\bibitem[{de~Zeeuw {et~al.}(2002)de~Zeeuw, Bureau, Emsellem, Bacon, Carollo,
  Copin, Davies, Kuntschner, Miller, Monnet, Peletier, \&
  Verolme}]{de_zeeuw_sauron_2002}
de~Zeeuw, P.~T., Bureau, M., Emsellem, E., {et~al.} 2002,
  {\textbackslash}mnras, 329, 513

\bibitem[{{Draine}(2011)}]{Draine2011}
{Draine}, B.~T. 2011, {Physics of the Interstellar and Intergalactic Medium},
  Princeton Series in Astrophysics (Princeton University Press)

\bibitem[{{Egusa} {et~al.}(2018){Egusa}, {Hirota}, {Baba}, \&
  {Muraoka}}]{Egusa2018}
{Egusa}, F., {Hirota}, A., {Baba}, J., \& {Muraoka}, K. 2018, \apj, 854, 90

\bibitem[{{Ellison} {et~al.}(2021){Ellison}, {Lin}, {Thorp}, {Pan}, {Scudder},
  {S{\'a}nchez}, {Bluck}, \& {Maiolino}}]{Ellison2021}
{Ellison}, S.~L., {Lin}, L., {Thorp}, M.~D., {et~al.} 2021, \mnras, 501, 4777

\bibitem[{{Emsellem} {et~al.}(2019){Emsellem}, {van der Burg}, {Fensch},
  {Je{\v{r}}{\'a}bkov{\'a}}, {Zanella}, {Agnello}, {Hilker}, {M{\"u}ller},
  {Rejkuba}, {Duc}, {Durrell}, {Habas}, {Lelli}, {Lim}, {Marleau}, {Peng}, \&
  {S{\'a}nchez-Janssen}}]{Emsellem2019}
{Emsellem}, E., {van der Burg}, R. F.~J., {Fensch}, J., {et~al.} 2019, \aap,
  625, A76

\bibitem[{{Erroz-Ferrer} {et~al.}(2019){Erroz-Ferrer}, {Carollo}, {den Brok},
  {Onodera}, {Brinchmann}, {Marino}, {Monreal-Ibero}, {Schaye}, {Woo},
  {Cibinel}, {Debattista}, {Inami}, {Maseda}, {Richard}, {Tacchella}, \&
  {Wisotzki}}]{Erroz-Ferrer2019}
{Erroz-Ferrer}, S., {Carollo}, C.~M., {den Brok}, M., {et~al.} 2019, \mnras,
  484, 5009

\bibitem[{{Eskew} {et~al.}(2012){Eskew}, {Zaritsky}, \& {Meidt}}]{Eskew2012}
{Eskew}, M., {Zaritsky}, D., \& {Meidt}, S. 2012, \aj, 143, 139

\bibitem[{{Evans} {et~al.}(2018){Evans}, {Riello}, {De Angeli}, {Carrasco},
  {Montegriffo}, {Fabricius}, {Jordi}, {Palaversa}, {Diener}, {Busso},
  {Cacciari}, {van Leeuwen}, {Burgess}, {Davidson}, {Harrison}, {Hodgkin},
  {Pancino}, {Richards}, {Altavilla}, {Balaguer-N{\'u}{\~n}ez}, {Barstow},
  {Bellazzini}, {Brown}, {Castellani}, {Cocozza}, {De Luise}, {Delgado},
  {Ducourant}, {Galleti}, {Gilmore}, {Giuffrida}, {Holl}, {Kewley}, {Koposov},
  {Marinoni}, {Marrese}, {Osborne}, {Piersimoni}, {Portell}, {Pulone},
  {Ragaini}, {Sanna}, {Terrett}, {Walton}, {Wevers}, \&
  {Wyrzykowski}}]{Evans2018}
{Evans}, D.~W., {Riello}, M., {De Angeli}, F., {et~al.} 2018, \aap, 616, A4

\bibitem[{{Fujimoto} {et~al.}(2019){Fujimoto}, {Chevance}, {Haydon},
  {Krumholz}, \& {Kruijssen}}]{Fujimoto2019}
{Fujimoto}, Y., {Chevance}, M., {Haydon}, D.~T., {Krumholz}, M.~R., \&
  {Kruijssen}, J.~M.~D. 2019, \mnras, 487, 1717

\bibitem[{{Fusco} {et~al.}(2020){Fusco}, {Bacon}, {Kamann}, {Conseil},
  {Neichel}, {Correia}, {Beltramo-Martin}, {Vernet}, {Kolb}, \&
  {Madec}}]{Fusco2020}
{Fusco}, T., {Bacon}, R., {Kamann}, S., {et~al.} 2020, \aap, 635, A208

\bibitem[{{Gadotti} {et~al.}(2019){Gadotti}, {S{\'a}nchez-Bl{\'a}zquez},
  {Falc{\'o}n-Barroso}, {Husemann}, {Seidel}, {P{\'e}rez}, {de
  Lorenzo-C{\'a}ceres}, {Martinez-Valpuesta}, {Fragkoudi}, {Leung}, {van de
  Ven}, {Leaman}, {Coelho}, {Martig}, {Kim}, {Neumann}, \&
  {Querejeta}}]{2019MNRAS.482..506G}
{Gadotti}, D.~A., {S{\'a}nchez-Bl{\'a}zquez}, P., {Falc{\'o}n-Barroso}, J.,
  {et~al.} 2019, \mnras, 482, 506

\bibitem[{{Gaia Collaboration} {et~al.}(2018){Gaia Collaboration}, {Brown},
  {Vallenari}, {Prusti}, {de Bruijne}, {Babusiaux}, {Bailer-Jones}, {Biermann},
  {Evans}, \& al.}]{GAIA2018}
{Gaia Collaboration}, {Brown}, A.~G.~A., {Vallenari}, A., {et~al.} 2018, \aap,
  616, A1

\bibitem[{{Gensior} {et~al.}(2020){Gensior}, {Kruijssen}, \&
  {Keller}}]{Gensior2020}
{Gensior}, J., {Kruijssen}, J.~M.~D., \& {Keller}, B.~W. 2020, \mnras, 495, 199

\bibitem[{{Genzel} {et~al.}(2010){Genzel}, {Tacconi}, {Gracia-Carpio},
  {Sternberg}, {Cooper}, {Shapiro}, {Bolatto}, {Bouch{\'e}}, {Bournaud},
  {Burkert}, {Combes}, {Comerford}, {Cox}, {Davis}, {F{\"o}rster Schreiber},
  {Garcia-Burillo}, {Lutz}, {Naab}, {Neri}, {Omont}, {Shapley}, \&
  {Weiner}}]{Genzel2010}
{Genzel}, R., {Tacconi}, L.~J., {Gracia-Carpio}, J., {et~al.} 2010, \mnras,
  407, 2091

\bibitem[{{Gil de Paz} {et~al.}(2007){Gil de Paz}, {Boissier}, {Madore},
  {Seibert}, {Joe}, {Boselli}, {Wyder}, {Thilker}, {Bianchi}, {Rey}, {Rich},
  {Barlow}, {Conrow}, {Forster}, {Friedman}, {Martin}, {Morrissey}, {Neff},
  {Schiminovich}, {Small}, {Donas}, {Heckman}, {Lee}, {Milliard}, {Szalay}, \&
  {Yi}}]{de_paz2007}
{Gil de Paz}, A., {Boissier}, S., {Madore}, B.~F., {et~al.} 2007, \apjs, 173,
  185

\bibitem[{Grisdale {et~al.}(2017)Grisdale, Agertz, Romeo, Renaud, \&
  Read}]{grisdale_impact_2017}
Grisdale, K., Agertz, O., Romeo, A.~B., Renaud, F., \& Read, J.~I. 2017,
  Monthly Notices of the Royal Astronomical Society, 466, 1093

\bibitem[{{Haas} {et~al.}(2013){Haas}, {Schaye}, {Booth}, {Dalla Vecchia},
  {Springel}, {Theuns}, \& {Wiersma}}]{Haas2013}
{Haas}, M.~R., {Schaye}, J., {Booth}, C.~M., {et~al.} 2013, \mnras, 435, 2955

\bibitem[{{Haffner} {et~al.}(2009){Haffner}, {Dettmar}, {Beckman}, {Wood},
  {Slavin}, {Giammanco}, {Madsen}, {Zurita}, \& {Reynolds}}]{Haffner2009}
{Haffner}, L.~M., {Dettmar}, R.~J., {Beckman}, J.~E., {et~al.} 2009, Reviews of
  Modern Physics, 81, 969

\bibitem[{Harris {et~al.}(2020)Harris, Millman, van~der Walt, Gommers,
  Virtanen, Cournapeau, Wieser, Taylor, Berg, Smith, Kern, Picus, Hoyer, van
  Kerkwijk, Brett, Haldane, Fernández~del Río, Wiebe, Peterson,
  Gérard-Marchant, Sheppard, Reddy, Weckesser, Abbasi, Gohlke, \&
  Oliphant}]{numpy}
Harris, C.~R., Millman, K.~J., van~der Walt, S.~J., {et~al.} 2020, Nature, 585,
  357–362

\bibitem[{{Henshaw} {et~al.}(2020{\natexlab{a}}){Henshaw}, {Ginsburg}, \&
  {Riener}}]{Henshaw2020b}
{Henshaw}, J., {Ginsburg}, A., \& {Riener}, M. 2020{\natexlab{a}}, {scousepy:
  Semi-automated multi-COmponent Universal Spectral-line fitting Engine}

\bibitem[{{Henshaw} {et~al.}(2020{\natexlab{b}}){Henshaw}, {Kruijssen},
  {Longmore}, {Riener}, {Leroy}, {Rosolowsky}, {Ginsburg}, {Battersby},
  {Chevance}, {Meidt}, {Glover}, {Hughes}, {Kainulainen}, {Klessen},
  {Schinnerer}, {Schruba}, {Beuther}, {Bigiel}, {Blanc}, {Emsellem}, {Henning},
  {Herrera}, {Koch}, {Pety}, {Ragan}, \& {Sun}}]{Henshaw2020}
{Henshaw}, J.~D., {Kruijssen}, J.~M.~D., {Longmore}, S.~N., {et~al.}
  2020{\natexlab{b}}, Nature Astronomy, 4, 1064

\bibitem[{{Herrera} {et~al.}(2020){Herrera}, {Pety}, {Hughes}, {Meidt},
  {Kreckel}, {Querejeta}, {Saito}, {Lang}, {Jim{\'e}nez-Donaire}, {Pessa},
  {Cormier}, {Usero}, {Sliwa}, {Faesi}, {Blanc}, {Bigiel}, {Chevance}, {Dale},
  {Grasha}, {Glover}, {Hygate}, {Kruijssen}, {Leroy}, {Rosolowsky},
  {Schinnerer}, {Schruba}, {Sun}, \& {Utomo}}]{Herrera2020}
{Herrera}, C.~N., {Pety}, J., {Hughes}, A., {et~al.} 2020, \aap, 634, A121

\bibitem[{{Hirota} {et~al.}(2018){Hirota}, {Egusa}, {Baba}, {Kuno}, {Muraoka},
  {Tosaki}, {Miura}, {Nakanishi}, \& {Kawabe}}]{Hirota2018}
{Hirota}, A., {Egusa}, F., {Baba}, J., {et~al.} 2018, \pasj, 70, 73

\bibitem[{{Ho} {et~al.}(2019){Ho}, {Kreckel}, {Meidt}, {Groves}, {Blanc},
  {Bigiel}, {Dale}, {Emsellem}, {Glover}, {Grasha}, {Kewley}, {Kruijssen},
  {Lang}, {McElroy}, {Kudritzki}, {Sanchez-Blazquez}, {Sand strom}, {Santoro},
  {Schinnerer}, \& {Schruba}}]{Ho2019}
{Ho}, I.~T., {Kreckel}, K., {Meidt}, S.~E., {et~al.} 2019, \apjl, 885, L31

\bibitem[{{Ho} {et~al.}(2016){Ho}, {Medling}, {Groves}, {Rich}, {Rupke},
  {Hampton}, {Kewley}, {Bland-Hawthorn}, {Croom}, {Richards}, {Schaefer},
  {Sharp}, \& {Sweet}}]{Ho2016}
{Ho}, I.~T., {Medling}, A.~M., {Groves}, B., {et~al.} 2016, \apss, 361, 280

\bibitem[{{Ho} {et~al.}(2018){Ho}, {Meidt}, {Kudritzki}, {Groves}, {Seibert},
  {Madore}, {Schinnerer}, {Rich}, {Kobayashi}, \& {Kewley}}]{Ho2018}
{Ho}, I.~T., {Meidt}, S.~E., {Kudritzki}, R.-P., {et~al.} 2018, \aap, 618, A64

\bibitem[{{Ho} {et~al.}(2017){Ho}, {Seibert}, {Meidt}, {Kudritzki},
  {Kobayashi}, {Groves}, {Kewley}, {Madore}, {Rich}, {Schinnerer},
  {D'Agostino}, \& {Poetrodjojo}}]{Ho2017}
{Ho}, I.~T., {Seibert}, M., {Meidt}, S.~E., {et~al.} 2017, \apj, 846, 39

\bibitem[{{Hopkins} {et~al.}(2014){Hopkins}, {Kere{\v{s}}}, {O{\~n}orbe},
  {Faucher-Gigu{\`e}re}, {Quataert}, {Murray}, \& {Bullock}}]{Hopkins2014}
{Hopkins}, P.~F., {Kere{\v{s}}}, D., {O{\~n}orbe}, J., {et~al.} 2014, \mnras,
  445, 581

\bibitem[{{Hopkins} {et~al.}(2013){Hopkins}, {Narayanan}, \&
  {Murray}}]{Hopkins2013}
{Hopkins}, P.~F., {Narayanan}, D., \& {Murray}, N. 2013, \mnras, 432, 2647

\bibitem[{{Hsieh} {et~al.}(2017){Hsieh}, {Lin}, {Lin}, {Pan}, {Hsu},
  {S{\'a}nchez}, {Cano-D{\'\i}az}, {Zhang}, {Yan}, {Barrera-Ballesteros},
  {Boquien}, {Riffel}, {Brownstein}, {Cruz-Gonz{\'a}lez}, {Hagen}, {Ibarra},
  {Pan}, {Bizyaev}, {Oravetz}, \& {Simmons}}]{Hsieh2017}
{Hsieh}, B.~C., {Lin}, L., {Lin}, J.~H., {et~al.} 2017, \apjl, 851, L24

\bibitem[{{Hughes} {et~al.}(2016){Hughes}, {Meidt}, {Colombo}, {Schruba},
  {Schinnerer}, {Leroy}, \& {Wong}}]{Hughes2016}
{Hughes}, A., {Meidt}, S., {Colombo}, D., {et~al.} 2016, in From Interstellar
  Clouds to Star-Forming Galaxies: Universal Processes?, ed. P.~{Jablonka},
  P.~{Andr{\'e}}, \& F.~{van der Tak}, Vol. 315, 30--37

\bibitem[{{Hunter} {et~al.}(1998){Hunter}, {Elmegreen}, \&
  {Baker}}]{Hunter1998}
{Hunter}, D.~A., {Elmegreen}, B.~G., \& {Baker}, A.~L. 1998, \apj, 493, 595

\bibitem[{Hunter(2007)}]{matplotlib}
Hunter, J.~D. 2007, Computing in Science \& Engineering, 9, 90

\bibitem[{{Husser} {et~al.}(2016){Husser}, {Kamann}, {Dreizler}, {Wendt},
  {Wulff}, {Bacon}, {Wisotzki}, {Brinchmann}, {Weilbacher}, {Roth}, \&
  {Monreal-Ibero}}]{Husser2016}
{Husser}, T.-O., {Kamann}, S., {Dreizler}, S., {et~al.} 2016, \aap, 588, A148

\bibitem[{{Jeffreson} {et~al.}(2020){Jeffreson}, {Kruijssen}, {Keller},
  {Chevance}, \& {Glover}}]{Jeffreson2020}
{Jeffreson}, S. M.~R., {Kruijssen}, J.~M.~D., {Keller}, B.~W., {Chevance}, M.,
  \& {Glover}, S. C.~O. 2020, \mnras, 498, 385

\bibitem[{{Jim{\'e}nez-Donaire} {et~al.}(2019){Jim{\'e}nez-Donaire}, {Bigiel},
  {Leroy}, {Usero}, {Cormier}, {Puschnig}, {Gallagher}, {Kepley}, {Bolatto},
  {Garc{\'\i}a-Burillo}, {Hughes}, {Kramer}, {Pety}, {Schinnerer}, {Schruba},
  {Schuster}, \& {Walter}}]{JimenezDonaire2019}
{Jim{\'e}nez-Donaire}, M.~J., {Bigiel}, F., {Leroy}, A.~K., {et~al.} 2019,
  \apj, 880, 127

\bibitem[{{Johansson} {et~al.}(2012){Johansson}, {Thomas}, \&
  {Maraston}}]{Johansson2012}
{Johansson}, J., {Thomas}, D., \& {Maraston}, C. 2012, \mnras, 421, 1908

\bibitem[{{Kalinova} {et~al.}(2017){Kalinova}, {van de Ven}, {Lyubenova},
  {Falc{\'o}n-Barroso}, {Colombo}, \& {Rosolowsky}}]{Kalinova2017}
{Kalinova}, V., {van de Ven}, G., {Lyubenova}, M., {et~al.} 2017, \mnras, 464,
  1903

\bibitem[{Kauffmann {et~al.}(2003)Kauffmann, Heckman, Tremonti, Brinchmann,
  Charlot, White, Ridgway, Brinkmann, Fukugita, Hall, Ivezić, Richards, \&
  Schneider}]{kauffmann_host_2003}
Kauffmann, G., Heckman, T.~M., Tremonti, C., {et~al.} 2003, Monthly Notices of
  the Royal Astronomical Society, 346, 1055

\bibitem[{{Kawamura} {et~al.}(2009){Kawamura}, {Mizuno}, {Minamidani},
  {Filipovi{\'c}}, {Staveley-Smith}, {Kim}, {Mizuno}, {Onishi}, {Mizuno}, \&
  {Fukui}}]{Kawamura2009}
{Kawamura}, A., {Mizuno}, Y., {Minamidani}, T., {et~al.} 2009, \apjs, 184, 1

\bibitem[{{Kennicutt}(1989)}]{Kennicutt1989}
{Kennicutt}, Robert~C., J. 1989, \apj, 344, 685

\bibitem[{{Kennicutt}(1998)}]{Kennicutt1998}
{Kennicutt}, Robert~C., J. 1998, \apj, 498, 541

\bibitem[{{Kennicutt} {et~al.}(2003){Kennicutt}, {Armus}, {Bendo}, {Calzetti},
  {Dale}, {Draine}, {Engelbracht}, {Gordon}, {Grauer}, {Helou}, {Hollenbach},
  {Jarrett}, {Kewley}, {Leitherer}, {Li}, {Malhotra}, {Regan}, {Rieke},
  {Rieke}, {Roussel}, {Smith}, {Thornley}, \& {Walter}}]{Kennicutt2003}
{Kennicutt}, Robert~C., J., {Armus}, L., {Bendo}, G., {et~al.} 2003, \pasp,
  115, 928

\bibitem[{{Kennicutt} {et~al.}(2011){Kennicutt}, {Calzetti}, {Aniano},
  {Appleton}, {Armus}, {Beir{\~a}o}, {Bolatto}, {Brandl}, {Crocker}, {Croxall},
  {Dale}, {Donovan Meyer}, {Draine}, {Engelbracht}, {Galametz}, {Gordon},
  {Groves}, {Hao}, {Helou}, {Hinz}, {Hunt}, {Johnson}, {Koda}, {Krause},
  {Leroy}, {Li}, {Meidt}, {Montiel}, {Murphy}, {Rahman}, {Rix}, {Roussel},
  {Sandstrom}, {Sauvage}, {Schinnerer}, {Skibba}, {Smith}, {Srinivasan},
  {Vigroux}, {Walter}, {Wilson}, {Wolfire}, \& {Zibetti}}]{Kennicutt2011}
{Kennicutt}, R.~C., {Calzetti}, D., {Aniano}, G., {et~al.} 2011, \pasp, 123,
  1347

\bibitem[{{Kim} {et~al.}(2021){Kim}, {Chevance}, {Kruijssen}, {Schruba},
  {Sandstrom}, {Barnes}, {Bigiel}, {Blanc}, {Cao}, {Dale}, {Faesi}, {Glover},
  {Grasha}, {Groves}, {Herrera}, {Klessen}, {Kreckel}, {Lee}, {Leroy}, {Pety},
  {Querejeta}, {Schinnerer}, {Sun}, {Usero}, {Ward}, \& {Williams}}]{Kim2021}
{Kim}, J., {Chevance}, M., {Kruijssen}, J.~M.~D., {et~al.} 2021, \mnras, 504,
  487

\bibitem[{{Kopsacheili} {et~al.}(2020){Kopsacheili}, {Zezas}, \&
  {Leonidaki}}]{Kopsacheili2020}
{Kopsacheili}, M., {Zezas}, A., \& {Leonidaki}, I. 2020, \mnras, 491, 889

\bibitem[{{Kreckel} {et~al.}(2016){Kreckel}, {Blanc}, {Schinnerer}, {Groves},
  {Adamo}, {Hughes}, \& {Meidt}}]{Kreckel2016}
{Kreckel}, K., {Blanc}, G.~A., {Schinnerer}, E., {et~al.} 2016, \apj, 827, 103

\bibitem[{{Kreckel} {et~al.}(2018){Kreckel}, {Faesi}, {Kruijssen}, {Schruba},
  {Groves}, {Leroy}, {Bigiel}, {Blanc}, {Chevance}, {Herrera}, {Hughes},
  {McElroy}, {Pety}, {Querejeta}, {Rosolowsky}, {Schinnerer}, {Sun}, {Usero},
  \& {Utomo}}]{Kreckel2018}
{Kreckel}, K., {Faesi}, C., {Kruijssen}, J.~M.~D., {et~al.} 2018, \apjl, 863,
  L21

\bibitem[{{Kreckel} {et~al.}(2017){Kreckel}, {Groves}, {Bigiel}, {Blanc},
  {Kruijssen}, {Hughes}, {Schruba}, \& {Schinnerer}}]{Kreckel2017}
{Kreckel}, K., {Groves}, B., {Bigiel}, F., {et~al.} 2017, \apj, 834, 174

\bibitem[{{Kreckel} {et~al.}(2020){Kreckel}, {Ho}, {Blanc}, {Glover}, {Groves},
  {Rosolowsky}, {Bigiel}, {Boqu{\'\i}en}, {Chevance}, {Dale}, {Deger},
  {Emsellem}, {Grasha}, {Kim}, {Klessen}, {Kruijssen}, {Lee}, {Leroy}, {Liu},
  {McElroy}, {Meidt}, {Pessa}, {Sanchez-Blazquez}, {Sandstrom}, {Santoro},
  {Scheuermann}, {Schinnerer}, {Schruba}, {Utomo}, {Watkins}, \&
  {Williams}}]{Kreckel2020}
{Kreckel}, K., {Ho}, I.~T., {Blanc}, G.~A., {et~al.} 2020, \mnras, 499, 193

\bibitem[{{Kreckel} {et~al.}(2019){Kreckel}, {Ho}, {Blanc}, {Groves},
  {Santoro}, {Schinnerer}, {Bigiel}, {Chevance}, {Congiu}, {Emsellem}, {Faesi},
  {Glover}, {Grasha}, {Kruijssen}, {Lang}, {Leroy}, {Meidt}, {McElroy}, {Pety},
  {Rosolowsky}, {Saito}, {Sandstrom}, {Sanchez-Blazquez}, \&
  {Schruba}}]{Kreckel2019}
{Kreckel}, K., {Ho}, I.~T., {Blanc}, G.~A., {et~al.} 2019, \apj, 887, 80

\bibitem[{{Kretschmer} \& {Teyssier}(2020)}]{Kretschmer2020}
{Kretschmer}, M. \& {Teyssier}, R. 2020, \mnras, 492, 1385

\bibitem[{{Krumholz} {et~al.}(2018){Krumholz}, {Burkhart}, {Forbes}, \&
  {Crocker}}]{Krumholz2018}
{Krumholz}, M.~R., {Burkhart}, B., {Forbes}, J.~C., \& {Crocker}, R.~M. 2018,
  \mnras, 477, 2716

\bibitem[{{Lang} {et~al.}(2020){Lang}, {Meidt}, {Rosolowsky}, {Nofech},
  {Schinnerer}, {Leroy}, {Emsellem}, {Pessa}, {Glover}, {Groves}, {Hughes},
  {Kruijssen}, {Querejeta}, {Schruba}, {Bigiel}, {Blanc}, {Chevance},
  {Colombo}, {Faesi}, {Henshaw}, {Herrera}, {Liu}, {Pety}, {Puschnig}, {Saito},
  {Sun}, \& {Usero}}]{Lang2020}
{Lang}, P., {Meidt}, S.~E., {Rosolowsky}, E., {et~al.} 2020, \apj, 897, 122

\bibitem[{{Lee} {et~al.}(2021){Lee}, {Whitmore}, {Thilker}, {Deger}, {Larson},
  {Ubeda}, {Anand}, {Boquien}, {Chandar}, {Dale}, {Emsellem}, {Leroy},
  {Rosolowsky}, {Schinnerer}, {Schmidt}, {Turner}, {Van Dyk}, {White},
  {Barnes}, {Belfiore}, {Bigiel}, {Blanc}, {Cao}, {Chevance}, {Congiu},
  {Egorov}, {Glover}, {Grasha}, {Groves}, {Henshaw}, {Hughes}, {Klessen},
  {Koch}, {Kreckel}, {Kruijssen}, {Liu}, {Lopez}, {Mayker}, {Meidt}, {Murphy},
  {Pan}, {Pety}, {Querejeta}, {Razza}, {Saito}, {Sanchez-Blazquez}, {Santoro},
  {Sardone}, {Scheuermann}, {Schruba}, {Sun}, {Usero}, {Watkins}, \&
  {Williams}}]{Lee2021}
{Lee}, J.~C., {Whitmore}, B.~C., {Thilker}, D.~A., {et~al.} 2021, arXiv
  e-prints, arXiv:2101.02855

\bibitem[{{Leroy} {et~al.}(2021{\natexlab{a}}){Leroy}, {Hughes}, {Liu}, {Pety},
  {Rosolowsky}, {Saito}, {Schinnerer}, {Schruba}, {Usero}, {Faesi}, {Herrera},
  {Chevance}, {Hygate}, {Kepley}, {Koch}, {Querejeta}, {Sliwa}, {Will},
  {Wilson}, {Anand}, {Barnes}, {Belfiore}, {Be{\v{s}}li{\'c}}, {Bigiel},
  {Blanc}, {Bolatto}, {Boquien}, {Cao}, {Chandar}, {Chastenet}, {Chiang},
  {Congiu}, {Dale}, {Deger}, {den Brok}, {Eibensteiner}, {Emsellem},
  {Garc{\'\i}a-Rodr{\'\i}guez}, {Glover}, {Grasha}, {Groves}, {Henshaw},
  {Jim{\'e}nez Donaire}, {Kim}, {Klessen}, {Kreckel}, {Kruijssen}, {Larson},
  {Lee}, {Mayker}, {McElroy}, {Meidt}, {Mok}, {Pan}, {Puschnig}, {Razza},
  {S{\'a}nchez-Bl'azquez}, {Sandstrom}, {Santoro}, {Sardone}, {Scheuermann},
  {Sun}, {Thilker}, {Turner}, {Ubeda}, {Utomo}, {Watkins}, \&
  {Williams}}]{Leroy2021}
{Leroy}, A.~K., {Hughes}, A., {Liu}, D., {et~al.} 2021{\natexlab{a}}, \apjs,
  255, 19

\bibitem[{{Leroy} {et~al.}(2019){Leroy}, {Sandstrom}, {Lang}, {Lewis}, {Salim},
  {Behrens}, {Chastenet}, {Chiang}, {Gallagher}, {Kessler}, \&
  {Utomo}}]{Leroy2019}
{Leroy}, A.~K., {Sandstrom}, K.~M., {Lang}, D., {et~al.} 2019, \apjs, 244, 24

\bibitem[{{Leroy} {et~al.}(2021{\natexlab{b}}){Leroy}, {Schinnerer}, {Hughes},
  {Rosolowsky}, {Pety}, {Schruba}, {Usero}, {Blanc}, {Chevance}, {Emsellem},
  {Faesi}, {Herrera}, {Liu}, {Meidt}, {Querejeta}, {Saito}, {Sandstrom}, {Sun},
  {Williams}, {Anand}, {Barnes}, {Behrens}, {Belfiore}, {Benincasa},
  {Be{\v{s}}li{\'c}}, {Bigiel}, {Bolatto}, {den Brok}, {Cao}, {Chandar},
  {Chastenet}, {Chiang}, {Congiu}, {Dale}, {Deger}, {Eibensteiner}, {Egorov},
  {Garc{\'\i}a-Rodr{\'\i}guez}, {Glover}, {Grasha}, {Henshaw}, {Ho}, {Kepley},
  {Kim}, {Klessen}, {Kreckel}, {Koch}, {Kruijssen}, {Larson}, {Lee}, {Lopez},
  {Machado}, {Mayker}, {McElroy}, {Murphy}, {Ostriker}, {Pan}, {Pessa},
  {Puschnig}, {Razza}, {S{\'a}nchez-Bl{\'a}zquez}, {Santoro}, {Sardone},
  {Scheuermann}, {Sliwa}, {Sormani}, {Stuber}, {Thilker}, {Turner}, {Utomo},
  {Watkins}, \& {Whitmore}}]{Leroy2021b}
{Leroy}, A.~K., {Schinnerer}, E., {Hughes}, A., {et~al.} 2021{\natexlab{b}},
  \apjs, 257, 43

\bibitem[{{Leroy} {et~al.}(2008){Leroy}, {Walter}, {Brinks}, {Bigiel}, {de
  Blok}, {Madore}, \& {Thornley}}]{Leroy2008}
{Leroy}, A.~K., {Walter}, F., {Brinks}, E., {et~al.} 2008, \aj, 136, 2782

\bibitem[{{Leroy} {et~al.}(2013){Leroy}, {Walter}, {Sandstrom}, {Schruba},
  {Munoz-Mateos}, {Bigiel}, {Bolatto}, {Brinks}, {de Blok}, {Meidt}, {Rix},
  {Rosolowsky}, {Schinnerer}, {Schuster}, \& {Usero}}]{Leroy2013}
{Leroy}, A.~K., {Walter}, F., {Sandstrom}, K., {et~al.} 2013, \aj, 146, 19

\bibitem[{{Leung} {et~al.}(2018){Leung}, {Leaman}, {van de Ven}, {Lyubenova},
  {Zhu}, {Bolatto}, {Falc{\'o}n-Barroso}, {Blitz}, {Dannerbauer}, {Fisher},
  {Levy}, {Sanchez}, {Utomo}, {Vogel}, {Wong}, \& {Ziegler}}]{Leung2018}
{Leung}, G. Y.~C., {Leaman}, R., {van de Ven}, G., {et~al.} 2018, \mnras, 477,
  254

\bibitem[{{Lin} {et~al.}(2020){Lin}, {Ellison}, {Pan}, {Thorp}, {Su},
  {S{\'a}nchez}, {Belfiore}, {Bothwell}, {Bundy}, {Chen}, {Concas}, {Hsieh},
  {Hsieh}, {Li}, {Maiolino}, {Masters}, {Newman}, {Rowlands}, {Shi},
  {Smethurst}, {Stark}, {Xiao}, \& {Yu}}]{Lin_2020}
{Lin}, L., {Ellison}, S.~L., {Pan}, H.-A., {et~al.} 2020, \apj, 903, 145

\bibitem[{{Lopez} {et~al.}(2011){Lopez}, {Krumholz}, {Bolatto}, {Prochaska}, \&
  {Ramirez-Ruiz}}]{Lopez2011}
{Lopez}, L.~A., {Krumholz}, M.~R., {Bolatto}, A.~D., {Prochaska}, J.~X., \&
  {Ramirez-Ruiz}, E. 2011, \apj, 731, 91

\bibitem[{{Lopez} {et~al.}(2014){Lopez}, {Krumholz}, {Bolatto}, {Prochaska},
  {Ramirez-Ruiz}, \& {Castro}}]{Lopez2014}
{Lopez}, L.~A., {Krumholz}, M.~R., {Bolatto}, A.~D., {et~al.} 2014, \apj, 795,
  121

\bibitem[{{Mac Low} \& {Klessen}(2004)}]{MacLow2004}
{Mac Low}, M.-M. \& {Klessen}, R.~S. 2004, Reviews of Modern Physics, 76, 125

\bibitem[{{Makarov} {et~al.}(2014){Makarov}, {Prugniel}, {Terekhova},
  {Courtois}, \& {Vauglin}}]{Makarov2014}
{Makarov}, D., {Prugniel}, P., {Terekhova}, N., {Courtois}, H., \& {Vauglin},
  I. 2014, \aap, 570, A13

\bibitem[{{Martin} \& {Kennicutt}(2001)}]{Martin2001}
{Martin}, C.~L. \& {Kennicutt}, Robert~C., J. 2001, \apj, 555, 301

\bibitem[{{McKee} \& {Ostriker}(2007)}]{McKee2007}
{McKee}, C.~F. \& {Ostriker}, E.~C. 2007, \araa, 45, 565

\bibitem[{{McLeod} {et~al.}(2020){McLeod}, {Kruijssen}, {Weisz}, {Zeidler},
  {Schruba}, {Dalcanton}, {Longmore}, {Chevance}, {Faesi}, \&
  {Byler}}]{McLeod2020}
{McLeod}, A.~F., {Kruijssen}, J.~M.~D., {Weisz}, D.~R., {et~al.} 2020, \apj,
  891, 25

\bibitem[{{Medling} {et~al.}(2018){Medling}, {Cortese}, {Croom}, {Green},
  {Groves}, {Hampton}, {Ho}, {Davies}, {Kewley}, {Moffett}, {Schaefer},
  {Taylor}, {Zafar}, {Bekki}, {Bland-Hawthorn}, {Bloom}, {Brough}, {Bryant},
  {Catinella}, {Cecil}, {Colless}, {Couch}, {Drinkwater}, {Driver},
  {Federrath}, {Foster}, {Goldstein}, {Goodwin}, {Hopkins}, {Lawrence},
  {Leslie}, {Lewis}, {Lorente}, {Owers}, {McDermid}, {Richards}, {Sharp},
  {Scott}, {Sweet}, {Taranu}, {Tescari}, {Tonini}, {van de Sande}, {Walcher},
  \& {Wright}}]{Medling2018}
{Medling}, A.~M., {Cortese}, L., {Croom}, S.~M., {et~al.} 2018, \mnras, 475,
  5194

\bibitem[{{Meidt} {et~al.}(2013){Meidt}, {Schinnerer}, {Garc{\'\i}a-Burillo},
  {Hughes}, {Colombo}, {Pety}, {Dobbs}, {Schuster}, {Kramer}, {Leroy}, {Dumas},
  \& {Thompson}}]{Meidt2013}
{Meidt}, S.~E., {Schinnerer}, E., {Garc{\'\i}a-Burillo}, S., {et~al.} 2013,
  \apj, 779, 45

\bibitem[{{Meidt} {et~al.}(2014){Meidt}, {Schinnerer}, {van de Ven},
  {Zaritsky}, {Peletier}, {Knapen}, {Sheth}, {Regan}, {Querejeta},
  {Mu{\~n}oz-Mateos}, {Kim}, {Hinz}, {Gil de Paz}, {Athanassoula}, {Bosma},
  {Buta}, {Cisternas}, {Ho}, {Holwerda}, {Skibba}, {Laurikainen}, {Salo},
  {Gadotti}, {Laine}, {Erroz-Ferrer}, {Comer{\'o}n}, {Men{\'e}ndez-Delmestre},
  {Seibert}, \& {Mizusawa}}]{Meidt2014}
{Meidt}, S.~E., {Schinnerer}, E., {van de Ven}, G., {et~al.} 2014, \apj, 788,
  144

\bibitem[{{Mingozzi} {et~al.}(2020){Mingozzi}, {Belfiore}, {Cresci}, {Bundy},
  {Bershady}, {Bizyaev}, {Blanc}, {Boquien}, {Drory}, {Fu}, {Maiolino},
  {Riffel}, {Schaefer}, {Storchi-Bergmann}, {Telles}, {Tremonti}, {Zakamska},
  \& {Zhang}}]{Mingozzi2020}
{Mingozzi}, M., {Belfiore}, F., {Cresci}, G., {et~al.} 2020, \aap, 636, A42

\bibitem[{{Momose} {et~al.}(2013){Momose}, {Koda}, {Kennicutt}, {Egusa},
  {Calzetti}, {Liu}, {Donovan Meyer}, {Okumura}, {Scoville}, {Sawada}, \&
  {Kuno}}]{Momose2013}
{Momose}, R., {Koda}, J., {Kennicutt}, Robert~C., J., {et~al.} 2013, \apjl,
  772, L13

\bibitem[{{O'Donnell}(1994)}]{ODonnell1994}
{O'Donnell}, J.~E. 1994, \apj, 422, 158

\bibitem[{{Oey} {et~al.}(2003){Oey}, {Parker}, {Mikles}, \& {Zhang}}]{Oey2003}
{Oey}, M.~S., {Parker}, J.~S., {Mikles}, V.~J., \& {Zhang}, X. 2003, \aj, 126,
  2317

\bibitem[{{Olivier} {et~al.}(2021){Olivier}, {Lopez}, {Rosen}, {Nayak},
  {Reiter}, {Krumholz}, \& {Bolatto}}]{Olivier2021}
{Olivier}, G.~M., {Lopez}, L.~A., {Rosen}, A.~L., {et~al.} 2021, \apj, 908, 68

\bibitem[{{Ostriker} {et~al.}(2010){Ostriker}, {McKee}, \&
  {Leroy}}]{Ostriker2010}
{Ostriker}, E.~C., {McKee}, C.~F., \& {Leroy}, A.~K. 2010, \apj, 721, 975

\bibitem[{{Ostriker} \& {Shetty}(2011)}]{Ostriker2011}
{Ostriker}, E.~C. \& {Shetty}, R. 2011, \apj, 731, 41

\bibitem[{{Padmanabhan} {et~al.}(2008){Padmanabhan}, {Schlegel}, {Finkbeiner},
  {Barentine}, {Blanton}, {Brewington}, {Gunn}, {Harvanek}, {Hogg},
  {Ivezi{\'c}}, {Johnston}, {Kent}, {Kleinman}, {Knapp}, {Krzesinski}, {Long},
  {Neilsen}, {Nitta}, {Loomis}, {Lupton}, {Roweis}, {Snedden}, {Strauss}, \&
  {Tucker}}]{Padmanabhan2008}
{Padmanabhan}, N., {Schlegel}, D.~J., {Finkbeiner}, D.~P., {et~al.} 2008, \apj,
  674, 1217

\bibitem[{{Pellegrini} {et~al.}(2011){Pellegrini}, {Baldwin}, \&
  {Ferland}}]{Pellegrini2011}
{Pellegrini}, E.~W., {Baldwin}, J.~A., \& {Ferland}, G.~J. 2011, \apj, 738, 34

\bibitem[{{Pellegrini} {et~al.}(2020){Pellegrini}, {Reissl}, {Rahner},
  {Klessen}, {Glover}, {Pakmor}, {Herrera-Camus}, \& {Grand}}]{Pellegrini2020}
{Pellegrini}, E.~W., {Reissl}, S., {Rahner}, D., {et~al.} 2020, \mnras, 498,
  3193

\bibitem[{{Pessa} {et~al.}(2021){Pessa}, {Schinnerer}, {Belfiore}, {Emsellem},
  {Leroy}, {Schruba}, {Kruijssen}, {Pan}, {Blanc}, {Sanchez-Blazquez},
  {Bigiel}, {Chevance}, {Congiu}, {Dale}, {Faesi}, {Glover}, {Grasha},
  {Groves}, {Ho}, {Jim{\'e}nez-Donaire}, {Klessen}, {Kreckel}, {Koch}, {Liu},
  {Meidt}, {Pety}, {Querejeta}, {Rosolowsky}, {Saito}, {Santoro}, {Sun},
  {Usero}, {Watkins}, \& {Williams}}]{Pessa2021}
{Pessa}, I., {Schinnerer}, E., {Belfiore}, F., {et~al.} 2021, \aap, 650, A134

\bibitem[{{Pietrinferni} {et~al.}(2004){Pietrinferni}, {Cassisi}, {Salaris}, \&
  {Castelli}}]{Pietrinferni2004}
{Pietrinferni}, A., {Cassisi}, S., {Salaris}, M., \& {Castelli}, F. 2004, \apj,
  612, 168

\bibitem[{{Poggianti} {et~al.}(2017){Poggianti}, {Moretti}, {Gullieuszik},
  {Fritz}, {Jaff{\'e}}, {Bettoni}, {Fasano}, {Bellhouse}, {Hau}, {Vulcani},
  {Biviano}, {Omizzolo}, {Paccagnella}, {D'Onofrio}, {Cava}, {Sheen}, {Couch},
  \& {Owers}}]{2017ApJ...844...48P}
{Poggianti}, B.~M., {Moretti}, A., {Gullieuszik}, M., {et~al.} 2017, \apj, 844,
  48

\bibitem[{{Querejeta} {et~al.}(2015){Querejeta}, {Meidt}, {Schinnerer},
  {Cisternas}, {Mu{\~n}oz-Mateos}, {Sheth}, {Knapen}, {van de Ven}, {Norris},
  {Peletier}, {Laurikainen}, {Salo}, {Holwerda}, {Athanassoula}, {Bosma},
  {Groves}, {Ho}, {Gadotti}, {Zaritsky}, {Regan}, {Hinz}, {Gil de Paz},
  {Menendez-Delmestre}, {Seibert}, {Mizusawa}, {Kim}, {Erroz-Ferrer}, {Laine},
  \& {Comer{\'o}n}}]{Querejeta2015}
{Querejeta}, M., {Meidt}, S.~E., {Schinnerer}, E., {et~al.} 2015, \apjs, 219, 5

\bibitem[{{Querejeta} {et~al.}(2021){Querejeta}, {Schinnerer}, {Meidt}, {Sun},
  {Leroy}, {Emsellem}, {Klessen}, {Munoz-Mateos}, {Salo}, {Laurikainen},
  {Beslic}, {Blanc}, {Chevance}, {Dale}, {Eibensteiner}, {Faesi},
  {Garcia-Rodriguez}, {Glover}, {Grasha}, {Henshaw}, {Herrera}, {Hughes},
  {Kreckel}, {Kruijssen}, {Liu}, {Murphy}, {Pan}, {Pety}, {Razza},
  {Rosolowsky}, {Saito}, {Schruba}, {Usero}, {Watkins}, \&
  {Williams}}]{Querejeta2021}
{Querejeta}, M., {Schinnerer}, E., {Meidt}, S., {et~al.} 2021, \aap, in press;
  arXiv:2109.04491, arXiv:2109.04491

\bibitem[{{Rathjen} {et~al.}(2021){Rathjen}, {Naab}, {Girichidis}, {Walch},
  {W{\"u}nsch}, {Dinnbier}, {Seifried}, {Klessen}, \& {Glover}}]{Rathjen2021}
{Rathjen}, T.-E., {Naab}, T., {Girichidis}, P., {et~al.} 2021, \mnras, 504,
  1039

\bibitem[{Renaud {et~al.}(2015)Renaud, Bournaud, Emsellem, Agertz,
  Athanassoula, Combes, Elmegreen, Kraljic, Motte, \&
  Teyssier}]{renaud_environmental_2015}
Renaud, F., Bournaud, F., Emsellem, E., {et~al.} 2015, Monthly Notices of the
  Royal Astronomical Society, 454, 3299

\bibitem[{{Riello} {et~al.}(2018){Riello}, {De Angeli}, {Evans}, {Busso},
  {Hambly}, {Davidson}, {Burgess}, {Montegriffo}, {Osborne}, {Kewley},
  {Carrasco}, {Fabricius}, {Jordi}, {Cacciari}, {van Leeuwen}, \&
  {Holland}}]{Riello2018}
{Riello}, M., {De Angeli}, F., {Evans}, D.~W., {et~al.} 2018, \aap, 616, A3

\bibitem[{{Romeo}(2020)}]{Romeo2020}
{Romeo}, A.~B. 2020, \mnras, 491, 4843

\bibitem[{{Rosales-Ortega} {et~al.}(2010){Rosales-Ortega}, {Kennicutt},
  {S{\'a}nchez}, {D{\'\i}az}, {Pasquali}, {Johnson}, \&
  {Hao}}]{Rosales-Ortega2010}
{Rosales-Ortega}, F.~F., {Kennicutt}, R.~C., {S{\'a}nchez}, S.~F., {et~al.}
  2010, \mnras, 405, 735

\bibitem[{{Rosolowsky} {et~al.}(2021){Rosolowsky}, {Hughes}, {Leroy}, {Sun},
  {Querejeta}, {Schruba}, {Usero}, {Herrera}, {Liu}, {Pety}, {Saito},
  {Be{\v{s}}li{\'c}}, {Bigiel}, {Blanc}, {Chevance}, {Dale}, {Deger}, {Faesi},
  {Glover}, {Henshaw}, {Klessen}, {Kruijssen}, {Larson}, {Lee}, {Meidt}, {Mok},
  {Schinnerer}, {Thilker}, \& {Williams}}]{Rosolowsky2021}
{Rosolowsky}, E., {Hughes}, A., {Leroy}, A.~K., {et~al.} 2021, \mnras, 502,
  1218

\bibitem[{{Rousseau-Nepton} {et~al.}(2019){Rousseau-Nepton}, {Martin},
  {Robert}, {Drissen}, {Amram}, {Prunet}, {Martin}, {Moumen}, {Adamo},
  {Alarie}, {Barmby}, {Boselli}, {Bresolin}, {Bureau}, {Chemin}, {Fernandes},
  {Combes}, {Crowder}, {Della Bruna}, {Duarte Puertas}, {Egusa}, {Epinat},
  {Ksoll}, {Girard}, {G{\'o}mez Llanos}, {Gouliermis}, {Grasha}, {Higgs},
  {Hlavacek-Larrondo}, {Ho}, {Iglesias-P{\'a}ramo}, {Joncas}, {Kam}, {Karera},
  {Kennicutt}, {Klessen}, {Lianou}, {Liu}, {Liu}, {de Amorim}, {Lyman},
  {Martel}, {Mazzilli-Ciraulo}, {McLeod}, {Melchior}, {Millan}, {Moll{\'a}},
  {Momose}, {Morisset}, {Pan}, {Pati}, {Pellerin}, {Pellegrini}, {P{\'e}rez},
  {Petric}, {Plana}, {Rahner}, {Ruiz Lara}, {S{\'a}nchez-Menguiano},
  {Spekkens}, {Stasi{\'n}ska}, {Takamiya}, {Vale Asari}, \&
  {V{\'\i}lchez}}]{Rousseau-Nepton2019}
{Rousseau-Nepton}, L., {Martin}, R.~P., {Robert}, C., {et~al.} 2019, \mnras,
  489, 5530

\bibitem[{{Saintonge} {et~al.}(2017){Saintonge}, {Catinella}, {Tacconi},
  {Kauffmann}, {Genzel}, {Cortese}, {Dav{\'e}}, {Fletcher},
  {Graci{\'a}-Carpio}, {Kramer}, {Heckman}, {Janowiecki}, {Lutz}, {Rosario},
  {Schiminovich}, {Schuster}, {Wang}, {Wuyts}, {Borthakur}, {Lamperti}, \&
  {Roberts-Borsani}}]{Saintonge2017}
{Saintonge}, A., {Catinella}, B., {Tacconi}, L.~J., {et~al.} 2017, \apjs, 233,
  22

\bibitem[{{Saintonge} {et~al.}(2011){Saintonge}, {Kauffmann}, {Wang}, {Kramer},
  {Tacconi}, {Buchbender}, {Catinella}, {Graci{\'a}-Carpio}, {Cortese},
  {Fabello}, {Fu}, {Genzel}, {Giovanelli}, {Guo}, {Haynes}, {Heckman},
  {Krumholz}, {Lemonias}, {Li}, {Moran}, {Rodriguez-Fernandez}, {Schiminovich},
  {Schuster}, \& {Sievers}}]{Saintonge2011}
{Saintonge}, A., {Kauffmann}, G., {Wang}, J., {et~al.} 2011, \mnras, 415, 61

\bibitem[{{Salim} {et~al.}(2007){Salim}, {Rich}, {Charlot}, {Brinchmann},
  {Johnson}, {Schiminovich}, {Seibert}, {Mallery}, {Heckman}, {Forster},
  {Friedman}, {Martin}, {Morrissey}, {Neff}, {Small}, {Wyder}, {Bianchi},
  {Donas}, {Lee}, {Madore}, {Milliard}, {Szalay}, {Welsh}, \& {Yi}}]{Salim2007}
{Salim}, S., {Rich}, R.~M., {Charlot}, S., {et~al.} 2007, \apjs, 173, 267

\bibitem[{{S{\'a}nchez} {et~al.}(2021){S{\'a}nchez}, {Barrera-Ballesteros},
  {Colombo}, {Wong}, {Bolatto}, {Rosolowsky}, {Vogel}, {Levy}, {Kalinova},
  {Alvarez-Hurtado}, {Luo}, \& {Cao}}]{Sanchez2021}
{S{\'a}nchez}, S.~F., {Barrera-Ballesteros}, J.~K., {Colombo}, D., {et~al.}
  2021, \mnras, 503, 1615

\bibitem[{{S{\'a}nchez} {et~al.}(2012){S{\'a}nchez}, {Kennicutt}, {Gil de Paz},
  {van de Ven}, {V{\'\i}lchez}, {Wisotzki}, {Walcher}, {Mast}, {Aguerri},
  {Albiol-P{\'e}rez}, {Alonso-Herrero}, {Alves}, {Bakos}, {Bart{\'a}kov{\'a}},
  {Bland-Hawthorn}, {Boselli}, {Bomans}, {Castillo-Morales}, {Cortijo-Ferrero},
  {de Lorenzo-C{\'a}ceres}, {Del Olmo}, {Dettmar}, {D{\'\i}az}, {Ellis},
  {Falc{\'o}n-Barroso}, {Flores}, {Gallazzi}, {Garc{\'\i}a-Lorenzo},
  {Gonz{\'a}lez Delgado}, {Gruel}, {Haines}, {Hao}, {Husemann},
  {Igl{\'e}sias-P{\'a}ramo}, {Jahnke}, {Johnson}, {Jungwiert}, {Kalinova},
  {Kehrig}, {Kupko}, {L{\'o}pez-S{\'a}nchez}, {Lyubenova}, {Marino},
  {M{\'a}rmol-Queralt{\'o}}, {M{\'a}rquez}, {Masegosa}, {Meidt},
  {Mendez-Abreu}, {Monreal-Ibero}, {Montijo}, {Mour{\~a}o}, {Palacios-Navarro},
  {Papaderos}, {Pasquali}, {Peletier}, {P{\'e}rez}, {P{\'e}rez}, {Quirrenbach},
  {Rela{\~n}o}, {Rosales-Ortega}, {Roth}, {Ruiz-Lara},
  {S{\'a}nchez-Bl{\'a}zquez}, {Sengupta}, {Singh}, {Stanishev}, {Trager},
  {Vazdekis}, {Viironen}, {Wild}, {Zibetti}, \& {Ziegler}}]{Sanchez2012}
{S{\'a}nchez}, S.~F., {Kennicutt}, R.~C., {Gil de Paz}, A., {et~al.} 2012,
  \aap, 538, A8

\bibitem[{{S{\'a}nchez} {et~al.}(2016{\natexlab{a}}){S{\'a}nchez}, {P{\'e}rez},
  {S{\'a}nchez-Bl{\'a}zquez}, {Garc{\'\i}a-Benito}, {Ibarra-Mede},
  {Gonz{\'a}lez}, {Rosales-Ortega}, {S{\'a}nchez-Menguiano}, {Ascasibar},
  {Bitsakis}, {Law}, {Cano-D{\'\i}az}, {L{\'o}pez-Cob{\'a}}, {Marino}, {Gil de
  Paz}, {L{\'o}pez-S{\'a}nchez}, {Barrera-Ballesteros}, {Galbany}, {Mast},
  {Abril-Melgarejo}, \& {Roman-Lopes}}]{Sanchez2016b}
{S{\'a}nchez}, S.~F., {P{\'e}rez}, E., {S{\'a}nchez-Bl{\'a}zquez}, P., {et~al.}
  2016{\natexlab{a}}, \rmxaa, 52, 171

\bibitem[{{S{\'a}nchez} {et~al.}(2016{\natexlab{b}}){S{\'a}nchez}, {P{\'e}rez},
  {S{\'a}nchez-Bl{\'a}zquez}, {Gonz{\'a}lez}, {Ros{\'a}les-Ortega},
  {Cano-D{\'\i}az}, {L{\'o}pez-Cob{\'a}}, {Marino}, {Gil de Paz}, {Moll{\'a}},
  {L{\'o}pez-S{\'a}nchez}, {Ascasibar}, \&
  {Barrera-Ballesteros}}]{Sanchez2016a}
{S{\'a}nchez}, S.~F., {P{\'e}rez}, E., {S{\'a}nchez-Bl{\'a}zquez}, P., {et~al.}
  2016{\natexlab{b}}, \rmxaa, 52, 21

\bibitem[{{S{\'a}nchez} {et~al.}(2014){S{\'a}nchez}, {Rosales-Ortega},
  {Iglesias-P{\'a}ramo}, {Moll{\'a}}, {Barrera-Ballesteros}, {Marino},
  {P{\'e}rez}, {S{\'a}nchez-Blazquez}, {Gonz{\'a}lez Delgado}, {Cid Fernandes},
  {de Lorenzo-C{\'a}ceres}, {Mendez-Abreu}, {Galbany}, {Falcon-Barroso},
  {Miralles-Caballero}, {Husemann}, {Garc{\'\i}a-Benito}, {Mast}, {Walcher},
  {Gil de Paz}, {Garc{\'\i}a-Lorenzo}, {Jungwiert}, {V{\'\i}lchez},
  {J{\'\i}lkov{\'a}}, {Lyubenova}, {Cortijo-Ferrero}, {D{\'\i}az}, {Wisotzki},
  {M{\'a}rquez}, {Bland-Hawthorn}, {Ellis}, {van de Ven}, {Jahnke},
  {Papaderos}, {Gomes}, {Mendoza}, \& {L{\'o}pez-S{\'a}nchez}}]{Sanchez2014}
{S{\'a}nchez}, S.~F., {Rosales-Ortega}, F.~F., {Iglesias-P{\'a}ramo}, J.,
  {et~al.} 2014, \aap, 563, A49

\bibitem[{Sanchez-Blazquez {et~al.}(2011)Sanchez-Blazquez, Ocvirk, Gibson,
  P{\'e}rez, \& Peletier}]{Sanchez-Blazquez2011}
Sanchez-Blazquez, P., Ocvirk, P., Gibson, B.~K., P{\'e}rez, I., \& Peletier,
  R.~F. 2011, \mnras, 415, 709

\bibitem[{{S{\'a}nchez-Menguiano} {et~al.}(2016){S{\'a}nchez-Menguiano},
  {S{\'a}nchez}, {Kawata}, {Chemin}, {P{\'e}rez}, {Ruiz-Lara},
  {S{\'a}nchez-Bl{\'a}zquez}, {Galbany}, {Anderson}, {Grand}, {Minchev}, \&
  {G{\'o}mez}}]{Sanchez-Menguiano2016}
{S{\'a}nchez-Menguiano}, L., {S{\'a}nchez}, S.~F., {Kawata}, D., {et~al.} 2016,
  \apjl, 830, L40

\bibitem[{{Sanders} {et~al.}(1985){Sanders}, {Scoville}, \&
  {Solomon}}]{Sanders1985}
{Sanders}, D.~B., {Scoville}, N.~Z., \& {Solomon}, P.~M. 1985, \apj, 289, 373

\bibitem[{{Santoro} {et~al.}(2021){Santoro}, {Kreckel}, {Belfiore}, {Groves},
  {Congiu}, {Thilker}, {Blanc}, {Schinnerer}, {Ho}, {Kruijssen}, {Meidt},
  {Klessen}, {Schruba}, {Querejeta}, {Pessa}, {Chevance}, {Kim}, {Emsellem},
  {McElroy}, {Barnes}, {Bigiel}, {Boquien}, {Dale}, {Glover}, {Grasha}, {Lee},
  {Leroy}, {Pan}, {Rosolowsky}, {Saito}, {Sanchez-Blazquez}, {Watkins}, \&
  {Williams}}]{Santoro2021}
{Santoro}, F., {Kreckel}, K., {Belfiore}, F., {et~al.} 2021, arXiv e-prints,
  arXiv:2111.09362

\bibitem[{{Sarzi} {et~al.}(2006){Sarzi}, {Falc{\'o}n-Barroso}, {Davies},
  {Bacon}, {Bureau}, {Cappellari}, {de Zeeuw}, {Emsellem}, {Fathi},
  {Krajnovi{\'c}}, {Kuntschner}, {McDermid}, \& {Peletier}}]{sarzi_sauron_2006}
{Sarzi}, M., {Falc{\'o}n-Barroso}, J., {Davies}, R.~L., {et~al.} 2006, \mnras,
  366, 1151

\bibitem[{{Scannapieco} {et~al.}(2012){Scannapieco}, {Wadepuhl}, {Parry},
  {Navarro}, {Jenkins}, {Springel}, {Teyssier}, {Carlson}, {Couchman}, {Crain},
  {Dalla Vecchia}, {Frenk}, {Kobayashi}, {Monaco}, {Murante}, {Okamoto},
  {Quinn}, {Schaye}, {Stinson}, {Theuns}, {Wadsley}, {White}, \&
  {Woods}}]{Scannapieco2012}
{Scannapieco}, C., {Wadepuhl}, M., {Parry}, O.~H., {et~al.} 2012, \mnras, 423,
  1726

\bibitem[{{Schlafly} \& {Finkbeiner}(2011)}]{Schlafly2011}
{Schlafly}, E.~F. \& {Finkbeiner}, D.~P. 2011, \apj, 737, 103

\bibitem[{{Schmidt}(1959)}]{Schmidt1959}
{Schmidt}, M. 1959, \apj, 129, 243

\bibitem[{{Schombert} {et~al.}(2019){Schombert}, {McGaugh}, \&
  {Lelli}}]{Schombert2019}
{Schombert}, J., {McGaugh}, S., \& {Lelli}, F. 2019, \mnras, 483, 1496

\bibitem[{{Schruba} {et~al.}(2011){Schruba}, {Leroy}, {Walter}, {Bigiel},
  {Brinks}, {de Blok}, {Dumas}, {Kramer}, {Rosolowsky}, {Sandstrom},
  {Schuster}, {Usero}, {Weiss}, \& {Wiesemeyer}}]{Schruba2011}
{Schruba}, A., {Leroy}, A.~K., {Walter}, F., {et~al.} 2011, \aj, 142, 37

\bibitem[{{Semenov} {et~al.}(2018){Semenov}, {Kravtsov}, \&
  {Gnedin}}]{Semenov2018}
{Semenov}, V.~A., {Kravtsov}, A.~V., \& {Gnedin}, N.~Y. 2018, \apj, 861, 4

\bibitem[{{Semenov} {et~al.}(2021){Semenov}, {Kravtsov}, \&
  {Gnedin}}]{Semenov2021}
{Semenov}, V.~A., {Kravtsov}, A.~V., \& {Gnedin}, N.~Y. 2021, \apj, 918, 13

\bibitem[{{Serre} {et~al.}(2010){Serre}, {Villeneuve}, {Carfantan},
  {Jolissaint}, {Mazet}, {Bourguignon}, \& {Jarno}}]{Serre2010}
{Serre}, D., {Villeneuve}, E., {Carfantan}, H., {et~al.} 2010, in Society of
  Photo-Optical Instrumentation Engineers (SPIE) Conference Series, Vol. 7736,
  Adaptive Optics Systems II, ed. B.~L. {Ellerbroek}, M.~{Hart}, N.~{Hubin}, \&
  P.~L. {Wizinowich}, 773649

\bibitem[{{Sheth} {et~al.}(2010){Sheth}, {Regan}, {Hinz}, {Gil de Paz},
  {Men{\'e}ndez-Delmestre}, {Mu{\~n}oz-Mateos}, {Seibert}, {Kim},
  {Laurikainen}, {Salo}, {Gadotti}, {Laine}, {Mizusawa}, {Armus},
  {Athanassoula}, {Bosma}, {Buta}, {Capak}, {Jarrett}, {Elmegreen},
  {Elmegreen}, {Knapen}, {Koda}, {Helou}, {Ho}, {Madore}, {Masters},
  {Mobasher}, {Ogle}, {Peng}, {Schinnerer}, {Surace}, {Zaritsky},
  {Comer{\'o}n}, {de Swardt}, {Meidt}, {Kasliwal}, \& {Aravena}}]{Sheth2010}
{Sheth}, K., {Regan}, M., {Hinz}, J.~L., {et~al.} 2010, \pasp, 122, 1397

\bibitem[{{Shetty} {et~al.}(2020){Shetty}, {Bershady}, {Westfall},
  {Cappellari}, {Drory}, {Law}, {Yan}, \& {Bundy}}]{Shetty2020}
{Shetty}, S., {Bershady}, M.~A., {Westfall}, K.~B., {et~al.} 2020, \apj, 901,
  101

\bibitem[{{Sorai} {et~al.}(2019){Sorai}, {Kuno}, {Muraoka}, {Miyamoto},
  {Kaneko}, {Nakanishi}, {Nakai}, {Yanagitani}, {Tanaka}, {Sato}, {Salak},
  {Umei}, {Morokuma-Matsui}, {Matsumoto}, {Ueno}, {Pan}, {Noma}, {Takeuchi},
  {Yoda}, {Kuroda}, {Yasuda}, {Yajima}, {Oi}, {Shibata}, {Seta}, {Watanabe},
  {Kita}, {Komatsuzaki}, {Kajikawa}, {Yashima}, {Cooray}, {Baji}, {Segawa},
  {Tashiro}, {Takeda}, {Kishida}, {Hatakeyama}, {Tomiyasu}, \&
  {Saita}}]{Sorai2019}
{Sorai}, K., {Kuno}, N., {Muraoka}, K., {et~al.} 2019, \pasj, 71, S14

\bibitem[{{Soto} {et~al.}(2016){Soto}, {Lilly}, {Bacon}, {Richard}, \&
  {Conseil}}]{Soto2016}
{Soto}, K.~T., {Lilly}, S.~J., {Bacon}, R., {Richard}, J., \& {Conseil}, S.
  2016, \mnras, 458, 3210

\bibitem[{Strauss {et~al.}(2002)Strauss, Weinberg, Lupton, Narayanan, Annis,
  Bernardi, Blanton, Burles, Connolly, Dalcanton, Doi, Eisenstein, Frieman,
  Fukugita, Gunn, Ivezić, Kent, Kim, Knapp, Kron, Munn, Newberg, Nichol,
  Okamura, Quinn, Richmond, Schlegel, Shimasaku, SubbaRao, Szalay, Vanden~Berk,
  Vogeley, Yanny, Yasuda, York, \& Zehavi}]{strauss_spectroscopic_2002}
Strauss, M.~A., Weinberg, D.~H., Lupton, R.~H., {et~al.} 2002, The Astronomical
  Journal, 124, 1810

\bibitem[{{Sun} {et~al.}(2020{\natexlab{a}}){Sun}, {Leroy}, {Ostriker},
  {Hughes}, {Rosolowsky}, {Schruba}, {Schinnerer}, {Blanc}, {Faesi},
  {Kruijssen}, {Meidt}, {Utomo}, {Bigiel}, {Bolatto}, {Chevance}, {Chiang},
  {Dale}, {Emsellem}, {Glover}, {Grasha}, {Henshaw}, {Herrera},
  {Jimenez-Donaire}, {Lee}, {Pety}, {Querejeta}, {Saito}, {Sandstrom}, \&
  {Usero}}]{Sun2020b}
{Sun}, J., {Leroy}, A.~K., {Ostriker}, E.~C., {et~al.} 2020{\natexlab{a}},
  \apj, 892, 148

\bibitem[{{Sun} {et~al.}(2020{\natexlab{b}}){Sun}, {Leroy}, {Schinnerer},
  {Hughes}, {Rosolowsky}, {Querejeta}, {Schruba}, {Liu}, {Saito}, {Herrera},
  {Faesi}, {Usero}, {Pety}, {Kruijssen}, {Ostriker}, {Bigiel}, {Blanc},
  {Bolatto}, {Boquien}, {Chevance}, {Dale}, {Deger}, {Emsellem}, {Glover},
  {Grasha}, {Groves}, {Henshaw}, {Jimenez-Donaire}, {Kim}, {Klessen},
  {Kreckel}, {Lee}, {Meidt}, {Sandstrom}, {Sardone}, {Utomo}, \&
  {Williams}}]{Sun2020}
{Sun}, J., {Leroy}, A.~K., {Schinnerer}, E., {et~al.} 2020{\natexlab{b}},
  \apjl, 901, L8

\bibitem[{{Turner} {et~al.}(2021){Turner}, {Dale}, {Lee}, {Boquien}, {Chandar},
  {Deger}, {Larson}, {Mok}, {Thilker}, {Ubeda}, {Whitmore}, {Belfiore},
  {Bigiel}, {Blanc}, {Emsellem}, {Grasha}, {Groves}, {Klessen}, {Kreckel},
  {Kruijssen}, {Leroy}, {Rosolowsky}, {Sanchez-Blazquez}, {Schinnerer},
  {Schruba}, {Van Dyk}, \& {Williams}}]{Turner2021}
{Turner}, J.~A., {Dale}, D.~A., {Lee}, J.~C., {et~al.} 2021, \mnras, 502, 1366

\bibitem[{{Utreras} {et~al.}(2020){Utreras}, {Blanc}, {Escala}, {Meidt},
  {Emsellem}, {Bigiel}, {Glover}, {Henshaw}, {Hygate}, {Kruijssen},
  {Rosolowsky}, {Schinnerer}, \& {Schruba}}]{Utreras2020}
{Utreras}, J., {Blanc}, G.~A., {Escala}, A., {et~al.} 2020, \apj, 892, 94

\bibitem[{{Vazdekis} {et~al.}(2016){Vazdekis}, {Koleva}, {Ricciardelli},
  {R{\"o}ck}, \& {Falc{\'o}n-Barroso}}]{Vazdekis2016}
{Vazdekis}, A., {Koleva}, M., {Ricciardelli}, E., {R{\"o}ck}, B., \&
  {Falc{\'o}n-Barroso}, J. 2016, \mnras, 463, 3409

\bibitem[{{Venturi} {et~al.}(2018){Venturi}, {Nardini}, {Marconi}, {Carniani},
  {Mingozzi}, {Cresci}, {Mannucci}, {Risaliti}, {Maiolino}, {Balmaverde},
  {Bongiorno}, {Brusa}, {Capetti}, {Cicone}, {Ciroi}, {Feruglio}, {Fiore},
  {Gallazzi}, {La Franca}, {Mainieri}, {Matsuoka}, {Nagao}, {Perna},
  {Piconcelli}, {Sani}, {Tozzi}, \& {Zibetti}}]{Venturi2018}
{Venturi}, G., {Nardini}, E., {Marconi}, A., {et~al.} 2018, \aap, 619, A74

\bibitem[{{Verley} {et~al.}(2007){Verley}, {Combes}, {Verdes-Montenegro},
  {Bergond}, \& {Leon}}]{Verley2007}
{Verley}, S., {Combes}, F., {Verdes-Montenegro}, L., {Bergond}, G., \& {Leon},
  S. 2007, \aap, 474, 43

\bibitem[{{Viaene} {et~al.}(2014){Viaene}, {Fritz}, {Baes}, {Bendo},
  {Blommaert}, {Boquien}, {Boselli}, {Ciesla}, {Cortese}, {De Looze}, {Gear},
  {Gentile}, {Hughes}, {Jarrett}, {Karczewski}, {Smith}, {Spinoglio}, {Tamm},
  {Tempel}, {Thilker}, \& {Verstappen}}]{Viaene2014}
{Viaene}, S., {Fritz}, J., {Baes}, M., {et~al.} 2014, \aap, 567, A71

\bibitem[{Virtanen {et~al.}(2020)Virtanen, Gommers, Oliphant, Haberland, Reddy,
  Cournapeau, Burovski, Peterson, Weckesser, Bright, {van der Walt}, Brett,
  Wilson, Millman, Mayorov, Nelson, Jones, Kern, Larson, Carey, Polat, Feng,
  Moore, {VanderPlas}, Laxalde, Perktold, Cimrman, Henriksen, Quintero, Harris,
  Archibald, Ribeiro, Pedregosa, {van Mulbregt}, \& {SciPy 1.0
  Contributors}}]{scipy}
Virtanen, P., Gommers, R., Oliphant, T.~E., {et~al.} 2020, Nature Methods, 17,
  261

\bibitem[{{Vogt} {et~al.}(2017){Vogt}, {P{\'e}rez}, {Dopita},
  {Verdes-Montenegro}, \& {Borthakur}}]{Vogt2017}
{Vogt}, F.~P.~A., {P{\'e}rez}, E., {Dopita}, M.~A., {Verdes-Montenegro}, L., \&
  {Borthakur}, S. 2017, \aap, 601, A61

\bibitem[{{Vriend}(2015)}]{MuseWise2015}
{Vriend}, W.-J. 2015, in Science Operations 2015: Science Data Management, 1

\bibitem[{{Weilbacher} {et~al.}(2020{\natexlab{a}}){Weilbacher}, {Palsa},
  {Streicher}, {Bacon}, {Urrutia}, {Wisotzki}, {Conseil}, {Husemann}, {Jarno},
  {Kelz}, {P{\'e}contal-Rousset}, {Richard}, {Roth}, {Selman}, \&
  {Vernet}}]{Weilbacher2020}
{Weilbacher}, P.~M., {Palsa}, R., {Streicher}, O., {et~al.} 2020{\natexlab{a}},
  \aap, 641, A28

\bibitem[{{Weilbacher} {et~al.}(2020{\natexlab{b}}){Weilbacher}, {Palsa},
  {Streicher}, {Bacon}, {Urrutia}, {Wisotzki}, {Conseil}, {Husemann}, {Jarno},
  {Kelz}, {P{\'e}contal-Rousset}, {Richard}, {Roth}, {Selman}, \&
  {Vernet}}]{Weilbacher+20}
{Weilbacher}, P.~M., {Palsa}, R., {Streicher}, O., {et~al.} 2020{\natexlab{b}},
  \aap, 641, A28

\bibitem[{{Westfall} {et~al.}(2019){Westfall}, {Cappellari}, {Bershady},
  {Bundy}, {Belfiore}, {Ji}, {Law}, {Schaefer}, {Shetty}, {Tremonti}, {Yan},
  {Andrews}, {Brownstein}, {Cherinka}, {Coccato}, {Drory}, {Maraston},
  {Parikh}, {S{\'a}nchez-Gallego}, {Thomas}, {Weijmans}, {Barrera-Ballesteros},
  {Du}, {Goddard}, {Li}, {Masters}, {Ibarra Medel}, {S{\'a}nchez}, {Yang},
  {Zheng}, \& {Zhou}}]{Westfall2019}
{Westfall}, K.~B., {Cappellari}, M., {Bershady}, M.~A., {et~al.} 2019, \aj,
  158, 231

\bibitem[{{Williams} {et~al.}(2018){Williams}, {Lazzarini}, {Plucinsky},
  {Sasaki}, {Antoniou}, {Vulic}, {Eracleous}, {Long}, {Binder}, {Dalcanton},
  {Lewis}, \& {Weisz}}]{Williams2018}
{Williams}, B.~F., {Lazzarini}, M., {Plucinsky}, P.~P., {et~al.} 2018, \apjs,
  239, 13

\bibitem[{{Williams} {et~al.}(2022){Williams}, {Kreckel}, {Belfiore}, {Groves},
  {Sandstrom}, {Santoro}, {Blanc}, {Bigiel}, {Boquien}, {Chevance}, {Congiu},
  {Emsellem}, {Glover}, {Grasha}, {Klessen}, {Koch}, {Kruijssen}, {Leroy},
  {Liu}, {Meidt}, {Pan}, {Querejeta}, {Rosolowsky}, {Saito},
  {S{\'a}nchez-Bl{\'a}zquez}, {Schinnerer}, {Schruba}, \&
  {Watkins}}]{Williams2022}
{Williams}, T.~G., {Kreckel}, K., {Belfiore}, F., {et~al.} 2022, \mnras, 509,
  1303

\bibitem[{{Williams} {et~al.}(2021){Williams}, {Schinnerer}, {Emsellem},
  {Meidt}, {Querejeta}, {Belfiore}, {Be{\v{s}}li{\'c}}, {Bigiel}, {Chevance},
  {Dale}, {Glover}, {Grasha}, {Klessen}, {Kruijssen}, {Leroy}, {Pan}, {Pety},
  {Pessa}, {Rosolowsky}, {Saito}, {Santoro}, {Schruba}, {Sormani}, {Sun}, \&
  {Watkins}}]{Williams2021}
{Williams}, T.~G., {Schinnerer}, E., {Emsellem}, E., {et~al.} 2021, \aj, 161,
  185

\bibitem[{{Wong} \& {Blitz}(2002)}]{Wong2002}
{Wong}, T. \& {Blitz}, L. 2002, \apj, 569, 157

\bibitem[{{Zaritsky} {et~al.}(1994){Zaritsky}, {Kennicutt}, \&
  {Huchra}}]{Zaritsky1994}
{Zaritsky}, D., {Kennicutt}, Robert~C., J., \& {Huchra}, J.~P. 1994, \apj, 420,
  87

\bibitem[{{Zhang} {et~al.}(2017){Zhang}, {Yan}, {Bundy}, {Bershady}, {Haffner},
  {Walterbos}, {Maiolino}, {Tremonti}, {Thomas}, {Drory}, {Jones}, {Belfiore},
  {S{\'a}nchez}, {Diamond-Stanic}, {Bizyaev}, {Nitschelm}, {Andrews},
  {Brinkmann}, {Brownstein}, {Cheung}, {Li}, {Law}, {Roman Lopes}, {Oravetz},
  {Pan}, {Storchi Bergmann}, \& {Simmons}}]{Zhang2017}
{Zhang}, K., {Yan}, R., {Bundy}, K., {et~al.} 2017, \mnras, 466, 3217

\bibitem[{{Zurita} {et~al.}(2000){Zurita}, {Rozas}, \& {Beckman}}]{Zurita2000}
{Zurita}, A., {Rozas}, M., \& {Beckman}, J.~E. 2000, \aap, 363, 9

\end{thebibliography}

\begin{appendix}

\section{Contributions}
\label{sec:contribs}

The PHANGS--MUSE Survey campaign and the associated data products are the outcome of a large team effort, with major contributions from many and input from the entire team. The VLT/MUSE Large Programme was led by E. Schinnerer. The current paper has benefited from work and insights from many members of the team. In the following, we summarise the key contributions that brought us from the early discussions in 2015, to this point.

\smallskip

\noindent \textbf{Observation Design, Data Processing, and Quality Assurance of the MUSE Data:} The specific efforts required for designing the MUSE science strategy and observations, its implementation, monitoring, data gathering, data reduction and analysis, quality checks was channelled through a MUSE-focused Working Group within the broader PHANGS team, led by E. Emsellem since its implementation in 2017. The PHANGS--MUSE working group included F. Belfiore, G. Blanc, E. Congiu, E. Emsellem, I-T. Ho, B. Groves, K. Kreckel, R. McElroy, I. Pessa, P. Sanchez-Blazquez, F. Santoro, E. Schinnerer. The design and submission of the MUSE Observation Blocks to ESO Portal were handled by B. Groves, R. McElroy, and C. Faesi, with help from E. Emsellem. The observing campaign monitoring and book-keeping were conducted by R. McElroy and F. Santoro, who also led the uploading and management of the raw data on the Heidelberg computer nodes, with contributions from E. Emsellem and E. Schinnerer. An initial shell version of a MUSE dataflow was written by G. Blanc and B. Groves. The {\tt pymusepipe} package was written by E. Emsellem, with team contributions by E. Congiu and F. Santoro. The specific alignment process for the MUSE mosaicking was conducted in parallel by E. Emsellem, I-T. Ho, R. McElroy, K. Kreckel and F. Santoro. The \DAP was developed by F. Belfiore, initially based on the {\tt gist} package \citep{bittner2019}, with contributions by E. Emsellem and I. Pessa (an early version of the data analysis framework was written and implemented by I-T. Ho). Extensive tests pertaining to the extraction of the stellar population information and stellar extinction were conducted by I. Pessa with support from F. Belfiore, E. Emsellem, I-T. Ho and P. Sanchez-Blazquez. An early version of the datasets were reduced by R. McElroy, and analysed by I-T Ho and K. Kreckel using LZIFU \citep{Ho2016}. The subsequent releases were reduced mainly by E. Emsellem and F. Santoro, with the efforts on the analysis led by F. Belfiore and I. Pessa. All data organisation on the MPIA computers was led by R. McElroy from 2017 to 2018, and then by F. Santoro.

\smallskip

\noindent \textbf{Management of the PHANGS Collaboration:} E.~Schinnerer has served as the leader of the PHANGS collaboration since 2015. G.~Blanc, E.~Emsellem, A.~Leroy, and E.~Rosolowsky have acted as the PHANGS steering committee. E.~Rosolowsky has served as team manager since 2018. The PHANGS `core team' provides key input and oversight to all major collaboration decisions. The core team includes: F.~Bigiel, G.~Blanc, E.~Emsellem, A.~Escala, B.~Groves, A.~Hughes, K.~Kreckel, J.M.D.~Kruijssen, J.~Lee, A.~Leroy, S.~Meidt, M.~Querejeta, J.~Pety, E.~Rosolowsky, P.~Sanchez-Blazquez, K.~Sandstrom, E.~Schinnerer, A.~Schruba, and A.~Usero. Scientific exploitation of PHANGS--MUSE has taken place largely in the context of the `Ionised ISM and its Relation to Star Formation' (IonisedISM) and `Large Scale Dynamical Processes' (Dynamics) science working groups. The IonisedISM group was led by B. Groves and K. Sandstrom in 2019, and since 2020 by B. Groves and K.~Kreckel. The Dynamics group has been led by S.~Meidt and M.~Querejeta since 2019.

\noindent \textbf{Preparation of this Paper:} E. Emsellem has managed the preparation of the text and figures in this paper in close collaboration with E. Schinnerer. E. Emsellem wrote Sect 1, 3.1, 4.1, 4.2.1-4.2.5, 6.1.3-6.1.5, 6.3.1, 7, and 9, provided Figs. 2, 3, 6, 7, 8, 9, 10, 14, 16, 23, made heavy reviews of all sections until finalisation. Major contributions to the paper were provided by F. Belfiore, K. Kreckel and F. Santoro. F. Belfiore wrote Sects. 5.1, 5.2, 6.2.1, 6.2.2 and generated Figs. 2, 3, 4, 11, 12, 13, 17, 18, 19, 21, Tables 1, 2, 3. I. Pessa provided text and figures for stellar population related subsections (6.2.4, 6.3.2; Fig.~21, 22, 24). E. Congiu provided text for Sect 3.2, 4.2.6, 4.2.7, 4.2.8, 6.1.2 and the basis for Fig 7. B. Groves provided heavy input and edits for the sample and science goals and the quality assessment Sections (2.2), and generated and wrote the section about star masks (5.3). F. Santoro provided text in Sect 4, Fig.~5 and Table A.1. K. Kreckel provided text in 6.1.1, 6.1.3, 6.2.3, 8 and Fig. 15, 20, 25, 27, 28. Beyond these, many in the team provided significant input, text or figures, including: G. Blanc, O. Egorov (Fig. 26), R. Klessen, P.~Sanchez-Blazquez.

\smallskip

\noindent \textbf{Observatory and Community Support:} PHANGS--MUSE would not exist without the key contributions from many staff at the European Southern Observatory, and the expert and unfailing support from the Observatory. This includes all staff involved in receiving and dealing with OBs, the Observatory Staff in Chile, the User Support Department staff associated with MUSE as well as all the behind-the-scene supporting people whose efforts and dedication made the PHANGS-MUSE Large Programme a reality. We especially thank Elena Valenti of the ESO user-support department for her continuous support. We also wish to acknowledge specific discussions and superb software-related support from Ralf Palsa. Beyond ESO, our programme benefited greatly from insights and support from several members of the community. We acknowledge the specific contributions from individuals in the MUSE GTO team, including Roland Bacon, Simon Conseil, Laure Piqueras, and Peter Weilbacher, who were or are key actors in the development and writing of \mpdaf\ and \MUSEp, and provided invaluable insights. 

\clearpage
\onecolumn

        \begin{longtable}{lllllll}
                \caption{Table reporting galaxy and pointing ID (col1), sky coordinates of the pointing (Cols. 2 and 3), day and starting time of the OB (Col. 4), progressive number (increasing with the exposure observing time) of the science exposures part of the OB and included in the final mosaic (Col. 5), PSF FWHM estimated using the final OB data cube (Col. 6), and MUSE observation mode (Col. 7).}\\
            \label{table:ObsPointings} \\
        \hline
                        Galaxy $\&$ Pointing ID & RA & DEC & TPL start & Exposure $\#$ & PSF (FWHM) & MUSE mode \\
            $\mathrm{}$ & [$\mathrm{{}^{\circ}}$/] & [$\mathrm{{}^{\circ}}$] & $\mathrm{}$ & $\mathrm{}$ & [$\mathrm{{}^{\arcsec}}$] & WFM \\
                \hline
                \endfirsthead
                \caption{continued.}\\
        \label{toto} \\
                \hline
        \hline
                        Galaxy $\&$ Pointing ID & RA & DEC & TPL start & Exposure $\#$ & PSF (FWHM) & MUSE mode\\
            $\mathrm{}$ & $\mathrm{{}^{\circ}}$ & $\mathrm{{}^{\circ}}$ & $\mathrm{}$ & $\mathrm{}$ & $\mathrm{{}^{\arcsec}}$ & WFM \\
                \hline
                \endhead
                \hline
                \endfoot

IC5332 P01 & 353.622663 & -36.10046 & 2018-06-14T08:00:41 & 1-2-3-4 & 0.59 & noAO \\
IC5332 P02 & 353.603097 & -36.10029 & 2018-07-11T06:07:18 & 1-2-3-4 & 0.80 & noAO \\
IC5332 P03 & 353.622731 & -36.08404 & 2018-07-11T08:14:50 & 1-2-3-4 & 0.67 & noAO \\
IC5332 P04 & 353.60243 & -36.0847 & 2018-07-11T09:18:27 & 1-2-3-4 & 0.72 & noAO \\
IC5332 P05 & 353.612201 & -36.117 & 2018-07-12T07:19:22 & 1-2-3-4 & 0.75 & noAO \\
\hline
NGC0628 P01 & 24.179717 & 15.75473 & 2015-09-15T05:00:21 & 1-2-3 & 0.73 & noAO \\
NGC0628 P02 & 24.168492 & 15.76554 & 2017-07-22T07:36:21 & 1-2-3 & 0.77 & noAO \\
NGC0628 P03 & 24.157267 & 15.77634 & 2017-07-25T07:31:28 & 1-2 & 0.73 & noAO \\
 &  &  & 2017-11-13T03:43:40 & 1-2-3 &  &  \\
NGC0628 P04 & 24.146037 & 15.78714 & 2017-09-16T04:17:06 & 1-2-3 & 0.84 & noAO \\
NGC0628 P05 & 24.190942 & 15.76554 & 2016-12-30T01:01:19 & 1-2-3 & 0.74 & noAO \\
NGC0628 P06 & 24.179717 & 15.77634 & 2016-10-01T04:56:00 & 1 & 0.62 & noAO \\
 &  &  & 2016-10-01T05:21:15 & 1-2 &  &  \\
NGC0628 P07 & 24.168492 & 15.78714 & 2016-10-01T06:08:00 & 1-2-3 & 0.60 & noAO \\
NGC0628 P08 & 24.157262 & 15.79794 & 2017-07-21T08:25:39 & 1-2-3 & 0.69 & noAO \\
NGC0628 P09 & 24.202171 & 15.77634 & 2017-11-13T01:22:29 & 1-2-3 & 0.70 & noAO \\
NGC0628 P10 & 24.191146 & 15.78908 & 2014-10-31T03:39:46 & 1-2-3 & 0.75 & noAO \\
NGC0628 P11 & 24.175675 & 15.79605 & 2014-10-31T04:40:25 & 1-2-3 & 0.74 & noAO \\
NGC0628 P12 & 24.168492 & 15.80875 & 2017-11-13T02:32:55 & 1-2-3 & 0.66 & noAO \\
\hline
NGC1087 P01 & 41.596158 & -0.49892 & 2017-11-13T04:56:31 & 1-2-3-4 & 0.69 & noAO \\
NGC1087 P02 & 41.612722 & -0.4987 & 2017-12-21T02:05:40 & 1-2-3-4 & 0.79 & noAO \\
NGC1087 P03 & 41.612686 & -0.48263 & 2017-12-21T03:09:30 & 1-2-3 & 0.83 & noAO \\
 &  &  & 2017-12-21T03:56:29 & 1-2 &  &  \\
NGC1087 P04 & 41.596478 & -0.48255 & 2018-01-12T01:32:38 & 1-2-3-4 & 0.63 & noAO \\
NGC1087 P05 & 41.596292 & -0.51499 & 2018-01-10T01:43:24 & 1-2-3-4 & 0.84 & noAO \\
NGC1087 P06 & 41.612674 & -0.51478 & 2018-01-11T01:02:44 & 1-2-3-4 & 0.63 & noAO \\
\hline
NGC1300 P01 & 49.921565 & -19.41124 & 2019-02-03T01:41:13 & 1-2-3-4 & 0.78 & AO \\
NGC1300 P02 & 49.904626 & -19.41147 & 2019-08-29T09:19:34 & 1-2 & 0.66 & AO \\
NGC1300 P03 & 49.938663 & -19.41111 & 2019-09-25T07:57:43 & 1-2-3-4 & 0.75 & AO \\
NGC1300 P04 & 49.93881 & -19.42729 & 2019-10-08T07:39:28 & 1-2-3-4 & 0.81 & AO \\
NGC1300 P05 & 49.904584 & -19.39516 & 2019-12-02T04:41:28 & 1-2-3-4 & 0.54 & AO \\
NGC1300 P06 & 49.921554 & -19.39518 & 2019-12-03T05:12:50 & 1-2-3-4 & 0.50 & AO \\
NGC1300 P07 & 49.938568 & -19.39515 & 2019-12-21T00:55:54 & 1-2-3-4 & 0.63 & AO \\
NGC1300 P08 & 49.955684 & -19.40844 & 2019-12-23T01:39:45 & 1-2-3-4 & 0.56 & AO \\
NGC1300 P09 & 49.921481 & -19.42735 & 2019-12-22T01:34:49 & 1-2-3-4 & 0.58 & AO \\
NGC1300 P10 & 49.904615 & -19.4272 & 2019-12-22T02:42:19 & 1-2-3-4 & 0.62 & AO \\
NGC1300 P11 & 49.887401 & -19.42193 & 2020-01-16T01:26:09 & 1-2-3-4 & 0.54 & AO \\
NGC1300 P12 & 49.887328 & -19.4058 & 2020-01-16T02:41:23 & 1-2-3-4 & 0.69 & AO \\
\hline
NGC1365 P01 & 53.421733 & -36.14044 & 2018-01-10T02:49:23 & 1-2-3-4 & 0.71 & noAO \\
NGC1365 P02 & 53.381807 & -36.1409 & 2018-10-17T07:19:24 & 1-2-3-4 & 0.82 & noAO \\
NGC1365 P03 & 53.401334 & -36.12483 & 2018-01-20T01:16:28 & 1-2-3-4 & 0.83 & noAO \\
NGC1365 P04 & 53.381516 & -36.12465 & 2018-01-20T02:25:06 & 1-2-3-4 & 0.90 & noAO \\
NGC1365 P05 & 53.421647 & -36.12486 & 2018-10-16T05:26:10 & 1-2-3-4 & 0.92 & noAO \\
NGC1365 P06 & 53.381737 & -36.15673 & 2018-11-05T05:41:35 & 1-2-3-4 & 0.72 & noAO \\
NGC1365 P07 & 53.401428 & -36.15677 & 2018-11-06T05:32:56 & 1-2-3-4 & 0.64 & noAO \\
NGC1365 P08 & 53.421804 & -36.15702 & 2018-11-07T04:30:38 & 1-2-3-4 & 0.90 & noAO \\
NGC1365 P09 & 53.441205 & -36.14062 & 2018-12-04T03:58:31 & 1-2-3 & 0.90 & noAO \\
NGC1365 P10 & 53.441226 & -36.12449 & 2018-12-04T04:53:10 & 1-2-3-4 & 1.08 & noAO \\
NGC1365 P11 & 53.361474 & -36.14065 & 2018-12-05T04:08:13 & 1-2-3 & 0.58 & noAO \\
NGC1365 P12 & 53.361534 & -36.15674 & 2018-12-05T05:18:24 & 1-2-3-4 & 0.76 & noAO \\
NGC1365 P30 & 53.402083 & -36.14056 & 2014-10-12T04:31:28 & 1-2-3-4 & 0.82 & noAO \\
 &  &  & 2014-10-12T05:30:02 & 1-2-3-4 &  &  \\
 \hline
NGC1385 P01 & 54.378854 & -24.50028 & 2019-10-06T08:06:01 & 1-2-3-4 & 0.49 & AO \\
NGC1385 P02 & 54.360994 & -24.50053 & 2019-12-31T03:56:25 & 1-2 & 0.38 & AO \\
 &  &  & 2019-12-31T04:55:57 & 1-2 &  &  \\
NGC1385 P03 & 54.369803 & -24.48441 & 2020-01-20T01:12:41 & 1-2-3-4 & 0.57 & AO \\
 &  &  & 2020-01-20T02:12:43 & 1-2-3-4 &  &  \\
NGC1385 P04 & 54.369803 & -24.48441 & 2020-12-05T02:17:13 & 1-2-3-4 & 0.69 & AO \\
NGC1385 P05 & 54.378854 & -24.51639 & 2020-01-21T01:14:06 & 1-2-3-4 & 0.59 & AO \\
\hline
NGC1433 P01 & 55.506902 & -47.22178 & 2018-10-16T06:52:54 & 1-2-3-4 & 0.70 & AO \\
NGC1433 P02 & 55.530064 & -47.22185 & 2019-10-05T06:48:42 & 1-2-3-4 & 0.63 & AO \\
NGC1433 P03 & 55.553469 & -47.2217 & 2019-10-05T07:57:53 & 1-2-3-4 & 0.83 & AO \\
NGC1433 P04 & 55.483191 & -47.22189 & 2019-10-06T06:02:47 & 1-2-3-4 & 0.60 & AO \\
NGC1433 P05 & 55.459083 & -47.22203 & 2019-10-07T06:46:16 & 1-2-3 & 0.63 & AO \\
 &  &  & 2019-10-07T07:59:06 & 1 &  &  \\
NGC1433 P06 & 55.45893 & -47.2055 & 2019-11-02T04:33:56 & 1-2-3-4 & 0.65 & AO \\
NGC1433 P07 & 55.482862 & -47.20541 & 2019-11-20T02:08:18 & 1-2-3 & 0.70 & AO \\
 &  &  & 2019-11-20T03:09:00 & 1 &  &  \\
NGC1433 P08 & 55.506986 & -47.20549 & 2019-11-21T02:11:14 & 1-2-3-4 & 0.65 & AO \\
NGC1433 P09 & 55.530772 & -47.20538 & 2019-11-22T06:27:26 & 1-2-3-4 & 0.71 & AO \\
NGC1433 P10 & 55.554174 & -47.20554 & 2019-12-20T04:30:19 & 1-2-3-4 & 0.62 & AO \\
NGC1433 P11 & 55.553626 & -47.23813 & 2019-12-21T02:16:23 & 1-2-3-4 & 0.61 & AO \\
NGC1433 P12 & 55.529952 & -47.23809 & 2019-12-21T04:27:47 & 1-2-3-4 & 0.65 & AO \\
NGC1433 P13 & 55.506442 & -47.23789 & 2019-12-22T04:24:22 & 1-2 & 0.65 & AO \\
 &  &  & 2019-12-22T05:05:18 & 1-2 &  &  \\
NGC1433 P14 & 55.482729 & -47.23792 & 2019-12-23T03:48:53 & 1-2-3-4 & 0.51 & AO \\
NGC1433 P15 & 55.459171 & -47.23791 & 2019-12-30T03:38:48 & 1-2-3-4 & 0.64 & AO \\
\hline
NGC1512 P01 & 60.998425 & -43.34935 & 2018-12-30T01:11:52 & 1-2-3-4 & 0.73 & noAO \\
NGC1512 P02 & 60.997871 & -43.33288 & 2018-12-30T03:46:33 & 1-2-3-4 & 0.85 & noAO \\
NGC1512 P03 & 60.976057 & -43.3331 & 2018-02-17T01:02:45 & 1-2-3-4-5 & 1.18 & noAO \\
NGC1512 P04 & 60.953684 & -43.33286 & 2018-02-18T01:08:42 & 1-2-3-4 & 0.80 & noAO \\
NGC1512 P05 & 60.953722 & -43.34897 & 2018-02-19T01:04:07 & 1-2-3-4 & 0.68 & noAO \\
NGC1512 P06 & 60.954141 & -43.36546 & 2019-01-10T02:41:43 & 1-2-3-4 & 0.70 & noAO \\
NGC1512 P07 & 60.976389 & -43.36524 & 2019-01-10T03:47:10 & 1-2-3-4 & 0.83 & noAO \\
NGC1512 P08 & 60.998479 & -43.36538 & 2019-01-10T04:52:57 & 1-2-3-4 & 0.93 & noAO \\
NGC1512 P30 & 60.975987 & -43.34905 & 2017-09-21T06:53:05 & 1 & 0.64 & noAO \\
 &  &  & 2017-09-21T08:30:27 & 1-2 &  &  \\
 &  &  & 2017-09-22T08:39:40 & 1 &  &  \\
\hline
NGC1566 P01 & 65.030061 & -54.93785 & 2018-12-14T03:12:39 & 1-2-3-4 & 0.54 & AO \\
NGC1566 P02 & 64.974665 & -54.93714 & 2019-01-15T02:28:00 & 1-2-3-4 & 0.56 & AO \\
NGC1566 P03 & 65.013148 & -54.95397 & 2020-12-10T04:30:27 & 1-2-3-4 & 0.60 & AO \\
NGC1566 P04 & 64.985261 & -54.92184 & 2019-01-25T00:53:23 & 1-2-3-4 & 0.65 & AO \\
NGC1566 P05 & 65.013286 & -54.92178 & 2019-01-27T00:52:13 & 1-2-3-4 & 0.63 & AO \\
NGC1566 P06 & 64.985399 & -54.95382 & 2019-01-27T02:02:45 & 1-2-3-4 & 0.72 & AO \\
NGC1566 P07 & 65.011776 & -54.9057 & 2019-01-28T01:09:07 & 1-2-3-4 & 0.64 & AO \\
NGC1566 P30 & 65.001794 & -54.93786 & 2017-10-23T04:45:57 & 1-2-3-4 & 0.64 & AO \\
\hline
NGC1672 P01 & 71.444433 & -59.25258 & 2017-11-12T06:54:01 & 1-2-3-4 & 0.65 & noAO \\
NGC1672 P02 & 71.440988 & -59.23654 & 2017-12-23T04:11:46 & 1-2-3-4 & 0.89 & noAO \\
NGC1672 P03 & 71.410094 & -59.23802 & 2017-11-13T06:07:01 & 1-2-3-4 & 0.73 & noAO \\
NGC1672 P04 & 71.412987 & -59.25384 & 2017-11-25T05:07:09 & 1-2-3-4 & 0.65 & noAO \\
NGC1672 P05 & 71.475798 & -59.25114 & 2017-12-26T05:11:09 & 1-2-3-4 & 0.80 & noAO \\
NGC1672 P06 & 71.47238 & -59.23514 & 2017-12-19T04:31:59 & 1-2-3-4 & 0.77 & noAO \\
NGC1672 P07 & 71.377135 & -59.23943 & 2017-12-19T05:38:10 & 1-2-3-4 & 0.71 & noAO \\
NGC1672 P08 & 71.381706 & -59.25546 & 2018-01-11T02:26:31 & 1-2-3-4 & 0.68 & noAO \\
\hline
NGC2835 P01 & 139.47034 & -22.33869 & 2017-12-15T06:22:14 & 1-2-3-4 & 0.78 & noAO \\
NGC2835 P02 & 139.470252 & -22.37082 & 2018-01-16T07:38:48 & 1-2-3-4 & 1.08 & noAO \\
NGC2835 P03 & 139.487844 & -22.36179 & 2018-01-18T03:42:20 & 1-2-3-4 & 0.93 & noAO \\
NGC2835 P04 & 139.487928 & -22.3458 & 2018-01-23T03:26:36 & 1-2-3-4 & 0.67 & noAO \\
NGC2835 P05 & 139.452986 & -22.34596 & 2018-02-14T02:03:35 & 1-2-3-4 & 0.85 & noAO \\
NGC2835 P06 & 139.452841 & -22.36209 & 2018-02-20T01:20:57 & 1-2-3-4 & 0.71 & noAO \\
NGC2835 P30 & 139.470371 & -22.35446 & 2017-02-02T02:58:32 & 1-2-3-4 & 0.87 & noAO \\
\hline
NGC3351 P01 & 161.007339 & 11.7042 & 2019-02-10T04:59:15 & 1-2-3-4 & 0.71 & noAO \\
NGC3351 P02 & 160.974424 & 11.7042 & 2019-02-10T06:03:50 & 1-2-3-4 & 0.76 & noAO \\
NGC3351 P03 & 160.990893 & 11.68806 & 2019-03-02T03:17:26 & 1-2-3-4 & 0.66 & noAO \\
NGC3351 P04 & 160.990873 & 11.72028 & 2019-03-02T04:16:25 & 1-2-3-4 & 0.73 & noAO \\
NGC3351 P05 & 161.007392 & 11.68802 & 2019-03-03T03:51:18 & 1-2-3-4 & 0.82 & noAO \\
NGC3351 P06 & 160.974484 & 11.72025 & 2019-03-03T05:02:01 & 1-2-3-4 & 0.98 & noAO \\
NGC3351 P07 & 161.007289 & 11.72022 & 2019-03-11T02:49:48 & 1-2-3-4 & 0.84 & noAO \\
NGC3351 P08 & 160.974375 & 11.68819 & 2019-03-12T02:42:02 & 1-2-3-4 & 0.74 & noAO \\
NGC3351 P30 & 160.990417 & 11.70381 & 2016-03-30T00:04:22 & 1-2-3-4 & 0.61 & noAO \\
 &  &  & 2016-04-04T00:43:01 & 1-2-3-4 &  &  \\
 \hline
NGC3627 P01 & 170.072929 & 12.98949 & 2018-01-25T07:19:09 & 1-2-3-4 & 0.68 & noAO \\
NGC3627 P02 & 170.055847 & 12.98976 & 2018-05-13T23:25:01 & 1-2-3-4 & 0.78 & noAO \\
NGC3627 P03 & 170.054709 & 12.97342 & 2018-05-08T01:35:58 & 1-2-3-4 & 0.98 & noAO \\
NGC3627 P04 & 170.071261 & 12.97362 & 2018-05-14T00:35:00 & 1-2-3-4 & 0.75 & noAO \\
NGC3627 P05 & 170.072366 & 13.0058 & 2018-05-14T01:41:04 & 1-2-3-4 & 0.80 & noAO \\
NGC3627 P06 & 170.056162 & 13.00601 & 2018-05-14T23:25:02 & 1-2-3-4 & 0.77 & noAO \\
NGC3627 P07 & 170.054957 & 12.95769 & 2018-05-15T00:29:52 & 1-2-3-4 & 0.74 & noAO \\
NGC3627 P08 & 170.071501 & 12.95767 & 2018-05-15T01:34:18 & 1-2-3-4 & 0.81 & noAO \\
\hline
NGC4254 P01 & 184.713694 & 14.41518 & 2018-04-16T02:49:03 & 1-2-3-4-5 & 0.63 & AO \\
NGC4254 P02 & 184.697794 & 14.4153 & 2018-05-19T02:22:33 & 1-2-3-4 & 0.57 & AO \\
NGC4254 P03 & 184.714741 & 14.43107 & 2018-06-08T00:17:56 & 1-2-3-4 & 0.59 & AO \\
NGC4254 P04 & 184.697599 & 14.43081 & 2018-06-08T23:17:55 & 1-2-3-4 & 0.61 & AO \\
NGC4254 P05 & 184.708005 & 14.39908 & 2018-06-04T23:35:32 & 1-2 & 0.81 & AO \\
 &  &  & 2018-06-05T00:11:22 & 1-2-3 &  &  \\
NGC4254 P06 & 184.691941 & 14.39909 & 2018-06-05T01:06:43 & 1-3-4 & 0.77 & AO \\
NGC4254 P07 & 184.724544 & 14.39901 & 2018-06-09T23:26:07 & 1-2-3-4 & 0.62 & AO \\
NGC4254 P08 & 184.730498 & 14.41463 & 2018-06-06T23:44:42 & 1-2-3-4 & 0.48 & AO \\
NGC4254 P09 & 184.731384 & 14.43116 & 2018-06-13T00:04:09 & 1-2-3 & 0.45 & AO \\
 &  &  & 2018-06-13T00:51:40 & 1 &  &  \\
NGC4254 P10 & 184.698343 & 14.44673 & 2019-03-11T04:59:39 & 1-2-3-4 & 0.53 & AO \\
NGC4254 P11 & 184.714996 & 14.44694 & 2019-03-02T05:27:50 & 1-2-3-4 & 0.44 & AO \\
NGC4254 P12 & 184.731621 & 14.44696 & 2019-03-02T06:35:48 & 1-2-3-4 & 0.45 & AO \\
\hline
NGC4303 P01 & 185.478821 & 4.47383 & 2019-05-10T03:10:00 & 1 & 0.55 & AO \\
 &  &  & 2019-05-10T03:51:19 & 1-2-3 &  &  \\
NGC4303 P02 & 185.494958 & 4.47371 & 2019-05-27T23:39:52 & 1-2-3-4 & 0.64 & AO \\
NGC4303 P03 & 185.462592 & 4.47361 & 2019-06-29T23:25:58 & 1-2-3 & 0.59 & AO \\
 &  &  & 2019-06-30T00:26:58 & 1 &  &  \\
NGC4303 P04 & 185.478627 & 4.48972 & 2020-01-30T07:08:21 & 1-2 & 0.54 & AO \\
 &  &  & 2020-01-30T07:43:58 & 1-2 &  &  \\
NGC4303 P05 & 185.47875 & 4.4575 & 2020-02-03T06:23:31 & 1-2-3-4 & 0.58 & AO \\
NGC4303 P06 & 185.494912 & 4.48972 & 2020-02-03T07:35:27 & 1-2-3-4 & 0.53 & AO \\
NGC4303 P07 & 185.494788 & 4.45766 & 2020-02-28T07:39:20 & 1 & 0.51 & AO \\
 &  &  & 2020-02-28T08:05:55 & 1-2-3 &  &  \\
NGC4303 P08 & 185.462442 & 4.48974 & 2020-02-19T05:54:25 & 1-2-3-4 & 0.61 & AO \\
NGC4303 P09 & 185.462532 & 4.45765 & 2020-02-19T07:35:28 & 1-2-3-4 & 0.70 & AO \\
\hline
NGC4321 P01 & 185.734704 & 15.8219 & 2019-04-28T02:38:38 & 1-2-3-4 & 0.79 & AO \\
NGC4321 P02 & 185.717999 & 15.82332 & 2019-04-30T02:20:03 & 1-2-3-4 & 0.59 & AO \\
NGC4321 P03 & 185.750833 & 15.82194 & 2019-05-01T01:06:01 & 1-2-3-4 & 0.85 & AO \\
NGC4321 P04 & 185.73408 & 15.83803 & 2020-03-02T06:11:38 & 1-2-3-4 & 0.47 & AO \\
NGC4321 P05 & 185.717559 & 15.83938 & 2020-03-03T06:06:28 & 1-2-3-4 & 0.72 & AO \\
NGC4321 P06 & 185.734316 & 15.80579 & 2020-02-20T07:07:56 & 1-2-3 & 0.46 & AO \\
 &  &  & 2020-02-20T08:03:06 & 1 &  &  \\
NGC4321 P07 & 185.717498 & 15.80723 & 2020-03-18T05:09:33 & 1-2-3-4 & 0.61 & AO \\
NGC4321 P08 & 185.751212 & 15.80621 & 2021-02-12T06:35:52 & 1-2-3-4 & 1.00 & AO \\
NGC4321 P09 & 185.700805 & 15.82756 & 2020-03-22T04:56:36 & 1-2-3-4 & 0.50 & AO \\
NGC4321 P10 & 185.701823 & 15.84284 & 2020-03-23T04:43:09 & 1-2-3-4 & 0.67 & AO \\
NGC4321 P11 & 185.750833 & 15.83806 & 2020-03-24T04:28:48 & 1-2-3-4 & 1.09 & AO \\
\hline
NGC4535 P01 & 188.576744 & 8.19195 & 2018-04-09T03:18:03 & 1-2-3-4 & 0.47 & AO \\
NGC4535 P02 & 188.593278 & 8.19259 & 2018-04-09T04:46:39 & 1-2-3-4 & 0.43 & AO \\
NGC4535 P03 & 188.592956 & 8.20803 & 2018-04-10T02:42:01 & 1-2-3-4 & 0.43 & AO \\
NGC4535 P04 & 188.576548 & 8.20801 & 2018-04-14T04:31:10 & 1-2-3-4 & 0.47 & AO \\
NGC4535 P05 & 188.592752 & 8.17594 & 2018-04-16T04:53:52 & 1-2-3-4 & 0.44 & AO \\
NGC4535 P06 & 188.576717 & 8.1758 & 2018-05-17T00:00:23 & 1-2-3-4 & 0.43 & AO \\
\hline
NGC5068 P01 & 199.729433 & -21.04312 & 2018-05-14T02:48:05 & 1-2-3-4 & 0.67 & noAO \\
 &  &  & 2018-06-14T02:46:50 & 1-2-3 &  &  \\
NGC5068 P02 & 199.729986 & -21.02694 & 2018-05-14T04:20:06 & 1-3-4 & 0.88 & noAO \\
NGC5068 P03 & 199.711794 & -21.04348 & 2018-05-15T02:42:20 & 1-2-3-4 & 0.96 & noAO \\
NGC5068 P04 & 199.712242 & -21.02699 & 2018-05-20T02:58:19 & 1-2-3-4 & 0.69 & noAO \\
NGC5068 P05 & 199.745845 & -21.04327 & 2018-05-21T04:13:30 & 1-2-3-4 & 0.62 & noAO \\
NGC5068 P06 & 199.72313 & -21.05915 & 2018-06-15T02:09:06 & 1-2-3-4 & 0.83 & noAO \\
NGC5068 P07 & 199.705524 & -21.0592 & 2018-06-17T01:57:12 & 1-2-3-4 & 0.77 & noAO \\
NGC5068 P08 & 199.712023 & -21.01147 & 2018-07-10T23:50:45 & 1-2-3-4 & 0.56 & noAO \\
NGC5068 P09 & 199.695073 & -21.01376 & 2018-07-11T00:56:19 & 1-2-3-4 & 0.52 & noAO \\
NGC5068 P10 & 199.740055 & -21.05934 & 2018-07-14T00:44:22 & 1-2-3-4 & 0.90 & noAO \\
\hline
NGC7496 P01 & 347.4467 & -43.42833 & 2019-06-09T08:31:41 & 1 & 0.62 & AO \\
 &  &  & 2019-06-09T08:53:47 & 1-2-3 &  &  \\
NGC7496 P02 & 347.440551 & -43.41284 & 2019-07-04T08:15:45 & 1-2 & 0.81 & AO \\
 &  &  & 2019-07-04T09:23:58 & 1-2-3 &  &  \\
NGC7496 P03 & 347.452917 & -43.44361 & 2019-08-25T06:43:38 & 1-2-3-4 & 0.79 & AO \\
\hline
        \end{longtable}

\end{appendix}

\end{document}